\documentclass[10pt,aps,prl,twocolumn,showpacs,groupedaddress,notitlepage,preprintnumbers,nobibnotes,superscriptaddress,floatfix,longbibliography]{revtex4-2}

\usepackage{
    amsmath, amssymb, latexsym, amsfonts, dsfont,
    amscd, amsthm, thmtools, thm-restate, mathtools,
    microtype, bm
}

\usepackage{epsfig, graphicx, subfigure}
\usepackage{booktabs}

\usepackage{physics, slashed, bbm}

\usepackage[dvipsnames]{xcolor}
\usepackage{soul, comment, todonotes, setspace, indentfirst}
\usepackage[T1]{fontenc}
\usepackage{multirow}
\usepackage[normalem]{ulem}

\usepackage{algorithm2e}
\RestyleAlgo{ruled}

\usepackage{hyperref}
\hypersetup{
    colorlinks=true,
    linkcolor=blue,
    filecolor=magenta,
    urlcolor=blue,
    citecolor=blue
}
\usepackage[nameinlink,noabbrev]{cleveref}

\newtheorem{theorem}{Theorem}

\newtheorem{lemma}{Lemma}

\newtheorem{proposition}{Proposition}
\newtheorem{problem}{Problem}

\newcommand{\C}{\mathbb{C}}

\newcommand{\E}{\mathbb{E}}

\newcommand{\bbP}{\mathbb{P}}

\newcommand{\R}{\mathbb{R}}

\newcommand{\Z}{\mathbb{Z}}

\newcommand{\cD}{\mathcal{D}}

\newcommand{\cH}{\mathcal{H}}

\newcommand{\cK}{\mathcal{K}}

\newcommand{\cN}{\mathcal{N}}

\newcommand{\cP}{\mathcal{P}}

\newcommand{\cT}{\mathcal{T}}

\newcommand{\cZ}{\mathcal{Z}}

\newcommand{\eps}{\epsilon}
\newcommand{\id}{\mathbbm{1}}

\newcommand{\wt}{\widetilde}

\DeclarePairedDelimiter\ceil{\lceil}{\rceil}

\newcommand{\opnorm}[1]{\left\| #1 \right\|}
\newcommand{\pnorm}[2]{\left\| #2 \right\|_{#1}}
\newcommand{\triplenorm}[1]{{\left\vert\kern-0.25ex\left\vert\kern-0.25ex\left\vert #1
    \right\vert\kern-0.25ex\right\vert\kern-0.25ex\right\vert}}

\makeatletter
\newwrite\apptoc@out
\newcommand{\apptoc@section}[4]{%
    \noindent\hyperlink{#4}{#1\quad#2}\dotfill#3\par
}
\newcommand{\apptoc@subsection}[4]{%
    \noindent\hspace*{1.5em}\hyperlink{#4}{#1\quad#2}\dotfill#3\par
}
\newcommand{\apptoc@subsubsection}[4]{%
    \noindent\hspace*{3em}\hyperlink{#4}{#1\quad#2}\dotfill#3\par
}
\newcommand{\appendixtableofcontents}{%
    \begingroup
    \small
    \IfFileExists{\jobname.atoc}{\input{\jobname.atoc}}{\emph{Run LaTeX again to generate appendix contents.}\par}%
    \endgroup
}
\newcommand{\apptoc@write}[3]{%
    \begingroup
    \edef\apptoc@number{#2}%
    \edef\apptoc@page{\thepage}%
    \edef\apptoc@href{\@currentHref}%
    \immediate\write\apptoc@out{\string\csname\space apptoc@#1\string\endcsname{\apptoc@number}{\detokenize{#3}}{\apptoc@page}{\apptoc@href}}%
    \endgroup
}
\newcommand{\startappendixtoc}{%
    \immediate\openout\apptoc@out=\jobname.atoc
    \let\apptoc@oldsection\section
    \let\apptoc@oldsubsection\subsection
    \let\apptoc@oldsubsubsection\subsubsection
    \RenewDocumentCommand{\section}{s o m}{%
        \IfBooleanTF{##1}{\apptoc@oldsection*{##3}}{%
            \IfNoValueTF{##2}{\apptoc@oldsection{##3}}{\apptoc@oldsection[##2]{##3}}%
            \apptoc@write{section}{\thesection}{##3}%
        }%
    }%
    \RenewDocumentCommand{\subsection}{s o m}{%
        \IfBooleanTF{##1}{\apptoc@oldsubsection*{##3}}{%
            \IfNoValueTF{##2}{\apptoc@oldsubsection{##3}}{\apptoc@oldsubsection[##2]{##3}}%
            \apptoc@write{subsection}{\thesection.\thesubsection}{##3}%
        }%
    }%
    \RenewDocumentCommand{\subsubsection}{s o m}{%
        \IfBooleanTF{##1}{\apptoc@oldsubsubsection*{##3}}{%
            \IfNoValueTF{##2}{\apptoc@oldsubsubsection{##3}}{\apptoc@oldsubsubsection[##2]{##3}}%
        }%
    }%
}
\AtEndDocument{\immediate\closeout\apptoc@out}
\makeatother

\newtheorem*{problem*}{Problem}

\newcommand{\MIT}{\affiliation{Center for Theoretical Physics - a Leinweber Institute, Massachusetts Institute of Technology, Cambridge, MA 02139, USA}}
\newcommand{\stanford}{\affiliation{Department of Electrical Engineering, Stanford University, Stanford, CA 94305, USA}}

\begin{document}
\preprint{MIT-CTP/6153}
\title{Optimal Quantum Sensing Algorithms for Discrete-Frequency Spectroscopy}

\author{Yu-Jie Liu}
\thanks{These authors contributed equally.\\}\MIT
\author{Richard R. Allen}
\thanks{These authors contributed equally.\\}\MIT
\author{Ruohan Shen}
\thanks{These authors contributed equally.\\}\MIT
\author{Bingtian Ye}\MIT
\author{Timothy T. Chang}\stanford
\author{Joonhee Choi}\stanford
\author{Soonwon Choi}\MIT

\begin{abstract}
Spectroscopy is a versatile tool that drives scientific discovery and underpins modern technologies. Quantum sensing offers promising opportunities to advance spectroscopy through enhanced sensitivity and spectral and spatial resolution. However, existing approaches largely focus on improving general measurement capabilities, often overlooking the structure of specific sensing tasks and leaving potential performance gains untapped.
Here, we develop an algorithmic approach to quantum sensing tailored to various spectroscopic tasks, providing new avenues for improvement.
Specifically, we consider general time-varying signals supported on a known set of $N$ candidate frequencies, a setting we call discrete-frequency spectroscopy.
We formulate three types of practically motivated sensing tasks—Detection, Search, and Learning—that require progressively more spectral information.
For all tasks, we design optimal quantum sensing algorithms achieving a $\sqrt{N}$ speedup over conventional frequency-scanning approaches.
Optimal Detection and Search need only a single-qubit sensor under $\pi$-pulse control, whereas optimal Learning provably requires coherent processing with a quantum computer.
End-to-end numerical simulations incorporating various experimental imperfections predict up to two-orders-of-magnitude speedup in Detection.
As an example, we analyze the feasibility of our Detection and Search algorithms for real-world molecular screening based on state-of-the-art nitrogen-vacancy centers. We also show how quantum-computing-enhanced Learning enables wireless communication in a weak-signal regime where conventional methods fail. Our results establish a practical route toward realizing quantum advantages via designing task-specific sensing algorithms.
\end{abstract}
\maketitle

\section{Introduction}

\begin{figure*}
    \centering
    \includegraphics[width=1\linewidth]{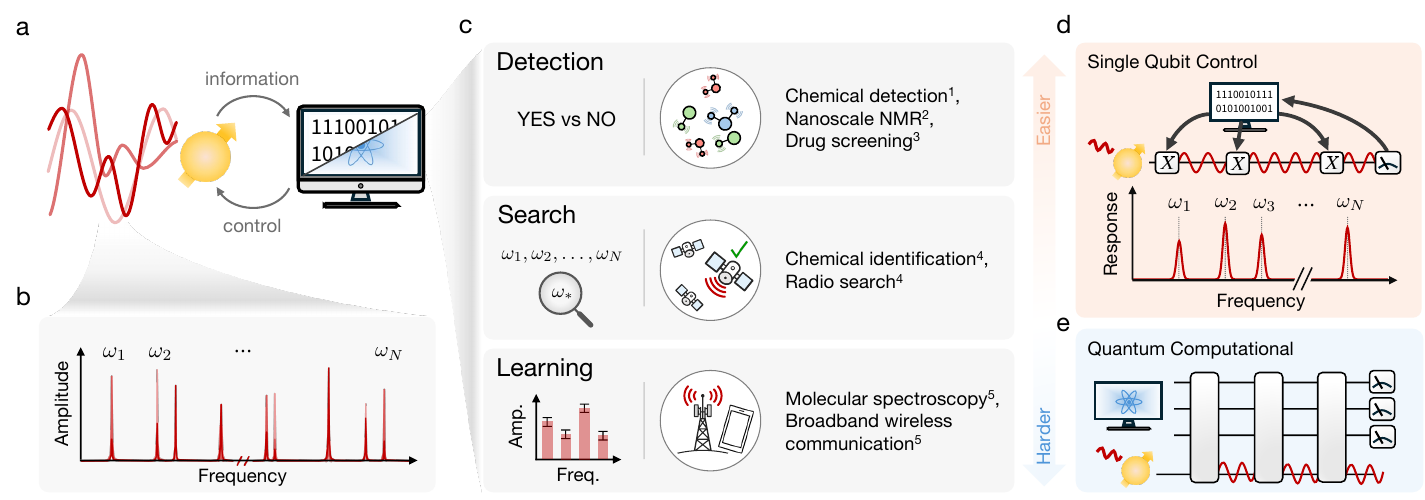}
    \caption{\textbf{Quantum sensing algorithms for discrete-frequency spectroscopy}. \textbf{a.} Our aim is to probe properties of a structured, time-dependent signal field (red) using a quantum sensor qubit (yellow) controlled by a classical or quantum computer in the minimal sensing time. \textbf{b.} The signal spectrum is supported on a discrete set of $N$ known frequencies $\omega_1,\omega_2,\ldots,\omega_N$. The number of candidate frequencies $N$ quantifies the complexity of the signal. \textbf{c.} Our model captures a variety of real-world sensing tasks, which fall into three categories of generally increasing difficulty (vertical color bar): Detection, Search, and Learning. The superscript of each example indicates its corresponding formulation in P1-P5. \textbf{d.} Detection and Search can be solved by our single-qubit algorithm in time $\widetilde{O}(\sqrt{N}/\epsilon)$, where $\epsilon$ is the minimum detectable value of $B$, the root-mean-square amplitude of the time-dependent signal. The algorithm consists of repeated four-step experiments: (i) preparation of the superposition state $\ket{+}=(\ket{0}+\ket{1})/\sqrt{2}$; (ii) evolution under the signal while applying a sequence of $X$ gates ($\pi$ pulses) at precomputed times, chosen to ensure roughly equal response to all candidate frequencies; (iii) measurement of the sensor qubit; and (iv) joint classical processing of the measurement outcomes across experiments. \textbf{e.} Although Learning is a more general problem, it can also be solved in time $\widetilde{O}(\sqrt{N}/\epsilon)$, where $\epsilon$ is the target accuracy for each amplitude component, using our quantum-computing-enhanced algorithm. We prove that joint processing with a quantum computer is necessary to achieve this optimal scaling.}
    \label{fig:overview}
\end{figure*}

From atomic-scale processes to astrophysical phenomena, time-dependent signals reveal the underlying structure of our physical world.
\textit{Spectroscopy} extracts useful information encoded in the spectrum of such signals, providing an indispensable tool for both scientific discovery and practical technologies.
The broad importance of spectroscopy has driven the development of increasingly powerful measurement techniques.
In particular, quantum sensing has advanced general measurement capabilities, with improvement along multiple axes, from enhanced sensitivity~\cite{ligo,giovannetti:2004quantum,pezze:2018,tsang:2011,tsang2012fundamental,berry2015quantum,gardner:2025} to improved spectral~\cite{rosskopf2017quantum,Schmitt2017} and spatial~\cite{mamin2013nanoscale,staudacher2013nuclear,rugar2015proton} resolution.

Real-world spectroscopic tasks, however, often require extracting targeted information from structured signals. Examples include chemical screening, where the goal is to detect one or more compounds from a known library, and wireless communication, where
one must recover digital information encoded in electromagnetic fields according to standardized protocols.
Such diverse objectives motivate a distinct, \emph{algorithmic} approach to quantum enhancement in spectroscopy---designing quantum sensing algorithms tailored to specific tasks in order to minimize resource requirements.
This approach has emerged in recent works under the name of quantum computational sensing.
Several studies show that this approach enables provable quantum advantages, albeit primarily for idealized or even contrived sensing tasks with limited connections to practical applications~\cite{allen2025quantumcomputingenhancedsensing,khan2025quantum,Cotler_2026,kannan2026exponential,mokeev2026enhancing,mele2026exponentialseparationsensingcontinuous,conlon2026superexponentialadvantagesqueezedlight}.

In this work, we develop quantum sensing algorithms for probing unknown, time-dependent signals supported on a known set of $N$ frequencies, which we call \textit{discrete-frequency spectroscopy}.
This setting encompasses a wide range of practical applications (Fig.~\ref{fig:overview}a-c).
To model diverse objectives across a number of different spectroscopic applications, we formulate five representative sensing tasks organized into three categories: Detection, Search, and Learning, which demand progressively more spectral information.
For each task, we construct an explicit quantum sensing algorithm that improves over the sensing time of conventional frequency-scanning approaches by a factor $\sqrt{N}$.
Further, we prove that our approaches are optimal---no algorithm can have an asymptotically shorter sensing time.
Despite demanding different amounts of spectral information, all tasks share the same  sensing-time scaling.
These differences instead manifest in the quantum resources needed to attain this scaling:
optimal Detection and Search can be achieved by a single sensor qubit controlled by a sequence of $\pi$-pulses (i.e., $X$ gates) applied at precomputed times
(Fig.~\ref{fig:overview}d), whereas optimal Learning provably requires coherent joint processing between the sensor and a quantum computer (Fig.~\ref{fig:overview}e).

The simplicity of our Detection and Search algorithms enables a detailed assessment under realistic experimental conditions, predicting up to two orders of magnitude reduction in sensing time with state-of-the-art single-qubit sensors, such as nitrogen-vacancy (NV) centers and trapped ions (Fig.~\ref{fig:main_numerics}). We further explore applications of our algorithms to molecular screening using realistic parameters. For broadband wireless communication, we show that our quantum-computing-enhanced Learning algorithm unlocks new regimes of sensitivity and data transmission rate inaccessible to conventional approaches (Fig.~\ref{fig:qg}), which, formally, can be translated to exponential quantum sensing advantage, similar to~\cite{mele2026exponentialseparationsensingcontinuous, Cotler_2026}. Together, our work opens a route toward practical advantages through algorithmic quantum sensing spectroscopy.

\section{Discrete-Frequency spectroscopy}

Our sensing tasks concern a general time-dependent signal of the form,
\begin{align}\label{eq:signal}
    f(t) =  B\sum_{n=1}^N \left(u_n \cos(\omega_n t) + u_{N+n} \sin(\omega_n t)\right),
\end{align}
where $B\geq 0$ denotes the signal strength, and the normalized coefficients $\vec{u}=(u_1,\dots,u_{2N})\in\R^{2N}$, which satisfy $\sum_{n=1}^{2N}u_n^2=1$, characterize the waveform of the signal. The frequencies $\{\omega_n\}_{n = 1}^N$ are exactly known (and need not be equally spaced), with the number of frequencies $N$ characterizing the complexity of our problems. For simplicity, we assume that all frequencies are separated from zero and from one another by at least $\Delta$ and focus on the high-sensitivity regime $\epsilon\leq\sqrt{N}\Delta$ in the main text. More general parameter regimes are addressed in the Supplemental Material (SM).

The signal is coupled to a single sensor qubit via the Hamiltonian $H(t) = \hbar \gamma f(t) Z$, where $Z$ is the Pauli-$Z$ operator and $\gamma$ is the coupling matrix element.
The unknown signal is ``always-on.'' That is, the signal evolves continuously and cannot be restarted or replayed from any point in time.
When multiple $m$ sensors are available, one can construct a single effective sensor with the collectively enhanced coupling $\gamma_\textrm{eff} = m\gamma$ by restricting the quantum state of the sensors to stay within the span of $\ket{0}^{\otimes m}$ and $\ket{1}^{\otimes m}$, for example, by preparing an entangled $m$-qubit GHZ state. This effective sensor can be viewed as a \textit{single} sensor with enhanced sensitivity.
We assume this approach has already been taken, since it leads to the optimal $m$-sensor protocols for all problems considered here. Throughout, we work in units where $\hbar \gamma_\textrm{eff} = 1$.

For the most general quantum sensing algorithm, we assume access to an arbitrary number of ancilla qubits that are not coupled to the signal, the ability to perform instantaneous entangling gates and measurements, and the absence of any control imperfections or noise.
These idealized conditions enable us to focus on the fundamental theoretical limits, to determine the minimum sensing time and optimal quantum sensing algorithm to accomplish a specific task.
We relax all of these assumptions when we discuss experimental relevance later in this work.

Motivated by the diversity of real-world applications (Fig.~\ref{fig:overview}c), we formulate five formal problems within discrete-frequency spectroscopy, P1-P5, grouped into three distinct types as follows.
Detection (P1-P3) asks whether a signal is absent ($B=0$) or has strength exceeding $\epsilon$, under progressively less prior knowledge: (P1) when the full waveform $\{u_n\}_{n=1}^{2N}$ is known; (P2) when only the spectral-weight pattern $\{u_n^2+u_{N+n}^2\}_{n=1}^N$ is known, but not the relative spectral phase between the cosine and sine components; (P3) when no prior information about the signal is available, except the set of candidate frequencies.
Search (P4) asks to identify the frequencies of a signal with only one (or a constant number of) nonzero frequency components, each with amplitude exceeding the threshold $\eps$.
Learning (P5) asks to estimate each of the spectral coefficients $\{Bu_n\}_{n=1}^{2N}$ up to precision $\epsilon$. For each task, we demand a high success probability, e.g., $>0.99$.

Each sensing task models a number of practical applications.
For example, Detection with known waveform (P1) models the detection of a chemical component under conventional NMR experiments with transverse polarization, whereas Detection with known spectral-weight pattern (P2) captures nanoscale NMR experiments~\cite{lovchinsky:2016,aslam:2017}, wherein the signal arises from statistical polarization lacking any phase coherence. These problems, together with general Detection (P3) and Search (P4), arise naturally in applications such as molecular screening, where one may only need to detect or identify a compound from a known family; see Application I for further discussion.
Learning (P5) is relevant, e.g., to broadband wireless communication, where transmitted digital data is encoded across $N$ different subcarrier frequencies; see Application II.

\section{Fundamental Hardness and optimal sensing time}
Before presenting our algorithms, we establish the formal hardness of all problems. We use the standard asymptotic notation $\Omega(\cdot)$ and $O(\cdot)$ to denote lower and upper bounds on how a quantity scales, respectively.
\begin{theorem}\label{thm: lower_bound_main_text} Any quantum sensing algorithm which solves any of Detection (P1-P3), Search (P4), or Learning (P5) must take total time
\begin{equation}\label{eq:lowerbound}
    T = \widetilde{\Omega}\bigg(\frac{\sqrt{N}}{\epsilon}\bigg).
\end{equation}
\end{theorem} \noindent
The tilde suppresses polylogarithmic factors in $N$ and $1/\eps$.
The $1/\epsilon$ scaling is the Heisenberg-limited sensing time required to detect a signal of strength $\epsilon$, while the $\sqrt{N}$ factor reflects the complexity arising from the multi-frequency structure of the signals.
This lower bound on sensing time holds regardless of the number of ancilla qubits or unitary gates available, thus establishing a fundamental limit on the efficiency of quantum sensing algorithms for spectroscopy; see SM for details.

To our knowledge, no existing quantum sensing algorithms achieve this bound~\footnote{Note that an optimal algorithm for P1 is very simple. Since the waveform is known, an optimal strategy applies $X$ gates to the sensor qubit whenever $f(t)$ crosses zero, allowing the signal to accumulate maximally before detection by a conventional Ramsey-type protocol used in DC sensing.}.
Conventional quantum sensing protocols perform spectroscopy by probing individual frequencies sequentially, using a sequence of $\pi$-pulses ($X$ gates) resonantly tuned to one candidate frequency at a time; this is known as Carr-Purcell-Meiboom-Gill (CPMG) spectroscopy~\cite{Kotler2011QuantumLockIn,qsensing:2017,magesan:2013,cooper:2014,xu:2016sensing,holzgrafe:2020,allert2022advances,kiang2024:quantum,Pgardner:2024}. Such an approach takes sensing time scaling as $\sim N/\epsilon$ to solve any of P1-P5.

We design quantum sensing algorithms which outperform these existing approaches and attain the optimal scaling in Theorem~\ref{thm: lower_bound_main_text} for all five problems, even in the worst case:
\begin{theorem}\label{thm: upper_bound_main_text}
   For each of Detection (P1-P3), Search (P4) and Learning (P5),
   there is a quantum sensing algorithm which solves the task in total time
   \begin{equation}
       T=\widetilde{O}\bigg(\frac{\sqrt{N}}{\epsilon}\bigg).
   \end{equation}
\end{theorem}
\noindent

Together with Theorem~\ref{thm: lower_bound_main_text}, this establishes the optimal sensing time scaling, up to polylogarithmic factors, for all five problems and yields a quadratic speedup in $N$ over conventional protocols.

Surprisingly, despite the apparent differences among P1-P5, Theorems~\ref{thm: lower_bound_main_text} and~\ref{thm: upper_bound_main_text} demonstrate that all problems require the same fundamental sensing time. Instead, these differences are reflected in the resource requirements among tasks. Our optimal algorithms for Detection and Search only use a single-qubit sensor undergoing $\pi$-pulse control. Meanwhile, we prove that optimal Learning requires ancilla qubits. In the absence of ancilla qubits, we find that:
\begin{theorem}\label{thm: ancilla_are_necessary}
   Any ancilla-free quantum sensing algorithm for Learning (P5) must take total time
\begin{equation}
    T = \Omega\bigg(\frac{N}{\epsilon}\bigg),
\end{equation}
when $\epsilon/\Delta = O(1)$.
\end{theorem}
\noindent
The bound holds even with arbitrary, adaptive state preparation and control unitaries on the sensor qubit; see SM for details.

\section{Quantum sensing algorithms for detection and search}\label{sec:singleq}

Having established the fundamental sensing time limits for all of our problems, we now present explicit algorithms that achieve them, beginning with Detection and Search.

The key idea is to represent the signal by a $2N$-dimensional vector $B\vec{u}$ of spectral coefficients and probe it through randomized projections. Specifically, we construct a single-qubit control sequence that encodes a randomized projection of the signal into a quantum phase, preparing the state $(|0\rangle+e^{i\Phi}|1\rangle)/\sqrt{2}$ with
\begin{align}\label{eq:random_proj_phase}
    \Phi=\mu B\tau\vec{R}\cdot\vec{u}.
\end{align}
Here, $\tau$ is the sensing duration, $\mu$ is a normalization factor to be chosen, and $\vec{R}$ is a randomly chosen, approximately normalized vector. Physically, $\vec{R}$ can be interpreted as the filter-function response of a sensor, determined by the applied $\pi$ pulses, as elaborated below.
This capability immediately enables Detection: in the absence of a signal ($\Phi=0$), a Pauli-$X$ measurement always yields $X=+1$, whereas observing $X=-1$ indicates the presence of a signal. For a randomly chosen $\vec{R}$ in $2N$ dimensions, its overlap with any fixed normalized $\vec{u}$ is typically of order $1/\sqrt{N}$. Consequently, choosing $\tau=\wt{O}(\sqrt{N}/\epsilon)$ produces a detectable phase whenever $B\geq\epsilon$. Repeating the randomized projection and measurement exponentially suppresses the failure probability. Search follows the same principle, but progressively restricts the support of $\vec{R}$ to subsets of candidate frequencies, identifying which frequencies are present via a binary-search procedure.

It remains to demonstrate that we can prepare the randomized projection states. We accomplish this using \emph{modulation functions}~\cite{qsensing:2017}, i.e., by engineering an effective $H(t)=f(t)Z \mapsto H_{\mathrm{eff}}(t)=g(t)f(t)Z$ for an appropriate choice of $g(t)$. We can engineer an arbitrary modulation function satisfying $|g(t)| \le 1$ using fast $\pi$-pulses, since conjugating by $\pi$-pulses reverses the sign of the coupling, $Z \mapsto XZX = -Z$. While we use continuous functions $g(t)$ for conceptual simplicity, practical pulse sequences accounting for finite pulse durations and control bandwidth can be designed using methods such as Delta-Sigma~\cite{schreier:2017} or sign modulator (see SM for detailed construction). 
Starting from the state $\ket{+}=(\ket{0}+\ket{1})/\sqrt{2}$ and evolving under the effective Hamiltonian yields the sensor state $(|0\rangle+e^{i\Phi}|1\rangle)/\sqrt{2}$ with phase
\begin{align}\label{eq:mod_fun_phase}
    \Phi = 2 \int_{t_0}^{t_0+\tau} g(t) f(t) dt,
\end{align}
where $\tau$ and $t_0$ are the interrogation and start times. For $g(t)$ supported entirely on the candidate frequencies, Eq.~\eqref{eq:mod_fun_phase} approximates the inner product between the spectral coefficients of $g(t)$ and $f(t)$ at large $\tau$.
Therefore, in order to realize the phase Eq.~\eqref{eq:random_proj_phase}, we define
\begin{equation}\label{eq: modulation_function}
    \tilde{g}_{\vec{x}}(t)
    =
   \frac{\mu}{\sqrt{N}}
    \sum_{n=1}^N
    \left(
        x_n\cos(\omega_n t)
        +
        x_{N+n} \sin(\omega_n t)
    \right),
\end{equation}
for $\vec x = (x_1,\dots,x_{2N})$ a randomly chosen vector with each $x_n$ an independent standard Gaussian. The vector $\vec R = \vec x/\sqrt{N}$ then defines the desired random projection.
In general, $\tilde{g}_{\vec{x}}(t)$ might not satisfy the requisite condition $|\tilde{g}_{\vec{x}}(t)| \leq 1$ at all times.
Nevertheless, one can introduce a proper modulation function $g_{\vec{x}}(t)$ simply by ``clipping'' $\tilde{g}_{\vec{x}}(t)$ to $\pm 1$ whenever its magnitude exceeds unity. We show in the SM that this clipping does not affect the resulting measurement statistics when $\mu=O(1/\sqrt{\log(N)})$, demonstrating the desired result.

\section{Quantum-computing-enhanced algorithms for signal learning}
\label{sec:estimation}

We now consider the most demanding of the problems, Learning, wherein we must estimate all $2N$ amplitudes of $B\vec{u}$.
The single-qubit approach above would be inefficient here, since learning each amplitude from randomized projections requires a  total sensing time linear in $N$.
To learn the signal more efficiently, we must devise a method of estimating multiple projections simultaneously.

We achieve this by using a quantum computer to coherently control the modulation function applied to the sensor, thereby computing all projections \emph{in superposition}.
Specifically, we consider a quantum computer consisting of $2N$ registers, each containing $M = O(\log (1/\epsilon))$ qubits, for a total of $2NM$ qubits. We use these ancilla qubits to
encode the projections as $| \vec{x} \rangle$ with $\vec{x} =  (x_1, x_2,\dots,x_{2N})$, where each $x_n\in [-1,1]$ is stored to $M$ bits of precision in the $n$-th register.
We initialize the quantum computer in the uniform superposition of all $| \vec{x} \rangle$ and the sensor in $\ket{0}$, then conditionally apply the modulation function $g_{\vec{x}}(t)$ and evolve for time $\tau = 2 \sqrt N /\mu \eps$. The state of the quantum computer is then approximately
\begin{equation}\label{eq: oracle_on_equal_super}
    \sum_{\vec{x}} e^{i \frac{B\vec{u}}{\epsilon}\cdot \vec{x}}|\vec{x}\rangle = \bigotimes_{n=1}^{2N} \sum_{x_n} e^{i \frac{Bu_n}{\epsilon} x_n} \ket{x_n},
\end{equation}
up to normalization. By taking the inverse quantum Fourier transform of each register, we obtain an $\eps$-accurate approximation of $Bu_n$, as desired.

We note that this algorithm relies on precisely the quantum gradient estimation algorithm applied to the linear function $B\vec{u} \cdot \vec{x}$~\cite{Bernstein1997Quantum,gradient}, where the quantum Fourier transform enables large dynamic range of sensing~\cite{vorobyov2021quantum}.
A crucial difference is the unitary preparing the state Eq.~\eqref{eq: oracle_on_equal_super} is synthesized from continuous sensing dynamics, rather than as a black-box oracle.
In other words, as in Ref.~\cite{allen2025quantumcomputingenhancedsensing}, we follow the paradigm of taking \textit{nature as an oracle} and leverage quantum algorithms to improve quantum sensing.

For periodic signals with equally spaced frequencies $\omega_n=n\Delta$, which are ubiquitous in practical applications, we show that the same optimal sensing time $\wt{O}(\sqrt{N}/\eps)$ is achieved by a drastically simpler algorithm, using only SWAP gates between the sensor and ancillary memory qubits, single-qubit measurements, and classical post-processing, without generating entanglement (see SM).

\begin{figure}[h!]
    \centering
    \includegraphics[width=0.98\linewidth]{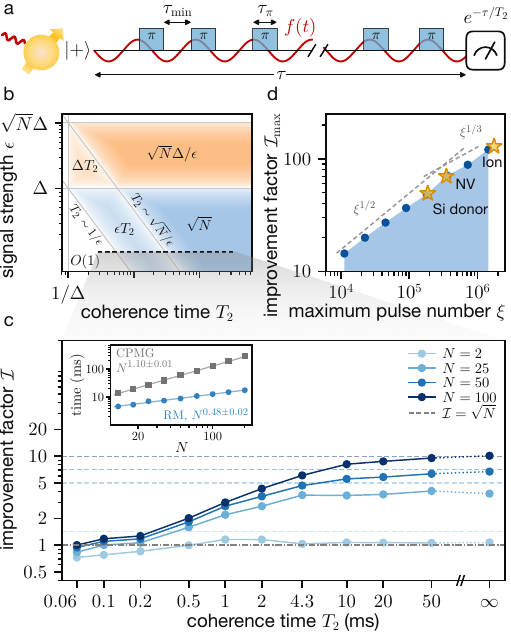}
    \caption{\textbf{Performance under practical constraints.}
\textbf{a.}~We consider three hardware constraints: finite $\pi$-pulse duration $\tau_{\pi}$, during which the signal continues to evolve; limited control bandwidth, which sets the minimum pulse spacing $\tau_{\mathrm{min}}$; and finite coherence time $T_2$, which attenuates the metrological signal by $e^{-\tau/T_2}$ and limits the interrogation time $\tau$. We investigate their effects on our Detection algorithm.
\textbf{b.}~Different scalings of the improvement factor $\mathcal{I}$, up to logarithmic factors, define the phase diagram in $T_2$ and signal strength $\epsilon$. The orange regions are affected by Fourier broadening, whereas the blue regions are not. Darker colors indicate weaker $T_2$-limitations. The uncolored regions, where the heuristic estimate does not apply, are discussed in the SM.
\textbf{c.}~Numerically simulated $\mathcal{I}$ versus $T_2$ for several $N$, with $\Delta/2\pi=40$~kHz and $\epsilon=0.05\Delta$, along the dashed line in \textbf{b}. Above the gray dash-dotted line $\mathcal{I}=1$, random modulation outperforms CPMG spectroscopy. Colored dashed lines indicate $\mathcal{I}=\sqrt{N}$ for each $N$. The inset shows total sensing times $T_{\mathrm{CPMG}}$ and $T_{\mathrm{RM}}$ versus $N$ at $T_2=\infty$, with lines showing least-squares fits to $AN^\alpha$ on logarithmic scales and labeled fitted exponents $\alpha$. The fitted exponents, $\alpha \approx 1.1$ (CPMG) and $0.48$ (RM), agree with the expected values of $1$ and $0.5$, respectively.
\textbf{d.}~Numerically simulated maximum improvement $\mathcal{I}_{\max}$ versus $\xi$, varying only $T_2$ with $\tau_\pi=8$~ns, $\tau_{\mathrm{min}}=4$~ns, and candidate frequencies between $2$ and $20$~MHz. Gray dashed lines guide the eye with slopes $\xi^{1/2}$ and $\xi^{1/3}$, the latter predicted asymptotically by theory. Stars indicate a phosphorus donor electron spin in silicon ($T_2=0.56$~s~\cite{Muhonen2014}, $\tau_\pi=3~\mu$s~\cite{Muhonen2015}, $\tau_{\mathrm{min}}=0$), an NV center ($T_2=4.2$~ms~\cite{Han2025}, $\tau_\pi=8$~ns, $\tau_{\mathrm{min}}=4$~ns~\cite{Lange2011}), and a trapped ion ($T_2=2.1$~s~\cite{Ruster2016}, $\tau_\pi=1.2~\mu$s~\cite{Leu2023}, $\tau_{\mathrm{min}}=0$).
Each star is positioned at its corresponding $\xi$, with $\mathcal{I}_{\max}$ interpolated logarithmically between simulated points, except for the trapped ion, whose value is extrapolated using $\xi^{1/3}$ beyond the simulated range.
    }
    \label{fig:main_numerics}
\end{figure}

\section{Performance under experimental imperfections}

Having established the fundamental limit and quantum algorithms under idealized settings, we now investigate the performance of our approach under realistic conditions. Specifically, we consider the performance in non-asymptotic scaling regimes reflecting the effects of various control imperfections and noises such as limited control bandwidth and speed as well as decoherence characterized by time scales $\tau_{\text{min}}, \tau_{\pi}$, and $T_2$ (Fig.~\ref{fig:main_numerics}a).
We focus on Detection (P3) and compare the total sensing time of our algorithm, $T_{\mathrm{RM}}$, against generously optimized sequential CPMG spectroscopy, $T_{\mathrm{CPMG}}$, as a baseline.
We quantify the advantage through the improvement factor
\begin{equation}
	  \mathcal{I} := \frac{T_{\mathrm{CPMG}}}{T_{\mathrm{RM}}}.
\end{equation}

We first give a heuristic scaling estimate of $\mathcal{I}$ that takes into account two effects: finite coherence time $T_2$ and Fourier broadening arising from the interplay between the target sensitivity $\epsilon$ and minimum frequency spacing $\Delta$.
The finite coherence time limits the number of candidate frequencies that our random-modulation algorithm can probe coherently in a single round.
Probing all $N$ candidate frequencies coherently requires an interrogation time $\sim\sqrt{N}/\epsilon$.
When $T_2\gtrsim\sqrt{N}/\epsilon$, all candidates can be probed in a single round, giving $T_{\mathrm{RM}}\sim\sqrt{N}/\epsilon$.
When $T_2\lesssim\sqrt{N}/\epsilon$, we divide the candidates into groups of size $N'\sim(\epsilon T_2)^2$, so that each group can be probed in time $\sim T_2$.
Testing all $N/N'$ groups therefore takes total sensing time $T_{\mathrm{RM}}\sim(N/N')T_2\sim N/(\epsilon^2T_2)$.
Fourier broadening can instead reduce the sensing time for sequential CPMG spectroscopy.
Each interrogation lasts $\sim1/\epsilon$, producing a frequency filter of width $\sim\epsilon$.
For $\epsilon\lesssim\Delta$, individual frequencies must be probed separately, giving $T_{\mathrm{CPMG}}\sim N/\epsilon$.
For $\epsilon\gtrsim\Delta$, each filter can probe at most $\sim\epsilon/\Delta$ candidate frequencies simultaneously, reducing the total sensing time proportionally.
For equally spaced frequencies, CPMG fully exploits Fourier broadening, giving $T_{\mathrm{CPMG}}\sim N\Delta/\epsilon^2$.
Combining the sensing time estimates above, we obtain
\begin{equation}
\mathcal{I} \sim \min\left\{\sqrt{N},\epsilon T_2\right\}
\times \min\left\{1,\frac{\Delta}{\epsilon}\right\}.
\end{equation}
Both effects reduce the advantage from its ideal scaling $\mathcal{I}\sim\sqrt{N}$, recovered when $T_2\gtrsim\sqrt{N}/\epsilon$ and $\epsilon\ll\Delta$.
The corresponding regimes are shown in Fig.~\ref{fig:main_numerics}b.

We now quantitatively evaluate $\mathcal{I}$, accounting for the effects discussed above as well as finite pulse durations and pulse spacing encountered in practical implementations.
Since these combined effects are difficult to capture analytically, we use end-to-end numerical simulations.
For each round of the random-modulation algorithm, we explicitly design the pulse sequence satisfying the pulse-duration and minimum-spacing constraints using a sign-modulator approach (see SM for details).
We numerically integrate the full time-dependent dynamics, including the simultaneous action of the control pulses and the signal, and sample the resulting measurement outcomes.
We then compute an explicit test statistic from the sampled outcomes across rounds and compare it against a predetermined threshold to decide whether a signal is present.
We similarly simulate CPMG spectroscopy, but with instantaneous pulses, which can only reduce its sensing time and thus underestimate our improvement factor.
For given sensor and signal parameters, we independently optimize each approach to minimize its total sensing time for the same detection task at a fixed failure probability of $0.1$.
Fig.~\ref{fig:main_numerics}c shows that as $T_2$ increases at fixed signal parameters, $\mathcal{I}$ starts from a plateau, then grows, and finally saturates.
We find meaningful improvement already for $N \ge 2$, with improvement approaching the expected scaling $\sqrt N$ for sufficiently large $T_2$ at fixed $\eps$ and $\Delta$ (see the inset of Fig.~\ref{fig:main_numerics}c).

Finally, to assess the practical potential of our algorithm on existing quantum sensing platforms, we evaluate the maximum achievable improvement factor $\mathcal{I}_{\max}$ over experimentally motivated parameter ranges.
For a sensor with fixed experimental capabilities, we optimize the improvement factor over $\epsilon$ and $N$ subject to limited control speed and bandwidth set by the finite pulse duration $\tau_\pi$ and the minimum pulse spacing $\tau_{\mathrm{min}}$.
Our theoretical analysis predicts the scaling $\mathcal{I}_{\max}\sim\xi^{1/3}$, where $\xi:=T_2/(\tau_\pi+\tau_{\mathrm{min}})$ is the maximum number of pulses that fit within one coherence time.
Interestingly, our numerical results exhibit faster growth at small $\xi$ (Fig.~\ref{fig:main_numerics}d), which we explain through finite-size analysis (see SM), leading to the prediction $\mathcal{I}_{\mathrm{max}} \sim \xi^{1/2}$.
Using state-of-the-art experimental parameters, we estimate $\mathcal{I}_{\max}\approx50$ for a phosphorus donor electron spin in silicon~\cite{Muhonen2014,Muhonen2015}, $70$ for an NV center~\cite{Han2025,Lange2011}, and $130$ for a trapped ion~\cite{Ruster2016,Leu2023}, as indicated by the stars in Fig.~\ref{fig:main_numerics}d.

\section{Applications}

Given the practical relevance of discrete-frequency spectroscopy, our algorithms offer opportunities for near- and long-term technological advances. We discuss this potential with two concrete applications.

\subsection{Application I: Molecular Screening}\label{sec:application_one}
Molecular screening offers a natural near-term application of our single-qubit algorithm. Fluorinated fragments are widely used in fragment-based drug discovery because their $^{19}$F NMR spectra provide characteristic molecular fingerprints that help identify different compounds within a mixture~\cite{Binas2021}.
For a given library of these known fingerprints, our algorithm can detect a signal (P1-P3) or locate active spectral lines (P4), identifying unknown chemical species without reconstructing the full spectrum.
To illustrate the experimental requirements, we envision a $^{19}$F NMR sample placed near an NV center in diamond. At a bias field of $50~\mathrm{mT}$, the $^{19}$F resonance lies near $2~\mathrm{MHz}$, so chemical-shift differences of $10$--$100~\mathrm{ppm}$ correspond to frequency separations of $20$--$200~\mathrm{Hz}$. For a known library, the spectral lines define a discrete set of $N$ candidate frequencies.
As an illustrative example, fluorinated-fragment screening experiments have used mixtures of $N \sim 20$ compounds~\cite{Binas2021}.

We model the signal from the $n$-th spectral component as a weak, decaying oscillating field,
$B_n(t)=B_n e^{-t/T_s}\cos(\omega_n t+\phi_n)$, where $B_n$ is the initial amplitude, $\omega_n$ and $\phi_n$ are the frequency and phase, and $T_s$ is the signal coherence time.
The signal is drawn from $N=20$ candidate frequencies.
For a uniformly hyperpolarized sample with $P\approx0.1\%$, the polarizations reported for fluoropyridines at a concentration of $100~\mathrm{mM}$ would yield fields of $B_n \sim 0.1~\mathrm{nT}$ at the sensor~\cite{Ariyasingha2019}.

Our single-qubit sensing sequence distributes its response over these frequencies. The accumulated root-mean-square phase then scales as $\theta_{\mathrm{rms}}(T)\sim\frac{\gamma_{\mathrm s}B_nT_s}{\sqrt{N}}\left(1-e^{-T/T_s}\right)$, up to a sequence-dependent numerical prefactor, where $\gamma_{\mathrm s}/2\pi=28~\mathrm{GHz/T}$. For $T_s=1~\mathrm{s}$ and $T=300~\mathrm{ms}$, setting this prefactor to unity results in a sufficiently strong response $\theta_{\mathrm{rms}}\sim1$ at $B_n\sim0.1~\mathrm{nT}$, which is comparable to the hyperpolarized-sample estimate above. The required interrogation time is also compatible with experimentally demonstrated coherence in NV ensembles, where dynamical decoupling has extended $T_2 \approx 0.5~\mathrm{s}$ at $77~\mathrm{K}$~\cite{BarGill2013}. These estimates motivate the deployment of our methods for realistic molecular screening using concentrated, hyperpolarized samples.

\subsection{Application II: Wireless Communications}\label{sec:application_two}

\begin{figure}[t]
    \centering
    \includegraphics[width=\linewidth]{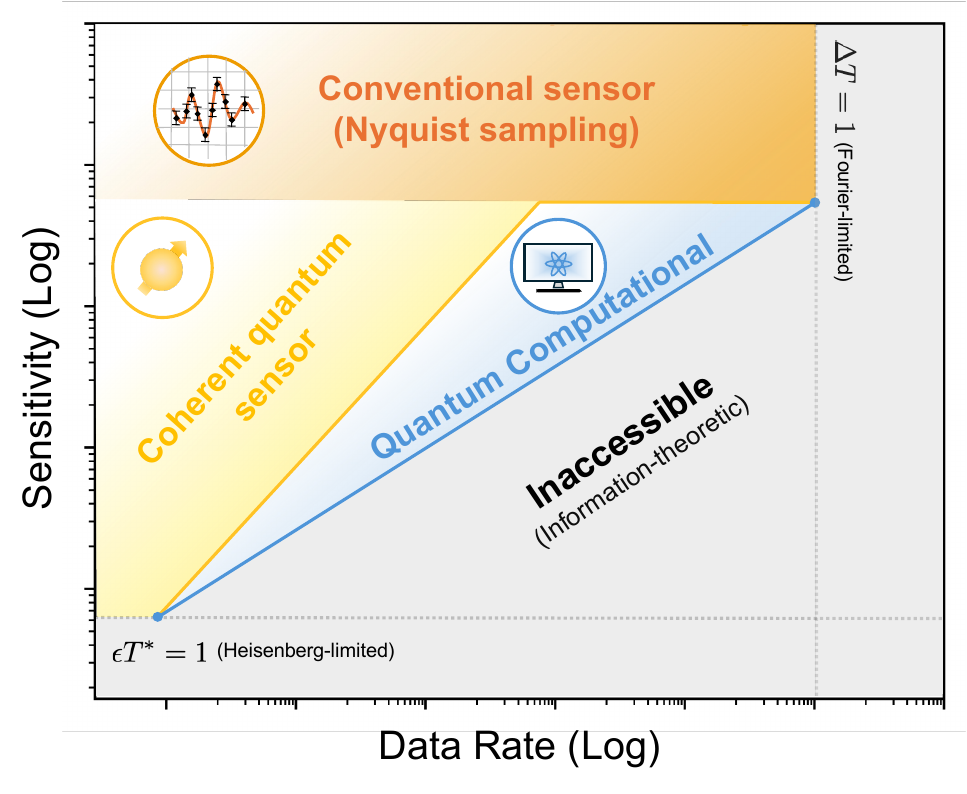}
    \caption{
    \textbf{Signal learning for wireless communication.}
We apply Learning (P5) to decode digital information encoded in the amplitudes of a multi-frequency signal. We compare achievable sensitivity and data rate for Nyquist-rate sampling, coherent single-sensor spectroscopy, and quantum-computing-enhanced sensing as symbol duration $T$ and frequency spacing $\Delta$ vary, with $N \ge 1$ subcarriers and fixed bandwidth $W$. Channel fluctuations and environmental effects impose a maximum usable symbol duration $T^*$ which we assume to be shorter than $T_2$ (see SM for $T_2$-limited analysis). Together with the subcarrier orthogonality requirement $T\geq1/\Delta$, this setting restricts the operating regime to $1/\Delta\leq T\leq T^*$. Nyquist-rate time sampling achieves a sensitivity $\epsilon_{\text{samp}}\sim \sqrt{W/T}$, defining the orange regime, with the dashed boundary marking the Fourier limit $\Delta T=1$ and Heisenberg limit $\epsilon T^* = 1$. The yellow regime additionally requires coherent single-sensor CPMG spectroscopy, while the blue regime requires quantum-computing-enhanced sensing, achieving $\epsilon_{\text{QC}}\sim \sqrt{W/\Delta T^2}$. The color gradients indicate that each scheme also supports less stringent sensitivity or lower data rates, while improved sensitivity and higher rates are subject to fundamental limits. The yellow solid boundary marks the fundamental ancilla-free sensing limit (a generalization of Theorem~\ref{thm: ancilla_are_necessary}; see SM), beyond which no ancilla-free single-sensor protocol can operate. The gray region is impossible for all protocols by a generalization of Theorem~\ref{thm: lower_bound_main_text} (see SM).}
    \label{fig:qg}
\end{figure}

Our Learning algorithm naturally applies to broadband, multicarrier communication schemes such as orthogonal frequency-division multiple access (OFDMA). Such protocols are widely used in modern wireless networks to enable simultaneous communication between a base station (e.g., cell tower) and multiple users (e.g., mobile devices)~\cite{tse2005fundamentals}. For fixed sensor hardware capabilities, our approach enables reliable data transmission at signal strengths too weak for conventional receivers to accurately decode.

In OFDMA, digital information is encoded and transmitted via electromagnetic waves over a bandwidth $W$, using $N$ frequency tones (subcarriers) equally spaced by $\Delta$. Each user encodes binary information in the discrete amplitudes of the sine and cosine modes of its assigned subcarrier over time duration $T$, as in Eq.~\eqref{eq:signal}.
In the simplest protocol, the information is encoded in the amplitude pairs $(Bu_n,B u_{N+n}) = (\pm \eps, \pm \eps)$, encoding two bits in their relative phases; this is known as quadrature phase-shift keying (QPSK). The data transmission rate is $R=2N/T$.
Decoding requires estimating all quadrature amplitudes to a sufficient accuracy $\sim \epsilon$, precisely the Learning (P5) task.
The communication performance (e.g., by the cell tower) is characterized by the rate-sensitivity pair $(R, \epsilon)$.

To assess the advantage of our quantum-computing-enhanced algorithm, we compare the regimes of achievable $(R,\epsilon)$ by three different receiver models utilizing the same
sensor.
\textit{Time-sampling receiver:} a single sensor particle samples at the
Nyquist rate, and a classical Fourier transform estimates the subcarrier amplitudes, as in conventional antenna receivers (see SM).
\textit{Coherent quantum receiver:} this class covers all single-particle protocols without ancillary quantum memory, including time sampling and protocols that exploit long-time coherent accumulation.
\textit{Quantum computational receiver:} our quantum algorithm estimates the subcarrier amplitudes.
Fig.~\ref{fig:qg} illustrates the achievable regimes. In the weak-signal regime, the quantum computational receiver supports data rates inaccessible to the other two receiver types. Such an improvement offers a concrete and new motivation for building quantum computers, because of their potential for future wireless communication technology in the regime of weak signals, such as over long distances or in outer space.

These improvements can be formally translated to an exponential quantum sensing advantage, as in other works~\cite{mele2026exponentialseparationsensingcontinuous, Cotler_2026}. Consider a random data stream subject to a hidden parity check, i.e., a subset of bits which must sum to an even number. A quantum computational receiver using our Learning algorithm can identify the parity check exponentially faster in the number of transmitted bits than any coherent quantum receiver without ancillas (see SM for details). 
This problem and speedup are deliberately crafted to emphasize a formal separation between the Learning capabilities of receivers with and without ancilla qubits; we find it more instructive to illustrate the benefits of quantum computing through improvements in fundamental receiver capabilities, such as data transmission rate and sensitivity (Fig.~\ref{fig:qg}).

\section{Discussion}

In this work, we have developed a framework for optimal quantum sensing spectroscopy that fully leverages the advantages of quantum sensors. Although we focused on signals supported on a finite set of sinusoidal modes, the underlying principles extend naturally to finite subsets of other orthogonal bases, such as Walsh functions~\cite{magesan:2013,cooper:2014} and wavelets~\cite{xu:2016sensing,kiang2024:quantum}, as well as to signals that admit accurate approximations in such bases.

Beyond these extensions, it is instructive to ask what physical intuition underlies our quantum-computing-enhanced Learning algorithm and how their advantages can be realized experimentally. From an algorithmic perspective, the unknown signal acts as a quantum phase oracle, enabling algorithms such as gradient estimation. From a physical perspective, quantum dynamics coherently process the signal before measurement, much like an optical prism that separates frequencies through the collective response of electrons in a material, mapping them to distinct spatial locations. This analogy motivates a simple analog implementation of Learning using coupled bosonic modes.
We consider a sensor mode coupled to the signal through $H(t)=f(t)(a+a^\dagger)$, where $a$ ($a^\dagger$) is its bosonic annihilation (creation) operator. With $N-1$ ancillary modes uncoupled to the signal, our protocol achieves the same sensing time scaling $\widetilde{O}(\sqrt{N}/\epsilon)$ as in the qubit-sensor setting. It requires only vacuum-state initialization and evolution under a time-independent control Hamiltonian with nearest-neighbor hopping. After the evolution, the spectral coefficients are then separately encoded in the coherent parameters of $N$ decoupled modes, accessible through heterodyne measurements (see SM). Unlike a physical prism, our approach realizes spectral separation algorithmically through programmable quantum control, potentially across diverse sensing platforms and operating regimes.

Finally, a natural question is whether our algorithms can be directly applied to the continuous-frequency setting. The naive approach of approximating a continuous band with densely spaced candidate frequencies need not yield optimal algorithms, as continuous-frequency spectroscopy requires sensitivity across an entire interval rather than a set of sharply defined frequencies. This distinction is already reflected in Detection and Search: while our algorithms achieve a quadratic speedup using only a single sensor qubit, analogous speedups in continuous-frequency search are believed to require ancilla-assisted coherent quantum processing~\cite{allen2025quantumcomputingenhancedsensing}. Developing optimal, easily implementable algorithms for continuous-frequency spectroscopy remains an important direction for future research.

\section{Acknowledgments}

We thank Dmitry Abanin, Ronald Fernando Garcia Ruiz, James Gardner, Nick Gharabaghi, Dongseob Kim, Dan Stamper-Kurn, Kavish Trivedi, Matteo Votto, Kaito Wada, Jun Ye, and Xueyue Zhang for fruitful discussions. We thank Tian-Xing Zheng for helpful comments on the manuscript. We are especially grateful to Paola Cappellaro and Takuya Isogawa for various discussions concerning potential experimental realization using NV centers.
ChatGPT 6.1 and Claude Opus 5.5 were used to assist with manuscript proofreading and numerical calculations, as well as to prove the lower bound for ancilla-free Learning. 
This work is supported by the U.S. Department of Energy, Genesis Mission (No. DOE-GM-RFA-P1-2026-00070).
Y.L., R.R.A., R.S., and S.C. acknowledge support from the the NSF through the Center for Ultracold Atoms, a Physics Frontiers Center (PHY-2317134) and QuSeC-TAQS Award 2326787, as well as a gift from Google.
S.C. also acknowledges support from the Alfred P. Sloan Research Fellowship. 
T.T.C. and J.C. acknowledge support from the AFOSR Young Investigator Program (YIP No. FA9550-25-1-0147), the U.S. Department of Energy, Office of Science, Early Career Research Program (ECRP No. DE-SC0026258), and the Office of Naval Research (No. N000142612398).
The authors acknowledge the MIT Office of Research Computing and Data for providing high performance computing resources that have contributed to the research results reported within this paper.

\bibliography{references.bib}

@article{childs_theory_2021,
    title = {Theory of {Trotter} {Error} with {Commutator} {Scaling}},
    volume = {11},
    issn = {2160-3308},
    url = {https://link.aps.org/doi/10.1103/PhysRevX.11.011020},
    doi = {10.1103/PhysRevX.11.011020},
    number = {1},
    urldate = {2025-08-25},
    journal = {Physical Review X},
    author = {Childs, Andrew M. and Su, Yuan and Tran, Minh C. and Wiebe, Nathan and Zhu, Shuchen},
    month = feb,
    year = {2021},
    pages = {011020},
}

@misc{allen2025quantumcomputingenhancedsensing,
      title={Quantum Computing Enhanced Sensing},
      author={Richard R. Allen and Francisco Machado and Isaac L. Chuang and Hsin-Yuan Huang and Soonwon Choi},
      year={2025},
      eprint={2501.07625},
      archivePrefix={arXiv},
      primaryClass={quant-ph},
      url={https://arxiv.org/abs/2501.07625},
}

@article{qsensing:2017,
  title = {Quantum sensing},
  author = {Degen, C. L. and Reinhard, F. and Cappellaro, P.},
  journal = {Rev. Mod. Phys.},
  volume = {89},
  issue = {3},
  pages = {035002},
  numpages = {39},
  year = {2017},
  month = {Jul},
  publisher = {American Physical Society},
  doi = {10.1103/RevModPhys.89.035002},
  url = {https://link.aps.org/doi/10.1103/RevModPhys.89.035002}
}

@article{tsang:2011,
  title = {Fundamental Quantum Limit to Waveform Estimation},
  author = {Tsang, Mankei and Wiseman, Howard M. and Caves, Carlton M.},
  journal = {Phys. Rev. Lett.},
  volume = {106},
  issue = {9},
  pages = {090401},
  numpages = {4},
  year = {2011},
  month = {Mar},
  publisher = {American Physical Society},
  doi = {10.1103/PhysRevLett.106.090401},
  url = {https://link.aps.org/doi/10.1103/PhysRevLett.106.090401}
}

@article{xu:2016sensing,
  title = {Wavelet-based fast time-resolved magnetic sensing with electronic spins in diamond},
  author = {Xu, Nanyang and Jiang, Fengjian and Tian, Yu and Ye, Jianfeng and Shi, Fazhan and Lv, Haijiang and Wang, Ya and Wrachtrup, J\"org and Du, Jiangfeng},
  journal = {Phys. Rev. B},
  volume = {93},
  issue = {16},
  pages = {161117(R)},
  numpages = {5},
  year = {2016},
  month = {Apr},
  publisher = {American Physical Society},
  doi = {10.1103/PhysRevB.93.161117},
  url = {https://link.aps.org/doi/10.1103/PhysRevB.93.161117}
}

@book{schreier:2017,
  author    = {Richard Schreier and Shanthi Pavan and Gabor C. Temes},
  title     = {Understanding Delta-Sigma Data Converters},
  edition   = {2},
  publisher = {John Wiley \& Sons},
  year      = {2017},
  series    = {IEEE Press Series on Microelectronic Systems},
  isbn       = {9781119258278},
  doi        = {10.1002/9781119258308}
}

@article{gardner:2025,
  title = {Stochastic Waveform Estimation at the Fundamental Quantum Limit},
  author = {Gardner, James W. and Gefen, Tuvia and Haine, Simon A. and Hope, Joseph J. and Preskill, John and Chen, Yanbei and McCuller, Lee},
  journal = {PRX Quantum},
  volume = {6},
  issue = {3},
  pages = {030311},
  numpages = {42},
  year = {2025},
  month = {Jul},
  publisher = {American Physical Society},
  doi = {10.1103/h91r-4ws9},
  url = {https://link.aps.org/doi/10.1103/h91r-4ws9}
}

@article{magesan:2013,
  title = {Compressing measurements in quantum dynamic parameter estimation},
  author = {Magesan, Easwar and Cooper, Alexandre and Cappellaro, Paola},
  journal = {Phys. Rev. A},
  volume = {88},
  issue = {6},
  pages = {062109},
  numpages = {13},
  year = {2013},
  month = {Dec},
  publisher = {American Physical Society},
  doi = {10.1103/PhysRevA.88.062109},
  url = {https://link.aps.org/doi/10.1103/PhysRevA.88.062109}
}

@article{lovchinsky:2016,
  author  = {I. Lovchinsky and A. O. Sushkov and E. Urbach and N. P. de Leon and S. Choi and K. De Greve and R. Evans and R. Gertner and E. Bersin and C. M{\"u}ller and L. McGuinness and F. Jelezko and R. L. Walsworth and H. Park and M. D. Lukin},
  title   = {Nuclear Magnetic Resonance Detection and Spectroscopy of Single Proteins Using Quantum Logic},
  journal = {Science},
  volume  = {351},
  number  = {6275},
  pages   = {836--841},
  year    = {2016},
  doi     = {10.1126/science.aad8022}
}

@article{aslam:2017,
  author  = {N. Aslam and M. Pfender and P. Neumann and R. Reuter and A. Zappe and F. Favaro de Oliveira and A. Denisenko and H. Sumiya and S. Onoda and J. Isoya and J. Wrachtrup},
  title   = {Nanoscale Nuclear Magnetic Resonance with Chemical Resolution},
  journal = {Science},
  volume  = {357},
  number  = {6346},
  pages   = {67--71},
  year    = {2017},
  doi     = {10.1126/science.aam8697}
}

@article{gradient,
  title = {Fast Quantum Algorithm for Numerical Gradient Estimation},
  author = {Jordan, Stephen P.},
  journal = {Phys. Rev. Lett.},
  volume = {95},
  issue = {5},
  pages = {050501},
  numpages = {4},
  year = {2005},
  month = {Jul},
  publisher = {American Physical Society},
  doi = {10.1103/PhysRevLett.95.050501},
  url = {https://link.aps.org/doi/10.1103/PhysRevLett.95.050501}
}

@article{Pgardner:2024,
  title = {Achieving the Fundamental Quantum Limit of Linear Waveform Estimation},
  author = {Gardner, James W. and Gefen, Tuvia and Haine, Simon A. and Hope, Joseph J. and Chen, Yanbei},
  journal = {Phys. Rev. Lett.},
  volume = {132},
  issue = {13},
  pages = {130801},
  numpages = {7},
  year = {2024},
  month = {Mar},
  publisher = {American Physical Society},
  doi = {10.1103/PhysRevLett.132.130801},
  url = {https://link.aps.org/doi/10.1103/PhysRevLett.132.130801}
}

@article{tsang2012fundamental,
  title   = {Fundamental Quantum Limits to Waveform Detection},
  author  = {Tsang, Mankei and Nair, Ranjith},
  journal = {Physical Review A},
  volume  = {86},
  number  = {4},
  pages   = {042115},
  year    = {2012},
  doi     = {10.1103/PhysRevA.86.042115}
}

@article{kiang2024:quantum,
  title   = {Quantum Sensing of Fast Time-Varying Magnetic Field With Daubechies Wavelets},
  author  = {Kiang, Chou-Wei and Ding, Jian-Jiun and Kiang, Jean-Fu},
  journal = {IEEE Access},
  volume  = {12},
  pages   = {23181--23189},
  year    = {2024},
  doi     = {10.1109/ACCESS.2024.3364817}
}

@article{berry2015quantum,
  title   = {Quantum {Bell--Ziv--Zakai} Bounds and Heisenberg Limits
             for Waveform Estimation},
  author  = {Berry, Dominic W. and Tsang, Mankei and Hall, Michael J. W.
             and Wiseman, Howard M.},
  journal = {Physical Review X},
  volume  = {5},
  number  = {3},
  pages   = {031018},
  year    = {2015},
  doi     = {10.1103/PhysRevX.5.031018}
}

@article{ligo,
  title = {Broadband Quantum Enhancement of the LIGO Detectors with Frequency-Dependent Squeezing},
  author = {Ganapathy, D. and Jia, W. and Nakano, M. and Xu, V. and Aritomi, N. and Cullen, T. and Kijbunchoo, N. and Dwyer, S. E. and Mullavey, A. and McCuller, L. and others},
  collaboration = {LIGO O4 Detector Collaboration},
  journal = {Phys. Rev. X},
  volume = {13},
  issue = {4},
  pages = {041021},
  numpages = {14},
  year = {2023},
  month = {Oct},
  publisher = {American Physical Society},
  doi = {10.1103/PhysRevX.13.041021},
  url = {https://link.aps.org/doi/10.1103/PhysRevX.13.041021}
}

@article{pezze:2018,
  title = {Quantum metrology with nonclassical states of atomic ensembles},
  author = {Pezz\`e, Luca and Smerzi, Augusto and Oberthaler, Markus K. and Schmied, Roman and Treutlein, Philipp},
  journal = {Rev. Mod. Phys.},
  volume = {90},
  issue = {3},
  pages = {035005},
  numpages = {70},
  year = {2018},
  month = {Sep},
  publisher = {American Physical Society},
  doi = {10.1103/RevModPhys.90.035005},
  url = {https://link.aps.org/doi/10.1103/RevModPhys.90.035005}
}

@article{giovannetti:2004quantum,
  title   = {Quantum-Enhanced Measurements: Beating the Standard Quantum Limit},
  author  = {Giovannetti, Vittorio and Lloyd, Seth and Maccone, Lorenzo},
  journal = {Science},
  volume  = {306},
  number  = {5700},
  pages   = {1330--1336},
  year    = {2004},
  doi     = {10.1126/science.1104149}
}

@article{cooper:2014,
  title   = {Time-resolved magnetic sensing with electronic spins in diamond},
  author  = {Cooper, Alexandre and Magesan, Easwar and Yum, H. N. and Cappellaro, Paola},
  journal = {Nature Communications},
  volume  = {5},
  pages   = {3141},
  year    = {2014},
  doi     = {10.1038/ncomms4141}
}

@article{allert2022advances,
  title   = {Advances in nano- and microscale {NMR} spectroscopy using diamond quantum sensors},
  author  = {Allert, Robin D. and Briegel, Karl D. and Bucher, Dominik B.},
  journal = {Chemical Communications},
  volume  = {58},
  number  = {59},
  pages   = {8165--8181},
  year    = {2022},
  doi     = {10.1039/D2CC01546C}
}

@misc{khan2025quantum,
      title={Quantum Computational Sensing}, 
      author={Saeed A. Khan and Sridhar Prabhu and Logan G. Wright and Peter L. McMahon},
      year={2026},
      eprint={2507.16918},
      archivePrefix={arXiv},
      primaryClass={quant-ph},
      url={https://arxiv.org/abs/2507.16918}, 
}

@article{Cotler_2026,
   title={Quantum Advantage for Sensing Properties of Classical Fields},
   volume={16},
   ISSN={2160-3308},
   url={http://dx.doi.org/10.1103/6ps1-mw9l},
   DOI={10.1103/6ps1-mw9l},
   number={4},
   journal={Physical Review X},
   publisher={American Physical Society (APS)},
   author={Cotler, Jordan and Danielson, Daine L. and Kannan, Ishaan},
   year={2026},
   month=Oct }

@misc{kannan2026exponential,
      title={Exponential quantum advantage for learning signals with a single qubit}, 
      author={Ishaan Kannan and Sridhar Prabhu and Saeed A. Khan and Mandar M. Sohoni and Xingrui Song and Saswata Roy and Alen Senanian and Valla Fatemi and Peter L. McMahon and Jordan Cotler},
      year={2026},
      eprint={2608.13521},
      archivePrefix={arXiv},
      primaryClass={quant-ph},
      url={https://arxiv.org/abs/2608.13521}, 
}

@article{Bernstein1997Quantum,
  author  = {Bernstein, Ethan and Vazirani, Umesh},
  title   = {Quantum Complexity Theory},
  journal = {SIAM Journal on Computing},
  volume  = {26},
  number  = {5},
  pages   = {1411--1473},
  year    = {1997},
  doi     = {10.1137/S0097539796300921}
}

@article{Kotler2011QuantumLockIn,
  author  = {Kotler, Shlomi and Akerman, Nitzan and Glickman, Yinnon and Keselman, Anna and Ozeri, Roee},
  title   = {Single-ion quantum lock-in amplifier},
  journal = {Nature},
  volume  = {473},
  number  = {7345},
  pages   = {61--65},
  year    = {2011},
  doi     = {10.1038/nature10010}
}

@article{holzgrafe:2020,
  title = {Nanoscale NMR Spectroscopy Using Nanodiamond Quantum Sensors},
  author = {Holzgrafe, Jeffrey and Gu, Qiushi and Beitner, Jan and Kara, Dhiren M. and Knowles, Helena S. and Atat\"ure, Mete},
  journal = {Phys. Rev. Appl.},
  volume = {13},
  issue = {4},
  pages = {044004},
  numpages = {8},
  year = {2020},
  month = {Apr},
  publisher = {American Physical Society},
  doi = {10.1103/PhysRevApplied.13.044004},
  url = {https://link.aps.org/doi/10.1103/PhysRevApplied.13.044004}
}

@article{rugar2015proton,
  title   = {Proton magnetic resonance imaging using a nitrogen--vacancy spin sensor},
  author  = {Rugar, D. and Mamin, H. J. and Sherwood, M. H. and Kim, M. and Rettner, C. T. and Ohno, K. and Awschalom, D. D.},
  journal = {Nature Nanotechnology},
  volume  = {10},
  number  = {2},
  pages   = {120--124},
  year    = {2015},
  doi     = {10.1038/nnano.2014.288}
}

@article{mamin2013nanoscale,
  title   = {Nanoscale Nuclear Magnetic Resonance with a Nitrogen-Vacancy Spin Sensor},
  author  = {Mamin, H. J. and Kim, M. and Sherwood, M. H. and
             Rettner, C. T. and Ohno, K. and Awschalom, D. D. and Rugar, D.},
  journal = {Science},
  volume  = {339},
  number  = {6119},
  pages   = {557--560},
  year    = {2013},
  doi     = {10.1126/science.1231540}
}

@article{vanacoleyen2013entanglement,
  title   = {Entanglement Rates and Area Laws},
  author  = {Van Acoleyen, Karel and Mari{\"e}n, Micha{\"e}l
             and Verstraete, Frank},
  journal = {Physical Review Letters},
  volume  = {111},
  pages   = {170501},
  year    = {2013},
  doi     = {10.1103/PhysRevLett.111.170501}
}

@article{rosskopf2017quantum,
  title   = {A quantum spectrum analyzer enhanced by a nuclear spin memory},
  author  = {Rosskopf, Tobias and Zopes, Jonathan and Boss, Jens M. and Degen, Christian L.},
  journal = {npj Quantum Information},
  volume  = {3},
  pages   = {33},
  year    = {2017},
  doi     = {10.1038/s41534-017-0030-6}
}

@article{staudacher2013nuclear,
  title   = {Nuclear Magnetic Resonance Spectroscopy on a (5-Nanometer)$^3$ Sample Volume},
  author  = {Staudacher, T. and Shi, F. and Pezzagna, S. and Meijer, J. and Du, J. and Meriles, C. A. and Reinhard, F. and Wrachtrup, J.},
  journal = {Science},
  volume  = {339},
  number  = {6119},
  pages   = {561--563},
  year    = {2013},
  doi     = {10.1126/science.1231675}
}

@article{Binas2021,
  author = {Binas, Oliver and de Jesus, Vanessa and Landgraf, Tom and
            V{\"o}lklein, Albrecht Eduard and Martins, Jason and Hymon, Daniel and
            Bains, Jasleen Kaur and Berg, Hannes and Biedenb{\"a}nder, Thomas and
            F{\"u}rtig, Boris and Gande, Santosh Lakshmi and Niesteruk, Anna and
            Oxenfarth, Andreas and Qureshi, Nusrat Shahin and Schamber, Tatjana and
            Schnieders, Robbin and Tr{\"o}ster, Alix and Wacker, Anna and
            Wirmer-Bartoschek, Julia and Wirtz Martin, Maria Alexandra and
            Stirnal, Elke and Azzaoui, Kamal and Richter, Christian and
            Sreeramulu, Sridhar and Blommers, Marcel Jules Jos{\'e} and
            Schwalbe, Harald},
  title = {{$^{19}$F} {NMR}-Based Fragment Screening for 14 Different
           Biologically Active {RNAs} and 10 {DNA} and Protein Counter-Screens},
  journal = {ChemBioChem},
  year = {2021},
  volume = {22},
  number = {2},
  pages = {423--433},
  doi = {10.1002/cbic.202000476}
}

@article{Schmitt2017,
  author = {Schmitt, Simon and Gefen, Tuvia and St{\"u}rner, Felix M. and
            Unden, Thomas and Wolff, Gerhard and M{\"u}ller, Christoph and
            Scheuer, Jochen and Naydenov, Boris and Markham, Matthew and
            Pezzagna, Sebastien and Meijer, Jan and Schwarz, Ilai and
            Plenio, Martin B. and Retzker, Alex and McGuinness, Liam P. and
            Jelezko, Fedor},
  title = {Submillihertz magnetic spectroscopy performed with a nanoscale quantum sensor},
  journal = {Science},
  year = {2017},
  volume = {356},
  number = {6340},
  pages = {832--837},
  doi = {10.1126/science.aam5532}
}

@article{Ariyasingha2019,
  author = {Ariyasingha, Nuwandi M. and Lindale, Jacob R. and
            Eriksson, Shannon L. and Clark, Grayson P. and Theis, Thomas and
            Shchepin, Roman V. and Chukanov, Nikita V. and Kovtunov, Kirill V. and
            Koptyug, Igor V. and Warren, Warren S. and Chekmenev, Eduard Y.},
  title = {Quasi-Resonance Fluorine-19 Signal Amplification by Reversible Exchange},
  journal = {The Journal of Physical Chemistry Letters},
  year = {2019},
  volume = {10},
  number = {15},
  pages = {4229--4236},
  doi = {10.1021/acs.jpclett.9b01505}
}

@article{BarGill2013,
  author = {Bar-Gill, Nir and Pham, Linh M. and Jarmola, Andrejs and
            Budker, Dmitry and Walsworth, Ronald L.},
  title = {Solid-state electronic spin coherence time approaching one second},
  journal = {Nature Communications},
  year = {2013},
  volume = {4},
  pages = {1743},
  doi = {10.1038/ncomms2771}
}

@book{tse2005fundamentals,
  author = {Tse, David and Viswanath, Pramod},
  title = {Fundamentals of Wireless Communication},
  publisher = {Cambridge University Press},
  year = {2005},
  doi = {10.1017/CBO9780511807213}
}

@article{
Han2025,
author = {Shuo Han  and Xiangyu Ye  and Xu Zhou  and Zhaoxin Liu  and Yuhang Guo  and Mengqi Wang  and Wentao Ji  and Ya Wang  and Jiangfeng Du },
title = {Solid-state spin coherence time approaching the physical limit},
journal = {Science Advances},
volume = {11},
number = {9},
pages = {eadr9298},
year = {2025},
doi = {10.1126/sciadv.adr9298},
URL = {https://www.science.org/doi/abs/10.1126/sciadv.adr9298}}

@article{Lange2011,
  title = {Single-Spin Magnetometry with Multipulse Sensing Sequences},
  author = {de Lange, G. and Rist{\`e}, D. and Dobrovitski, V. V. and Hanson, R.},
  journal = {Phys. Rev. Lett.},
  volume = {106},
  issue = {8},
  pages = {080802},
  numpages = {4},
  year = {2011},
  month = {Feb},
  publisher = {American Physical Society},
  doi = {10.1103/PhysRevLett.106.080802},
  url = {https://link.aps.org/doi/10.1103/PhysRevLett.106.080802}
}

@article{Zopes2017,
  title = {High-Resolution Quantum Sensing with Shaped Control Pulses},
  author = {Zopes, J. and Sasaki, K. and Cujia, K. S. and Boss, J. M. and Chang, K. and Segawa, T. F. and Itoh, K. M. and Degen, C. L.},
  journal = {Phys. Rev. Lett.},
  volume = {119},
  issue = {26},
  pages = {260501},
  numpages = {5},
  year = {2017},
  month = {Dec},
  publisher = {American Physical Society},
  doi = {10.1103/PhysRevLett.119.260501},
  url = {https://link.aps.org/doi/10.1103/PhysRevLett.119.260501}
}

@misc{Nomura2023,
      title={Imaging of high-frequency electromagnetic field by multipulse sensing using nitrogen vacancy centers in diamond},
      author={Shintaro Nomura and Hideyuki Watanabe and Satoshi Kashiwaya},
      year={2023},
      eprint={2307.02089},
      archivePrefix={arXiv},
      primaryClass={quant-ph},
      url={https://arxiv.org/abs/2307.02089},
}

@book{Walter2004,
    author = {Gautschi, Walter},
    title = {Orthogonal Polynomials: Computation and Approximation},
    publisher = {Oxford University Press},
    year = {2004},
    month = {04},
    isbn = {9780198506720},
    doi = {10.1093/oso/9780198506720.001.0001},
    url = {https://doi.org/10.1093/oso/9780198506720.001.0001},
}

@article{Christandl2004,
  title = {Perfect State Transfer in Quantum Spin Networks},
  author = {Christandl, Matthias and Datta, Nilanjana and Ekert, Artur and Landahl, Andrew J.},
  journal = {Phys. Rev. Lett.},
  volume = {92},
  issue = {18},
  pages = {187902},
  numpages = {4},
  year = {2004},
  month = {May},
  publisher = {American Physical Society},
  doi = {10.1103/PhysRevLett.92.187902},
  url = {https://link.aps.org/doi/10.1103/PhysRevLett.92.187902}
}

@article{Weedbrook2012,
  title = {Gaussian quantum information},
  author = {Weedbrook, Christian and Pirandola, Stefano and Garc{\'{\i}}a-Patr{\'o}n, Ra{\'u}l and Cerf, Nicolas J. and Ralph, Timothy C. and Shapiro, Jeffrey H. and Lloyd, Seth},
  journal = {Rev. Mod. Phys.},
  volume = {84},
  issue = {2},
  pages = {621--669},
  numpages = {0},
  year = {2012},
  month = {May},
  publisher = {American Physical Society},
  doi = {10.1103/RevModPhys.84.621},
  url = {https://link.aps.org/doi/10.1103/RevModPhys.84.621}
}

@book{Golub2010,
url = {https://doi.org/10.1515/9781400833887},
title = {Matrices, Moments and Quadrature with Applications},
author = {Gene H. Golub and Gérard Meurant},
publisher = {Princeton University Press},
address = {Princeton},
doi = {doi:10.1515/9781400833887},
isbn = {9781400833887},
year = {2010},
lastchecked = {2026-10-21}
}

@article{Boley1987,
doi = {10.1088/0266-5611/3/4/010},
url = {https://doi.org/10.1088/0266-5611/3/4/010},
year = {1987},
month = {nov},
publisher = {},
volume = {3},
number = {4},
pages = {595},
author = {D Boley and G H Golub},
title = {A survey of matrix inverse eigenvalue problems},
journal = {Inverse Problems},
}

@article{Muhonen2014,
  title={Storing quantum information for 30 seconds in a nanoelectronic device},
  author={Muhonen, Juha T and Dehollain, Juan P and Laucht, Arne and Hudson, Fay E and Kalra, Rachpon and Sekiguchi, Takeharu and Itoh, Kohei M and Jamieson, David N and McCallum, Jeffrey C and Dzurak, Andrew S and others},
  journal={Nature Nanotechnology},
  volume={9},
  number={12},
  pages={986--991},
  year={2014},
  publisher={Nature Publishing Group UK, London},
  doi = {https://doi.org/10.1038/nnano.2014.211}
}

@article{Muhonen2015,
doi = {10.1088/0953-8984/27/15/154205},
url = {https://doi.org/10.1088/0953-8984/27/15/154205},
year = {2015},
month = {mar},
publisher = {IOP Publishing},
volume = {27},
number = {15},
pages = {154205},
author = {Muhonen, J T and Laucht, A and Simmons, S and Dehollain, J P and Kalra, R and Hudson, F E and Freer, S and Itoh, K M and Jamieson, D N and McCallum, J C and Dzurak, A S and Morello, A},
title = {Quantifying the quantum gate fidelity of single-atom spin qubits in silicon by randomized benchmarking},
journal = {Journal of Physics: Condensed Matter},
}

@article{Ruster2016,
  title={A long-lived Zeeman trapped-ion qubit},
  author={Ruster, Thomas and Schmiegelow, Christian T and Kaufmann, Henning and Warschburger, Claudia and Schmidt-Kaler, Ferdinand and Poschinger, Ulrich G},
  journal={Applied Physics B},
  volume={122},
  number={10},
  pages={254},
  year={2016},
  publisher={Springer},
  doi = {https://doi.org/10.1007/s00340-016-6527-4}
}

@article{Leu2023,
  title = {Fast, High-Fidelity Addressed Single-Qubit Gates Using Efficient Composite Pulse Sequences},
  author = {Leu, A. D. and Gely, M. F. and Weber, M. A. and Smith, M. C. and Nadlinger, D. P. and Lucas, D. M.},
  journal = {Phys. Rev. Lett.},
  volume = {131},
  issue = {12},
  pages = {120601},
  numpages = {6},
  year = {2023},
  month = {Sep},
  publisher = {American Physical Society},
  doi = {10.1103/PhysRevLett.131.120601},
  url = {https://link.aps.org/doi/10.1103/PhysRevLett.131.120601}
}

@article{Wang2022,
  title = {Sensing of Arbitrary-Frequency Fields Using a Quantum Mixer},
  author = {Wang, Guoqing and Liu, Yi-Xiang and Schloss, Jennifer M. and Alsid, Scott T. and Braje, Danielle A. and Cappellaro, Paola},
  journal = {Phys. Rev. X},
  volume = {12},
  issue = {2},
  pages = {021061},
  numpages = {22},
  year = {2022},
  month = {Jun},
  publisher = {American Physical Society},
  doi = {10.1103/PhysRevX.12.021061},
  url = {https://link.aps.org/doi/10.1103/PhysRevX.12.021061}
}

@article{mokeev2026enhancing,
  title = {Enhancing Optical Imaging via Quantum Computation},
  author = {Mokeev, Aleksandr and Saif, Babak and Lukin, Mikhail D. and Borregaard, Johannes},
  journal = {PRX Quantum},
  volume = {7},
  issue = {1},
  pages = {010318},
  numpages = {18},
  year = {2026},
  month = {Jan},
  publisher = {American Physical Society},
  doi = {10.1103/s94k-929p},
  url = {https://link.aps.org/doi/10.1103/s94k-929p}
}

@misc{mele2026exponentialseparationsensingcontinuous,
      title={Exponential separation in sensing continuous signals via squeezing}, 
      author={Francesco Anna Mele and Nadine Meister and Haimeng Zhao and Senrui Chen and Hsin-Yuan Huang},
      year={2026},
      eprint={2609.38678},
      archivePrefix={arXiv},
      primaryClass={quant-ph},
      url={https://arxiv.org/abs/2609.38678}, 
}

@misc{conlon2026superexponentialadvantagesqueezedlight,
      title={Super-Exponential Advantage of Squeezed Light in Phase Estimation under Discrete Phase Randomisation}, 
      author={Lorcan O. Conlon and V Vijendran and Aritra Das and Yu-Xin Wang and Erfan Abbasgholinejad and Anthony J. Brady and Jacob Bringewatt and Sean R. Muleady and Syed M. Assad and Alexey V. Gorshkov},
      year={2026},
      eprint={2610.03657},
      archivePrefix={arXiv},
      primaryClass={quant-ph},
      url={https://arxiv.org/abs/2610.03657}, 
}

@article{vorobyov2021quantum,
  title={Quantum Fourier transform for nanoscale quantum sensing},
  author={Vorobyov, Vadim and Zaiser, Sebastian and Abt, Nikolas and Meinel, Jonas and Dasari, Durga and Neumann, Philipp and Wrachtrup, J{\"o}rg},
  journal={npj Quantum Information},
  volume={7},
  number={1},
  pages={124},
  year={2021},
  publisher={Nature Publishing Group UK, London}
}
\clearpage
\onecolumngrid
\appendix
\LinesNumbered

\section*{Appendices}

\addcontentsline{toc}{section}{Appendix}
\setcounter{secnumdepth}{3}

\appendixtableofcontents
\startappendixtoc

\section{Preliminaries}\label{app:setting}

We begin by describing the model of quantum sensing algorithms used in this work and by formally defining the five sensing problems P1--P5 considered in the main text. The problems differ both in what information is initially known about the signal and what information we must determine about the signal, leading to the hierarchy of Detection, Search, and Learning protocols analyzed in subsequent appendices.

\subsection{Quantum sensing algorithms for spectroscopy}\label{sec:sensing-model}

We consider time-dependent signals $f(t)$ which couple to a single sensor qubit via $H(t)=\hbar\gamma f(t)Z$, where $Z$ denotes the Pauli-$Z$ operator and $\gamma$ is the coupling strength. We work in units such that $\hbar\gamma=1$. For notational convenience, we write $v_i\equiv Bu_i$ for the signal coefficients denoted by $Bu_i$ in the main text, as these coefficients appear frequently in the following analysis. The signals we consider are therefore of the form $f(t)=f_{\vec v}(t)$, where
\begin{align}
 f_{\vec v}(t)=\sum_{n=1}^N\bigl(v_n\cos(\omega_nt)+v_{n+N}\sin(\omega_nt)\bigr).
 \label{eq:appendix-signal}
\end{align}
Here $N\ge1$ is the number of distinct frequencies, $\vec v\in\R^{2N}$ is unknown but fixed throughout the experiment, and the known frequencies satisfy
\begin{align}
 \omega_1\ge\Delta>0,\qquad \omega_{n+1}-\omega_n\ge\Delta\quad(1\le n<N).
 \label{eq:frequency-separation}
\end{align}
We write $H_{\vec{v}}(t) = f_{\vec{v}}(t) Z$. It is convenient to collect the known modes in the vector
\begin{align}
 \phi(t)=\bigl(\cos(\omega_1t),\ldots,\cos(\omega_Nt),\sin(\omega_1t),\ldots,\sin(\omega_Nt)\bigr)^\top,
 \qquad f_{\vec v}(t)=\vec v\cdot\phi(t).
 \label{eq:fourier-vector}
\end{align}
The signal is ``always on,'' i.e., it is not possible to restart time evolution under $H_{\vec{v}}(t)$. We allow noiseless instantaneous control unitaries, as well as arbitrary state preparation and measurement. A protocol may also use ancillary qubits that do not couple directly to the signal, and control unitaries may act jointly on the sensor and ancillas. Concretely, a protocol using $n_Q$ ancilla qubits can be specified by a sequence of times $0=t_0<\cdots<t_P=T$ and control unitaries $V_0,\ldots,V_P$ acting on all $n_Q+1$ qubits. The final state is
\begin{align}
 V_P\left(\prod_{p=1}^{P-1}\left(\exp\left(-i\int_{t_p}^{t_{p+1}} H_{\vec{v}}(t)\,\mathrm dt\right)\otimes\id_Q\right)V_p\right)\ket{0}^{\otimes(n_Q+1)},
 \label{eq:sensing-protocol-state}
\end{align}
where $\id_Q$ denotes the identity matrix on all $Q$ qubits. Finally, a POVM at time $T$ records the classical output of the protocol. We note that this framework includes protocols which perform intermediate measurements and adaptively apply different control unitaries conditional on the outcomes, by increasing $n_Q$ and applying the principle of deferred measurement. Since $P$ may be arbitrarily large, with $t_p$ spaced infinitesimally apart, this model is equivalent to an analog, continuous control model, in which the Hamiltonian is modified to $H_{\mathrm{total}}(t) = H_{\vec{v}}(t) + H_c(t)$, for $H_c(t)$ an arbitrary control Hamiltonian acting on all $n_Q + 1$ qubits. In this setting, the choice of $H_c(t)$ specifies a quantum sensing algorithm.

\subsection{Sensing problems}\label{sec:sensing-problems}

We now formally define the five spectroscopy problems (P1--P5) considered in this work. For each, we let $\eps>0$ denote the precision parameter. We also let $\delta\in(0,1/2)$ denote the allowed failure probability.

The first three problems, P1--P3, are all \emph{detection} problems, where we must decide whether a signal exists or not, given some prior knowledge about the potential signal. The problems are distinguished by the strength of this prior knowledge. For P1, we know the \emph{waveform} of the signal, i.e., the unit vector $\vec u=\vec v/\pnorm{2}{\vec v}$. For P2, we only know the \emph{spectral-weight pattern} of the signal, i.e., $\{u_n^2+ u_{N+n}^2\}_{n=1}^N$. For P3, we have no prior information about the signal. We formally define P1--P3 as follows:

\begin{problem}[Detection with known waveform (P1)]\label{problem:waveform-detection}
Let $\vec u\in\R^{2N}$ be a known vector with $\| \vec u \|_2=1$. Given that $\vec v=\| \vec v \|_2 \vec u$, decide whether $\vec v=0$ or $\| \vec v \|_2\ge\eps$ with probability at least $1-\delta$.
\end{problem}

\begin{problem}[Detection with known spectral-weight pattern (P2)]\label{problem:spectral-detection}
Let $\vec p\in\R^N_{\ge 0}$ be a known vector with $\pnorm{1}{\vec p}=1$. Given that $v_n^2+v_{n+N}^2=\| \vec v \|_2^2\cdot p_n$, decide whether $\vec v=0$ or $\| \vec v \|_2\ge\eps$ with probability at least $1-\delta$.
\end{problem}

\begin{problem}[Detection for unknown signal (P3)]\label{problem:signal-detection}
Decide whether $\vec v=0$ or $\| \vec v \|_2\ge\eps$ with probability at least $1-\delta$.
\end{problem}

P1--P3 are ordered by how much prior knowledge we have about the signal; we have the most information in P1 and the least in P3. As such, any protocol for P3 also solves P1 and P2, simply by ignoring the prior information. It therefore suffices to give a detection algorithm for P3, which we do in Appendix~\ref{app:random-modulation}.

The next problem, P4, is a qualitatively different \emph{Search} problem, where the goal is to determine which frequency components contribute to the unknown signal. We assume the signal is sparse, i.e., there are only a constant number of contributing frequencies (though our protocol in Appendix~\ref{app:signal-search} works for non-sparse signals, as well). Let $\cK=\{n\in[N]:(v_n^2+v_{n+N}^2)^{1/2}>0\}$ denote the set of contributing frequencies. We formally define P4 below:

\begin{problem}[Search, P4]\label{problem:signal-search}
Assume $|\cK|\le K =  O(1)$ and assume $(v_n^2+v_{n+N}^2)^{1/2}\ge\eps$ for every $n\in\cK$. Determine $\cK$ with probability at least $1-\delta$.
\end{problem}

In Appendix~\ref{app:signal-search}, we show how to solve Search by recursively performing Detection on subsets of candidate frequencies. We adapt the modulation function used for P3 so that each subset test is independent of the signal outside that subset.

Finally, P5 is the most difficult of the problems and requires us to learn a complete description of the signal. We assume the signal amplitudes are bounded by a constant to constrain the dynamic range (though our protocol in Appendix~\ref{app:signal-profile-estimation} works for any bound $v_{\max}$ such that $\| \vec v \|_\infty \le v_{\max}$, incurring only sublogarithmic overhead). Formally, we have:

\begin{problem}[Learning, P5]\label{problem:signal_learning}
Assume $\| \vec v \|_\infty \le v_{\max} = O(1)$. Construct an estimate $\wt{\vec v}$ of $\vec v$ such that $\| \wt{\vec v}-\vec v \|_\infty\le\eps$ with probability at least $1-\delta$.
\end{problem}

In Appendix~\ref{app:signal-profile-estimation} we give a quantum-computing-enhanced sensing protocol for Learning. In addition, in Appendix~\ref{app:periodic-signal-learning}, we give a simpler Learning protocol which uses ancilla qubits but never generates entanglement, so long as the signal is periodic. Further, in Appendix~\ref{sec:boson}, we describe an alternative, analog approach to Learning when the signal is coupled to a bosonic sensor.

\section{Detection algorithm}\label{app:random-modulation}

We now solve the most general Detection problem, Problem~\ref{problem:signal-detection}: distinguish $\vec v=0$ from $\pnorm{2}{\vec v}\ge\eps$. The same protocol solves Problems~\ref{problem:waveform-detection} and~\ref{problem:spectral-detection} by ignoring the additional prior information.

Our main result is an ancilla-free algorithm for Detection which achieves the near-optimal total evolution time:

\begin{theorem}[Detection]\label{thm:signal-detection}
There is an ancilla-free quantum algorithm which solves Problem~\ref{problem:signal-detection} using total sensing time
\begin{align}
    T=O\!\left(
      \left(\frac{\sqrt{N\log(N)}}{\eps}
      +\frac{\log(N)}{\Delta}\right)
      \log\frac1\delta
    \right).\label{eq:detection-time}
\end{align}
\end{theorem}
\noindent
\emph{Proof Sketch.}
We construct a random modulation function $g(t)$ by taking a linear combination of all Fourier modes with independent Gaussian coefficients, then clipping the resulting function to the allowed range $[-1,1]$. Over a sufficiently long interval, different Fourier modes are nearly orthogonal, while each mode has mean square close to $1/2$.
The phase before clipping is therefore Gaussian, with a standard deviation controlled by the signal norm. We choose the interrogation time so that a signal of norm $\eps$ already gives a phase of constant size. The clipping error is small compared with this standard deviation, even for arbitrarily large amplitudes. This ensures a constant probability of detecting a present signal, which we can amplify to $1-\delta$ by repetition.

We now describe the algorithm in detail. We use the Fourier vector $\phi(t)$ defined in Eq.~\eqref{eq:fourier-vector}. Draw $\zeta\sim\cN(0,\id_{2N})$ and define
\begin{align}
    Y(t)=\frac{\mu}{\sqrt N}\zeta^\top\phi(t),\qquad
    g(t)=\operatorname{Clip}(Y(t)),\qquad
    \operatorname{Clip}(x)=\max\{-1,\min\{x,1\}\}.
    \label{eq:random-modulation}
\end{align}
For each fixed $t$, $Y(t)$ has variance $\mu^2$, since $\pnorm{2}{\phi(t)}^2=N$.
Starting from $\ket{+}$, evolving under the unknown signal with modulation function $g(t)$ over time interval $I$ produces the state $\exp(-i Z \int_I g(t) f(t) \mathrm{d}t) \ket{+}$. An $X$ basis measurement of this state has mean $\cos(2\int_I g(t) f(t) \mathrm{d}t)$. Write $h_N=\sum_{n=1}^N n^{-1}$ for the $N$-th harmonic number.

Our Detection algorithm is given in Algorithm~\ref{alg:signal-detection}:

\begin{algorithm}[H]
\caption{Detection}\label{alg:signal-detection}
\KwIn{Access to $H_{\vec v}(t)$; known frequencies and separation bound $\Delta$; $\eps>0$ and $\delta\in(0,1/2)$.}
Set $L\gets\ceil{\ln(128N)}$, $\mu\gets1/(2\sqrt L)$,
$\tau\gets\max\{16\sqrt N/(\eps\mu),\,28h_N/\Delta\}$,
$r\gets\ceil{16\ln(1/\delta)}$\;
\For{$a=1,\ldots,r$}{
    Draw fresh $\zeta\sim\cN(0,\id_{2N})$ and independently $\lambda\sim\operatorname{Unif}[0,1]$\;
    Prepare $\ket{+}$ and interrogate for time $\tau$ with modulation function
    $\lambda\operatorname{Clip}(\mu\zeta^\top\phi(t)/\sqrt N)$\;
    Measure in the $X$ basis\;
    \If{the outcome is $-1$}{\KwRet{signal present}\;}
}
\KwRet{signal absent}\;
\end{algorithm}

We first prove two estimates used in the analysis. Both hold uniformly over the starting time of an interrogation, so the signal need not be restarted. They will also be used for signal learning.

\begin{lemma}[Finite-time Gram matrix]\label{lem:gram}
For an interval $I$ of length $\tau$, let $G_I=\int_I\phi(t)\phi(t)^\top\,\mathrm dt$ denote the Gram matrix of the Fourier modes. Then
\begin{align}
    \pnorm{\infty\to\infty}{G_I-\tfrac\tau2\id_{2N}}
    \le\frac{7h_N}{\Delta},\label{eq:gram-bounds}
\end{align}
where $\pnorm{\infty\to\infty}{A}=\max_i\sum_j|A_{ij}|$ is the maximum absolute row sum. If $\tau\ge28h_N/\Delta$, then, for every $\vec v\in\R^{2N}$,
\begin{align}
    \pnorm{2}{G_I\vec v}\ge\frac\tau4\pnorm{2}{\vec v},\qquad
    \int_I f_{\vec v}(t)^2\,\mathrm dt
    \le\frac{3\tau}{4}\pnorm{2}{\vec v}^2.
    \label{eq:gram-quadratic}
\end{align}
\end{lemma}
\begin{proof}
Fix the row for the cosine or sine at frequency $\omega_n$. For $n'\ne n$, product-to-sum identities and $|\int_I e^{i\nu t}\,\mathrm dt|\le2/|\nu|$ bound each of the two entries at frequency $\omega_{n'}$ by
\begin{align}
    \frac1{|\omega_n-\omega_{n'}|}+\frac1{\omega_n+\omega_{n'}}.
    \label{eq:gram-cross-term}
\end{align}
After subtracting $\tau/2$ from the diagonal, the two entries at frequency $\omega_n$ have total absolute value at most $1/\omega_n$. Since $\omega_n\ge n\Delta$ and $|\omega_n-\omega_{n'}|\ge|n-n'|\Delta$, the row sum is at most
\begin{align}
    \frac1{\omega_n}
    +2\sum_{n'\ne n}\left(\frac1{|\omega_n-\omega_{n'}|}+\frac1{\omega_n+\omega_{n'}}\right)
    \le\frac{1+6h_N}{\Delta}\le\frac{7h_N}{\Delta}.
    \label{eq:gram-row-sum}
\end{align}
This proves Eq.~\eqref{eq:gram-bounds}. Every eigenvalue of a matrix has absolute value at most its maximum absolute row sum, as follows by applying the eigenvalue equation at a largest coordinate of an eigenvector. Applying this fact to the real symmetric matrix $G_I-\tau\id_{2N}/2$ shows that the eigenvalues of $G_I$ lie in $[\tau/2-7h_N/\Delta,\tau/2+7h_N/\Delta]$. If $\tau\ge28h_N/\Delta$, this interval is contained in $[\tau/4,3\tau/4]$. Diagonalizing $G_I$ and using $\int_I f_{\vec v}^2(t)~\mathrm{d} t=\vec v^\top G_I\vec v$ proves Eq.~\eqref{eq:gram-quadratic}.
\end{proof}

Next, we bound the effect of clipping. The key idea is to bound the error in the overall phase, rather than the probability that the function $Y(t)$ is clipped at any given time.

\par\addvspace{\topsep}
\noindent\begin{minipage}{\linewidth}
\begin{lemma}[Clipping]\label{lem:clipping}
Let $Y(t)$ be a jointly measurable real random function on an interval $I$ of length $\tau$, and suppose
\begin{align}
    \bbP[|Y(t)|\ge x]\le2e^{-x^2/(2\mu^2)}
    \qquad(t\in I,\ x\ge0)
    \label{eq:clipping-tail}
\end{align}
for some $\mu>0$. For deterministic real $f\in L^2(I)$, set
\begin{align}
    \Phi=\int_IY(t)f(t)\,\mathrm dt,\qquad
    \wt\Phi=\int_I\operatorname{Clip}(Y(t))f(t)\,\mathrm dt.
    \label{eq:clipping-phases}
\end{align}
Then
\begin{align}
    \E[(\wt\Phi-\Phi)^2]
    \le4\tau\pnorm{L^2(I)}{f}^2\mu^4e^{-1/(2\mu^2)}.
    \label{eq:clipping-second-moment}
\end{align}
\end{lemma}
\end{minipage}\par\addvspace{\topsep}
Lemma~\ref{lem:clipping}, together with $|\cos(2\Phi) - \mathrm{cos}(2\wt\Phi)| \le 2 | \Phi - \wt\Phi|$ and Cauchy-Schwarz, imply that the expected values of an $X$-measurement on the states $\exp(-i \Phi Z) \ket{+}$ and $\mathrm{exp}(-i \wt \Phi Z) \ket{+}$ differ by at most $2\big(\E[(\wt\Phi-\Phi)^2]\big)^{1/2}$.

\begin{proof}
The tail bound implies $\E\int_IY(t)^2\,\mathrm dt<\infty$, so both phases are defined almost surely. Integrating the same bound gives
\begin{align}
    \E\big[(|Y(t)|-1)_+^2\big]
    =2\int_0^\infty u\,\bbP[|Y(t)|>1+u]\,\mathrm du \le4e^{-1/(2\mu^2)}\int_0^\infty u e^{-u/\mu^2}\,\mathrm du
    =4\mu^4e^{-1/(2\mu^2)},
    \label{eq:clipping-pointwise-error}
\end{align}
where $x_+=\max\{x,0\}$. Cauchy--Schwarz in time and Tonelli's theorem imply
\begin{align}
    \E[(\wt\Phi-\Phi)^2]
    \le\pnorm{L^2(I)}{f}^2
       \int_I\E\big[(|Y(t)|-1)_+^2\big]\,\mathrm dt,
    \label{eq:clipping-integrated-error}
\end{align}
which proves Eq.~\eqref{eq:clipping-second-moment}.
\end{proof}

We can now apply these results to prove Theorem~\ref{thm:signal-detection}.

\begin{proof}[Proof of Theorem~\ref{thm:signal-detection}]
Use Algorithm~\ref{alg:signal-detection}. On an arbitrary interrogation interval $I$, the phase before clipping is Gaussian, with variance
\begin{align}
    \sigma^2=\frac{\mu^2}{N}\pnorm{2}{G_I\vec v}^2.
    \label{eq:gaussian-phase-variance}
\end{align}
The chosen duration satisfies
\begin{align}
    \frac{7h_N}{\Delta}\le\frac\tau4,
    \qquad \tau\ge\frac{16\sqrt N}{\eps\mu}.
    \label{eq:detection-gram-duration}
\end{align}
Lemma~\ref{lem:gram} therefore gives $\sigma\ge\mu\tau\pnorm{2}{\vec v}/(4\sqrt N)\ge4$ if the signal is present. For each fixed $t$, the Gaussian variable $Y(t)$ has moment-generating function $\E [e^{sY(t)}]=e^{s^2\mu^2/2}$. Markov's inequality implies the tail bound in Lemma~\ref{lem:clipping}. The two lemmas give
\begin{align}
    \E[(\wt\Phi-\Phi)^2]
    \le3\tau^2\pnorm{2}{\vec v}^2\mu^4e^{-1/(2\mu^2)} \le48N\mu^2e^{-1/(2\mu^2)}\sigma^2
     \le\frac{\sigma^2}{64}.
    \label{eq:detection-relative-phase-error}
\end{align}
The last inequality uses $\mu\le1$ and $e^{-1/(2\mu^2)}=e^{-2L}\le(128N)^{-2}$.

A standard normal random variable has probability greater than $1/2$ of having absolute value at least $1/2$: its density is at most $1/\sqrt{2\pi}<1/2$, so the complementary interval has probability less than $1/2$. Markov's inequality consequently yields
\begin{align}
    \bbP\left[|\wt\Phi|\ge\frac\sigma4\right]
    \ge\bbP\left[|\Phi|\ge\frac\sigma2\right]
       -\bbP\left[|\wt\Phi-\Phi|>\frac\sigma4\right] >\frac12-\frac14=\frac14.
    \label{eq:detection-phase-probability}
\end{align}
In particular, $|\wt\Phi|\ge1$ with probability at least $1/4$. Conditional on $\wt\Phi=x$, averaging over the independent attenuation factor $\lambda$ gives
\begin{align}
    \bbP[X=-1\mid\wt\Phi=x]
    &=\int_0^1\sin^2(\lambda x)\,\mathrm d\lambda
    =\frac12-\frac{\sin(2x)}{4x}\ge\frac14
    \qquad(|x|\ge1).
    \label{eq:detection-attenuation}
\end{align}
Thus each round detects a present signal with probability at least $1/16$, uniformly over its starting time and the signal amplitude. If the signal is not present, every round returns $+1$. Fresh randomness and reinitialization imply that the probability of no detection after $r$ rounds is at most $(15/16)^r\le e^{-r/16}\le\delta$. Finally, $\mu^{-1}=O(\sqrt{\log(N)})$ and $h_N=O(\log(N))$, so $r\tau$ satisfies Eq.~\eqref{eq:detection-time}.
\end{proof}

\section{Search algorithm}\label{app:signal-search}

We now solve Search, Problem~\ref{problem:signal-search}. Recall that $K\in[N]$ denotes a known sparsity bound such that $|\cK|\le K$, where $\cK$ denotes the set of active frequencies. Though the problem is formulated for $K = O(1)$, our algorithm works for arbitrary $K$, and we keep the dependence explicit in the analysis.

The algorithm successively divides the candidate frequencies into two
sets and tests which sets contain signal. Applying the modulation function used for Detection directly can produce a response from
frequencies outside that subset; if such a frequency has large amplitude, this can lead to an error. Therefore, we will use a new modulation function construction, so that all outside frequencies contribute exactly zero phase.

Our main result is an ancilla-free algorithm for Search achieving the near-optimal total evolution time:

\begin{theorem}[Search]\label{thm:signal-search} There is an ancilla-free quantum algorithm which solves Problem~\ref{problem:signal-search} using total sensing time
\begin{align}
 T=O\!\left(
 \left(\frac{\sqrt{N\log(N)}}{\eps}
       +\frac{\log^2(N)}{\Delta}\right)
 \log\!\left(\frac{\log(N)}{\delta}\right)
 \right).\label{eq:search-time}
\end{align}
\end{theorem}
\noindent
\emph{Proof sketch.}
We modify the random modulation function used for Detection using the known Gram matrix of the Fourier modes, so that we can test whether a subset contains any active frequencies without interference from outside modes. We then use this subset test to binary search over the possible frequencies in $\cK$.

We first specify the subset test. Set
\begin{align}
 L=\ceil{\ln(128N)},\qquad \mu=\frac1{4\sqrt L}, \qquad J=\ceil{\log_2N},\qquad Q=1+2KJ,\qquad
 r=\ceil{16\ln(Q/\delta)}.
 \label{eq:search-parameters}
\end{align}
Here, $Q$ bounds the total number of subset tests. In particular, $Q=1$
when $N=1$.

For a nonempty tested set $S$ of size $m$, each trial uses an interval $I$ of length
\begin{align}
 \tau_m=\max\left\{\frac{8\sqrt m}{\mu\eps},\frac{28h_N}{\Delta}\right\},
 \label{eq:search-trial-duration}
\end{align}
where $h_N$ again denotes the $N$-th harmonic number. We again define the Gram matrix $G_I$ of the Fourier modes over $I$, as well as a normalized version:
\begin{align}
    G_I=\int_I\phi(t)\phi(t)^\top\,\mathrm dt,
 \qquad A_I=\frac{2G_I}{\tau_m}.
\end{align}
Draw a vector $\zeta\in\R^{2N}$ whose
coordinates $\zeta_n,\zeta_{n+N}$ are independent standard Gaussians
for $n\in S$, and are zero for $n\notin S$. Define
\begin{align}
 Y(t)=\frac{\mu}{\sqrt m}\zeta^\top A_I^{-1}\phi(t),\qquad
 b(t)=\operatorname{Clip}(Y(t)),
 \label{eq:search-random-modulation}\\
 \wt Y(t)=b(t)+\phi(t)^\top G_I^{-1}
       \int_I\phi(u)(Y(u)-b(u))\,\mathrm du,
 \label{eq:search-corrected-modulation}\\
 c=1+\frac{8N}{\tau_m}\int_I|Y(t)-b(t)|\,\mathrm dt,
 \qquad g(t)=\frac{\wt Y(t)}c.
 \label{eq:search-bounded-modulation}
\end{align}
The additional term in Eq.~\eqref{eq:search-corrected-modulation} ensures that $\int_I \phi(t) \wt{Y}(t)~\mathrm{d}t = \int_I \phi(t) Y(t)~\mathrm{d}t$ in spite of the clipping. The denominator
in Eq.~\eqref{eq:search-bounded-modulation} ensures $|g(t)|\le1$.

We now present our subset testing subroutine in Algorithm~\ref{alg:search-subset} and our full Search algorithm in Algorithm~\ref{alg:signal-search}:

\begin{algorithm}[H]
\caption{Subset test}\label{alg:search-subset}
\KwIn{Nonempty $S\subseteq[N]$; $\eps>0$; parameters
in Eq.~\eqref{eq:search-parameters}.}
\For{$a=1,\ldots,r$}{
 Let $I$ be the next interval of length $\tau_{|S|}$ from Eq.~\eqref{eq:search-trial-duration}\;
 Draw fresh Gaussian $\zeta$ as specified above, and construct $g(t)$ by Eqs.~\eqref{eq:search-random-modulation}--\eqref{eq:search-bounded-modulation}\;
 Independently draw $\lambda\sim\operatorname{Unif}[0,1]$\;
 Prepare $\ket{+}$, interrogate over $I$ with modulation function $\lambda g(t)$, and measure $X$\;
 \If{the outcome is $-1$}{\KwRet{signal present}\;}
}
\KwRet{signal absent}\;
\end{algorithm}

\begin{algorithm}[H]
\caption{Search}\label{alg:signal-search}
\KwIn{Access to $H_{\vec v}(t)$; known frequencies and separation bound $\Delta$; sparsity bound $K \in [N]$; $\eps>0$ and $\delta\in(0,1/2)$.}
Choose parameters as in Eq.~\eqref{eq:search-parameters}, initialize a queue
with $[N]$, and set $\wt{\cK}\gets\varnothing$\;
\While{the queue is nonempty}{
 Remove the next set $S$ and run Algorithm~\ref{alg:search-subset}\;
 \If{signal is reported present}{
  \eIf{$|S|=1$}{Add its element to $\wt{\cK}$\;}{
   Divide $S$ (nearly) in half and append both subsets to the queue\;
  }
 }
}
\KwRet{$\wt{\cK}$}\;
\end{algorithm}

Here, by nearly in half, we mean that $S$ is partitioned into two disjoint subsets whose sizes differ by at most one.
We prove that the subset test never retains an empty set and rarely discards an active one.

\begin{lemma}[Subset test]\label{lem:search-subset}
Algorithm~\ref{alg:search-subset} always reports signal absent if
$S\cap\cK=\varnothing$. If $S\cap\cK\ne\varnothing$, it reports signal absent
with probability at most $\delta/Q$.
\end{lemma}
\begin{proof}
First, we verify that the modulation function is valid and satisfies the desired orthogonality properties. Since
$\tau_m\ge28h_N/\Delta$, Lemma~\ref{lem:gram} gives
$\pnorm{\infty\to\infty}{A_I-\id_{2N}}\le1/2$. The triangle inequality
implies $\pnorm{\infty}{A_Iz}\ge\pnorm{\infty}{z}/2$ for all
$z\in\R^{2N}$, so $A_I$ is invertible and
$\| A_I^{-1} \|_{\infty\to\infty}\le2$.
Because $\pnorm{\infty}{\phi(t)}\le1$,
the correction satisfies
\begin{align}
 |\wt Y(t)-b(t)|
 &\le\pnorm{\infty}{G_I^{-1}\phi(t)}
       \pnorm{1}{\int_I\phi(u)(Y(u)-b(u))\,\mathrm du}  \le\frac{8N}{\tau_m}\int_I|Y(u)-b(u)|\,\mathrm du=c-1.
 \label{eq:search-correction-size}
\end{align}
Thus $|\wt Y(t)|\le1+(c-1)=c$, which proves $|g(t)|\le1$.
Integrating Eq.~\eqref{eq:search-corrected-modulation} and using
$G_IA_I^{-1}=(\tau_m/2)\id_{2N}$ gives
\begin{align}
 \int_I\phi(t)g(t)\,\mathrm dt
 &=\frac{1}{c} \int_I\phi(t)Y(t)\,\mathrm dt
 =\frac{1}{c}\frac{\tau_m\mu}{2\sqrt m}\zeta.
 \label{eq:search-exact-moments}
\end{align}
In particular, every frequency outside $S$ contributes zero phase on every
trial. If $S$ contains no signal, the $X$ outcome is therefore always $+1$.

We next show that $c\le2$ with probability at least $15/16$.
The bound $\| A_I^{-1}\phi(t) \|_{\infty}\le2$ implies that the
Gaussian variable $Y(t)$ has variance at most $8\mu^2=1/(2L)$ at
each fixed $t$. Its moment-generating function and Markov's inequality
give $\bbP[|Y(t)|>u]\le2e^{-Lu^2}$. Hence
\begin{align}
 \E[|Y(t)-b(t)|]
 \le2\int_1^{\infty}e^{-Lu^2}\,\mathrm du
 \le2\int_1^{\infty}u e^{-Lu^2}\,\mathrm du
 =\frac{e^{-L}}L,\label{eq:search-clipping-average}
\end{align}
so $\bbP[c>2] \le\E[c-1]\le 8Ne^{-L}/L
 \le1/16$.

For every Gaussian draw define the target phase
\begin{align}
 \Phi=\frac{\tau_m\mu}{2\sqrt m}\zeta^\top\vec v.
 \label{eq:search-target-phase}
\end{align}
If $S$ contains an active frequency, its coefficient vector has norm
at least $\eps$. Thus $\Phi$ is a centered Gaussian with standard
deviation at least $\tau_m\mu\eps/(2\sqrt m)\ge4$.
A standard normal variable lies in $(-1/2,1/2)$ with probability less
than $1/2$, since its density is at most $1/\sqrt{2\pi}<1/2$.
Consequently $\bbP[|\Phi|\ge2]>1/2$.
The events
$|\Phi|\ge2$ and $c\le2$ hold together with probability at least
$7/16$ by the above estimate. On their intersection, the implemented phase
before attenuation is $\Phi/c$ and has absolute value at least one.
The attenuation calculation Eq.~\eqref{eq:detection-attenuation} then gives
an $X$ outcome of $-1$ with probability at least $1/4$.
Every active trial therefore detects signal with probability at least
$7/64\ge1/16$. All estimates hold uniformly over the interval's
starting time.  Fresh randomness and reinitialization imply that the probability of no detection after $r$ rounds is at most $(15/16)^r\le e^{-r/16}\le\delta/Q$, as desired.
\end{proof}

Having demonstrated that the subset test succeeds with large probability, we can now analyze Algorithm~\ref{alg:signal-search} to prove Theorem~\ref{thm:signal-search}:

\begin{proof}[Proof of Theorem~\ref{thm:signal-search}]
Use Algorithm~\ref{alg:signal-search}. Every subset reported present intersects $\cK$, by
Lemma~\ref{lem:search-subset}. Subsets at a fixed depth are disjoint, so at most $K$ are retained. Thus at most $2K$ are tested at each depth. The tree of tested subsets has depth at most $J = \ceil{\log_2(N)}$, giving at most $Q=1+2KJ$ tests. Since each of the $Q$ test fails with probability at most $\delta/Q$, the overall failure probability is at most $\delta$, as desired. If no test fails, we indeed add every active frequency and no inactive frequencies to $\wt{\cK}$, solving Search.

We now bound the overall time taken by the algorithm. At depth
$1\le\ell\le J$, at most $\min\{2^\ell,2K\}$ sets are tested,
each of size at most $\ceil{N/2^\ell}\le2N/2^\ell$.
Writing $\cT$ for the tested sets, we obtain
\begin{align}
 \sum_{S\in\cT}\sqrt{|S|}
 &\le\sqrt N+\sqrt{2N}\sum_{\ell=1}^{J}
     \min\{2^{\ell/2},2K\,2^{-\ell/2}\}
 =O(\sqrt{NK}),
 \label{eq:search-tree-cost}\\
 |\cT|
 &\le1+\sum_{\ell=1}^{J}\min\{2^\ell,2K\}
 =O\!\left(K\log\frac{N}{K}\right).
 \label{eq:search-tree-size}
\end{align}
Each test uses at most $r$ sensing intervals, so
\begin{align}
 T\le r\left(
 \frac8{\mu\eps}\sum_{S\in\cT}\sqrt{|S|}
 +\frac{28h_N}{\Delta}|\cT|
 \right).\label{eq:search-total-time}
\end{align}
Finally, $\mu^{-1}=O(\sqrt{\log(N)})$,
$h_N=O(\log(N))$, and
$r=O(\log(K\log(N)/\delta))$. Substitution gives
\begin{align}
 T=O\!\left(
 \left(\frac{\sqrt{NK\log(N)}}{\eps}
 +\frac{K\log(N/K)\log(N)}{\Delta}\right)
 \log\!\left(\frac{K\log(N)}{\delta}\right)
 \right).\label{eq:search-time-general}
\end{align}
This bound holds for any $K \in [N]$; substituting $K = O(1)$ yields Eq.~\eqref{eq:search-time}, proving the result.
\end{proof}

\section{Learning algorithm}\label{app:signal-profile-estimation}

We now solve the Learning problem, Problem~\ref{problem:signal_learning}. Recall that $v_{\max} > 0$ denotes a bound on the maximum amplitude $\|\vec{v}\|_\infty$. Though the problem is formulated for $v_{\max} = O(1)$, our algorithm works for any $v_{\max}$, and incurs only sublogarithmic overhead. We keep the dependence on $v_{\max}$ explicit in our analysis. We assume $\eps < v_{\max}$, because otherwise the zero vector is trivially a solution.

To estimate all the signal amplitudes simultaneously, we use a quantum computer to coherently control the modulation function of the signal, thereby sensing in superposition. Before clipping, the signal encodes the entries of $G_I\vec v$ as phases on separate registers of the quantum computer, where $G_I$ denotes the Gram matrix of the Fourier modes as in Lemma~\ref{lem:gram}. An inverse quantum Fourier transform reads out these phases, and classical inversion of the known Gram matrix provides our estimate $\wt{\vec v}$.

Our main result is a quantum-computing-enhanced algorithm for Learning with near-optimal total evolution time:
\par\addvspace{\topsep}
\noindent\begin{minipage}{\linewidth}
\begin{theorem}[Learning]
\label{thm:signal_learning_qft}
There is a quantum-computing-enhanced sensing algorithm which solves Problem~\ref{problem:signal_learning} using total sensing time
\begin{align}
    T
    =O\!\left(
        \frac{\sqrt N}{\eps}
        \sqrt{\log\left(\frac{N}{\delta}\right)}
        \sqrt{\log\left(\frac{N}{\delta}\right) + \log \left(\frac{1}{\eps}\right)}
        +\frac{\log(N)}{\Delta}
    \right)\label{eq:learning_qft_time}
\end{align}
and $O\bigl(N\log(N/\delta)\log(1/\eps)\bigr)$ ancilla qubits.
\end{theorem}
\end{minipage}\par\addvspace{\topsep}
\noindent
\emph{Proof Sketch.}
Before clipping, the accumulated phase is linear in the modulation function coefficients, so the state of the quantum computer is a tensor product over different registers. Fourier readout estimates each phase coefficient to constant accuracy with constant success probability. We use $r=O(\log(N/\delta))$ registers for each coefficient and take
their median, so all coordinates are accurate except with probability $\delta$.
We again use Lemma~\ref{lem:clipping} to bound error caused by clipping the modulation functions.

We now describe the Learning protocol in detail. We again let $h_N$ denote the $N$-th harmonic number. We set
\begin{align}
    a=\frac{16\pi}{\eps},\qquad
    r=\ceil{8\ln(8N/\delta)},\qquad
    L=\ln(512a^2N^2rv_{\max}^2/\delta^2),
    \label{eq:learning_parameters}
\end{align}
as well as
\begin{align}
    \mu_0=\frac{1}{\sqrt{2L}},\qquad
    \tau=\max\left\{\frac{2a\sqrt{Nr}}{\mu_0},\frac{28h_N}{\Delta}\right\},
    \qquad \mu=\frac{2a\sqrt{Nr}}{\tau}\le\mu_0.
    \label{eq:learning_scales}
\end{align}
For each coefficient, we use $r$ registers of $n_{\mathrm{reg}}$ qubits, where
\begin{align}
    n_{\mathrm{reg}}=\ceil{\log_2\bigl(32(av_{\max}+1)\bigr)},\qquad M=2^{n_{\mathrm{reg}}}.
    \label{eq:learning_register_size}
\end{align}
The basis label $\ell\in\{0,\ldots,M-1\}$ represents the centered value $x(\ell) = (2\ell+1-M)/M$. The uniform distribution on these values is symmetric about zero and
supported in $(-1,1)$. Write $\vec x^{(s)}$ for the $2N$ values in copy
$s \in [r]$. On an interval $I$ of length $\tau$, we use the controlled modulation function
\begin{align}
    Y_{\vec x}(t)=\frac{\mu}{\sqrt{Nr}}
       \sum_{s=1}^r\vec x^{(s)}\cdot\phi(t),\qquad
    g_{\vec x}(t)=\operatorname{Clip}(Y_{\vec x}(t)).
    \label{eq:learning_modulation}
\end{align}
For the inverse quantum Fourier transform, we use the following convention, with corresponding mapping between un-signed and signed Fourier indices:
\begin{align}
    \operatorname{QFT}_M^{-1}\ket{\ell}
    =\frac{1}{\sqrt M}\sum_{y=0}^{M-1}
       e^{-2\pi i y\ell/M}\ket y, \qquad s_M(y)=
    \begin{cases}
        y,&0\le y<M/2,\\
        y-M,&M/2\le y<M
    \end{cases}.
\end{align}
Finally, for the classical reconstruction step, we define the normalized Gram matrix $A_I=2G_I/\tau$.

We can now present our full protocol for Learning, Algorithm~\ref{alg:signal_learning}:

\begin{algorithm}[H]
\caption{Learning}
\label{alg:signal_learning}
\KwIn{Access to $H_{\vec v}(t)$; known frequencies and amplitude bound $v_{\max}$; $\eps>0$ and $\delta\in(0,1/2)$.}
Choose parameters as in Eqs.~\eqref{eq:learning_parameters}--\eqref{eq:learning_register_size}\;
Prepare $2Nr$ registers of $n_{\mathrm{reg}}$ qubits in uniform superposition, and
prepare the sensor in $\ket{1}$\;
Apply the controlled modulation Eq.~\eqref{eq:learning_modulation} over
the next interval $I$ of length $\tau$\;
Apply $\operatorname{QFT}_M^{-1}$ to every register
and measure outcomes $y_{n,s}$ and $y_{n+N,s}$ for $n \in [N]$ and $s \in [r]$\;
Let $\tilde{\beta}_n$ and $\tilde{\beta}_{n+N}$ equal the medians of
$\{\pi s_M(y_{n,s})\}_{s=1}^r$ and $\{\pi s_M(y_{n+N,s})\}_{s=1}^r$, respectively, for each $n \in [N]$\;
\KwRet{$\wt{\vec v}=A_I^{-1}\widetilde{\vec\beta}/a$}\;
\end{algorithm}

We first show that each Fourier measurement estimates its phase
coefficient to constant accuracy.

\par\addvspace{\topsep}
\noindent\begin{minipage}{\linewidth}
\begin{lemma}[Fourier readout]\label{lem:learning_fourier_readout}
Let $M\ge32$ be a power of two and $|\beta|\le\pi M/4$.
Apply $\operatorname{QFT}_M^{-1}$ to
\begin{align}\label{eq:learning_readout_state}
    \ket{\chi_M(\beta)}
    =\frac1{\sqrt M}\sum_{\ell=0}^{M-1}
      e^{i\beta x(\ell)}\ket\ell
\end{align}
and measure an outcome $Y$. Then
\begin{align}\label{eq:learning_readout_accuracy}
    \bbP\bigl[|\pi s_M(Y)-\beta|\le4\pi\bigr]\ge\frac34.
\end{align}
\end{lemma}
\end{minipage}\par\addvspace{\topsep}

\begin{proof}
Write $q=\beta/\pi$. Apart from a global phase, the phase factor before
the transform is $e^{2\pi iq\ell/M}$. The probability of obtaining measurement outcome $y$ is obtained by summing a geometric series, yielding
\begin{align}\label{eq:learning_readout_distribution}
    \bbP[Y=y]
    =\frac{\sin^2(\pi(q-y))}
           {M^2\sin^2(\pi(q-y)/M)},
\end{align}
Let $d_M(q,y)=\min_{k\in\Z}|q-y-kM|$ denote circular distance.
For $0<d_M(q,y)\le M/2$, the inequality
$\sin(\pi d_M(q,y)/M)\ge2d_M(q,y)/M$ implies
\begin{align}\label{eq:learning_readout_pointwise}
    \bbP[Y=y]\le\frac{1}{4d_M(q,y)^2}.
\end{align}
For each integer $k\ge4$, at most two outcomes have circular distance
in $(k,k+1]$. Consequently,
\begin{align}\label{eq:learning_readout_tail}
    \bbP[d_M(q,Y)>4]
    \le\frac12\sum_{k=4}^\infty\frac1{k^2}<\frac14.
\end{align}
Since $|q|\le M/4$ and $M\ge32$, the interval $[q-4,q+4]$ lies
strictly inside $(-M/2,M/2)$. On the complementary event, the signed
representative therefore satisfies $|s_M(Y)-q|\le4$.
\end{proof}

We can now analyze Algorithm~\ref{alg:signal_learning} to prove Theorem~\ref{thm:signal_learning_qft}:

\begin{proof}[Proof of Theorem~\ref{thm:signal_learning_qft}]
Use Algorithm~\ref{alg:signal_learning}. The assumptions $\eps<v_{\max}$ and $\delta<1/2$ give $L>1$ and
hence $0<\mu\le\mu_0<1$. By Eq.~\eqref{eq:learning_scales}, the sensing time satisfies $\tau\ge28h_N/\Delta.$ Lemma~\ref{lem:gram} therefore bounds the normalized Gram matrix by
\begin{align}
    \pnorm{\infty\to\infty}{A_I-\id_{2N}}\le\frac12,
    \qquad
    \pnorm{\infty\to\infty}{A_I}\le\frac32,
    \qquad
    \| A_I^{-1} \|_{\infty \to \infty}\le2,\label{eq:learning_gram_conditioning}
\end{align}
as in our analysis for Search.

First consider the state obtained without clipping. On a basis state,
the phase is exactly
\begin{align}\label{eq:learning_unclipped_phase}
    \Phi_{\vec x}
    &=\frac{\mu}{\sqrt{Nr}}
      \sum_{s=1}^r (\vec x^{(s)})^\top G_I\vec v
      =a\sum_{s=1}^r\vec x^{(s)}\cdot A_I\vec v.
\end{align}
Since $Z\ket1=-\ket1$, the sensor gives the registers the phase
$e^{i\Phi_{\vec x}}$.
Define $\vec\beta=aA_I\vec v$. The register state factorizes as
\begin{align}\label{eq:learning_factorization}
    \ket{\Psi_{\mathrm{unclip}}}
    =\left(\bigotimes_{n=1}^{N}\bigotimes_{s=1}^r
      \ket{\chi_M(\beta_n)}\right)\otimes
      \left(\bigotimes_{n=1}^{N}\bigotimes_{s=1}^r
      \ket{\chi_M(\beta_{n+N})}\right).
\end{align}
Moreover, $\| \vec\beta \|_{\infty}\le3av_{\max}/2\le\pi M/4$ by
Eqs.~\eqref{eq:learning_gram_conditioning} and~\eqref{eq:learning_register_size}.
Lemma~\ref{lem:learning_fourier_readout} applies to every register,
and the outcomes in distinct copies are independent. For a fixed
coordinate, let $F$ count the estimates more than $4\pi$ from the true
value. Independence and the failure bound $1/4$ give
$\E [3^F]\le(3/2)^r$. The median can be inaccurate only if $F\ge r/2$,
so Markov's inequality bounds the failure probability by
$(\sqrt3/2)^r\le e^{-r/8}$. A union bound now gives
\begin{align}
    \bbP\bigl[\| \wt{\vec\beta}-\vec\beta\|_{\infty}>4\pi\bigr]
    \le2N e^{-r/8}\le\frac\delta4.\label{eq:learning_ideal_success}
\end{align}
On this successful event, we have
\begin{align}\label{eq:learning_reconstruction_error}
    \| \wt{\vec v}-\vec v \|_{\infty}
    &\le \| A_I^{-1} \|_{\infty \to \infty}
      \frac{\| \wt{\vec\beta}-\vec\beta\|_{\infty}}{a}
      \le\frac{8\pi}{a}=\frac\eps2.
\end{align}

We next bound the effect of clipping. Under the uniform distribution
on all register labels, the grid values $x(\ell)$ are independent, centered,
and bounded in absolute value by one. For every fixed $t$, the sum
of squared coefficients in the unclipped modulation equals
\begin{align}\label{eq:learning_control_variance}
    \frac{\mu^2}{Nr}\sum_{s=1}^r\sum_{n=1}^{N}\bigl(\phi_n(t)^2+\phi_{n+N}(t)^2\bigr)
    =\mu^2.
\end{align}
For a symmetric random variable $X\in[-1,1]$,
$\E [e^{uX}]=\E[\cosh(uX)]\le\cosh u\le e^{u^2/2}$; the last
inequality follows by comparing the power series. Independence thus
gives $\E [e^{\theta Y_{\vec x}(t)}]\le e^{\theta^2\mu^2/2}$.
Applying Markov's inequality with $\theta=z/\mu^2$ yields
$\bbP[|Y_{\vec x}(t)|\ge z]\le2e^{-z^2/(2\mu^2)}$ for $z\ge0$,
as required by Lemma~\ref{lem:clipping}. If
$d_{\vec x}=\int_I(g_{\vec x}(t)-Y_{\vec x}(t))f_{\vec v}(t)\,\mathrm{d}t$,
that lemma yields
\begin{align}\label{eq:learning_clipping_phase_bound}
    \E_{\vec x}[d_{\vec x}^2]
    \le4\tau\pnorm{L^2(I)}{f_{\vec v}}^2
          \mu^4e^{-1/(2\mu^2)}.
\end{align}
Further, Lemma~\ref{lem:gram} gives
\begin{align}\label{eq:learning_energy}
    \pnorm{L^2(I)}{f_{\vec v}}^2
    =\frac\tau2\vec v^\top A_I\vec v
    \le\frac{3\tau}{4}\pnorm{2}{\vec v}^2
    \le2Nv_{\max}^2\tau.
\end{align}
Write $\ket{\Psi_{\mathrm{clip}}}$ for the state prepared by the clipped
modulation. Both register states have uniform computational-basis probabilities, and
$|e^{iu}-e^{iv}|\le|u-v|$. Using
Eqs.~\eqref{eq:learning_clipping_phase_bound} and~\eqref{eq:learning_energy},
we obtain
\begin{align}
    \pnorm{2}{\ket{\Psi_{\mathrm{clip}}}
             -\ket{\Psi_{\mathrm{unclip}}}}^2
    \le\E_{\vec x}[d_{\vec x}^2] \le8Nv_{\max}^2\tau^2\mu^4e^{-1/(2\mu^2)}\le32a^2N^2rv_{\max}^2\mu_0^2e^{-L} \le\frac{\delta^2}{16}.
      \label{eq:learning_clipping_state_bound}
\end{align}
The inequalities use $\mu\le\mu_0$, $1/(2\mu_0^2)=L$, and
\eqref{eq:learning_parameters}, with $\mu_0<1$. The trace distance between these
pure states is at most their Euclidean distance, hence at most $\delta/4$.
Every subsequent measurement distribution consequently changes by
at most $\delta/4$ in total variation distance.
Combining this with Eq.~\eqref{eq:learning_ideal_success} proves that the
failure probability is at most $\delta/2\le \delta$.

Finally, Eq.~\eqref{eq:learning_parameters} gives
$r=O(\log(N/\delta))$ and
$L=O(\log(N/\delta)+\log(v_{\max}/\eps))$. The single sensing interval has
length $\tau$, so Eq.~\eqref{eq:learning_scales} and
$h_N=O(\log(N))$ give the time bound Eq.~\eqref{eq:learning_qft_time}. Each of the $2Nr$ registers
uses $n_{\mathrm{reg}}=O(\log(v_{\max}/\eps))$ qubits by Eq.~\eqref{eq:learning_register_size},
which proves the ancilla bound.
\end{proof}

\subsection{Simplified algorithm for periodic signal learning}\label{app:periodic-signal-learning}

We have presented an approach for learning signals with arbitrary frequencies, but it involves applying many qubit entangling gates to the quantum computer, which makes it difficult to realize in practice. In this section, we consider protocols for learning \emph{periodic} signals, where all frequencies are equally spaced. We present an algorithm which is considerably simpler than the protocol for learning arbitrary signals.

Formally, we define the \emph{periodic signal learning problem} as Problem~\ref{problem:signal_learning} with $\omega_n = n \Delta$ for all $n \in [N]$. We again assume $\| \vec{v} \|_\infty \le v_{\max} = O(1)$, but our algorithm works for any $v_{\max}$ with polylogarithmic overhead. We again assume $\eps < v_{\max}$, because otherwise the zero vector is trivially a solution.

Our main result is a simple quantum-computing-enhanced sensing for periodic signal learning with near-optimal total evolution time:

\par\addvspace{\topsep}
\noindent\begin{minipage}{\linewidth}
\begin{theorem}[Periodic signal learning]\label{thm:periodic-learning} There is a quantum-computing-enhanced sensing algorithm which solves Problem~\ref{problem:signal_learning} for periodic signals using total sensing time
\begin{align}
 O\!\left(
 \left(\frac{\sqrt N}{\eps}+\frac1\Delta\right)
 \left(\log\frac{N}{\delta}+\log\frac{1}{\eps}\right)^{3/2}
 \right).\label{eq:periodic-time}
\end{align}
The algorithm consists of SWAP gates between the sensor qubit and a quantum memory of $O\!\left(N\left(\log(N/\delta)+\log(1/\eps)\right)\right)$ qubits, plus single-qubit gates and measurements.
\end{theorem}
\end{minipage}\par\addvspace{\topsep}
\ \\
\noindent
\emph{Proof Sketch.}
We subdivide each period into $O(N)$ bins and use SWAP gates to store the phase accumulated in each bin in separate memory qubits. The key observation is that we can then coherently accumulate phase over multiple periods to obtain a Heisenberg limited estimate of $f_{\vec{v}}(t)$ at various time points (using a nearly unbiased estimator of arcsine). We perform a discrete Fourier transform on the resulting classical data to estimate $\vec{v}$. Because the errors in the measurements of different time bins are independent, the Fourier transform averages down the overall error by $\sqrt{N}$, leading to the desired scaling of the overall evolution time.

We now describe the periodic signal learning algorithm in detail, beginning with the Fourier reconstruction step.
Let $T_0=2\pi/\Delta$ denote the period of the signal, $M=4N$ the number of time bins, and $\tau=T_0/M$ the width of each bin. For
$m=0,\ldots,M-1$, set $I_m=[m\tau,(m+1)\tau]$ and
$t_m=(m+1/2)\tau$. For an integrable periodic function $r$, write
\begin{align}
 g_m(r)=\frac1\tau\int_{I_m}r(t)\,\mathrm{d}t,
 \qquad \vec g(r)=(g_0(r),\ldots,g_{M-1}(r)).\label{eq:periodic-bin-averages}
\end{align}
If $r(t)=\sum_{n=1}^N(x_n\cos(n\Delta t)+x_{n+N}\sin(n\Delta t))$ and
$\vec x=(x_1,\ldots,x_N,x_{N+1},\ldots,x_{2N})$, then
$\vec g(r)=\Phi\vec x$, where
\begin{align}
 \Phi_{m,n}=\kappa_n\cos(n\Delta t_m),\qquad
 \Phi_{m,n+N}=\kappa_n\sin(n\Delta t_m),\qquad
 \kappa_n=\mathrm{sinc}(\pi n/M)\ge\kappa_*:=\frac{2\sqrt2}{\pi}.\label{eq:periodic-bin-matrix}
\end{align}
The identity
$\sum_{m=0}^{M-1}e^{2\pi i n(m+1/2)/M}=0$ for every integer
$n$ with $0<|n|<M$ implies that
\begin{align}
 \Phi^\top\Phi=\frac M2 D^2,\qquad
 D=\mathrm{diag}(\kappa_1,\ldots,\kappa_N,\kappa_1,\ldots,\kappa_N).\label{eq:periodic-bin-orthogonality}
\end{align}
Consequently, the \emph{reconstruction matrix}
$R=(\Phi^\top\Phi)^{-1}\Phi^\top$ satisfies $R\Phi=\id_{2N}$, and
\begin{align}
 (R\vec z)_n&=\frac{2}{M\kappa_n}\sum_{m=0}^{M-1}z_m\cos(n\Delta t_m),&
 (R\vec z)_{n+N}&=\frac{2}{M\kappa_n}\sum_{m=0}^{M-1}z_m\sin(n\Delta t_m).\label{eq:periodic-reconstruction}
\end{align}
We also use the orthogonal projector $\Pi=\Phi R$ onto the image of $\Phi$.

The next lemma quantifies how independent bounded errors affect the
reconstruction. Its proof uses only the row norms of $R$ and $\Pi$ and
Hoeffding's inequality.

\par\addvspace{\topsep}
\noindent\begin{minipage}{\linewidth}
\begin{lemma}[Fourier concentration]\label{lem:periodic-fourier}
Suppose the coordinates of $\vec\xi\in\R^M$ are independent and have
mean zero, and each lies in an interval of length at most $2\nu$, where
$\nu>0$. Then, for $t>0$,
\begin{align}
 \bbP\bigl[\| R\vec\xi \|_{\infty} \ge t\bigr]
 &\le M e^{-Mt^2/(6\nu^2)},&
 \bbP\bigl[\| \Pi\vec\xi\|_{\infty}\ge t\bigr]
 &\le2M e^{-t^2/(2\nu^2)}.\label{eq:periodic-concentration}
\end{align}
Moreover, for every $\vec b\in\R^M$,
\begin{align}
 \| R\vec b \|_{\infty}\le\frac2{\kappa_*}\| \vec b \|_{\infty},
 \qquad
\| \Pi\vec b \|_{\infty}\le\sqrt M\| \vec b\|_{\infty}.\label{eq:periodic-bias-bounds}
\end{align}
\end{lemma}
\end{minipage}\par\addvspace{\topsep}
\begin{proof}
Hoeffding's inequality gives
\begin{align}
 \bbP\bigg[\Big|\sum_m c_m\xi_m\Big|\ge t\bigg]
 \le2\exp\left(-\frac{t^2}{2\nu^2\sum_m c_m^2}\right)\label{eq:periodic-hoeffding}
\end{align}
for every nonzero real vector $\vec c$. Each row of $R$ has squared Euclidean norm
$2/(M\kappa_n^2)\le3/M$. Each row of $\Pi$ has squared Euclidean norm
$\Pi_{mm}\le1$, since $\Pi$ is an orthogonal projector. Applying Eq.~\eqref{eq:periodic-hoeffding} and taking a union bound over $2N$ and $M$
coordinates, respectively, proves the concentration statements. Eq.~\eqref{eq:periodic-reconstruction} bounds its row sums by $2/\kappa_*$. Finally,
$\| \Pi\vec b\|_{\infty} \le \| \Pi\vec b\|_{2}
\le \| \vec{b} \|_2\le\sqrt M  \| \vec{b} \|_{\infty}$.
\end{proof}

Next, we describe how to prepare the single memory qubit states recording the accumulated phases under the unknown signal. For a known estimate $\wt{\vec v}$, let $\wt f=f_{\wt{\vec v}}$ and
$h_m=g_m(f-\wt f)$. Fix an integer $Q\ge1$, a scale parameter $\lambda>0$,
and known numbers $c_m$. To prepare $Q$ copies of
$e^{-i\lambda(h_m-c_m)Z}\ket{+}$ for every bin, set
$J=\ceil{Q\lambda/\tau}$ and $a=\lambda/(J\tau)$. Assign $Q$ memory
qubits to each bin and initialize them in $\ket{+}$. We choose the SWAP gate timing so that each copy receives the fraction $a$ of total integrated phase over the bin; this is possible for all copies because $Qa \le 1$. Any remaining time in the final period is consumed as idle.
After $J$ periods, each copy has acquired phase
$Ja\int_{I_m}f(t)\,\mathrm{d}t=\lambda g_m(f)$. A known single qubit rotation then prepares the desired state. This procedure uses $MQ$ memory qubits and total time
\begin{align}
 T_0\left \lceil \frac{MQ\lambda}{T_0} \right\rceil\le MQ\lambda+T_0.\label{eq:periodic-batch-time}
\end{align}

Next, we describe how, via a series of single-qubit measurements of phase states of this form, we can refine an estimate $\wt{\vec{v}}$ of the coefficient vector $\vec{v}$ (with associated estimate $\wt{f}(t)$ of $f(t) = f_{\vec{v}}(t)$). Specifically, we will take as input an estimate $\wt{\vec{v}}$, with the promise that $\| \vec{h} \|_\infty \le h_{\scriptscriptstyle \mathrm{max}}$, for $\vec{h} = \vec{g}(f - \wt{f})$ the bin averages of the residual and $h_{\scriptscriptstyle \mathrm{max}} > 0$ some bound, and return an improved estimate $\wt{\vec{h}}$ of $\vec{h}$ which satisfies a tighter bound with high probability. We will subsequently use this refinement subroutine to hierarchically estimate the coefficients in $\vec{v}$. Our subroutine is described in Algorithm~\ref{alg:periodic-refine}.

\begin{algorithm}[H]
\caption{$\mathsf{RefineEstimate}(\wt{\vec v},h_{\max},s,L)$}
\label{alg:periodic-refine}
\KwIn{Estimate $\wt{\vec v}$; bound $h_{\max}$ such that
$\| \vec h \|_\infty \le h_{\max}$;
$s>0$ and integer $L\ge1$.}
Set $q_*\gets\max\{0,\ceil{\log_2(8h_{\max}/s)}\}$,
$c_m\gets0$ for all $m$, and $C\gets h_{\max}$\;
\For{$q=1,\ldots,q_*$}{
 Set $\lambda\gets1/(3C)$\;
 Prepare $64L$ copies of $e^{-i\lambda(h_m-c_m)Z}\ket{+}$
 for every bin $m$\;
 Measure all copies in the $Y$ basis; let $\overline Y_m$ be the mean\;
 Set $c_m\gets c_m+\overline Y_m/(2\lambda)$ , $C\gets C/2$\;
}
Set $a_l\gets(2l)!/(4^l(l!)^2(2l+1))$ for $0\le l\le L$,
$A_L\gets\sum_{l=0}^L a_l$, and $\lambda_*\gets1/s$\;
Independently sample $\pi_m\in\{0,\ldots,L\}$ with
$\bbP[\pi_m=l]=a_l/A_L$ for each $m$\;
Prepare $2\pi_m+1$ copies of $e^{-i\lambda_*(h_m-c_m)Z}\ket{+}$ for each $m$\;
Measure all copies in the $Y$ basis and let $P_m$ denote the product for each $m$\;
\KwRet{$\wt h_m=c_m+(A_Ls/2)P_m$ for $m=0,\ldots,M-1$.}
\end{algorithm}

The following bounds explain why we use Algorithm~\ref{alg:periodic-refine}: its bias decays
exponentially in $L$, while its random error remains bounded by a multiple
of $s$.

\par\addvspace{\topsep}
\noindent\begin{minipage}{\linewidth}
\begin{lemma}[Refining estimator]\label{lem:periodic-refinement}
Suppose $q_*\le L$. The localization
steps in Algorithm~\ref{alg:periodic-refine} give
$|h_m-c_m|\le s/8$ for every $m$, except with probability at most
$2ML e^{-2L}$. Conditional on this
property, the final estimates have the form
\begin{align}
 \wt h_m=h_m+b_m+\xi_m,
 \qquad |b_m|\le s16^{-L}/100,\label{eq:periodic-refinement-error}
\end{align}
where the $\xi_m$ are
independent, mean zero random variables which each lie in an interval of length at most
$\pi s/2$. The sensing time is
\begin{align}
 O\!\left(\frac{ML}{s}+(q_*+1)T_0\right).\label{eq:periodic-refine-time}
\end{align}
\end{lemma}
\end{minipage}\par\addvspace{\topsep}
\begin{proof}
Fix a bin and write $u=h_m-c_m$. If $|u|\le C$, a $Y$ measurement
of $e^{-i\lambda uZ}\ket{+}$ has mean $\sin(2\lambda u)$, with
$\lambda=1/(3C)$. If the empirical mean differs from this mean by at most
$1/4$, then
\begin{align}
 \left|u-\frac{\overline Y_m}{2\lambda}\right|
 &\le\frac1{8\lambda}
   +\frac{|2\lambda u|^3}{12\lambda}
 \le\frac{3C}{8}+\frac{2C}{27}<\frac C2.\label{eq:periodic-localization-step}
\end{align}
Here we used $|\sin x-x|\le|x|^3/6$. Hoeffding's inequality for
$64L$ independent outcomes in $[-1,1]$ bounds the probability of an
empirical error exceeding $1/4$ by $2e^{-2L}$. This bound holds
conditional on every earlier successful localization step. A union bound
over $M q_*\le ML$ steps gives the claimed failure probability. On
success, $|h_m-c_m|\le h_{\max}2^{-q_*}\le s/8$; if $q_*=0$,
the same inequality follows from $8h_{\max}\le s$.

Condition on successful localization. Then
$u=h_m-c_m$ is fixed and $|2u/s|\le1/4$. Put $\mu=\sin(2u/s)$.
The fresh measurements satisfy $\E[P_m\mid\pi_m=l]=\mu^{2l+1}$, and hence
\begin{align}
 \E[\wt h_m]=c_m+\frac s2\sum_{l=0}^L a_l\mu^{2l+1}.\label{eq:periodic-estimator-mean}
\end{align}
We have $\arcsin x=\sum_{l\ge0}a_lx^{2l+1}$ for $|x|<1$. Since
$\arcsin\mu=2u/s$, $0<a_l\le1$, and $|\mu|\le1/4$, the bias is bounded by
\begin{align}
 \left|\E[\wt h_m]-h_m\right|
 &\le\frac s2\sum_{l=L+1}^\infty 4^{-(2l+1)}
 =\frac{s16^{-L}}{120}\le\frac{s16^{-L}}{100}.\label{eq:periodic-estimator-bias}
\end{align}
We have $A_L\le\arcsin(1)=\pi/2$. Thus $\wt h_m-c_m$ lies in
$[-\pi s/4,\pi s/4]$ and
$\xi_m=\wt h_m-\E[\wt h_m]$ lies in an interval of length at
most $\pi s/2$.

Finally, the phase multipliers are
$\lambda_q=2^{q-1}/(3h_{\max})$. For $q_*>0$,
\begin{align}
 \sum_{q=1}^{q_*}\lambda_q
 =\frac{2^{q_*}-1}{3h_{\max}}<\frac{16}{3s},\label{eq:periodic-localization-scales}
\end{align}
and when $q_*=0$, their sum is zero. There are $64L$ copies at each
localization step and $2L+1$ copies with phase multiplier $1/s$ in the final
step. Summing Eq.~\eqref{eq:periodic-batch-time} proves
Eq.~\eqref{eq:periodic-refine-time}.
\end{proof}

Finally, we use our refinement subroutine to hierarchically estimate the coefficients in $\vec{v}$, halving the precision at each step. We set
\begin{align}
 K=\ceil{\log_2(v_{\max}/\eps)},\qquad
 L=\ceil{4\ln\left(10^6 N K v_{\max} / (\delta\eps)\right)}.\label{eq:periodic-parameters}
\end{align}
Our full periodic signal learning algorithm is described in Algorithm~\ref{alg:periodic-learning}.

\begin{algorithm}[H]
\caption{Periodic signal learning}\label{alg:periodic-learning}
\KwIn{Access to $H_{\vec v}(t)$; known frequencies and amplitude bound $v_{\max}$;
$\eps>0$ and $\delta \in (0,1/2)$.}
Set $M\gets4N$, choose $K,L$ as in Eq.~\eqref{eq:periodic-parameters},
and construct $R$ using Eq.~\eqref{eq:periodic-reconstruction}\;
Set $\wt{\vec v}^{(0)}\gets0$\;
\For{$k=0,\ldots,K-1$}{
 Set $\alpha_k\gets v_{\max}2^{-k}$ and $s_k\gets\alpha_k\sqrt M/(12\sqrt L)$\;
 Set $h_{\max}^{(k)}\gets2Nv_{\max}$ if $k=0$, and
 $h_{\max}^{(k)}\gets\alpha_k\sqrt N$ otherwise\;
 Run $\wt{\vec h}^{(k)}\gets
 \mathsf{RefineEstimate}(\wt{\vec v}^{(k)},h_{\max}^{(k)},s_k,L)$\;
 Set $\wt{\vec v}^{(k+1)}\gets\wt{\vec v}^{(k)}+R\wt{\vec h}^{(k)}$\;
}
\KwRet{$\wt{\vec v}^{(K)}$.}
\end{algorithm}

\begin{proof}[Proof of Theorem~\ref{thm:periodic-learning}]
We prove inductively that, except on the failure events bounded below,
\begin{align}
 \| \vec v-\wt{\vec v}^{(k)} \|_{\infty}\le\alpha_k,
 \qquad
 \| \vec g(f-f_{\wt{\vec v}^{(k)}})  \|_{\infty}\le h_{\max}^{(k)}.\label{eq:periodic-invariant}
\end{align}
For $k=0$, the first inequality is the coefficient bound. The second
follows from $|f(t)|\le2Nv_{\max}$.

Suppose Eq.~\eqref{eq:periodic-invariant} holds for some $k$. The numbers
of localization steps are
\begin{align}
 q_0=\ceil{\log_2(96\sqrt{NL})},\qquad
 q_k=\ceil{\log_2(48\sqrt L)}\quad(k\ge1).\label{eq:periodic-localization-counts}
\end{align}
They are at most $L$. Indeed, Eq.~\eqref{eq:periodic-parameters} gives
$L\ge56$ and $N\le e^{L/4}$; also $\sqrt L\le e^{L/8}$ for $L\ge56$.
Thus $96\sqrt{NL}\le96e^{L/4}\le2^L$, proving both bounds.
Lemma~\ref{lem:periodic-refinement} therefore applies, so we have the following decomposition, with independent mean zero  errors each in an interval of length $2\nu_k$:
\begin{align}
 \wt{\vec h}^{(k)}
 =\Phi(\vec v-\wt{\vec v}^{(k)})+\vec b^{(k)}+\vec\xi^{(k)},
 \qquad
 \nu_k=\frac{\pi s_k}{4},\qquad
 \| \vec b^{(k)}  \|_{\infty} \le\frac{s_k16^{-L}}{100}.\label{eq:periodic-round-decomposition}
\end{align}
Since $R\Phi=\id_{2N}$ and $\Phi R=\Pi$, the updated errors are
\begin{align}
 \wt{\vec v}^{(k+1)}-\vec v
 =R(\vec b^{(k)}+\vec\xi^{(k)}), \qquad \vec g(f-f_{\wt{\vec v}^{(k+1)}})
 =-\Pi(\vec b^{(k)}+\vec\xi^{(k)}).\label{eq:periodic-update-errors}
\end{align}
The bias terms satisfy
\begin{align}
 \| R\vec b^{(k)} \|_{\infty}
 \le\frac{\alpha_k\sqrt N}{300\kappa_*\sqrt L}\,16^{-L}
 \le\frac{\alpha_k}{4}, \qquad \| \Pi\vec b^{(k)} \|_{\infty}
 \le\frac{\alpha_k N}{300\sqrt L}\,16^{-L}
 \le\frac{\alpha_k\sqrt N}{4}.\label{eq:periodic-round-bias}
\end{align}
For the last inequalities we used
$\sqrt N16^{-L}\le e^{L/8}16^{-L}\le1$, $L\ge1$, and
$\kappa_*>2/3$. Moreover,
$\nu_k^2=\pi^2\alpha_k^2M/(2304L)$. Lemma~\ref{lem:periodic-fourier}
therefore gives
\begin{align}
 \bbP\left[\| R\vec\xi^{(k)} \|_{\infty} \ge\frac{\alpha_k}{4}\right]
 \le M e^{-\frac{24L}{\pi^2}}\le M e^{-2L}, \quad \bbP\bigg[\| \Pi\vec\xi^{(k)}\|_{\infty}
            \ge\frac{\alpha_k\sqrt N}{4}\bigg]
 \le2M e^{-\frac{18L}{\pi^2}}\le2M e^{-L}.\label{eq:periodic-round-tails}
\end{align}
Outside these events, Eq.~\eqref{eq:periodic-update-errors} imply the desired condition at round $k+1$, because $\alpha_{k+1}=\alpha_k/2$.
The total
failure probability is at most
\begin{align}
 KM\bigl[(2L+1)e^{-2L}+2e^{-L}\bigr]
 &\le5KML e^{-L}
 \le5KM e^{-L/2}<\delta,\label{eq:periodic-total-failure}
\end{align}
using the expressions for $K,L,M$.
At the last round, $\alpha_K\le\eps$, so the algorithm succeeds with probability at least $1-\delta$.

It remains to bound the resources used by the algorithm. Each batch uses at most $O(NL)$ memory
qubits, which are measured and reused. By Eq.~\eqref{eq:periodic-refine-time},
\begin{align}
 T
 &=O\!\left(
 \sum_{k=0}^{K-1}\frac{ML}{s_k}
 +T_0\left(1+\sum_{k=0}^{K-1}(q_k+1)\right)\right) =O\!\left(
 \frac{\sqrt N L^{3/2}}{\eps}
 +\frac1\Delta\bigl(\log(NL)+K\log L\bigr)\right).\label{eq:periodic-total-time}
\end{align}
The first term follows from
$\sum_{k=0}^{K-1}\alpha_k^{-1}=(2^K-1)/v_{\max}<2/\eps$.
The choice of $L$ gives $K=O(L)$ and $\log N=O(L)$; also
$\log L=O(\sqrt L)$. Consequently the second term is
$O(L^{3/2}/\Delta)$. Finally,
$\log K=O(\log(v_{\max}/\eps))$, so
$L=O(\log(N/\delta)+\log(v_{\max}/\eps))$. For $v_{\max} = O(1)$, this proves the desired ancilla and evolution time bounds.
\end{proof}

\section{Fundamental sensing time limits}

In this section, we consider three specific sensing problems and prove a lower bound on the sensing time to solve them with the most general quantum algorithms.
These three problems are chosen such that they can be also solved by any general algorithm designed to solve P1-P5 in discrete-frequency spectroscopy.  Thus, lower bounds for these specific tasks imply lower bounds for P1–P5. The three problems are the following.
First, for P1--P3, we construct an instance of a periodic signal (i.e., a signal with equally spaced candidate frequencies) for which even detecting the signal, even when its waveform is exactly known, requires $\Omega(\sqrt N/(\epsilon\log N))$ sensing time.
Second, for P3--P5, we consider, for any fixed set of candidate frequencies sufficiently separated from each other and from zero, the minimal average distinguishability needed to distinguish the null signal from the family of $N$ monochromatic signals with an unknown frequency, forcing $\Omega(\sqrt N/\epsilon)$ sensing time. Third, we prove the same lower bound for P5 (up to logarithmic factors), applicable to a broader range of $\eps$ and $\Delta$, by bounding the information required to decode the signs of the coefficients of random vectors encoded in the signal.

\subsection{Fundamental limits for detection of a fully known candidate signal and P1-P3}\label{app:known-waveform-detection}
We begin by introducing an even ``easier'' problem in discrete-frequency spectroscopy, which we call P0. Unlike P1, where only the signal shape is known, in P0 we assume complete knowledge of the candidate signal, including both its shape and strength, and ask whether the true signal equals this candidate or the signal is absent. We formulate the problem precisely as follows.

\begin{problem}[Detection of a fully known candidate signal (P0)]
Let $0<\epsilon = o(\sqrt{N}\Delta/\log(2N))$, let the known frequencies obey
$\omega_1\geq\Delta$ and
$\omega_{n+1}-\omega_n\geq\Delta$, and fix
$\vec a\in S^{N-1}$ and $\vec\phi\in[0,2\pi)^N$.  Given evolution under
$H(t)=f(t)Z$, decide between
\begin{equation}
    f(t)=0
    \qquad\text{and}\qquad
    f(t)=\epsilon\sum_{n=1}^N
    a_n\cos(\omega_nt+\phi_n)
    \label{eq:ksd-hypotheses}
\end{equation}
with success probability at least $2/3$.
\end{problem}
\noindent
P0 is no harder than P1, P2, or P3: a protocol for any of these problems
must in particular distinguish the null signal from a fixed signal
of norm $\epsilon$.

\begin{theorem}[Worst-case lower bound for P1--P3]
\label{thm: harmonic_lb}
Take equally spaced frequencies $\omega_n=n\Delta$, and choose
$a_n=1/\sqrt N$ and $\phi_n=0$ for every $n\in[N]$.  Any protocol that solves
the resulting P0 instance with success probability at least $2/3$ requires
\begin{equation}
    T=\Omega\!\left(
        \frac{\sqrt N}{\epsilon\log(N)}
    \right).
    \label{eq:ksd-harmonic-lower-bound}
\end{equation}
Consequently, the same worst-case lower bound holds for P1, P2, and P3.
\end{theorem}

\noindent
The hard waveform is a normalized Dirichlet kernel.  Its modes interfere
constructively only in narrow periodic peaks: each peak has height
$\Theta(\epsilon\sqrt N)$ but width $\Theta(1/(N\Delta))$.  The integrated
absolute signal in one period is therefore only
$O(\epsilon\log N/(\Delta\sqrt N))$.  Even a protocol that knows the waveform
and always aligns the sensor with its sign must accumulate contributions from
$\Omega(\Delta\sqrt N/(\epsilon\log N))$ periods.

The proof uses the following proposition.

\begin{proposition}[Implicit Instance Optimal Time Complexity]\label{prop: imp_inst_opt} Let $T_{\mathrm{det}}$ be such that
\begin{align}\label{eq: main_integral}
    \epsilon \int_0^{T_{\mathrm{det}}} dt \left|\sum_{n=1}^N a_n \cos(\omega_n t + \phi_n) \right| = 1.
\end{align}
The sensing time complexity of P0 is $\Theta(T_{\mathrm{det}})$.
\end{proposition}
\noindent
The upper bound is achieved by Ramsey interferometry with the modulation function above, $\mathrm{sgn}(\sum_{n=1}^N a_n \cos(\omega_n t + \phi_n))$, while the lower bound follows by invoking Corollary A.5 in~\cite{childs_theory_2021} and analyzing the integrated operator norm difference between the candidate Hamiltonians.

Proposition~\ref{prop: imp_inst_opt} means that, to lower bound the time complexity of P0, it suffices to \emph{upper bound} the integral Eq.~\eqref{eq: main_integral}. This directly allows us to prove a general worst-case lower bound on the time complexity of P0.

\begin{proof}[Proof of Theorem~\ref{thm: harmonic_lb}]
Let
\begin{equation}
    S_N(u)=\sum_{n=1}^N\cos(nu).
\end{equation}
The Dirichlet-kernel identity gives
\begin{equation}
    2S_N(u)+1
    =
    \frac{\sin((N+1/2)u)}{\sin(u/2)}.
    \label{eq:ksd-dirichlet-identity}
\end{equation}
After the substitution $u=2x$ and symmetry about $x=\pi/2$,
\begin{align}
    \int_0^{2\pi}|2S_N(u)+1|\,du
    &=4\int_0^{\pi/2}
    \left|\frac{\sin((2N+1)x)}{\sin x}\right|dx
    \nonumber\\
    &\leq
    4\int_0^{\pi/2}
    \min\!\left(2N+1,\frac{\pi}{2x}\right)dx
    \nonumber\\
    &=2\pi\bigl[1+\log(2N+1)\bigr].
    \label{eq:ksd-dirichlet-l1-bound}
\end{align}
Here we used $|\sin((2N+1)x)|\leq(2N+1)\sin x$ and
$\sin x\geq2x/\pi$ on $[0,\pi/2]$.  Eq.~\eqref{eq:ksd-dirichlet-l1-bound}
and the triangle inequality imply
\begin{equation}
    \int_0^{2\pi}|S_N(u)|\,du
    \leq C\log(2N)
    \label{eq:ksd-cosine-sum-l1-bound}
\end{equation}
for a universal constant $C$.

The function $|S_N(\Delta t)|$ has period $2\pi/\Delta$.  Dividing
$[0,T]$ into complete periods and at most one partial period, and bounding the
partial period by a complete one, gives
\begin{equation}
    \frac{\epsilon}{\sqrt N}
    \int_0^T|S_N(\Delta t)|\,dt
    \leq
    C\frac{\epsilon\log(2N)}{\sqrt N}
    \left(T+\frac1\Delta\right).
    \label{eq:ksd-integrated-dirichlet-bound}
\end{equation}
Combining Proposition~\ref{prop: imp_inst_opt} with
Eq.~\eqref{eq:ksd-integrated-dirichlet-bound} yields
\begin{equation}
    T
    \geq
    c\frac{\sqrt N}{\epsilon\log(2N)}-\frac1\Delta
    \label{eq:ksd-lower-bound-with-boundary-term}
\end{equation}
for a universal $c>0$.  Since $\epsilon = o(\sqrt{N}\Delta/\log(2N))$, the second term is
negligible for all sufficiently large $N$, proving
Eq.~\eqref{eq:ksd-harmonic-lower-bound}.
\end{proof}

\subsection{Fundamental limits for P3-P5}

We now prove a lower bound for the three tasks with the least prior
information: Detection of an unknown signal (P3), Search (P4), and Learning
(P5).  Unlike the worst-case signal constructed in the preceding
subsection, the argument here applies to every fixed set of separated
candidate frequencies.  It does so by bounding the average distinguishability
of a family in which one unknown frequency is active.

We use the same signal model as in the rest of the paper.  The known
frequencies satisfy
\begin{equation}
    0<\omega_1<\cdots<\omega_N,
    \qquad
    \omega_1\geq\Delta,
    \qquad
    \omega_{k+1}-\omega_k\geq\Delta.
    \label{eq:arbitrary-separated-frequency-assumption}
\end{equation}
We allow the most general sensing protocol in our model: arbitrary ancillas,
signal-independent controls, intermediate measurements, classical
randomness, and adaptive feedback.  By purifying all operations and deferring
all measurements, such a protocol can be represented by a pure initial state,
a signal-independent coherent control $V(t)$ acting on the sensor and all
ancillas, and one final measurement.  In the toggling frame defined by this
control, the signal couples through
\begin{equation}
    \wt Z(t)=V(t)^\dagger(Z\otimes I_{\mathrm{anc}})V(t),
    \qquad \|\wt Z(t)\|=1,
    \label{eq:lower-bound-toggling-operator}
\end{equation}
while the null signal generates the identity evolution.

\begin{theorem}[Lower bounds for P3--P5]
\label{thm:p3-p5-lower-bound}
Let $N\geq3$, let the candidate frequencies satisfy
Eq.~\eqref{eq:arbitrary-separated-frequency-assumption}, and assume
$0<\epsilon<\Delta$ and $\epsilon\leq1/3$.  Any quantum sensing protocol that
solves P3, P4, or P5 with success probability at least $2/3$ has worst-case
total sensing time
\begin{equation}
    T=\Omega\!\left(\frac{\sqrt N}{\epsilon}\right).
    \label{eq:p3-p5-lower-bound}
\end{equation}
The bound permits arbitrary ancilla systems and fully adaptive controls.
\end{theorem}

\noindent
The restriction $\epsilon\leq1/3$ only ensures that the hard instances used
for P5 lie within the normalized amplitude range $\|\vec v\|_\infty\leq1$;
for any other constant amplitude bound, the same proof applies after a
constant rescaling.

The intuition follows Ref.~\cite{allen2025quantumcomputingenhancedsensing}. Consider the null signal and $N$ alternatives, each containing a single tone of amplitude $\alpha$ at a distinct candidate frequency. Any protocol solving P3–P5 must distinguish every alternative from the null.
The key limitation is that a quantum sensor cannot respond strongly to all $N$ possible frequencies within a short sensing time. Physically, oscillations at different frequencies tend to average out unless the sensor’s control is suitably matched to them. Although quantum control can make the sensor sensitive to multiple frequencies simultaneously, this sensitivity must be distributed among the $N$ possibilities. Over a sensing interval of duration $O(1/\alpha)$, the average change in the sensor–ancilla state across the $N$ alternatives is at most $O(1/\sqrt N)$, regardless of the control strategy or ancillas used.
For successful Detection or Search, however, every alternative must produce an appreciable change in the quantum state. Since these changes can accumulate at most linearly with the number of sensing intervals, at least $\Omega(\sqrt N)$ intervals are required. Each interval takes time $O(1/\alpha)$, giving $T=\Omega(\sqrt N/\alpha)$. Taking $\alpha=\Theta(\epsilon)$ yields the claimed lower bound $T=\Omega(\sqrt N/\epsilon)$.

We now make this argument precise.  For an amplitude $\alpha>0$, consider the
null signal $f_0(t)=0$ and the one-sparse alternatives
\begin{equation}
    f_k^{(\alpha)}(t)=\alpha\cos(\omega_k t),
    \qquad k\in[N].
    \label{eq:arbitrary-separated-hard-family}
\end{equation}
On an interval $I$, let
\begin{equation}
    \wt U_{k,\alpha}(I)
    =\mathcal T\exp\!\left(
        -i\alpha\int_I\cos(\omega_k t)\wt Z(t)\,dt
    \right)
    \label{eq:one-sparse-interaction-evolution}
\end{equation}
denote the corresponding toggling-frame evolution (or interaction picture).

\subsubsection{Short-time average distinguishability}

\begin{lemma}[Separated-frequency short-time bound]
\label{lem:separated-frequency-short-time}
For every initial sensor--ancilla state $\ket\psi$, every signal-independent
control sequence, every interval $I$, and $x:=\alpha\tau\leq1$,
\begin{equation}
    \frac1N\sum_{k=1}^N
    \left\|
    \bigl(\wt U_{k,\alpha}(I)-I\bigr)\ket\psi
    \right\|_2^2
    \leq
    \frac{2e^2}{N}
    \left[
        x^2+2\pi\frac{\alpha}{\Delta}x
    \right].
    \label{eq:separated-frequency-short-time}
\end{equation}
\end{lemma}

\begin{proof}
We use the following observation: if $J$ is an interval of
length $L$, $g(y)\in \cH$ is a sensor-ancilla state that depends smoothly on $y\in J$, and the frequencies obey
Eq.~\eqref{eq:arbitrary-separated-frequency-assumption}, then
\begin{equation}
    \sum_{k=1}^N
    \left\|
        \int_J g(y)e^{i\omega_k y}dy
    \right\|_2^2
    \leq
    \left(L+\frac{2\pi}{\Delta}\right)
    \int_J\|g(y)\|_2^2dy.
    \label{eq:hilbert-valued-large-sieve}
\end{equation}
To see the scalar estimate underlying Eq.~\eqref{eq:hilbert-valued-large-sieve},
expand the squared norm.  The diagonal terms contribute
$L\sum_k|c_k|^2$, while the off-diagonal terms contain
\begin{equation}
    \int_J e^{i(\omega_j-\omega_k)y}\,dy
    =
    \frac{e^{i(\omega_j-\omega_k)b}
          -e^{i(\omega_j-\omega_k)a}}
         {i(\omega_j-\omega_k)},
    \qquad J=[a,b].
\end{equation}
Applying the generalized Hilbert inequality
\begin{equation}
    \left|
    \sum_{j\neq k}
    \frac{c_j\overline{c_k}}{\omega_j-\omega_k}
    \right|
    \leq
    \frac{\pi}{\Delta}\sum_k|c_k|^2
\end{equation}
separately to the two endpoints gives the scalar estimate
\begin{equation}
    \int_J
    \left|\sum_{k=1}^N c_ke^{i\omega_k y}\right|^2dy
    \leq
    \left(L+\frac{2\pi}{\Delta}\right)
    \sum_{k=1}^N|c_k|^2.
\end{equation}
Eq.~\eqref{eq:hilbert-valued-large-sieve} can be shown by evaluating it using an orthonormal basis for $\cH$, then bounding the result using the scalar estimate.

Write $I=[t_0,t_0+\tau]$ and denote the set of dummy time variables
\begin{equation}
\mathcal D_m(I)
=\{(t_1,\ldots,t_m):t_0\le t_m\le\cdots\le t_1\le t_0+\tau\}.
\end{equation}
The $m$-th term in Dyson series acting on $\ket\psi$ is
\begin{equation}
    \ket{V_m(\omega_k)}
    =
    (-i\alpha)^m
    \int_{\cD_m(I)}d\vec t\,
    \cZ_1\cdots\cZ_m\ket\psi
    \prod_{\ell=1}^m\cos(\omega_k t_\ell),
    \label{eq:separated-frequency-dyson-term}
\end{equation}
where $\cZ_\ell=\wt Z(t_\ell)$ and
$\|\cZ_1\cdots\cZ_m\ket\psi\|_2=1$.  Expanding the cosines gives
\begin{equation}
    \ket{V_m(\omega_k)}
    =
    \frac{(-i\alpha)^m}{2^m}
    \sum_{\vec b\in\{\pm1\}^m}
    \int_{\cD_m(I)}d\vec t\,
    e^{i\omega_k\vec b\cdot\vec t}
    \cZ_1\cdots\cZ_m\ket\psi.
    \label{eq:separated-frequency-sign-expansion}
\end{equation}
To re-write the expression in the form of Eq.~\eqref{eq:hilbert-valued-large-sieve}, we denote the density $g_{m,\vec b}(y)$ for a fixed sign vector $\vec b$ by
\begin{equation}
    g_{m,\vec b}(y)
    =
    \int_{\cD_m(I)}d\vec t\,
    \delta(y-\vec b\cdot\vec t)
    \cZ_1\cdots\cZ_m\ket\psi. \label{sm:eq:density}
\end{equation}
The support of $g_{m,\vec b}(y)$ has length at most $m\tau$.
For state-valued functions, let
$\|h\|_{1}=\int\|h(y)\|_2dy$ and
$\|h\|_{\infty}=\operatorname*{sup}_y\|h(y)\|_2$. Integrating over the scalar variable $t_m$ in Eq.~\eqref{sm:eq:density} removes the delta function, we can bound the remaining integration and obtain
\begin{equation}
    \|g_{m,\vec b}\|_{\infty}
    \leq\frac{\tau^{m-1}}{(m-1)!},
    \qquad
    \|g_{m,\vec b}\|_{1}
    \leq\frac{\tau^m}{m!}.
\end{equation}
Consequently,
\begin{equation}
    \int\|g_{m,\vec b}(y)\|_2^2dy
    \leq
    \frac{\tau^{2m-1}}{m!(m-1)!}.
    \label{eq:pushforward-density-bound}
\end{equation}
Applying Eq.~\eqref{eq:hilbert-valued-large-sieve} and then the triangle
inequality to the $2^m$ sign vectors yields
\begin{equation}
    \left(
    \sum_{k=1}^N\|\ket{V_m(\omega_k)}\|_2^2
    \right)^{1/2}
    \leq
    \alpha^m
    \left[
        \left(m\tau+\frac{2\pi}{\Delta}\right)
        \frac{\tau^{2m-1}}{m!(m-1)!}
    \right]^{1/2}.
    \label{eq:separated-frequency-dyson-l2-bound}
\end{equation}

Since
$(\wt U_{k,\alpha}(I)-I)\ket\psi=\sum_{m\geq1}\ket{V_m(\omega_k)}$,
another application of the triangle inequality gives
\begin{align}
    &\left[
    \sum_{k=1}^N
    \left\|
    (\wt U_{k,\alpha}(I)-I)\ket\psi
    \right\|_2^2
    \right]^{1/2}\leq
    \sum_{m=1}^\infty
    x^m
    \sqrt{\frac{m+q}{m!(m-1)!}},
    \qquad
    q:=\frac{2\pi}{\Delta\tau}.
    \label{eq:separated-frequency-dyson-sum}
\end{align}
Using $\sqrt{m!(m-1)!}=\sqrt m\,(m-1)!$ and
$\sqrt{m+q}\leq\sqrt m+\sqrt q$, we find, for $x\leq1$,
\begin{align}
    \sum_{m=1}^\infty
    x^m\sqrt{\frac{m+q}{m!(m-1)!}}
    \leq
    xe^x+\sqrt q\,xe^x
    \leq
    e\left[
        x+\sqrt{2\pi\frac{\alpha}{\Delta}x}
    \right].
\end{align}
Squaring and using $(a+b)^2\leq2(a^2+b^2)$ proves
Eq.~\eqref{eq:separated-frequency-short-time}.
\end{proof}

\subsubsection{Accumulation over the full protocol}

\begin{theorem}[Lower bound for arbitrary separated frequencies]
\label{thm:arbitrary-separated-one-sparse-lower-bound}
Let the candidate frequencies satisfy
Eq.~\eqref{eq:arbitrary-separated-frequency-assumption}, let $N\geq3$, and
consider the null signal together with the family in
Eq.~\eqref{eq:arbitrary-separated-hard-family}.  Any protocol which either
distinguishes the null signal from every member of this family with error at
most $1/3$, or identifies $k$ with probability at least $2/3$, must satisfy
\begin{equation}
    T
    \geq
    \frac1\alpha
    \left[
        \sqrt{
        \frac{N}{18e^2(1+2\pi\alpha/\Delta)}
        }
        -1
    \right].
    \label{eq:arbitrary-separated-one-sparse-lower-bound}
\end{equation}
In particular, if $\alpha/\Delta=O(1)$, then
$T=\Omega(\sqrt N/\alpha)$.
\end{theorem}

\begin{proof}
Using the purified representation described above, partition a protocol of
worst-case duration $T$ into $R=\lceil\alpha T\rceil$ intervals of length at
most $1/\alpha$.  Let $\wt U_{r,k}$ be the toggling-frame propagator on
interval $r$ under $f_k^{(\alpha)}$.  Starting from the global purified state
$\lvert\psi_0\rangle$, define
\begin{equation}
    \lvert\psi_r^{(k)}\rangle
    =
    \wt U_{r,k}\cdots\wt U_{1,k}\lvert\psi_0\rangle,
    \qquad
    \lvert\delta_r^{(k)}\rangle
    =
    \lvert\psi_r^{(k)}\rangle-\lvert\psi_0\rangle,
\end{equation}
and
\begin{equation}
    D_r
    :=
    \frac1N\sum_{k=1}^N\|\lvert\delta_r^{(k)}\rangle\|_2^2,
    \qquad
    d_r
    :=
    \frac1N\sum_{k=1}^N
    \|(\wt U_{r,k}-I)\lvert\psi_0\rangle\|_2^2.
\end{equation}
The identity
\begin{equation}
    \lvert\delta_r^{(k)}\rangle
    =
    (\wt U_{r,k}-I)\lvert\psi_0\rangle
    +\wt U_{r,k}\lvert\delta_{r-1}^{(k)}\rangle
\end{equation}
together with the triangle and Cauchy--Schwarz inequalities gives
\begin{equation}
    \sqrt{D_r}
    \leq
    \sqrt{D_{r-1}}+\sqrt{d_r}.
    \label{eq:arbitrary-separated-composition-recursion}
\end{equation}
Lemma~\ref{lem:separated-frequency-short-time}, applied with
$x_r=\alpha\tau_r\leq1$, implies
\begin{equation}
    d_r
    \leq
    \frac{2e^2}{N}
    \left(1+2\pi\frac{\alpha}{\Delta}\right).
\end{equation}
Iterating Eq.~\eqref{eq:arbitrary-separated-composition-recursion} from
$D_0=0$ therefore gives
\begin{equation}
    D_R
    \leq
    \frac{2e^2R^2}{N}
    \left(1+2\pi\frac{\alpha}{\Delta}\right).
    \label{eq:arbitrary-separated-distance-upper-bound}
\end{equation}

It remains to lower-bound $D_R$ using the success condition.  For
Search, let $E_k$ be the final POVM event that the protocol outputs
$k$, and write $q_k=\Pr(E_k\mid f_0)$. Since the difference in probability of any measurement event cannot exceed the trace distance (TD):
\begin{equation}
    \operatorname{TD}\!\left(
        \lvert\psi_R^{(k)}\rangle\!\langle\psi_R^{(k)}\rvert,
        \lvert\psi_0\rangle\!\langle\psi_0\rvert
    \right)
    \geq
    \frac23-q_k.
\end{equation}
Since the events $E_k$ are mutually exclusive,
$\sum_k q_k\leq1$.  For pure states, trace distance is at most the Euclidean
distance between the corresponding state vectors.  Averaging over $k$ and
then using the root-mean-square--arithmetic-mean inequality therefore gives
\begin{equation}
    \sqrt{D_R}
    \geq
    \frac1N\sum_{k=1}^N\|\lvert\delta_R^{(k)}\rangle\|_2
    \geq
    \frac23-\frac1N,
\end{equation}
and hence
\begin{equation}
    D_R\geq\left(\frac23-\frac1N\right)^2\geq\frac19,
    \qquad N\geq3.
    \label{eq:arbitrary-separated-identification-distance-lower-bound}
\end{equation}
For Detection, let $E$ be the event ``signal present.''  The assumed error
bound gives, for every $k$,
\begin{equation}
    \operatorname{TD}\!\left(
        \lvert\psi_R^{(k)}\rangle\!\langle\psi_R^{(k)}\rvert,
        \lvert\psi_0\rangle\!\langle\psi_0\rvert
    \right)
    \geq
    \Pr(E\mid f_k^{(\alpha)})-\Pr(E\mid f_0)
    \geq\frac13,
\end{equation}
which again implies $D_R\geq1/9$.  Combining either lower bound with
Eq.~\eqref{eq:arbitrary-separated-distance-upper-bound} and using
$R\leq\alpha T+1$ proves
Eq.~\eqref{eq:arbitrary-separated-one-sparse-lower-bound}.
\end{proof}

\begin{proof}[Proof of Theorem~\ref{thm:p3-p5-lower-bound}]
For P3 and P4, apply Theorem~\ref{thm:arbitrary-separated-one-sparse-lower-bound} with $\alpha=\epsilon$: solving these tasks requires, respectively, Detection and Search of the unknown monochromatic signal. For P5, take $\alpha=3\epsilon\le1$. Successful Learning identifies the active frequency, since its estimated cosine coefficient is at least $2\epsilon$, whereas every inactive cosine coefficient has magnitude at most $\epsilon$. In each case, $\alpha=\Theta(\epsilon)$ and $\alpha/\Delta=O(1)$, so the theorem gives $T=\Omega(\sqrt N/\epsilon)$.
\end{proof}

\subsection{Fundamental limit for periodic signal learning}
\label{app:periodic-learning-lower-bound}

For periodic signals, an information-theoretic argument removes the restriction $\eps<\Delta$ from the P5 lower bound from the previous section, at the cost of a logarithmic factor. This lower bound is particularly useful for bounding the fundamental limit of wireless communication, as explained in Appendix~\ref{app:qg}.

\begin{theorem}[Lower bound for periodic signal learning]
\label{thm:periodic-learning-lower-bound}
Let $N\ge3$, $\omega_n=n\Delta$ with $\Delta>0$, and $0<\eps<v_{\max}$.
Any protocol that solves Problem~\ref{problem:signal_learning} with success probability at least $2/3$ for every signal satisfying $\|\vec v\|_\infty\le v_{\max}$ requires worst-case elapsed sensing time
\begin{equation}
    T=\Omega\!\left(
        \frac{\sqrt N}{\eps\sqrt{\log(N)}}
    \right).
    \label{eq:periodic-learning-lower-bound}
\end{equation}
The bound permits arbitrary nonsensing ancillas, signal-independent controls, intermediate measurements, and adaptive feedback, with no restriction on $\eps/\Delta$.
\end{theorem}

\begin{proof}
Set $\alpha=\min\{2\eps,v_{\max}\}$ and consider the signals
\begin{equation}
    f_x(t)=\alpha p_x(\Delta t),
    \qquad
    p_x(\theta)=\sum_{n=1}^N x_n\cos(n\theta),
    \qquad
    x\in\{-1,1\}^N.
\end{equation}
Their coefficients are $v_n=\alpha x_n$ and $v_{n+N}=0$.
Since $\eps<\alpha\le v_{\max}$, successful learning determines every sign $x_n$ from the estimated cosine coefficients.

We first restrict to an exponentially large family whose waveforms are uniformly bounded.
For independent uniform signs $X_n$, Hoeffding's inequality gives
\begin{equation}
    \Pr\!\left(|p_X(\theta)|>u\right)
    \le 2e^{-u^2/(2N)}
\end{equation}
at every fixed $\theta$.
Choose
\begin{equation}
    m=\lceil2\pi N^2\rceil,
    \qquad
    u=\sqrt{2N\log(4m)}.
\end{equation}
A union bound over $\theta_j=2\pi j/m$, $0\le j<m$, shows that at least half of all sign vectors satisfy $|p_x(\theta_j)|\le u$ at every grid point.
Moreover, $\|p_x'\|_\infty\le\sum_{n=1}^N n\le N^2$, while the grid spacing is at most $N^{-2}$.
Consequently, there is a set $\mathcal C\subseteq\{-1,1\}^N$ such that
\begin{equation}
    |\mathcal C|\ge2^{N-1},
    \qquad
    \sup_{\theta\in\mathbb R}|p_x(\theta)|
    \le\Lambda_N:=u+1
    =O\!\left(\sqrt{N\log(N)}\right)
\end{equation}
for every $x\in\mathcal C$.
Periodicity makes this bound uniform over all sensing times.

Purify the sensing protocol, including its measurement records and classical randomness, and denote the full sensor--ancilla system by $Q=sA$, where $s$ is the sensor qubit and $A$ contains all ancillary systems.
Introduce an auxiliary label register $L$ which models the nature that stores the ground truth, and consider the joint state
\begin{equation}
    |\Psi(t)\rangle
    =\frac1{\sqrt{|\mathcal C|}}
      \sum_{x\in\mathcal C}|x\rangle_L|\psi_x(t)\rangle_Q,
\end{equation}
where $|\psi_x(t)\rangle_Q$ is the purified sensing state under the signal $f_x$, and $|\psi_x(0)\rangle=|\psi_0\rangle$ is independent of $x$.
The signal interaction in this enlarged system is
\begin{equation}
    K_{Ls}(t)
    =\alpha\sum_{x\in\mathcal C}
      p_x(\Delta t)|x\rangle\langle x|_L\otimes Z_s,
    \qquad
    \|K_{Ls}(t)\|\le\alpha\Lambda_N.
\end{equation}
All sensing controls, including purified adaptive feedback, act only on $Q$ and therefore preserve the entanglement entropy across $L|Q$.
Because $K_{Ls}(t)$ couples the label register only to the two-dimensional sensor, the small incremental entangling theorem~\cite{vanacoleyen2013entanglement} gives
\begin{equation}
    \left|\frac{d}{dt}S(\rho_Q(t))\right|
    \le C_{\mathrm{SIE}}\alpha\Lambda_N\log2,
\end{equation}
where $\rho_Q(t)=\operatorname{Tr}_L|\Psi(t)\rangle\langle\Psi(t)|$ and $C_{\mathrm{SIE}}$ is a universal constant independent of the ancillary dimensions.
Since $S(\rho_Q(0))=0$, integration yields
\begin{equation}
    S(\rho_Q(T))
    \le C_{\mathrm{SIE}}\alpha\Lambda_N T\log2.
    \label{eq:periodic-learning-information-upper}
\end{equation}

Now take $X$ uniform on $\mathcal C$.
A successful coefficient estimate determines all signs $X_n$, since $\eps<\alpha$.
Thus the sensing protocol produces an estimated sign vector $\widehat X$ with $\Pr(\widehat X\ne X)\le1/3$.
Outputs outside $\mathcal C$ may be mapped to a fixed element of $\mathcal C$ without increasing this error probability.
The conditional sensing states $|\psi_x(T)\rangle_Q$ are pure, so the Holevo bound and Fano's inequality imply
\begin{align}
    S(\rho_Q(T))
    &\ge I(X;\widehat X) \nonumber\\
    &\ge \frac23\log|\mathcal C|-\log2
     \ge \frac{2N-5}{3}\log2.
\end{align}
Combining this with Eq.~\eqref{eq:periodic-learning-information-upper} gives
\begin{equation}
    T\ge
    \frac{2N-5}{3C_{\mathrm{SIE}}\alpha\Lambda_N}
    =\Omega\!\left(
        \frac{\sqrt N}{\eps\sqrt{\log(2N)}}
    \right),
\end{equation}
where we used $\alpha\le2\eps$.
\end{proof}

\subsection{Fundamental limit for ancilla-free signal learning}\label{app:ancilla-free-learning-bound}

In this section, we prove an $\Omega(N/\varepsilon)$ lower bound on the sensing time for any ancilla-free Learning algorithms
We use the same notation as before:
Norms of vectors are Euclidean, and matrix norms are the induced operator norms. We write
$[N]=\{1,\ldots,N\}$.
Recall that $N\ge 1$ and $\Delta>0$. The known frequencies satisfy
\begin{align}
\omega_1\ge\Delta,\qquad \omega_{n+1}-\omega_n\ge\Delta\quad (1\le n<N).
\end{align}
The unknown vector is $v\in[-1,1]^N$, and the Hamiltonian considered in this section is
\begin{align}
H_{\vec{v}}(t)=f_{\vec{v}}(t)Z,\qquad f_{\vec{v}}(t)=\sum_{n=1}^N v_n\cos(\omega_n t).
\end{align}
An ancilla-free sensing protocol uses at most $m<\infty$ experiments. Each experiment prepares a qubit state, evolves
it over an interval of duration $\tau$ with arbitrary intermediate signal-independent unitary controls, and
measures a discrete POVM $\{M_y\}_y$. Thus $M_y\succeq 0$ and $\sum_y M_y=\id$. The preparation, controls, interval, and POVM may depend on previous outcomes and
private classical randomness. Only classical memory is retained between experiments.

The total sensing time is $\cT=\sum_{j=1}^m\tau_j$. Early stopping is represented by zero-duration dummy
experiments. The number $m$ may be arbitrary; our bound will not depend on it.

\begin{theorem}\label{thm:main}
Suppose $0<\eps<1$. If such a protocol outputs $\wt v$ satisfying
\begin{align}
\bbP(\pnorm{\infty}{\wt v-v}\le\eps)\ge\frac{2}{3}\qquad\text{for every }v\in[-1,1]^N,
\end{align}
then its worst-case expected sensing time satisfies
\begin{align}
T:=\sup_{v\in[-1,1]^N}\E_v\cT\ge c \frac{N}{\eps(1+2\eps/\Delta)}\label{eq:main-bound}
\end{align}
for a universal constant $c>0$. When $\epsilon/\Delta = O(1)$, we have $T = \Omega(N/\epsilon)$.
\end{theorem}
We first outline the argument before presenting its technical ingredients. Set $a=\min\{2\eps,1\}$ and consider amplitudes $aX$, where $X\in\{-1,1\}^N$. Since $a>\eps$, successful amplitude estimation determines every sign. Under a uniform prior, this requires a constant amount of information per sign.

To bound the information gained from measurements, we imagine providing the observer with an additional noisy copy of the unknown amplitudes, introduced solely as a proof device. Although this copy leaves most individual signs unresolved, it already suffices to predict the measurement statistics of short experiments. We therefore focus on the information that actual measurements provide beyond this partial knowledge. Successful learning still requires additional information proportional to the number of unknown signs, whereas the noisy copy captures collective features of the signal that can be learned quickly. This allows us to isolate the information that is harder to acquire for successful learning.  The following lemma bounds the information gained from a single experiment conditioned on this noisy copy.

\begin{lemma}[Information gained in one experiment]\label{lem:information}
Let $0<a\le 1$, and give $X\in\{-1,1\}^N$ any
probability distribution. Fix one qubit experiment of duration $\tau$, and let $Y$ be its measurement
outcome when the amplitudes are $aX$. Draw a standard Gaussian vector $G$ independently of $X$ and
the experiment, and put $W=aX+4aG$. Then the mutual information between $X$ and $Y$, conditioned on $W$, satisfies
\begin{align}
I(X;Y\mid W)\le C\,a\left(1+\frac{a}{\Delta}\right)\tau\label{eq:one-experiment}
\end{align}
for a universal constant $C$. The experiment may use any qubit preparation, unitary controls, and
discrete POVM.
\end{lemma}

The allowance of \emph{every} distribution of $X$ is what handles adaptivity. After a history is fixed, both
the next experiment and the conditional distribution of $X$ are fixed. The lemma applies to precisely
that pair. The same auxiliary Gaussian is independent of every actual history. Consequently the
information chain rule adds \eqref{eq:one-experiment} over the experiments. Since $a\le 2\eps$, this will give an upper
bound proportional to $\eps(1+2\eps/\Delta)T$ for information that must be proportional to $N$.

We prove the one-experiment lemma first. Its only substantive input is an estimate of the
final state from noisy amplitudes. Frequency separation controls the coefficients of the evolution
expansion, and elementary Gaussian moments control the error of that estimate. We then convert
the vector error directly into an information bound.

\begin{lemma}[An elementary frequency estimate]\label{lem:frequency}
Let $I$ be an interval of length $\tau>0$, and let
$g:I\to\C^d$ have $\int_I\opnorm{g(t)}^2\,dt<\infty$. Then
\begin{align}
\sum_{n=1}^N\opnorm{\int_I\cos(\omega_n t)g(t)\,dt}_2^2
\le L\int_I\opnorm{g(t)}_2^2\,dt,\qquad L=\tau+\frac{8}{\Delta}.\label{eq:frequency}
\end{align}
The constant does not depend on $N$, $d$, or the location of $I$.
\end{lemma}

\begin{proof}
This is essentially a formal statement of Eq.~\eqref{eq:arbitrary-separated-frequency-assumption}.
We first bound a sum of oscillations. Let $t_0$ be the midpoint of $I$, set $s=\tau+4/\Delta$, and
introduce the triangular weight
\begin{align}
w(t)=\begin{cases}
1-|t-t_0|/s,& |t-t_0|\le s,\\
0,& |t-t_0|>s.
\end{cases}
\end{align}
It satisfies $w\ge 1/2$ on $I$ and $\int_\R w(t)\,dt=s$. Direct integration of the two linear pieces gives, for
$\xi\ne 0$,
\begin{align}
\int_\R w(t)e^{i\xi t}\,dt=e^{i\xi t_0}\frac{2(1-\cos(\xi s))}{s\xi^2},\qquad
\left|\int_\R w(t)e^{i\xi t}\,dt\right|\le\frac{4}{s\xi^2}.
\end{align}

Order the signed frequencies $\{-\omega_N,\ldots,-\omega_1,\omega_1,\ldots,\omega_N\}$ as $\lambda_1<\cdots<\lambda_{2N}$. They are $\Delta$-separated, so $|\lambda_j-\lambda_k|\ge\Delta|j-k|$. For any fixed $j$, the sum of the off-diagonal integral bounds
is
\begin{align}
\sum_{k\ne j}\frac{4}{s|\lambda_j-\lambda_k|^2}
\le\frac{8}{s\Delta^2}\sum_{r\ge 1}\frac{1}{r^2}\le\frac{16}{s\Delta^2}.
\end{align}
The last series is at most $2$: its first term is $1$, and for $r\ge 2$ we use $r^{-2}\le(r-1)^{-1}-r^{-1}$.

For vectors $z_1,\ldots,z_{2N}$, expand the squared norm of $\sum_j e^{i\lambda_jt}z_j$. Bound each off-diagonal term using $2|\braket{z_j}{z_k}|\leq \opnorm{z_j}_2^2+\opnorm{z_k}_2^2$. The preceding row-sum estimate gives
\begin{align}
\int_{\R}w(t)\opnorm{\sum_j e^{i\lambda_jt}z_j}_2^2\,dt
&\leq \left(s+\frac{16}{s\Delta^2}\right)\sum_j\opnorm{z_j}_2^2
\leq L\sum_j\opnorm{z_j}_2^2.
\end{align}
Indeed, $s\geq 4/\Delta$, so $s+16/(s\Delta^2)\leq \tau+8/\Delta=L$.

Write each cosine as the sum of its two exponentials divided by $2$. Assigning the vector $z_n/2$ to each of $\omega_n$ and $-\omega_n$ gives
\begin{align}
\int_I\opnorm{\sum_n z_n\cos(\omega_nt)}_2^2\,dt
&\leq 2\int_{\R}w(t)\opnorm{\sum_n z_n\cos(\omega_nt)}_2^2\,dt
\leq L\sum_n\opnorm{z_n}_2^2.
\label{eq:cosine-synthesis}
\end{align}
To obtain the claimed inequality, put $c_n=\int_I\cos(\omega_nt)g(t)\,dt$ and $A=\sum_n\opnorm{c_n}_2^2$. Cauchy--Schwarz and \eqref{eq:cosine-synthesis} give
\begin{align}
A&=\int_I\left\langle\sum_n c_n\cos(\omega_nt),g(t)\right\rangle\,dt
\leq \sqrt{LA}\left(\int_I\opnorm{g(t)}_2^2\,dt\right)^{1/2}.
\end{align}
If $A>0$, divide by $\sqrt{A}$ and square; if $A=0$, the claim already holds.
\end{proof}

\begin{lemma}[Gaussian moment counting]\label{lem:gaussian}
Let $\vec{G},\vec{W}\in\R^N$ have independent standard Gaussian coordinates, with $\vec{G}$ and $\vec{W}$ independent, and write $\vec{\zeta}=\vec{G}+i\vec{W}$. For an integer $k\geq 1$ and arbitrary vectors $c_{\mathbf{i}}\in\C^d$ indexed by $\mathbf{i}=(i_1,\ldots,i_k)\in[N]^k$,
\begin{align}
\E\opnorm{\sum_{\mathbf{i}}c_{\mathbf{i}}\zeta_{i_1}\cdots\zeta_{i_k}}_2^2
&\leq 2^kk!\sum_{\mathbf{i}}\opnorm{c_{\mathbf{i}}}_2^2.
\label{eq:gaussian-moments}
\end{align}
\end{lemma}

\begin{proof}
For one coordinate and nonnegative integers $p,q$, rotational symmetry gives $\E\zeta_j^p\overline{\zeta_j}^{\,q}=0$ if $p\neq q$. For $p=q$, polar coordinates and integration by parts give
\begin{align}
\E|\zeta_j|^{2p}&=\int_0^{\infty}r^{2p+1}e^{-r^2/2}\,dr=2^pp!.
\end{align}
Thus independence gives
\begin{align}
\E\left[\prod_{\ell=1}^k\zeta_{i_\ell}\prod_{\ell=1}^k\overline{\zeta_{j_\ell}}\right]
&=2^k\sum_{\pi\in S_k}\prod_{\ell=1}^k\mathbf{1}_{\{i_\ell=j_{\pi(\ell)}\}},
\end{align}
where $S_k$ is the set of permutations of $[k]$. Indeed, the expectation vanishes unless the two index lists have the same multiplicities. If index $n$ occurs $r_n$ times in both lists, the expectation is $2^k\prod_n r_n!$, and $\prod_n r_n!$ is exactly the number of permutations matching the two lists.

Expand the squared norm in \eqref{eq:gaussian-moments}. For each of the $k!$ permutations, Cauchy--Schwarz bounds the resulting inner product of two permuted coefficient arrays by $\sum_{\mathbf{i}}\opnorm{c_{\mathbf{i}}}_2^2$. Summing these bounds proves the claim.
\end{proof}

The next lemma constructs the state estimate used in the proof. Its error bound is uniform in the true parameter: the estimator is a single function of the noisy observation, not a function chosen separately for each possible true parameter.

\begin{lemma}[Estimating a final state from noisy amplitudes]\label{lem:state-estimate}
Fix an experiment with a pure initial state and duration $\tau>0$. Let $\psi({\vec{v}})\in\C^2$ be its final unit vector for real amplitudes ${\vec{v}}$, and put
\begin{align}
R^2&=\tau\left(\tau+\frac{8}{\Delta}\right).
\end{align}
For $\sigma>0$ with $\sigma R\leq 1/4$, there is a function $F_\sigma:\R^N\to\C^2$, depending only on the experiment and $\sigma$, such that for every real ${\vec{v}}$,
\begin{align}
\E_{\vec{G}}\opnorm{F_\sigma({\vec{v}}+\sigma \vec{G})-\psi(v)}_2^4&\leq(4\sigma R)^4.
\label{eq:state-estimate}
\end{align}
Here $\vec{G}$ is a standard Gaussian vector.
\end{lemma}

\begin{proof}
Following the known controls puts the Hamiltonian in the form $f_{\vec{v}}(t)A(t)$, with $A(t)$ a known Hermitian unitary. A final known unitary can be absorbed into the measurement. Write $I=[c,d]$ and let $U_v(t,c)$ be the propagator. The prepared unit vector is $\psi_0$, so $\psi(\vec{v})=U_{\vec{v}}(d,c)\psi_0$.

We also use the same evolution equation with complex coefficients in $f_{\vec{v}}$. These coefficients are an algebraic device in the proof. The proposed estimator is
\begin{align}
F_\sigma(\vec{w})&=\E_{\vec{W}}\psi(\vec{w}+i\sigma \vec{W}),
\label{eq:estimator}
\end{align}
where $\vec{W}$ is an independent standard Gaussian vector. The introduced imaginary part is used to subtract the bias in the second moment induced by the Gaussian noise $\vec{G}$.

Fix a real $\vec{v}$ and a perturbation $\vec{z}\in\C^N$, now we want to find an explicit expression for $\psi(\vec{v}+\vec{z})$. For convenience of calculation, we make a change of basis
\begin{align}
B_{\vec{v}}(t)&=U_{\vec{v}}(t,c)^\dagger A(t)U_{\vec{v}}(t,c),\\
\chi_{\vec{z}}(t)&=U_{\vec{v}}(t,c)^\dagger U_{{\vec{v}}+{\vec{z}}}(t,c)\psi_0.
\end{align}
Differentiating $\chi_{\vec{v}}(t)$ with respect to $t$ reveals the relation (Dyson series)
\begin{align}
\chi_{\vec{z}}(t)&=\psi_0-i\int_c^t f_{\vec{z}}(r)B_{\vec{v}}(r)\chi_{\vec{z}}(r)\,dr.
\label{eq:integral-evolution}
\end{align}
Repeated substitution in \eqref{eq:integral-evolution} gives
\begin{align}
\psi(\vec{v}+\vec{z})&=\psi(\vec{v})+\sum_{k\geq 1}P_{k,\vec{v}}(\vec{z}),
&P_{k,\vec{v}}(\vec{z})&=\sum_{\mathbf{i}\in[N]^k}Q_{\vec{v},\mathbf{i}}z_{i_1}\cdots z_{i_k},
\label{eq:evolution-expansion}
\end{align}
with
\begin{align}
Q_{\vec{v},\mathbf{i}}&=(-i)^kU_{\vec{v}}(d,c)\int_{d>t_1>\cdots>t_k>c}B_{\vec{v}}(t_1)\cdots B_{\vec{v}}(t_k)\psi_0\prod_{\ell=1}^k\cos(\omega_{i_\ell}t_\ell)\,dt_1\cdots dt_k.
\end{align}
To bound the integral, note that the vector $B_{\vec{v}}(t_1)\cdots B_{\vec{v}}(t_k)\psi_0$ has norm one. Set the integrand to zero whenever the times are not ordered as $t_1\ge\cdots\ge t_k$, so the integral can be taken over the full cube $I^k$.
Apply \eqref{eq:frequency} successively in the $k$ time variables, including the remaining finite indices in the vector at
each step. The outer unitary $U_{\vec{v}}(d,c)$ does not affect any norm. We obtain
\begin{align}
\sum_{\mathbf{i}\in[N]^k}\opnorm{Q_{\vec{v},\mathbf{i}}}_2^2
\le L^k\frac{\tau^k}{k!}=\frac{R^{2k}}{k!}.
\label{eq:coefficient-bound}
\end{align}
Here $L=\tau+8/\Delta$; the ordered region has volume $\tau^k/k!$ because the $k!$ possible time orderings
divide the cube $I^k$ equally.

Set $\vec{\zeta}=\vec{G}+i\vec{W}$ as in Lemma~\ref{lem:gaussian}. To bound the fourth moment, use $\opnorm{P}_2^4=\opnorm{P\otimes P}_2^2$.
The degree-$2k$ polynomial $P_{k,\vec{v}}(\sigma\vec{\zeta})\otimes P_{k,\vec{v}}(\sigma\vec{\zeta})$ has coefficients $\sigma^{2k}Q_{\vec{v},\mathbf{i}}\otimes Q_{\vec{v},\mathbf{j}}$. Eqs.~\eqref{eq:gaussian-moments} and~\eqref{eq:coefficient-bound}
give
\begin{align}
\E\opnorm{P_{k,\vec{v}}(\sigma\vec{\zeta})}_2^4
&\le 2^{2k}(2k)!\,\sigma^{4k}\left(\sum_{\mathbf{i}}\opnorm{Q_{\vec{v},\mathbf{i}}}_2^2\right)^2\\
&\le 2^{2k}\frac{(2k)!}{(k!)^2}(\sigma R)^{4k}\le(2\sigma R)^{4k}.
\end{align}
The last step uses $\binom{2k}{k}\le4^k$. The triangle inequality for the fourth root of the fourth moment
therefore gives
\begin{align}
\left(\E\opnorm{\psi(\vec{v}+\sigma \vec{G}+i\sigma \vec{W})-\psi(\vec{v})}_2^4\right)^{1/4}
\le\sum_{k\ge1}(2\sigma R)^k=\frac{2\sigma R}{1-2\sigma R}\le4\sigma R,
\label{eq:fourth-moment}
\end{align}
where we apply repeatedly the inequality $
(\mathbb E\|X+Y\|_2^4)^{1/4}
\le
(\mathbb E\|X\|_2^4)^{1/4}
+
(\mathbb E\|Y\|_2^4)^{1/4}.
$ for random variables $X,Y$.

Since \(F_\sigma\) averages the perturbed state over \(\vec{W}\), Jensen’s inequality gives
$\mathbb E_{\vec{G}}\|F_\sigma(\vec{v}+\sigma \vec{G})-\psi(\vec{v})\|_2^4
\le
\mathbb E_{\vec{G},\vec{W}}\|\psi(\vec{v}+\sigma \vec{G}+i\sigma \vec{W})-\psi(\vec{v})\|_2^4$. Apply this to \eqref{eq:fourth-moment} to
obtain \eqref{eq:state-estimate}.
\end{proof}

We now prove Lemma~\ref{lem:information}. The next calculation explains exactly why a fourth-moment bound is useful. The approximate vector (state) $F_\sigma(\vec{w}) = \ket{F}$ need not be normalized.

\begin{proof}[Proof of Lemma~\ref{lem:information}]
For distributions $p,q$, write $\mathrm{KL}(p\|q)=\sum_y p(y)\log(p(y)/q(y))$. Let $p(\vec{x},y)$ be the probability distribution over random signal vector $X$ and the corresponding measurement outcome $Y$, $
p_{\vec{x}}(y)= p(y|\vec{x})
$ and $\pi_{\vec{x}} = \sum_y p(\vec{x},y)$. We will use the relative-entropy identity
\begin{align}
\sum_{\vec{x}}\pi_{\vec{x}}\mathrm{KL}(p_{\vec{x}}\|q)=I(X;Y)+\mathrm{KL}\left(\sum_{\vec{x}}\pi_{\vec{x}}p_{\vec{x}}\middle\|q\right),
\label{eq:relative-entropy}
\end{align}
Any reference distribution
$q$ therefore gives an upper bound on mutual information. We also use this identity conditionally and, for
Gaussian densities, with integrals in place of sums.

We first establish a measurement estimate. Let $\psi$ be a unit vector, let $F\in\C^2$ be any vector,
and let $\eta>0$. Define
\begin{align}
p_y =\expval{M_y}{\psi},\qquad
r_y=\expval{M_y}{F}+\eta\Tr M_y,\qquad
q_y=\frac{r_y}{\sum_z r_z}.
\end{align}
These comparison probabilities depend only on $F,\eta$ and the POVM. They are well-defined even if
$F=0$, since $\sum_y r_y=\opnorm{F}_2^2+2\eta>0$. We claim that
\begin{align}
\mathrm{KL}(p\|q)\le2\opnorm{\psi-F}_2^2+\frac{\opnorm{\psi-F}_2^4}{\eta}+2\eta.
\label{eq:measurement-estimate}
\end{align}
To prove the claim, put $d=\sum_y r_y$. Using $\log t\le t-1$, \begin{align}
\mathrm{KL}(p\|q)=\sum_y p_y\log(p_y/r_y)+\log d
\le\sum_y p_y^2/r_y+d-2=\sum_y\frac{(p_y-r_y)^2}{r_y}.
\end{align}
Omit zero $M_y$, set
\begin{align}
e=\psi-F,\qquad A=\expval{M_y}{F},\quad
B=\expval{M_y}{e},\quad t=\Tr M_y>0.
\end{align}
Then $r_y=A+\eta t$ and $p_y-r_y=2\operatorname{Re}\bra{F}M_y\ket{e}+B-\eta t$. Cauchy--Schwarz gives $|\bra{F}M_y\ket{e}|\le\sqrt{AB}$.
Apply Cauchy--Schwarz once more, to the vectors $(\sqrt A,\sqrt{\eta t})$ and $(2\sqrt B,|B-\eta t|/\sqrt{\eta t})$. This yields
\begin{align}
\frac{(p_y-r_y)^2}{r_y}
\le\frac{(2\sqrt{AB}+|B-\eta t|)^2}{A+\eta t} \le4B+\frac{(B-\eta t)^2}{\eta t}=2B+\frac{B^2}{\eta t}+\eta t.
\end{align}
Since $M_y\preceq t\id$, we have $B\le t\opnorm{e}^2$. Summing over outcomes, using $\sum_y B=\opnorm{e}^2$ and $\sum_y t=2$,
proves \eqref{eq:measurement-estimate}.

We apply this bound to the physical experiment. First suppose the initial state is a pure state. Let
\begin{align}
R^2=\tau(\tau+8/\Delta),\qquad\alpha=a^2R^2.
\end{align}
If $\tau=0$, the outcome is independent of $X$ and the claim holds. Note that $\alpha$ is determined by $\tau,\Delta$ and $a$. Next, we bound the mutual information separately when $\alpha$ lies in different intervals. If $0<\alpha\le1/256$, set $\sigma=4a$ and
$\eta=256\alpha$. Then $\sigma R\le1/4$, and Lemma~\ref{lem:state-estimate} gives, for every real $\vec{v}$,
\begin{align}
\E_{\vec{G}}\opnorm{\psi(\vec{v})-F_\sigma(\vec{v}+4a\vec{G})}_2^4\le\eta^2,\qquad
\E_{\vec{G}}\opnorm{\psi(\vec{v})-F_\sigma(\vec{v}+4a\vec{G})}_2^2\le\eta.
\end{align}
The second inequality follows from applying the Cauchy--Schwarz to the first. Let $p_{\vec{x}}$ be the measurement outcome distribution when $\vec{v}=a\vec{x}$.
Construct $q_{\vec{w}}$ from $F_\sigma(\vec{w})$ as above. Averaging \eqref{eq:measurement-estimate} over the Gaussian gives
\begin{align}
\E_{\vec{G}}\mathrm{KL}(p_{\vec{x}}\|q_{a\vec{x}+4a\vec{G}})\le5\eta=1280\alpha\qquad\text{for every }\vec{x}.
\label{eq:averaged-divergence}
\end{align}
Applying Eq.~\eqref{eq:relative-entropy} at each fixed $\vec{w}$ then take the average over $\vec{w}$, we get
\begin{align}
I(X;Y\mid W) \le \E_{X,W}\mathrm{KL}(p_X\|q_W) \le 1280\alpha.
\end{align}
For every duration we also have $I(X;Y\mid W)\le\log 2$ from Holevo's bound.
If $\alpha>1/256$, this implies $I(X;Y\mid W)\le 256(\log 2)\alpha\le1280\alpha$ as
well. Thus, in all cases,
\begin{align}
I(X;Y\mid W) \le \min\{\log 2,\ 1280a^2\tau(\tau+8/\Delta)\}.
\label{eq:one-experiment-min}
\end{align}
The bound Eq.~\eqref{eq:one-experiment-min} holds for any experiments with pure initial states, so it holds for mixed initial states too.

Finally, $\min\{1,x^2\}\le x$ for $x\ge0$. Since $\sqrt{1280}<36$,
\begin{align}
\min\{\log 2,1280a^2\tau(\tau+8/\Delta)\}
&\le \min\{1,1280a^2\tau^2\}+\frac{10240a^2\tau}{\Delta} \\
&\le \left(36a+\frac{10240a^2}{\Delta}\right)\tau \\
&\le 10240a(1+a/\Delta)\tau.
\end{align}
This proves Eq.~\eqref{eq:one-experiment}, for example with $C=10240$.
\end{proof}

\begin{proof}[Proof of Theorem~\ref{thm:main}]
We may suppose $T<\infty$. Collect all private classical randomness into a
parameter-independent seed $R_0$. It may be discrete or continuous. The transcript is $S=
(R_0,Y_1,\ldots,Y_m)$; the estimator and all classical decisions are functions of this transcript.

Set $a=\min\{2\eps,1\}$, take $X$ uniform on $\{-1,1\}^N$, and let the true amplitudes be $aX$. Then
$\eps<a\le1$. Independently of $X$ and the entire protocol, draw a standard Gaussian vector $\vec{G}$ and set
$W=aX+4a\vec{G}$.

We first quantify the information required. Decode each sign from $\wt v$. On the success event, every sign is correct because $a>\eps$. Hence every decoded bit has
error probability at most $1/3$. Let $E_i$ indicate whether bit $i$ was decoded incorrectly. Since $X_i$ is
determined by $(S,E_i)$,
\begin{align}
H(X\mid S)\le\sum_i H(X_i\mid S)\le\sum_i H(E_i)\le Nh_2(1/3),
\end{align}
where $h_2(p)=-p\log p-(1-p)\log(1-p)$. The last step uses $h_2'(p)=\log((1-p)/p)\ge0$ on $(0,1/2]$.
The function $\log 2-h_2(p)$ has second derivative $1/[p(1-p)]\ge4$, and both its value and first
derivative vanish at $1/2$. Integrating twice gives $\log 2-h_2(p)\ge2(p-1/2)^2$. Therefore
\begin{align}
I(X;S)=N\log 2-H(X\mid S)\ge N/18.
\end{align}

The noisy fake signal reveals at most $N/32$ information. To verify this, let $q$ be the Gaussian
density with mean zero and covariance $16a^2\id_N$. The log density ratio for $W$ given $X=\vec{x}$ is
\begin{align}
\log\frac{p_{W\mid \vec{x}}(\vec{w})}{q(\vec{w})}=\frac{2a\vec{x}\cdot \vec{w}-\opnorm{a\vec{x}}_2^2}{32a^2}.
\end{align}
Since $\E[W\mid X=\vec{x}]=a\vec{x}$ and $\opnorm{\vec{x}}_2^2=N$, its expectation is $N/32$. Eq.~\eqref{eq:relative-entropy} implies
$I(X;W)\le N/32$. The information chain rule now gives
\begin{align}
I(X;S\mid W)=I(X;S)+I(X;W\mid S)-I(X;W)\ge\frac{7N}{288}.
\label{eq:information-required}
\end{align}

We next bound what the experiments can supply. Write $H_j=(R_0,Y_1,\ldots,Y_{j-1})$ for the history
before experiment $j$. Fix a seed value and an outcome history for that
seed, and call their combined value $h$. The next experiment is now fixed. Apply Lemma~\ref{lem:information} to this
experiment and the conditional distribution of $X$ given $H_j=h$.
Conditioning on $H_j=h$ may correlate the signs, but Lemma~\ref{lem:information} allows any distribution of $X$. Moreover, since $\vec{G}$ was drawn independently of $X$ and all protocol randomness, fixing the history leaves it standard Gaussian and independent of $X$ and the next measurement’s randomness, as required by the lemma. Thus
\begin{align}
I(X;Y_j\mid W,H_j=h)\le Ca(1+a/\Delta)\tau_j(h).
\end{align}
The conditional information includes the average over $W$ at that history. Averaging over histories,
including the seed, and using $I(X;R_0\mid W)=0$, the finite information chain rule yields
\begin{align}
I(X;S\mid W) =\sum_{j=1}^m I(X;Y_j\mid W,H_j) \le Ca(1+a/\Delta)\E\sum_{j=1}^m\tau_j\le 2C\eps(1+2\eps/\Delta) T.
\label{eq:information-supplied}
\end{align}
The last inequality uses $a\le2\eps$. Combining Eqs.~\eqref{eq:information-required} and~\eqref{eq:information-supplied} gives
\begin{align}
T\ge\frac{7}{576C}\frac{N}{\eps(1+2\eps/\Delta)},
\end{align}
which proves Eq.~\eqref{eq:main-bound}.
\end{proof}

\section{Learning algorithm with bosonic sensors}\label{sec:boson}

In this section, we describe an analog approach to realize an efficient Learning of the signal when the signal is coupled to a bosonic sensor. Before we go into details, we summarize the key findings in plain words:
\begin{enumerate}
    \item We consider the signal to be coupled to a bosonic mode $H(t) = f(t)(a+a^\dag)$, where $f(t)$ is a time-dependent signal whose spectrum is supported on a set of $N$ candidate frequencies. The goal is to learn the Fourier components of $f(t)$ within accuracy $\epsilon$. We show that this can be done in time $\widetilde{O}(\sqrt{N}/\epsilon)$ using time evolution of a time-independent, quadratic bosonic Hamiltonian involving the sensor bosonic mode and other $N-1$ bosonic modes that are not coupled to the signal.
    \item
    Given a fixed target estimation accuracy, at any given time, the physical bosonic modes form a product of coherent states. They have bounded total boson number and its variance $\widetilde{O}(N)$. Each mode has bounded time-averaged boson number and its variance $\widetilde{O}(1)$.
    \item The Hamiltonian can be realized with nearest neighbour two-site coupling for a chain and a star-like geometry.
\end{enumerate}
We now present the detailed setting and analysis.

\subsection{Setting}

We consider the same time-dependent signals as in Eq.~\eqref{eq:signal}, now coupled to a single bosonic mode instead of a sensor qubit.
Let $\{\omega_1,\dots,\omega_N\}$ be a set of known frequencies with $\omega_1\ge\Delta$ and $\omega_{n+1}-\omega_n\ge\Delta$ for all $n\in[N-1]$.
The signal is
\begin{align}
    f_{\vec v}(t)=\sum_{n=1}^N\left(v_n\cos(\omega_n t)+v_{N+n}\sin(\omega_n t)\right),
\end{align}
where $\vec v\in\R^{2N}$ contains the unknown cosine and sine amplitudes.
The signal couples to the bosonic mode with annihilation operator $a$ via the Hamiltonian
\begin{align}\label{eq:hamiltonian}
    H(t)=f_{\vec v}(t)\,(a+a^\dagger).
\end{align}

We study the Learning problem P5 (Problem~\ref{problem:signal_learning}) in this setting.

\begin{problem}[Learning with a bosonic sensor]\label{problem:analog-learning}
Assume $\|\vec v\|_\infty\le v_\text{max}=O(1)$.
Using time evolution under the Hamiltonian Eq.~\eqref{eq:hamiltonian}, construct an estimate $\widetilde{\vec v}$ of $\vec v$ such that $\|\widetilde{\vec v}-\vec v\|_\infty\le\epsilon$ with probability at least $1-\delta$.
\end{problem}

\subsection{Learning the signal with $N$ modes}

We give a protocol for Problem~\ref{problem:analog-learning} that uses $N-1$ ancilla bosonic modes $a_2,\dots,a_N$, which do not couple to the signal, together with a signal-independent quadratic control Hamiltonian, a vacuum initial state, and heterodyne measurement.
Set $a_1=a$ and let $U$ be an $N\times N$ unitary matrix whose first column has all entries equal to $1/\sqrt N$.
The modes $b_n=\sum_k U_{nk}a_k$ satisfy the canonical commutation relations, and inverting this relation gives $a=\frac1{\sqrt N}\sum_n b_n$.
During the interval $[0,T]$ we apply the control Hamiltonian $H_c=\sum_n\omega_n b_n^\dagger b_n$.
Rewriting the signal Hamiltonian Eq.~\eqref{eq:hamiltonian} in terms of the modes $b_n$, the total Hamiltonian becomes a sum of $N$ independent driven oscillators,
\begin{align}\label{eq:decoupled}
    H_c+H(t)=\sum_{n=1}^N\left[\omega_n b_n^\dagger b_n+\frac{f_{\vec v}(t)}{\sqrt N}\,(b_n+b_n^\dagger)\right].
\end{align}
In what follows, we show that this simple control Hamiltonian suffices to solve the learning problem. Although the Hamiltonian generally requires all-to-all couplings in the original basis, we demonstrate in a later section that the same task can be accomplished using only local couplings by choosing appropriate $U$.

The Heisenberg picture for each mode is $\dot b_n=-i\omega_n b_n-if_{\vec v}(t)/\sqrt N$, whose exact solution is
\begin{align}\label{eq:displacement}
    b_n(t)=e^{-i\omega_n t}\bigl(b_n(0)+\beta_n(t)\bigr),\qquad
    \beta_n(t)=-\frac{i}{\sqrt N}\int_0^t f_{\vec v}(s)\,e^{i\omega_n s}\,ds.
\end{align}
Here $\beta_n(t)$ is a complex displacement, and its addition to $b_n(0) = b_n$ implicitly includes the identity operator.
Starting from the vacuum, the state at time $t$ is therefore the product of coherent states with amplitudes $e^{-i\omega_n t}\beta_n(t)$.
We recover the signal amplitudes by inverting their exact finite-time response in the coherent displacements.
At time $T$ we perform local heterodyne measurements~\cite{Weedbrook2012} on the physical modes $a_k$.
The outcomes $\hat\alpha_k$ have means $\alpha_k(T)=\langle a_k\rangle(T)$ and independent Gaussian noise of variance $1/2$ in each real and imaginary component.
We then compute
\begin{align}\label{eq:boson-classical-readout}
    \hat\beta_n=e^{i\omega_nT}\sum_{k=1}^N U_{nk}\hat\alpha_k.
\end{align}
Since $b_n=\sum_k U_{nk}a_k$ and $\langle b_n\rangle(T)=e^{-i\omega_nT}\beta_n(T)$, these processed outcomes have means $\beta_n(T)$.
The classical transformation in Eq.~\eqref{eq:boson-classical-readout} is unitary, so it preserves the isotropic Gaussian noise, including independence and variance $1/2$ in each real and imaginary component.
Define the real basis vector and response matrix by
\begin{align}\label{eq:boson-response}
    \phi(t)&=\bigl(\cos(\omega_1t),\dots,\cos(\omega_Nt),\sin(\omega_1t),\dots,\sin(\omega_Nt)\bigr)^T,\nonumber\\
    A(T)&=\frac2T\int_0^T \phi(t)\phi(t)^T\,dt.
\end{align}
Let $z_n=2i\sqrt N\,\hat\beta_n/T$ and $y=(\operatorname{Re}z_1,\dots,\operatorname{Re}z_N,\operatorname{Im}z_1,\dots,\operatorname{Im}z_N)^T$.
Since $f_{\vec v}(t)=\phi(t)^T\vec v$, the exact displacement Eq.~\eqref{eq:displacement} gives
\begin{align}\label{eq:boson-observation}
    y=A(T)\vec v+\xi,\qquad \xi\sim\mathcal N\!\left(0,\frac{2N}{T^2}I_{2N}\right).
\end{align}
The matrix $K(T)$ depends only on the known frequencies and $T$.
Whenever it is invertible, the estimator
\begin{align}\label{eq:boson-estimator}
    \widetilde{\vec v}=
    A(T)^{-1}y
\end{align}
is unbiased and has covariance matrix $(2N/T^2)A(T)^{-2}$.

We choose the sensing time to control the noise after this inversion.
Let $H_m=\sum_{j=1}^m1/j$ for $m\ge1$ and $H_0=0$.
For $0<\delta<1$ and $\epsilon>0$, set
\begin{align}\label{eq:boson-parameters}
    t_0=\frac{4H_{2N-1}}{\Delta},\qquad L=\ln\frac{4N}{\delta},
\end{align}
and take
\begin{align}\label{eq:time}
    T=t_0+\frac{2\sqrt{NL}}{\epsilon}.
\end{align}
We will show that this choice makes Eq.~\eqref{eq:boson-estimator} solve Problem~\ref{problem:analog-learning}.

We first bound finite-time overlaps between the positive and negative signal frequencies.
Let $\lambda_1<\dots<\lambda_{2N}$ be the ordered list $(-\omega_N,\dots,-\omega_1,\omega_1,\dots,\omega_N)$.
The assumptions on the frequencies imply $|\lambda_j-\lambda_l|\ge |j-l|\Delta$.
For $z\in\C^{2N}$, define $S_z(t)=\sum_jz_je^{-i\lambda_jt}$ and its complex Gram matrix
\begin{align}\label{eq:boson-fourier-gram}
    \Gamma_{jl}(T)=\frac1T\int_0^T e^{i(\lambda_j-\lambda_l)t}\,dt.
\end{align}
The diagonal entries equal one, while for $j\ne l$,
\begin{align}
    |\Gamma_{jl}(T)|\le\frac{2}{T|\lambda_j-\lambda_l|},\qquad
    \sum_{l\ne j}|\Gamma_{jl}(T)|\le\frac{2}{\Delta T}\left(\sum_{r=1}^{j-1}\frac1r+\sum_{r=1}^{2N-j}\frac1r\right)\le\frac{t_0}{T}.
\end{align}
Since $\Gamma(T)$ is Hermitian, the row-sum bound yields an operator-norm bound $\|\Gamma(T)-I\|\le t_0/T$ and hence
\begin{align}\label{eq:boson-fourier-bound}
    \left(1-\frac{t_0}{T}\right)\|z\|_2^2
    \le\frac1T\int_0^T|S_z(t)|^2\,dt
    \le\left(1+\frac{t_0}{T}\right)\|z\|_2^2.
\end{align}
This inequality holds for every $T>0$.

We use Eq.~\eqref{eq:boson-fourier-bound} to bound the eigenvalues of the real response matrix $A(T)$.
Set $c_n=v_n+iv_{N+n}$ and define $d\in\C^{2N}$ by $d_{N+n}=c_n$ and $d_{N+1-n}=\overline{c_n}$ for $n\in[N]$.
Then $f_{\vec v}(t)=S_d(t)/2$ and $\|d\|_2^2=2\|\vec v\|_2^2$.
Consequently,
\begin{align}
    \vec v^{\,T}A(T)\vec v=\frac2T\int_0^T f_{\vec v}(t)^2\,dt=\frac1{2T}\int_0^T|S_d(t)|^2\,dt,
\end{align}
and applying Eq.~\eqref{eq:boson-fourier-bound} for arbitrary real $\vec v$ gives
\begin{align}\label{eq:boson-response-spectrum}
    \left(1-\frac{t_0}{T}\right)I_{2N}\preceq A(T)\preceq\left(1+\frac{t_0}{T}\right)I_{2N}.
\end{align}
For $T>t_0$, the estimator Eq.~\eqref{eq:boson-estimator} is therefore well defined, and each of its real components has variance at most $2N/(T-t_0)^2$.
The Gaussian tail bound and a union bound over all $2N$ components yield
\begin{align}\label{eq:boson-learning-tail}
    \Pr\!\left(\|\widetilde{\vec v}-\vec v\|_\infty>\epsilon\right)
    \le4N\exp\!\left[-\frac{\epsilon^2(T-t_0)^2}{4N}\right].
\end{align}
Substituting Eq.~\eqref{eq:time} makes the right side equal to $\delta$.
Thus a single run solves the learning problem with error probability $\delta$ in sensing time
\begin{align}\label{eq:boson-time-scaling}
    T=O\!\left(\frac{\ln(2N)}{\Delta}+\frac{\sqrt{N\ln(4N/\delta)}}{\epsilon}\right).
\end{align}
For fixed $\delta$, this is $\widetilde O(\sqrt N/\epsilon)$ whenever $\epsilon=O(\sqrt N\Delta)$.
The estimator and its error guarantee use the exact dynamics, including all nonresonant terms.

\subsection{Photon number during the protocol}

For the protocol for be physical, the number of bosonic particles (say, photon) needs to grow slowly with the parameters of the sensing problems. In this section, we establish that this is indeed the case by bounding the instantaneous photon number and fluctuation, and the time-averaged photon number and fluctuation in each physical mode $a_k$.

We begin with the time-average photon number.
Because the dynamics preserve products of coherent states, the amplitude and time-averaged photon number of mode $a_k$ are
\begin{align}\label{eq:boson-physical-photons}
    \alpha_k(t)&=\langle a_k\rangle(t)=\sum_{n=1}^N\overline{U_{nk}}e^{-i\omega_nt}\beta_n(t),\nonumber\\
    \overline n_k(T)&=\frac1T\int_0^T\langle a_k^\dagger a_k\rangle(t)\,dt=\frac1T\int_0^T|\alpha_k(t)|^2\,dt.
\end{align}
We will prove a bound that holds for every allowed $U$.

We first separate the linearly growing resonant contribution from the remaining terms in the exact amplitude.
Using the signed frequencies $\lambda_j$ and coefficients $d_j$ defined above, introduce the real $N\times2N$ matrix
\begin{align}\label{eq:boson-denominators}
    R_{nj}=\begin{cases}
        (\omega_n-\lambda_j)^{-1},&\lambda_j\ne\omega_n,\\
        0,&\lambda_j=\omega_n.
    \end{cases}
\end{align}
Every row and every column is a subset of a row or column of the matrix of inverse differences between the $2N$ separated frequencies.
The same harmonic sums used for Eq.~\eqref{eq:boson-fourier-bound} therefore give
\begin{align}\label{eq:boson-denominator-bound}
    \max_n\sum_j|R_{nj}|\le\frac{2H_{2N-1}}{\Delta}=\frac{t_0}{2},\qquad
    \max_j\sum_n|R_{nj}|\le\frac{t_0}{2},\qquad
    \|R\|\le\frac{t_0}{2}.
\end{align}
The operator-norm bound follows by taking the square root of the product of the maximum absolute row and column sums.
Let $\eta=Rd$.
Since $|c_n|\le\sqrt2v_\text{max}$ and $\|d\|_\infty\le\sqrt2v_\text{max}$, the row bound implies
\begin{align}\label{eq:boson-transient-bound}
    \|\eta\|_\infty\le\frac{\sqrt2v_\text{max}\,t_0}{2}.
\end{align}
Fix $k$ and write $u_n=\overline{U_{nk}}$, so that $\|u\|_2=1$.
Integrating each exponential in Eq.~\eqref{eq:displacement} gives the exact identity
\begin{align}\label{eq:boson-amplitude-expanded}
    \alpha_k(t)=\frac1{2\sqrt N}\left[-it\sum_nu_nc_ne^{-i\omega_nt}
    +\sum_nu_n\eta_ne^{-i\omega_nt}
    -\sum_jd_j(R^Tu)_je^{-i\lambda_jt}\right].
\end{align}
Indeed, for $\lambda_j\ne\omega_n$, the corresponding integral contributes
$-u_nd_j(e^{-i\lambda_jt}-e^{-i\omega_nt})/[2\sqrt N(\omega_n-\lambda_j)]$, while $\lambda_j=\omega_n$ gives the first term in Eq.~\eqref{eq:boson-amplitude-expanded}.
Define vectors $s,p,q\in\C^{2N}$ by
\begin{align}\label{eq:boson-amplitude-coefficients}
    s_{N+n}=u_nc_n,\qquad p_{N+n}=u_n\eta_n\quad(n\in[N]),\qquad
    s_j=p_j=0\quad(j\le N),\qquad q_j=d_j(R^Tu)_j,
\end{align}
and set $r=p-q$.
In terms of the exponential sums $S_z(t)$, Eq.~\eqref{eq:boson-amplitude-expanded} becomes
\begin{align}\label{eq:boson-amplitude-exact}
    \alpha_k(t)=\frac{-itS_s(t)+S_r(t)}{2\sqrt N}.
\end{align}
The first term contains the resonant growth, and the second retains both the nonresonant response and the terms from the lower endpoint of the time integral.

The coefficients in Eq.~\eqref{eq:boson-amplitude-exact} have bounds independent of the physical mode.
Using $\|u\|_2=1$, $|c_n|\le\sqrt2v_\text{max}$, and Eq.~\eqref{eq:boson-transient-bound}, we obtain
\begin{align}
    \|s\|_2^2=\sum_n|u_n|^2|c_n|^2\le2v_\text{max}^2,\qquad
    \|p\|_2\le\|\eta\|_\infty\le\frac{\sqrt2v_\text{max}t_0}{2}.
\end{align}
Also, Eq.~\eqref{eq:boson-denominator-bound} gives
\begin{align}
    \|q\|_2\le\|d\|_\infty\|R^Tu\|_2\le\frac{\sqrt2v_\text{max}t_0}{2}.
\end{align}
Since $R$ is real, its transpose has the same operator norm.
The triangle inequality therefore yields
\begin{align}\label{eq:boson-coefficient-bound}
    \|s\|_2\le\sqrt2v_\text{max},\qquad \|r\|_2\le\sqrt2v_\text{max}t_0.
\end{align}

We now bound the time integral in Eq.~\eqref{eq:boson-physical-photons}.
For a scalar function $g$, write $\|g\|_T=(T^{-1}\int_0^T|g(t)|^2dt)^{1/2}$.
By Eq.~\eqref{eq:boson-amplitude-exact}, the triangle inequality, and $t\le T$,
\begin{align}
    \sqrt{\overline n_k(T)}
    &\le\frac{T\|S_s\|_T+\|S_r\|_T}{2\sqrt N}\nonumber\\
    &\le\frac1{2\sqrt N}\sqrt{1+\frac{t_0}{T}}\left(T\|s\|_2+\|r\|_2\right)\nonumber\\
    &\le\frac{v_\text{max}}{\sqrt{2N}}\sqrt{1+\frac{t_0}{T}}(T+t_0),
\end{align}
where the second and third inequalities use Eqs.~\eqref{eq:boson-fourier-bound} and~\eqref{eq:boson-coefficient-bound}.
Squaring gives, for every $k\in[N]$ and $T>0$,
\begin{align}\label{eq:boson-average-bound}
    \overline n_k(T)\le\frac{v_\text{max}^2}{2N}\left(1+\frac{t_0}{T}\right)(T+t_0)^2.
\end{align}
For the learning time Eq.~\eqref{eq:time}, we have $T>t_0$, so
\begin{align}\label{eq:boson-average-learning}
    \overline n_k(T)\le\frac{4v_\text{max}^2T^2}{N}
    \le\frac{32v_\text{max}^2L}{\epsilon^2}+\frac{128v_\text{max}^2H_{2N-1}^2}{N\Delta^2}
    =O\!\left(v_\text{max}^2\left[\frac{\ln(4N/\delta)}{\epsilon^2}+\frac{\ln^2(2N)}{N\Delta^2}\right]\right).
\end{align}
If $t_0\le2\sqrt{NL}/\epsilon$, equivalently $\epsilon\le\Delta\sqrt{NL}/(2H_{2N-1})$, then
\begin{align}\label{eq:boson-resource-regime}
    T=O\!\left(\frac{\sqrt{N\ln(4N/\delta)}}{\epsilon}\right),\qquad
    \overline n_k(T)=O\!\left(\frac{v_\text{max}^2\ln(4N/\delta)}{\epsilon^2}\right).
\end{align}
For fixed $\delta$, a sufficient condition for Eq.~\eqref{eq:boson-resource-regime} is $\epsilon=O(\Delta\sqrt{N/\ln(2N)})$.
Under only $\epsilon=O(\sqrt N\Delta)$, the sensing time retains its $\widetilde O(\sqrt N/\epsilon)$ scaling, while the uniform photon-number bound is Eq.~\eqref{eq:boson-average-learning}.

We finally derive the upper bound of the photon number at a fixed time and upper bounds for the photon-number variances.
The exact displacement integral and Eq.~\eqref{eq:boson-denominator-bound} imply
\begin{align}
    |\beta_n(t)|\le\frac1{2\sqrt N}\left(t|c_n|+2\sum_j|R_{nj}||d_j|\right)
    \le\frac{v_\text{max}}{\sqrt{2N}}(t+t_0).
\end{align}
Since $U$ preserves total photon number,
\begin{align}\label{eq:total-photons}
    \langle b_n^\dagger b_n\rangle(t)\le\frac{v_\text{max}^2}{2N}(t+t_0)^2,\qquad
    \sum_k\langle a_k^\dagger a_k\rangle(t)=\sum_n|\beta_n(t)|^2\le\frac{v_\text{max}^2}{2}(t+t_0)^2.
\end{align}
Each physical-mode photon number is also bounded by this total.
At each fixed time, the state is a product of coherent states in both mode bases, so the individual photon numbers are Poisson variables and the total is a sum of independent Poisson variables.
Consequently,
\begin{align}\label{eq:photon-variance}
    \operatorname{Var}(a_k^\dagger a_k)=\langle a_k^\dagger a_k\rangle,\qquad
    \operatorname{Var}(b_n^\dagger b_n)=\langle b_n^\dagger b_n\rangle,\qquad
    \operatorname{Var}\!\left(\sum_ka_k^\dagger a_k\right)=\sum_k\langle a_k^\dagger a_k\rangle.
\end{align}
Thus Eq.~\eqref{eq:total-photons} also bounds the corresponding variances at each time.
For every physical mode, Eqs.~\eqref{eq:total-photons}, \eqref{eq:photon-variance}, and~\eqref{eq:boson-average-bound} give
\begin{align}\label{eq:boson-variance-bounds}
    \operatorname{Var}_t(a_k^\dagger a_k)&\le\frac{v_\text{max}^2}{2}(t+t_0)^2,\nonumber\\
    \frac1T\int_0^T\operatorname{Var}_t(a_k^\dagger a_k)dt
    &=\overline n_k(T)\le\frac{v_\text{max}^2}{2N}\left(1+\frac{t_0}{T}\right)(T+t_0)^2,
\end{align}
where $\operatorname{Var}_t$ denotes the quantum variance in the state at time $t$.
For the learning time Eq.~\eqref{eq:time}, the time-averaged variance therefore obeys Eq.~\eqref{eq:boson-average-learning}.

\subsection{A nearest-neighbor control Hamiltonian}

The protocol specifies the control Hamiltonian $H_c=\sum_n\omega_n b_n^\dagger b_n$ through the modes $b_n$, each of which is in general spread over all modes $a_1,\dots,a_N$.
This section shows that, for any given set of candidate frequencies, there exists a choice of $U$ that turns $H_c$ into a chain of the physical modes $a_k$ with nearest neighbor interaction, and the sensor mode $a_1$ at one end.
Substituting $b_n=\sum_kU_{nk}a_k$ gives $H_c=\sum_{k,l}M_{kl}\,a_k^\dagger a_l$ with $M=U^\dagger DU$ and $D=\mathrm{diag}(\omega_1,\dots,\omega_N)$.
The protocol constrains $U$ only through its first column, $U_{n1}=1/\sqrt N$ for all $n$, so the question is whether some unitary with this first column makes $M$ tridiagonal.

\begin{proposition}\label{prop:chain}
For any frequencies $\omega_1<\dots<\omega_N$, there is a real orthogonal matrix $U$ with $U_{n1}=1/\sqrt N$ for all $n$ such that
\begin{align}\label{eq:chain}
    H_c=\sum_{k=1}^N\nu_k\,a_k^\dagger a_k+\sum_{k=1}^{N-1}J_k\left(a_k^\dagger a_{k+1}+a_{k+1}^\dagger a_k\right)
\end{align}
with real $\nu_k$ and $J_k>0$.
Among all unitaries $U$ with this first column and leading to Eq.~\eqref{eq:chain}, the coefficients $\nu_k$ and $J_k>0$ in Eq.~\eqref{eq:chain} are unique~\cite{Boley1987}.
\end{proposition}

\begin{proof}
We first construct $U$ by the Lanczos iteration~\cite{Golub2010}.
Let $q_1=(1,\dots,1)^T/\sqrt N\in\R^N$, $q_0=0$ and $J_0=0$.
For $k=1,\dots,N-1$, set $\nu_k=q_k^TDq_k$, let $r_k=Dq_k-\nu_kq_k-J_{k-1}q_{k-1}$, and set $J_k=\|r_k\|$ and $q_{k+1}=r_k/J_k$; finally set $\nu_N=q_N^TDq_N$.
The vectors $q_1,\dots,q_k$ produced in this way are the Gram--Schmidt orthonormalization of $q_1,Dq_1,\dots,D^{k-1}q_1$.
The $n$th entry of $D^jq_1$ is $\omega_n^j/\sqrt N$, so the matrix with columns $q_1,Dq_1,\dots,D^{N-1}q_1$ is a Vandermonde matrix in the distinct $\omega_n$ divided by $\sqrt N$, and it is invertible.
Hence $r_k\neq0$ and $J_k>0$ for every $k\le N-1$, and $q_1,\dots,q_N$ is an orthonormal basis of $\R^N$.
By construction $Dq_k=J_{k-1}q_{k-1}+\nu_kq_k+J_kq_{k+1}$ for $k\le N-1$, and $Dq_N=J_{N-1}q_{N-1}+\nu_Nq_N$ because $Dq_N$ has zero overlap with $q_1,\dots,q_{N-2}$ by the symmetry of $D$.
Let $U$ be the matrix with columns $q_1,\dots,q_N$, so $U_{nk}=(q_k)_n$.
Then $U$ is real orthogonal, $U_{n1}=(q_1)_n=1/\sqrt N$, and $M_{kl}=q_k^TDq_l$ is tridiagonal with diagonal entries $\nu_k$ and off-diagonal entries $J_k$, which is Eq.~\eqref{eq:chain}.

We now show uniqueness.
Let $U$ be any unitary with first column $q_1$ for which $M=U^\dagger DU$ is tridiagonal with real diagonal entries $\nu_k'$ and off-diagonal entries $J_k'>0$, and let $u_k$ denote its $k$th column.
Column $k$ of the identity $DU=UM$ reads $Du_k=J_{k-1}'u_{k-1}+\nu_k'u_k+J_k'u_{k+1}$.
Given $u_1,\dots,u_k$, this identity fixes $\nu_k'=u_k^\dagger Du_k$, then $J_k'=\|Du_k-\nu_k'u_k-J_{k-1}'u_{k-1}\|$, and then $u_{k+1}$.
Starting from $u_1=q_1$, this is the recursion above, so $\nu_k'=\nu_k$, $J_k'=J_k$, and $u_k=q_k$ for all $k$.
\end{proof}

For equally spaced frequencies the coefficients have a closed form.
Let $\omega_n=\omega_1+(n-1)\Delta$.
The vector $q_k$ constructed in the proof equals $p_{k-1}(D)q_1$, where $p_0,p_1,\dots$ are the orthonormal polynomials for the uniform weight on the points $\omega_1,\dots,\omega_N$, and $\nu_k$ and $J_k$ are the coefficients of their three-term recurrence.
These polynomials are the discrete Chebyshev (Gram) polynomials~\cite{Walter2004}, whose recurrence gives
\begin{align}\label{eq:chain-equal}
    \nu_k=\frac{1}{N}\sum_{n=1}^N \omega_n,\qquad
    J_k=\frac{k\Delta}{2}\sqrt{\frac{N^2-k^2}{4k^2-1}}.
\end{align}
All modes $a_k$ therefore have the same on-site coupling.
Both factors $k/\sqrt{4k^2-1}$ and $\sqrt{N^2-k^2}$ decrease with $k$, so the couplings decrease along the chain from $J_1=\Delta\sqrt{(N^2-1)/12}$ at the sensor end to $J_{N-1}=(N-1)\Delta/(2\sqrt{2N-3})$ at the far end.
The chain with $J_k=\frac\Delta2\sqrt{k(N-k)}$ used for perfect state transfer~\cite{Christandl2004} also has the equally spaced spectrum $\omega_1,\dots,\omega_N$, but the squared overlap of its end site with its $n$th eigenmode is $\binom{N-1}{n-1}/2^{N-1}$, which is nonuniform for $N\ge3$, so in that case it does not satisfy $U_{n1}=1/\sqrt N$.
For $N=2$, both overlaps equal $1/2$, and the perfect-state-transfer chain coincides with the uniform-weight chain constructed above.
Together with local heterodyne measurements and the classical postprocessing in Eq.~\eqref{eq:boson-classical-readout}, Proposition~\ref{prop:chain} realizes the protocol using only nearest-neighbor couplings during sensing.

\subsection{A star-shaped control Hamiltonian}

In the previous section, we consider a chain geometry, where the sensor mode $a_1$ is located at one end of the chain and couples only to $a_2$, and all other modes couple to their nearest neighbors.
This section gives a second geometry, in which $a_1$ couples to every other mode and the modes $a_2,\dots,a_N$ do not couple to each other.
Write $\bar\omega=\frac1N\sum_n\omega_n$ and $p(z)=\prod_{n=1}^N(z-\omega_n)$.

We first restate the requirement on $U$ as a condition on $M$ alone.
For a Hermitian $N\times N$ matrix $M$, let $G(z)=\bra{e_1}(z-M)^{-1}\ket{e_1}$, where $e_1$ is the first standard basis vector.
If $M=U^\dagger DU$ with $U_{n1}=1/\sqrt N$, then $(z-M)^{-1}=U^\dagger(z-D)^{-1}U$ gives
\begin{align}\label{eq:resolvent}
    G(z)=\sum_{n=1}^N\frac{|U_{n1}|^2}{z-\omega_n}=\frac1N\sum_{n=1}^N\frac1{z-\omega_n}=\frac{p'(z)}{N\,p(z)}.
\end{align}
Conversely, suppose $M$ satisfies Eq.~\eqref{eq:resolvent}, and write $M=\sum_m\lambda_mP_m$ with distinct eigenvalues $\lambda_m$ and spectral projectors $P_m$, so that $G(z)=\sum_m\bra{e_1}P_m\ket{e_1}/(z-\lambda_m)$.
Matching the poles of the two expressions for $G(z)$ shows that each $\omega_n$ is an eigenvalue of $M$ with $\bra{e_1}P_m\ket{e_1}=1/N$.
These are $N$ distinct eigenvalues of an $N\times N$ matrix, so each is simple, with a normalized eigenvector $u_n$ whose phase we fix so that $(u_n)_1=1/\sqrt N$.
The matrix $U$ with rows $u_n^\dagger$ is unitary, satisfies $U_{n1}=1/\sqrt N$, and gives $M=U^\dagger DU$.
Hence a control Hamiltonian $H_c=\sum_{k,l}M_{kl}\,a_k^\dagger a_l$ can be used in the protocol if and only if $M$ satisfies Eq.~\eqref{eq:resolvent}.

\begin{proposition}\label{prop:star}
For any frequencies $\omega_1<\dots<\omega_N$, the control Hamiltonian
\begin{align}\label{eq:star}
    H_c=\bar\omega\,a_1^\dagger a_1+\sum_{j=1}^{N-1}\left[d_j\,a_{j+1}^\dagger a_{j+1}+g_j\left(a_1^\dagger a_{j+1}+a_{j+1}^\dagger a_1\right)\right]
\end{align}
satisfies Eq.~\eqref{eq:resolvent}, where $d_1<\dots<d_{N-1}$ are the zeros of $p'$, with $\omega_j<d_j<\omega_{j+1}$, and
\begin{align}\label{eq:star-couplings}
    g_j=\left(\frac{N}{\sum_{n=1}^N(d_j-\omega_n)^{-2}}\right)^{1/2}>0.
\end{align}
Up to relabeling of $a_2,\dots,a_N$, this is the only Hamiltonian of the form Eq.~\eqref{eq:star} with all $g_j>0$ that satisfies Eq.~\eqref{eq:resolvent}~\cite{Boley1987}.
\end{proposition}

\begin{proof}
We first locate the zeros of $p'$.
By Rolle's theorem, $p'$ has a zero in each of the $N-1$ intervals $(\omega_j,\omega_{j+1})$.
Since $p'$ has degree $N-1$, these are all of its zeros, each is simple, and $p'(z)=N\prod_{j=1}^{N-1}(z-d_j)$.

We next compute $G(z)$ for the Hamiltonian Eq.~\eqref{eq:star}.
Its matrix $M$ has $M_{11}=\bar\omega$, $M_{1,j+1}=M_{j+1,1}=g_j$, $M_{j+1,j+1}=d_j$, and all other entries zero.
The Schur complement formula for the $(1,1)$ entry of $(z-M)^{-1}$ gives
\begin{align}\label{eq:star-resolvent}
    G(z)=\left(z-\bar\omega-\sum_{j=1}^{N-1}\frac{g_j^2}{z-d_j}\right)^{-1}.
\end{align}
Comparing with Eq.~\eqref{eq:resolvent}, it suffices to show that
\begin{align}\label{eq:star-target}
    \frac{N\,p(z)}{p'(z)}=z-\bar\omega-\sum_{j=1}^{N-1}\frac{g_j^2}{z-d_j}.
\end{align}

We now expand the left side of Eq.~\eqref{eq:star-target} in partial fractions.
Polynomial division gives $N\,p(z)=(z-\alpha)\,p'(z)+r(z)$ for some constant $\alpha$ and some polynomial $r$ of degree at most $N-2$.
Write $s=\sum_n\omega_n$, so that $p(z)=z^N-s\,z^{N-1}+\dots$ and $p'(z)=Nz^{N-1}-(N-1)s\,z^{N-2}+\dots$.
The coefficient of $z^{N-1}$ is $-Ns$ on the left side and $-(N-1)s-N\alpha$ on the right side, so $\alpha=s/N=\bar\omega$.
Since the zeros $d_j$ of $p'$ are simple and $\deg r<\deg p'$, the partial fraction expansion of $r/p'$ is
\begin{align}
    \frac{r(z)}{p'(z)}=\sum_{j=1}^{N-1}\frac{r(d_j)}{p''(d_j)\,(z-d_j)}.
\end{align}
Setting $z=d_j$ in the division identity gives $r(d_j)=N\,p(d_j)$, so Eq.~\eqref{eq:star-target} holds with $g_j^2=-N\,p(d_j)/p''(d_j)$.

It remains to show that this value of $g_j^2$ equals Eq.~\eqref{eq:star-couplings}.
Write $L(z)=p'(z)/p(z)=\sum_n(z-\omega_n)^{-1}$.
Then $p''/p=L'+L^2$.
At $z=d_j$ we have $p'(d_j)=0$ and $p(d_j)\neq0$, so $L(d_j)=0$ and
\begin{align}
    \frac{p''(d_j)}{p(d_j)}=L'(d_j)=-\sum_{n=1}^N\frac1{(d_j-\omega_n)^2}.
\end{align}
Substituting into $g_j^2=-N\,p(d_j)/p''(d_j)$ gives Eq.~\eqref{eq:star-couplings}, and in particular $g_j^2>0$.

Finally we show uniqueness.
Suppose a Hamiltonian of the form Eq.~\eqref{eq:star} with diagonal entries $\alpha',d_1',\dots,d_{N-1}'$ and couplings $g_j'>0$ satisfies Eq.~\eqref{eq:resolvent}.
By the converse argument above, every normalized eigenvector $u$ of its matrix satisfies $|\braket{e_1|u}|^2=1/N$.
If $d_j'=d_l'$ for some $j\neq l$, the vector $g_l'e_{j+1}-g_j'e_{l+1}$ is an eigenvector orthogonal to $e_1$, which is impossible, so the $d_j'$ are distinct.
The right side of Eq.~\eqref{eq:star-target} with primed parameters then has simple poles exactly at the $d_j'$ with residues $-g_j'^2\neq0$.
The partial fraction expansion of $N\,p/p'$ is unique, so after relabeling $d_j'=d_j$, $g_j'=g_j$, and $\alpha'=\bar\omega$.
\end{proof}

The chain and the star give the sensor the same onsite coupling and the same sum of squared couplings to the other modes. Indeed, $M=U^\dagger DU$ and $|U_{n1}|^2=1/N$ imply $M_{11}=\bar\omega$ and $(M^2)_{11}=N^{-1}\sum_n\omega_n^2$. Hermiticity gives $(M^2)_{11}=M_{11}^2+\sum_{k=2}^N|M_{1k}|^2$, so the sum of squared couplings from the sensor is fixed by $\sum_{k=2}^N|M_{1k}|^2 = \sum_{n=1}^N (\omega_n-\bar{\omega})^2/N$. In the chain, this sum contains only the link $J_1$, whereas in the star it contains all $N-1$ links $g_j$. Therefore,
\begin{equation}
J_1^2=\sum_{j=1}^{N-1}g_j^2
=\frac1N\sum_{n=1}^N(\omega_n-\bar\omega)^2.
\end{equation}
For equally spaced frequencies, this squared strength is $\Delta^2(N^2-1)/12$, giving a two-site coupling $J_1=\Delta\sqrt{(N^2-1)/12}$ on the chain and a root-mean-square star coupling $\Delta\sqrt{(N+1)/12}$ spreading over $N-1$ two-site couplings on a star geometry.

\section{Experimental proposal and numerical simulations}
In this section we give a concrete implementation of the random-modulation protocol for the Detection problem (Problem~\ref{problem:signal-detection}) and study how experimental imperfections affect its performance.
Appendix~\ref{ssec: problem_setup} formally specifies the specific problem we consider in this section, together with the model of the sensor, the control, and how we model the experimental implementations.
Appendices~\ref{ssec: our_protocol} and~\ref{ssec: conventional_protocol} present the concrete version of the random-modulation protocol and CPMG spectroscopy that serves as the baseline.
We denote the total sensing times of the random-modulation protocol and of CPMG spectroscopy for the same detection problem by $T_{\mathrm{RM}}$ and $T_{\mathrm{CPMG}}$, defined in Eqs.~\eqref{eq: T_RM} and~\eqref{eq: T_CPMG}.
We quantify the improvement of the random-modulation protocol over the baseline by the \textit{improvement factor}, which is defined to be the ratio
\begin{equation}
  \mathcal{I} := \frac{T_{\mathrm{CPMG}}}{T_{\mathrm{RM}}}.
\end{equation}
Appendix~\ref{ssec: phase_diagram} estimates heuristically how $\mathcal{I}$ depends on the signal parameters and sensor parameters, and Appendix~\ref{app: end_to_end_numerics} conducts an end-to-end numerical simulation to study the performance under realistic imperfections.

\subsection{Problem setup}\label{ssec: problem_setup}
We consider a special case of Detection (Problem~\ref{problem:signal-detection}), which suffices to demonstrate the protocol of Appendix~\ref{ssec: our_protocol}.
The candidate frequencies are equally spaced,
\begin{equation}
    \omega_n = \omega_0 + n\,\Delta, \qquad n=1,2,\dots,N,
\end{equation}
where $\omega_0\ge 0$ is the carrier frequency.
As in the main text, the signal couples to the sensor through the signal Hamiltonian $H_s(t)=f(t)Z$, where $f(t)$ is the signal waveform.
There are $N+1$ hypotheses $\mathsf{H}_0,\mathsf{H}_1,\dots,\mathsf{H}_N$.
Under $\mathsf{H}_0$, $f(t)=0$.
Under $\mathsf{H}_n$ with $n\geq 1$, the waveform is
\begin{equation}
    f(t) = f_n(t) = \epsilon \cos(\omega_n t + \phi),
\end{equation}
a monochromatic signal at frequency $\omega_n$ with unknown initial phase $\phi\in[0,2\pi)$.
The amplitude $\epsilon$ is known.
Exactly one hypothesis holds.
The protocol must decide whether it is $\mathsf{H}_0$ or some $\mathsf{H}_n$ with $n\geq 1$, with success probability at least $1-\delta$ under every hypothesis.

We consider the following control model.
The protocol consists of a sequence of rounds.
In each round, the sensor is prepared in an arbitrary pure state, evolves under $H_s(t) + H_c(t)$, and is measured in an arbitrary basis, with preparation and measurement treated as instantaneous.
The control Hamiltonian $H_c(t)$ consists of $\pi$ pulses with a fixed Rabi frequency $\Omega_R$,
\begin{equation}
    H_c(t) = \frac{\Omega_R}{2} X
\end{equation}
during a pulse and $H_c(t) = 0$ otherwise.
A pulse centered at time $t$ therefore occupies $[t-\tau_\pi/2,\,t+\tau_\pi/2]$, where $\tau_\pi = \pi/\Omega_R$, and implements an $X$ gate up to a global phase.
The interval between the end of one pulse and the start of the next must be at least $\tau_{\mathrm{min}}$.
Sequences of $\pi$ pulses at a fixed Rabi frequency are the standard control in experiments, and the minimum interval $\tau_{\mathrm{min}}$ models the finite control bandwidth.
We assume $\tau_\pi+\tau_{\mathrm{min}}\ll 2\pi/\omega_N$, so that the pulse duration and the minimum interval are much shorter than the period of every candidate signal.
We call evolution under $H_s(t)+H_c(t)$ alone the ideal dynamics.

The sensor has a finite coherence time $T_2$, which we model as dephasing applied after the ideal dynamics of each round.
If a round lasts for time $\tau$ from the state preparation to the measurement, the measured state is obtained by running the ideal dynamics over the whole round and then multiplying each off-diagonal element in the $\{\ket{0},\ket{1}\}$ basis by $e^{-\tau/T_2}$.
We assume $T_2\gg 1/\Delta$.

\subsection{Random-modulation protocol}\label{ssec: our_protocol}

This subsection presents an end-to-end protocol that implements the Detection algorithm of the main text under the control model above.
The main text protocol assumes an infinite coherence time, so the evolution time of each round can be made long enough for the accumulated phase to be of order one, and an $X$ basis measurement then reads it out directly regardless of the sign of the phase.
With a finite $T_2$, the evolution time of each round is limited by $T_2$, and the evolution time needed to detect the signal grows with $N$, so shortening it for given $N$ and $\epsilon$ extends the range of $N$ that the protocol reaches within $T_2$.
We therefore measure in the $Y$ basis instead of the $X$ basis, since a $Y$ basis measurement reaches the maximal signal at half the accumulated phase, and hence half the evolution time, that an $X$ basis measurement needs.
The $Y$ basis measurement brings one difficulty.
Under $\mathsf{H}_n$ with $n\geq 1$, the protocol knows neither the active frequency nor its initial phase, so for a given random modulation it can determine neither the magnitude nor the sign of the accumulated phase, and the outcomes of different random modulations cannot be averaged as in the $X$ basis protocol, where the sign does not matter.
We resolve this with a test statistic that sums the squares of the averages taken within each random modulation, instead of averaging the outcomes across modulations.

\subsubsection{Protocol overview}\label{sssec: protocol_overview}
The random-modulation protocol draws at most $R$ random modulations and runs $M$ rounds of experiment with each of them.
Every round has the same evolution time $\tau_{\mathrm{evo}}$, so the total sensing time is at most $RM\tau_{\mathrm{evo}}$.
We label the rounds by $(r,m)$ with $r\in[R]$ and $m\in[M]$, and round $(r,m)$ interrogates the signal during the time window $[t_{r,m},\,t_{r,m}+\tau_{\mathrm{evo}}]$, where
\begin{equation}\label{eq: window_start}
    t_{r,m} = \bigl((r-1)M+m-1\bigr)\tau_{\mathrm{evo}} .
\end{equation}
Each random modulation is handled in four steps.
First, we generate the random modulation
\begin{equation}\label{eq: modulation_function_experiment}
    Y^{(r)}(t)
    =
    \sum_{n=1}^{N}
    \left(
        a^{(r)}_n\cos(\omega_n t)
        +
        b^{(r)}_n\sin(\omega_n t)
    \right),
\end{equation}
with coefficients $a^{(r)}_n=\cos\varphi^{(r)}_n$ and $b^{(r)}_n=\sin\varphi^{(r)}_n$, where each phase $\varphi^{(r)}_n$ is independently sampled and uniformly distributed on $[0,2\pi)$ for $n=1,\dots,N$ and $r=1,\dots,R$.
Each term of Eq.~\eqref{eq: modulation_function_experiment} is then $\cos(\omega_n t-\varphi^{(r)}_n)$, a modulation with fixed amplitude and a uniformly random phase.
In our numerical simulations, this ensemble of random modulations gives a shorter total sensing time than the modulation function of Eq.~\eqref{eq: modulation_function}, whose coefficients $a_n$ and $b_n$ are independent Gaussian random variables.
Second, for each round $(r,m)$ we compile $Y^{(r)}(t)$ on the time window of that round into a pulse sequence $\cP^{(r,m)}$ that approximates the modulation function $\mathrm{sgn}(Y^{(r)}(t))$, as described in Appendix~\ref{sssec: pulse_sequence_generation}.
Third, in each round $(r,m)$, we prepare the sensor in $\ket{+}$, evolve it under $\cP^{(r,m)}$, measure it in the $Y$ basis, and record the measured value multiplied by a sign determined by $Y^{(r)}(t)$ as the outcome $y_{r,m}\in\{+1,-1\}$, see Appendix~\ref{sssec: pulse_sequence_generation}.
Since all rounds share one global clock, under $\mathsf{H}_n$ the responses of the $M$ rounds of the same random modulation add coherently for the candidate frequencies within $1/(M\tau_{\mathrm{evo}})$ of $\omega_n$ and cancel for the others, see Appendix~\ref{ssec: phase_diagram}.
Fourth, we average the outcomes of each random modulation,
\begin{equation}\label{eq: y_hat}
    \hat{y}_r = \frac{1}{M} \sum_{m=1}^M y_{r,m},
\end{equation}
and sum the squares of these averages over the random modulations drawn so far into the test statistic
\begin{equation}\label{eq: S_statistic}
    S_r = \sum_{r'=1}^r \hat{y}_{r'}^2 .
\end{equation}
Under $\mathsf{H}_0$ every outcome is an unbiased coin flip, so every $\hat{y}_r$ averages to zero and $S_r$ is small.
Under $\mathsf{H}_n$ with $n\geq 1$, when $r$ is large enough, some of the first $r$ random modulations have their component at $\omega_n$ nearly in phase or in antiphase with the signal, whatever $n$ and $\phi$ are.
The rounds of such a modulation accumulate a large phase, so its $\hat{y}_r$ is far from zero and $S_r$ is large.
Averaging over $M$ rounds suppresses the projection noise in each $\hat{y}_r$.
Finally, we compare $S_r$ with a threshold $S^*$, report that a signal is present if $S_r>S^*$, and otherwise return to the first step with the next random modulation.
If $S_R\leq S^*$ after all $R$ random modulations, we report that no signal is present.
Appendix~\ref{sssec: data_processing} gives a concrete algorithm that chooses $\tau_{\mathrm{evo}}$, $M$, $R$, and $S^*$ in advance for a given coherence time $T_2$ and target error probability $\delta$.

\subsubsection{Pulse sequence generation}\label{sssec: pulse_sequence_generation}

For each round $(r,m)$, we compile $Y^{(r)}(t)$ on the time window of that round into a pulse sequence $\cP^{(r,m)}$, namely the increasing list of times in the window at which $\pi$ pulses are centered.
The target modulation function is $\mathrm{sgn}(Y^{(r)}(t))$, because it accumulates phase with the same sign as $Y^{(r)}(t)$ at every time and at the maximal rate.
We place grid points in the window, starting from $t_{r,m}$, with a fixed grid step $\tau_\pi+\tau_{\mathrm{min}}$, and we choose $\tau_{\mathrm{evo}}$ as an integer multiple of the grid step, so that $t_{r,m}+\tau_{\mathrm{evo}}$ is the last grid point.
We evaluate the sign of $Y^{(r)}(t)$ at every grid point, and for each pair of neighboring grid points at which the signs differ we center one $\pi$ pulse at their midpoint.
The pulse sequence $\cP^{(r,m)}$ consists of these midpoints in increasing order.
Since the grid step is at least $\tau_\pi$, each pulse lies between two neighboring grid points and hence inside the window.
The resulting pulse sequence respects the minimum interval $\tau_{\mathrm{min}}$ that Appendix~\ref{ssec: problem_setup} requires, because the centers of two consecutive pulses are at least one grid step apart and each pulse lasts $\tau_\pi$.
Finally, the outcome $y_{r,m}$ is the value measured in the $Y$ basis multiplied by $\mathrm{sgn}(Y^{(r)}(t_{r,m}+\tau_{\mathrm{evo}}))$, the sign of $Y^{(r)}(t)$ at the last grid point of the window, to compensate two sign flips of the measured value.
The first occurs when $Y^{(r)}(t_{r,m})<0$, since the pulse sequence then realizes $-\mathrm{sgn}(Y^{(r)}(t))$ instead of $\mathrm{sgn}(Y^{(r)}(t))$.
The second occurs when the number of pulses is odd, since the $X$ gates then leave one $X$ gate that maps $Y$ to $-Y$.

The pulse sequence above targets $\mathrm{sgn}(Y^{(r)}(t))$ because Detection only needs the phase to accumulate at the maximal rate.
Search, however, tests whether a set $S$ of candidate frequencies contains an active frequency, and it requires in addition that the candidate frequencies outside $S$ contribute zero phase.
The function $\mathrm{sgn}(Y^{(r)}(t))$ does not guarantee this, since taking the sign of a modulation function supported on $S$ can produce a response at candidate frequencies outside $S$.
For Search we therefore compute numerically the modulation function $g(t)$ of Eq.~\eqref{eq:search-bounded-modulation}, which has independent Gaussian coefficients on $S$ and is corrected after clipping so that the frequencies outside $S$ contribute exactly zero phase.
This $g(t)$ takes continuous values in $[-1,1]$, whereas a sequence of $\pi$ pulses realizes only the values $\pm 1$, so we compile $g(t)$ into a pulse sequence by delta-sigma modulation.
We estimate that this compilation requires a control bandwidth of $O(\omega_N)$.

\subsubsection{Dynamics in each round of experiment}\label{sssec: round_dynamics}
This subsubsection gives the distribution of the outcome $y_{r,m}$ of round $(r,m)$.
In round $(r,m)$ the sensor starts in $\ket{+}$ and evolves under $H_s(t)+H_c(t)$ from $t_{r,m}$ to $t_{r,m}+\tau_{\mathrm{evo}}$, where $H_c(t)$ is the control Hamiltonian of the pulse sequence $\cP^{(r,m)}$.
The ideal dynamics is generated by the propagator
\begin{equation}\label{eq: round_propagator}
    U^{(r,m)}
    =
    \mathcal{T}\exp\left(
        -i\int_{t_{r,m}}^{t_{r,m}+\tau_{\mathrm{evo}}}
        \bigl(H_s(t)+H_c(t)\bigr)\,dt
    \right),
\end{equation}
where $\mathcal{T}$ denotes time ordering.
The $T_2$ model multiplies the off-diagonal elements of the final state by $e^{-\tau_{\mathrm{evo}}/T_2}$, and the expectation value of $Y$ depends only on these elements, so the value measured in the $Y$ basis has expectation $e^{-\tau_{\mathrm{evo}}/T_2}\bra{+}U^{(r,m)\dagger}\,Y\,U^{(r,m)}\ket{+}$.
The outcome $y_{r,m}$ is this value multiplied by $\mathrm{sgn}(Y^{(r)}(t_{r,m}+\tau_{\mathrm{evo}}))$, as defined in Appendix~\ref{sssec: pulse_sequence_generation}, so for $y\in\{+1,-1\}$
\begin{equation}\label{eq: outcome_distribution}
    \Pr[y_{r,m}=y\mid \mathsf{H}_n,\phi]
    =
    \frac{1+y\,e^{-\tau_{\mathrm{evo}}/T_2}\,
    \mathrm{sgn}\bigl(Y^{(r)}(t_{r,m}+\tau_{\mathrm{evo}})\bigr)\,
    \bra{+}U^{(r,m)\dagger}\,Y\,U^{(r,m)}\ket{+}}{2} .
\end{equation}
We write $p_{r,m}$ for the value of Eq.~\eqref{eq: outcome_distribution} at $y=+1$.
Given the random modulation $Y^{(r)}(t)$, the outcomes $y_{r,1},\dots,y_{r,M}$ are independent.
Under $\mathsf{H}_0$ the signal vanishes, $U^{(r,m)}$ is a product of rotations about $X$, and $\ket{+}$ is an eigenstate of every such rotation, so $y_{r,m}$ is an unbiased coin flip,
\begin{equation}\label{eq: outcome_distribution_H0}
    \Pr[y_{r,m}=y\mid \mathsf{H}_0]
    =
    \frac{1}{2} ,
\end{equation}
so $p_{r,m}=1/2$.
Both the parameter selection of Appendix~\ref{sssec: data_processing} and the end-to-end numerical simulation in Appendix~\ref{app: end_to_end_numerics} evaluate Eq.~\eqref{eq: outcome_distribution} by integrating Eq.~\eqref{eq: round_propagator} numerically.

\subsubsection{Data processing}\label{sssec: data_processing}
Here we characterize the distribution of $S_r$ of Eq.~\eqref{eq: S_statistic} under every hypothesis, and use it to define the total sensing time $T_{\mathrm{RM}}$ and to choose $\tau_{\mathrm{evo}}$, $M$, $R$, and the threshold $S^*$.

We first fix one index $r$ and consider $\hat{y}_r$ of Eq.~\eqref{eq: y_hat}.
Given the random modulation $Y^{(r)}(t)$ of Eq.~\eqref{eq: modulation_function_experiment}, the number $K_r$ of outcomes equal to $+1$ among the $M$ independent Bernoulli trials with success probabilities $p_{r,1},\dots,p_{r,M}$ of Eq.~\eqref{eq: outcome_distribution} determines $\hat{y}_r=(2K_r-M)/M$.
The success probabilities depend on $Y^{(r)}(t)$, so the distribution of $K_r$ is its distribution for fixed $Y^{(r)}(t)$ averaged over $Y^{(r)}(t)$,
\begin{equation}\label{eq: K_pmf}
    \Pr[K_r=k\mid\mathsf{H}_n,\phi]
    =
    \mathbb{E}\left[\Pr[K_r=k\mid Y^{(r)}]\,\Big|\,\mathsf{H}_n,\phi\right] ,
    \qquad k=0,\dots,M ,
\end{equation}
where $\mathbb{E}[\cdot\mid\mathsf{H}_n,\phi]$ is the expectation over $Y^{(r)}(t)$.
Under $\mathsf{H}_0$, $p_{r,m}=1/2$ for every $m$ and every $Y^{(r)}(t)$, and Eq.~\eqref{eq: K_pmf} is the binomial distribution with success probability $1/2$.
Under $\mathsf{H}_n$ with $n\geq 1$ and initial phase $\phi$, we evaluate the average in Eq.~\eqref{eq: K_pmf} by Monte Carlo.
For each sample, we draw $Y^{(r)}(t)$ from Eq.~\eqref{eq: modulation_function_experiment}, compile it on the $M$ windows of Eq.~\eqref{eq: window_start} into the pulse sequences $\cP^{(r,1)},\dots,\cP^{(r,M)}$ as in Appendix~\ref{sssec: pulse_sequence_generation}, obtain $p_{r,1},\dots,p_{r,M}$ from Eq.~\eqref{eq: outcome_distribution} by integrating Eq.~\eqref{eq: round_propagator} numerically, and compute $\Pr[K_r=k\mid Y^{(r)}]$ from these $M$ probabilities by adding the trials one at a time, each step updating the distribution of the running count with the next $p_{r,m}$.

We now turn to $S_r$ of Eq.~\eqref{eq: S_statistic}.
The random modulations $Y^{(1)}(t),\dots,Y^{(R)}(t)$ are drawn independently, so $K_1,\dots,K_R$ are independent.
Under $\mathsf{H}_n$ with $n\geq 1$, the distribution of $K_r$ in Eq.~\eqref{eq: K_pmf} depends on $r$, because the $M$ windows of the $r$-th random modulation start at $t_{r,1}$ of Eq.~\eqref{eq: window_start}, where the phase of the signal is $\omega_n t_{r,1}+\phi$.
The Monte Carlo evaluation above is therefore carried out for each $r$ with the $M$ windows of that $r$.
For every $r\in[R]$,
\begin{equation}
    M^2 S_r = \sum_{r'=1}^{r} \left(2K_{r'}-M\right)^2
\end{equation}
is a sum of $r$ independent random variables.
Each term takes integer values in $\{0,1,\dots,M^2\}$ with probabilities known from Eq.~\eqref{eq: K_pmf}, so the distribution of $M^2 S_r$ on the integers $0,1,\dots,rM^2$ is computed exactly by summing the terms one at a time.
This gives $\Pr[S_r\leq S^*\mid\mathsf{H}_0]$ and $\Pr[S_r\leq S^*\mid\mathsf{H}_n,\phi]$ for every $r\in[R]$ and every threshold $S^*$.

We now define the total sensing time $T_{\mathrm{RM}}$ of the random-modulation protocol, and use the above distributions to choose $\tau_{\mathrm{evo}}$, $M$, $R$, and $S^*$ in order to minimize $T_{\mathrm{RM}}$.
We define
\begin{equation}\label{eq: T_RM}
    T_{\mathrm{RM}}
    \coloneqq
    \min_{\tau_{\mathrm{evo}},M,R,S^*}\;
    \max_{n\geq 1,\,\phi}\;
    M\tau_{\mathrm{evo}}\sum_{r=0}^{R-1}\Pr[S_r\leq S^*\mid\mathsf{H}_n,\phi] ,
\end{equation}
as the mean total sensing time under a signal.
Each random modulation takes sensing time $M\tau_{\mathrm{evo}}$, and $\sum_{r=0}^{R-1}\Pr[S_r\leq S^*\mid\mathsf{H}_n,\phi]$, with $S_0=0$, is the mean number of random modulations that the protocol draws under $\mathsf{H}_n$ with initial phase $\phi$.
This holds because $S_r$ is nondecreasing in $r$, so the protocol draws the $(r+1)$-th random modulation if and only if $S_r\leq S^*$.
The maximum takes the worst case over all possible signals.
The minimum is over all choices of the protocol parameters for which both the false positive probability and the false negative probability are at most $\delta$.
Since $S_r$ is nondecreasing in $r$, the protocol reports that a signal is present if and only if $S_R>S^*$, so the false positive probability is $\Pr[S_R>S^*\mid\mathsf{H}_0]$ and the false negative probability is $\Pr[S_R\leq S^*\mid\mathsf{H}_n,\phi]$.
We find the minimum in Eq.~\eqref{eq: T_RM} by a search in three steps.
First, for fixed $\tau_{\mathrm{evo}}$, $M$, and $R$, we set $S^*$ to the smallest value in the support of $S_R$ under $\mathsf{H}_0$ for which the false positive probability is at most $\delta$, and we check whether the false negative probability at this $S^*$ is at most $\delta$ for every $n\geq 1$ and every $\phi$.
If it is, we record the maximum in Eq.~\eqref{eq: T_RM}.
Second, for fixed $\tau_{\mathrm{evo}}$, we compare all pairs $(M,R)$ that pass this check, since the smallest $R$ that passes need not minimize the mean total sensing time.
Third, we vary $\tau_{\mathrm{evo}}$ and keep the choice of $\tau_{\mathrm{evo}}$, $M$, and $R$ with the smallest recorded value.
In the numerical experiments $\phi$ is restricted to a finite set of values, as specified in Appendix~\ref{sssec: optimization_verification}.

\subsection{CPMG spectroscopy}\label{ssec: conventional_protocol}
CPMG spectroscopy is the sequential scheme commonly used in practice~\cite{Kotler2011QuantumLockIn,qsensing:2017}, and we take it as the baseline.
It probes the candidate frequencies one at a time.
For each candidate frequency, the $\pi$ pulse sequence realizes a modulation function that is resonant with that frequency, and two modulation functions in quadrature are used so that the signal is detected whatever its initial phase $\phi$.
We ignore all control imperfections of CPMG spectroscopy and keep only the finite coherence time $T_2$.
Its pulses are instantaneous and no minimum pulse interval is imposed, so the pulse sequence realizes its target modulation function exactly.
These simplifications can only shorten its sensing time, so the improvement factor of the random-modulation protocol reported below is a conservative estimate.
Accordingly, we do not simulate the full dynamics of CPMG spectroscopy, and all of its outcome probabilities are computed from the modulation function alone, both in the parameter selection below and in the end-to-end numerical simulation of Appendix~\ref{app: end_to_end_numerics}.

Here we present CPMG spectroscopy concretely.
It tests the candidate frequencies one at a time and stops at the first positive test.
We label the tests by $r\in[N]$, since at most $N$ tests are performed, and the $r$-th test proceeds in four steps.
First, we choose a candidate frequency $\omega_{n_r}$ uniformly at random from the candidate frequencies that have not been tested.
Second, we run $2M$ rounds for this candidate frequency, labeled $(r,m,C)$ and $(r,m,S)$ with $m\in[M]$.
Round $(r,m,C)$ interrogates the signal during the time window $[t^{C}_{r,m},\,t^{C}_{r,m}+\tau_{\mathrm{evo}}]$ and round $(r,m,S)$ during $[t^{S}_{r,m},\,t^{S}_{r,m}+\tau_{\mathrm{evo}}]$, where
\begin{equation}\label{eq: conventional_window_start}
    t^{C}_{r,m}=\bigl(2(r-1)M+m-1\bigr)\tau_{\mathrm{evo}},
    \qquad
    t^{S}_{r,m}=\bigl(2(r-1)M+M+m-1\bigr)\tau_{\mathrm{evo}},
\end{equation}
so that the $M$ rounds with label $C$ come first and the $M$ rounds with label $S$ follow.
In each round the qubit is initialized in $\ket{+}$, evolves for the window under a modulation function, and is measured in the $Y$ basis.
The modulation function of round $(r,m,C)$ is $h^{C}_{n_r}(t)$ and that of round $(r,m,S)$ is $h^{S}_{n_r}(t)$, where the two quadrature functions are
\begin{equation}\label{eq: quadrature_functions}
    h^{C}_n(t)=\mathrm{sgn}\bigl(\cos(\omega_n t)\bigr),
    \qquad
    h^{S}_n(t)=\mathrm{sgn}\bigl(\sin(\omega_n t)\bigr).
\end{equation}
With instantaneous pulses, under $\mathsf{H}_n$ with $n\geq 1$ the state at the end of round $(r,m,C)$ is $(\ket{0}+e^{i\theta^{C}_{r,m}}\ket{1})/\sqrt{2}$ up to a global phase, and likewise for round $(r,m,S)$ with $\theta^{S}_{r,m}$, where
\begin{equation}\label{eq: conventional_phase}
    \theta^{C}_{r,m}=2\epsilon\int_{t^{C}_{r,m}}^{t^{C}_{r,m}+\tau_{\mathrm{evo}}} h^{C}_{n_r}(t)\cos(\omega_n t+\phi)\,dt,
    \qquad
    \theta^{S}_{r,m}=2\epsilon\int_{t^{S}_{r,m}}^{t^{S}_{r,m}+\tau_{\mathrm{evo}}} h^{S}_{n_r}(t)\cos(\omega_n t+\phi)\,dt .
\end{equation}
The modulation frequency $\omega_{n_r}$ is the candidate frequency chosen in the $r$-th test and need not equal the signal frequency $\omega_n$.
Let $y^{C}_{r,m}$ and $y^{S}_{r,m}$ be the $Y$ basis outcomes of rounds $(r,m,C)$ and $(r,m,S)$.
With the $T_2$ model of Eq.~\eqref{eq: outcome_distribution}, for $y\in\{+1,-1\}$
\begin{equation}\label{eq: conventional_outcome_distribution}
    \Pr[y^{C}_{r,m}=y\mid\mathsf{H}_n,\phi]=\frac{1+y\,e^{-\tau_{\mathrm{evo}}/T_2}\sin\theta^{C}_{r,m}}{2},
    \qquad
    \Pr[y^{S}_{r,m}=y\mid\mathsf{H}_n,\phi]=\frac{1+y\,e^{-\tau_{\mathrm{evo}}/T_2}\sin\theta^{S}_{r,m}}{2},
\end{equation}
and under $\mathsf{H}_0$ the signal vanishes, so both outcomes are unbiased coin flips.
Third, we average the outcomes of the two modulation functions separately,
\begin{equation}\label{eq: IQ_averages}
    I_r=\frac{1}{M}\sum_{m=1}^{M} y^{C}_{r,m},
    \qquad
    Q_r=\frac{1}{M}\sum_{m=1}^{M} y^{S}_{r,m},
\end{equation}
and combine them into the amplitude
\begin{equation}\label{eq: IQ_amplitude}
    A_r=\sqrt{I_r^2+Q_r^2}.
\end{equation}
Fourth, if $A_r>A^*$, where $A^*$ is a threshold fixed in advance, the protocol reports a signal at frequency $\omega_{n_r}$ and stops.
Otherwise the protocol returns to the first step.
When all $N$ candidate frequencies have been tested without a positive test, the protocol reports no signal and stops.
The evolution time $\tau_{\mathrm{evo}}$, the number of rounds $M$, and the threshold $A^*$ are fixed below.

Under $\mathsf{H}_0$ the outcomes carry no signal, so $I_r$ and $Q_r$ average to zero and $A_r$ is small.
Under $\mathsf{H}_n$ with $n\geq 1$ and $n_r=n$, the phases of Eq.~\eqref{eq: conventional_phase} depend on $\phi$ through $\cos\phi$ and $\sin\phi$ respectively, so for every $\phi$ at least one of $I_r$ and $Q_r$ averages to a value away from zero and $A_r$ is large.
Taking the two averages before squaring suppresses the projection noise in each of them, and squaring removes the dependence of the sign of the $Y$ basis outcome on $\phi$.

We now define the total sensing time $T_{\mathrm{CPMG}}$ of CPMG spectroscopy, and choose $\tau_{\mathrm{evo}}$, $M$, and $A^*$ in order to minimize $T_{\mathrm{CPMG}}$.
We define
\begin{equation}\label{eq: T_CPMG}
    T_{\mathrm{CPMG}}
    \coloneqq
    \min_{\tau_{\mathrm{evo}},M}\;
    \max_{n\geq 1,\,\phi}\;
    2M\tau_{\mathrm{evo}}\sum_{r=1}^{N} r\,\Pr[\text{the protocol stops after the $r$-th test}\mid\mathsf{H}_n,\phi] ,
\end{equation}
as the mean total sensing time under a signal.
Each test takes sensing time $2M\tau_{\mathrm{evo}}$, and the sum is the mean number of tests that the protocol performs under $\mathsf{H}_n$ with initial phase $\phi$, where the protocol stops after the $r$-th test if that test is the first positive test or if $r=N$.
The maximum takes the worst case over all possible signals.
The minimum is over all choices of $\tau_{\mathrm{evo}}$ and $M$ for which the false negative probability is at most the target error probability $\delta$, with the threshold $A^*$ set as follows so that the false positive probability is at most $\delta$.
Under $\mathsf{H}_0$ every outcome is an unbiased coin flip, so $M I_r$ and $M Q_r$ are independent shifted binomial variables with success probability $1/2$, and the distribution of $A_r$ depends only on $M$ and can be computed exactly.
We set $A^*$ to the smallest value in the support of $A_r$ for which $\Pr[A_r>A^*\mid\mathsf{H}_0]\leq\delta/N$, so that the union bound over the at most $N$ tests gives a false positive probability at most $\delta$.
The false negative probability is the probability under $\mathsf{H}_n$ with initial phase $\phi$ that no test is positive.
The false negative probability and the sum in Eq.~\eqref{eq: T_CPMG} have no closed form, because the probability that a test is positive depends on the position of its time windows on the global clock, and hence on the random order of the tests.
For fixed $\tau_{\mathrm{evo}}$ and $M$, we therefore estimate both by a Monte Carlo simulation of the protocol under $\mathsf{H}_n$ for every $n\geq 1$ and every $\phi$.
We find the minimum in Eq.~\eqref{eq: T_CPMG} by comparing all pairs $(\tau_{\mathrm{evo}},M)$ for which the estimated false negative probability is at most $\delta$ for every $n\geq 1$ and every $\phi$, since the smallest $M$ that passes need not minimize the mean total sensing time.
In the numerical experiments $\phi$ is restricted to a finite set of values, as specified in Appendix~\ref{sssec: optimization_verification}.

\subsection{Phase diagram of the improvement factor}\label{ssec: phase_diagram}
In this section we estimate how large the improvement factor $\mathcal{I}$ of the random-modulation protocol over CPMG spectroscopy can be as a function of $T_2$, $\epsilon$, $\Delta$, and $N$.
We focus on the regime $T_2 \gg 1/\Delta$.
We fix $\Delta$ and $N$, and for each choice of $T_2$ and $\epsilon$ we find the optimal strategy of the random-modulation protocol and of CPMG spectroscopy separately and compare their total sensing times.
The analysis is heuristic.

Two effects shape the phase diagram, and the complexity of the analysis comes from their interplay.
\textit{Fourier broadening} decides how many modes respond to the signal within one round, and \textit{off-resonant precession} decides how the responses of consecutive rounds add up when they are averaged.
We derive both from the phase accumulated in a single round.
Consider one round whose evolution starts at time $t_0$ and ends at time $t_0+\tau_{\mathrm{evo}}$ under a modulation function $g(t)$, and let the signal be $\epsilon\cos(\omega t+\phi)$ at a single frequency $\omega$.
By the same computation as in Eq.~\eqref{eq: conventional_phase}, the accumulated phase of the round is
\begin{equation}\label{eq: phase_from_filter}
    \theta(t_0)
    =2\epsilon\int_{t_0}^{t_0+\tau_{\mathrm{evo}}} g(t)\cos(\omega t+\phi)\,dt
    =2\epsilon\,\mathrm{Re}\!\left[e^{-i\phi}\,\hat g_{t_0}(\omega)\right],
    \qquad
    \hat g_{t_0}(\omega)=\int_{t_0}^{t_0+\tau_{\mathrm{evo}}} g(t)\,e^{-i\omega t}\,dt,
\end{equation}
where the second equality uses that $g$ is real.
Hence $|\theta(t_0)|\leq 2\epsilon\,|\hat g_{t_0}(\omega)|$, so the response of one round to a signal at frequency $\omega$ is set by the Fourier transform of the modulation function truncated to the round, evaluated at $\omega$.
We call this truncated Fourier transform $\hat g_{t_0}$ the \textit{filter function}.
Both the random modulation function $Y^{(r)}(t)$ of the random-modulation protocol and the two quadrature modulation functions $\cos(\omega_n t)$ and $\sin(\omega_n t)$ of CPMG spectroscopy are linear combinations of $\cos(\omega_i t)$ and $\sin(\omega_i t)$ over the candidate frequencies, defined on the global clock and truncated to the evolution time of one round.
The protocols above realize these modulation functions through the nonlinear transformation $\mathrm{sgn}$, which maximizes the sensitivity, and we ignore this difference for simplicity of the analysis.
Since the filter function is linear in $g$, it suffices to compute it for a single mode.
For the mode $\cos(\omega_j t)$ and $\omega>0$,
\begin{equation}\label{eq: truncated_mode_ft}
\int_{t_0}^{t_0+\tau_{\mathrm{evo}}}\cos(\omega_j t)\,e^{-i\omega t}\,dt
\approx \frac{\tau_{\mathrm{evo}}}{2}\,e^{-i(\omega-\omega_j)(t_0+\tau_{\mathrm{evo}}/2)}\,
\operatorname{sinc}\!\left(\frac{(\omega-\omega_j)\tau_{\mathrm{evo}}}{2}\right),
\end{equation}
where $\operatorname{sinc}(x)=\sin(x)/x$ and the term centered at $-\omega_j$ has been dropped, which is valid when $\omega_0\tau_{\mathrm{evo}}\gg 1$.
Substituting Eq.~\eqref{eq: truncated_mode_ft} into Eq.~\eqref{eq: phase_from_filter}, the phase accumulated from this mode is
\begin{equation}\label{eq: single_mode_phase}
    \theta(t_0)
    =
    \epsilon\tau_{\mathrm{evo}}\,
    \underbrace{\operatorname{sinc}\!\left(\frac{(\omega-\omega_j)\tau_{\mathrm{evo}}}{2}\right)}_{\text{Fourier broadening}}\,
    \underbrace{\cos \bigl(\phi+(\omega-\omega_j)(t_0+\tau_{\mathrm{evo}}/2)\bigr)}_{\text{off-resonant precession}},
\end{equation}
and we take the two factors in turn.

The sinc factor comes from the truncation of the mode to the round.
Over an infinitely long evolution the Fourier transform of the mode would be a single sharp peak at $\omega_j$.
Truncation to a round of length $\tau_{\mathrm{evo}}$ broadens it into a peak of height $\tau_{\mathrm{evo}}/2$ and width $1/\tau_{\mathrm{evo}}$, and this broadening is Fourier broadening.
The sinc factor depends only on the distance $|\omega-\omega_j|$, and is of order one when $|\omega-\omega_j|\lesssim 1/\tau_{\mathrm{evo}}$ and decays as $1/(|\omega-\omega_j|\tau_{\mathrm{evo}})$ beyond.
By linearity, the filter function at a candidate frequency $\omega_i$ is a sum over all modes $j$, each contributing $\tau_{\mathrm{evo}}/2$ times its coefficient times a sinc factor and a phase factor.
When $\tau_{\mathrm{evo}}\gg 1/\Delta$, every other mode has $|\omega_i-\omega_j|\geq\Delta$ and hence a sinc factor at most $1/(\Delta\tau_{\mathrm{evo}})\ll 1$, so only the matched mode responds in one round.
When $\tau_{\mathrm{evo}}\ll 1/\Delta$, the roughly $1/(\Delta\tau_{\mathrm{evo}})$ modes within $1/\tau_{\mathrm{evo}}$ of $\omega_i$ respond in one round.

The cosine factor comes from the detuning $\omega-\omega_j$ between the signal and the mode.
For the matched mode $\omega_j=\omega$ it equals $\cos\phi$, the same in every round.
For a mode with $\omega_j\neq\omega$ the mode precesses relative to the signal at the detuning frequency, so the argument of the cosine advances by $(\omega-\omega_j)\tau_{\mathrm{evo}}$ from one round to the next, and this is off-resonant precession.
This precession matters because both protocols average the outcomes of $M$ consecutive rounds that share one modulation function, whose start times are $t_m=t_1+(m-1)\tau_{\mathrm{evo}}$ for $m=1,\dots,M$ by Eqs.~\eqref{eq: window_start} and~\eqref{eq: conventional_window_start}.
Over these rounds the argument of the cosine factor in Eq.~\eqref{eq: single_mode_phase} advances by $(\omega-\omega_j)M\tau_{\mathrm{evo}}$ in total.
When $|\omega-\omega_j|M\tau_{\mathrm{evo}}\ll 1$, the cosine factor barely changes, so the phase is the same constant in every round and the $M$ outcomes average to its response as if the mode were matched.
When $|\omega-\omega_j|M\tau_{\mathrm{evo}}\gg 1$, the cosine factor goes through many cycles, so the phase takes both signs equally often and the responses of the $M$ rounds cancel in the average.
Hence the $M$ rounds together respond only to signals within $1/(M\tau_{\mathrm{evo}})$ of $\omega_j$, even though each round alone responds to signals within $1/\tau_{\mathrm{evo}}$.
We call the set of candidate frequencies within $1/(M\tau_{\mathrm{evo}})$ of a mode a frequency bin, so a bin contains one candidate frequency when $M\tau_{\mathrm{evo}}\gg 1/\Delta$, about $1/(\Delta M\tau_{\mathrm{evo}})$ candidate frequencies when $1/(N\Delta)\ll M\tau_{\mathrm{evo}}\ll 1/\Delta$, and all $N$ candidate frequencies when $M\tau_{\mathrm{evo}}\ll 1/(N\Delta)$.
The optimal strategy for Detection balances the phase accumulated in a single round, which grows with $\tau_{\mathrm{evo}}$, against the number of candidate frequencies in a bin, which shrinks with $M\tau_{\mathrm{evo}}$.

In the following we analyze the performance of CPMG spectroscopy and the random-modulation protocol using the above argument.
For each protocol we go through the regimes of $\epsilon$ and $T_2$, analyze how the total sensing time depends on the evolution time $\tau_{\mathrm{evo}}$ of each round, determine the optimal $\tau_{\mathrm{evo}}^*$, and compare the two protocols.

\subsubsection{CPMG spectroscopy}
In CPMG spectroscopy, one test probes one modulation frequency with $2M$ rounds, $M$ for each quadrature, so the test responds to the candidate frequencies in the frequency bin of width $1/(M\tau_{\mathrm{evo}})$ around the modulation frequency.
Let $N_{\mathrm{bin}}$ be the number of bins needed to cover all $N$ candidate frequencies.
In this section the total sensing time of CPMG spectroscopy means its mean total sensing time under a signal, as in Appendix~\ref{ssec: conventional_protocol}.
The total sensing time scales as
\begin{equation}\label{eq: conventional_total_time_scaling}
    N_{\mathrm{bin}} \times M \times \tau_{\mathrm{evo}},
\end{equation}
since each test takes $2M$ rounds of evolution time $\tau_{\mathrm{evo}}$, and the protocol stops at the first test whose frequency bin contains the signal frequency, which takes of order $N_{\mathrm{bin}}$ tests on average because the modulation frequencies are drawn uniformly at random.
By the bin sizes above, $N_{\mathrm{bin}}=O(1)$ when $M\tau_{\mathrm{evo}}\ll 1/(N\Delta)$, $N_{\mathrm{bin}}=O(N\Delta M\tau_{\mathrm{evo}})$ when $1/(N\Delta)\ll M\tau_{\mathrm{evo}}\ll 1/\Delta$, and $N_{\mathrm{bin}}=O(N)$ when $M\tau_{\mathrm{evo}}\gg 1/\Delta$, so the total sensing time is $O(M\tau_{\mathrm{evo}})$, $O(N\Delta(M\tau_{\mathrm{evo}})^2)$, and $O(NM\tau_{\mathrm{evo}})$ in the three ranges.
The three expressions agree at the boundaries and all of them increase with $M\tau_{\mathrm{evo}}$, so the best strategy is the one with the shortest test.

The number of rounds $M$ is set by the response of one round whose modulation frequency lies in the bin of the signal, which by Eq.~\eqref{eq: conventional_outcome_distribution} is $e^{-\tau_{\mathrm{evo}}/T_2}\sin\theta$ with $\theta=O(\epsilon\tau_{\mathrm{evo}})$ from Eq.~\eqref{eq: single_mode_phase}.
Since the protocol stops at the first positive test, the threshold $A^*$ of Appendix~\ref{ssec: conventional_protocol} keeps the false positive probability of each test below $\delta/N$, so detecting this response from $M$ rounds requires $M=O\bigl(\ln(N/\delta)\,e^{2\tau_{\mathrm{evo}}/T_2}/\sin^2\theta\bigr)=O\bigl(\ln N\,e^{2\tau_{\mathrm{evo}}/T_2}/\sin^2\theta\bigr)$ rounds, where we take $\delta=\Theta(1)$ throughout this section.
When $\epsilon\tau_{\mathrm{evo}}\ll 1$, this gives $M=O(\ln N\,e^{2\tau_{\mathrm{evo}}/T_2}/(\epsilon\tau_{\mathrm{evo}})^2)$ and $M\tau_{\mathrm{evo}}=O(\ln N\,e^{2\tau_{\mathrm{evo}}/T_2}/(\epsilon^2\tau_{\mathrm{evo}}))$, which decreases with $\tau_{\mathrm{evo}}$ as long as $\tau_{\mathrm{evo}}\ll T_2$ and grows once $\tau_{\mathrm{evo}}$ exceeds $T_2$.
When $\epsilon\tau_{\mathrm{evo}}=O(1)$, $M=O(\ln N\,e^{2\tau_{\mathrm{evo}}/T_2})$ and $M\tau_{\mathrm{evo}}=O(\ln N\,e^{2\tau_{\mathrm{evo}}/T_2}/\epsilon)$.
When $\epsilon\tau_{\mathrm{evo}}\gg 1$, the phase wraps around, so the response does not grow while the dephasing gets stronger, and a shorter $\tau_{\mathrm{evo}}$ with $\epsilon\tau_{\mathrm{evo}}=O(1)$ is always better.
Hence the shortest test has $\tau_{\mathrm{evo}}^*=O(\min\{1/\epsilon,T_2\})$.
When $T_2\gg 1/\epsilon$, this gives $\tau_{\mathrm{evo}}^*=O(1/\epsilon)$, $M=O(\ln N)$, and $M\tau_{\mathrm{evo}}=O(\ln N/\epsilon)$.
When $T_2\ll 1/\epsilon$, it gives $\tau_{\mathrm{evo}}^*=O(T_2)$, $M=O(\ln N/(\epsilon T_2)^2)$, and $M\tau_{\mathrm{evo}}=O(\ln N/(\epsilon^2T_2))$, as in conventional sensing of a single frequency with a finite coherence time.
The regimes below differ only in which range $M\tau_{\mathrm{evo}}$ falls in, and we read off $N_{\mathrm{bin}}$ and the total sensing time in each.

The four regimes are listed in Table~\ref{tab: conventional_regimes}, which gives $\tau_{\mathrm{evo}}^*$, $M$, and $M\tau_{\mathrm{evo}}$ from Eq.~\eqref{eq: conventional_total_time_scaling}, $N_{\mathrm{bin}}$, and the total sensing time.
\begin{table}[h]
    \centering
    \small
    \begin{tabular*}{\textwidth}{@{\extracolsep{\fill}}cccccccc@{}}
        \toprule
        Regime & $\epsilon$ & $T_2$ & $\tau_{\mathrm{evo}}^*$ & $M$ & $M\tau_{\mathrm{evo}}^*$ & $N_{\mathrm{bin}}$ & Total sensing time \\
        \midrule
        1 & $\epsilon\ll\Delta\ln N$ & $T_2\gg 1/\epsilon$ & $1/\epsilon$ & $\ln N$ & $\ln N/\epsilon$ & $N$ & $N\ln N/\epsilon$ \\
        2 & $\epsilon\ll\Delta\ln N$ & $1/\Delta\ll T_2\ll 1/\epsilon$ & $T_2$ & $\ln N/(\epsilon T_2)^2$ & $\ln N/(\epsilon^2T_2)$ & $N$ & $N\ln N/(\epsilon^2T_2)$ \\
        3 & $\Delta\ln N\ll\epsilon\ll N\Delta\ln N$ & $T_2\gg 1/\Delta$ & $1/\epsilon$ & $\ln N$ & $\ln N/\epsilon$ & $N\Delta\ln N/\epsilon$ & $N\Delta\ln^2 N/\epsilon^2$ \\
        4 & $\epsilon\gg N\Delta\ln N$ & $T_2\gg 1/\Delta$ & $1/\epsilon$ & $\ln N$ & $\ln N/\epsilon$ & $1$ & $\ln N/\epsilon$ \\
        \bottomrule
    \end{tabular*}
    \caption{Optimal strategy of CPMG spectroscopy in each regime, with $\delta=\Theta(1)$. All entries hold up to constant factors, and $T_2\gg 1/\Delta$ throughout.}
    \label{tab: conventional_regimes}
\end{table}

\subsubsection{Random-modulation protocol}
In this section we analyze the random-modulation protocol in a simplified model of the protocol described in Appendix~\ref{ssec: our_protocol}, in which it always draws all $R$ random modulations and then compares $S\coloneqq S_R$ with the threshold $S^*$.
Since $S_r$ is nondecreasing in $r$, this changes neither the false positive probability nor the false negative probability, and the total sensing time becomes $RM\tau_{\mathrm{evo}}$.
This is at least the mean total sensing time in Eq.~\eqref{eq: T_RM} at the same parameters, and hence gives a conservative comparison with the baseline.
We choose $\tau_{\mathrm{evo}}$, $M$, and $R$ to minimize $RM\tau_{\mathrm{evo}}$, and all relations below hold up to constant factors.
The total sensing time depends on $\tau_{\mathrm{evo}}$ only through two quantities, the contribution to the signal of one pair of rounds whose phases coincide and the number of consecutive rounds whose phases coincide.
We first derive this dependence from the statistic $S$, and then, in each of three ranges of $\tau_{\mathrm{evo}}$, compute the two quantities from the covariance of the phases and minimize the total sensing time.
Only two kinds of rounds turn out to be useful, namely rounds shorter than $1/(N\Delta)$ and rounds whose length is close to an integer multiple of $2\pi/\Delta$.

The random-modulation protocol runs $M$ consecutive rounds of evolution time $\tau_{\mathrm{evo}}\leq T_2$ under each of $R$ random modulations.
As above, we ignore the transformation $\mathrm{sgn}$.
To simplify the analysis, we also replace the uniformly random phases of Eq.~\eqref{eq: modulation_function_experiment} by independent Gaussian coefficients, so that the accumulated phases are jointly Gaussian, and take the modulation function of one random modulation to be
\begin{equation}
    g(t)
    =
    \frac{1}{\sqrt{N}}\sum_{j=1}^{N}\left(a_j\cos(\omega_j t)+b_j\sin(\omega_j t)\right),
\end{equation}
where $a_j,b_j$ are independent standard Gaussian random variables.
We fix a hypothesis $\mathsf{H}_n$ with $n\geq 1$ and write $\theta_{r,m}$ for the phase accumulated in round $m$ of random modulation $r$.
By Eq.~\eqref{eq: phase_from_filter}, each $\theta_{r,m}$ is a linear combination of the coefficients $a_j,b_j$, so the phases of one random modulation are jointly Gaussian with mean zero.
Their covariance depends only on the distance $k$ between the two rounds, as Eq.~\eqref{eq: C_k_triangle} below shows, and we write
\begin{equation}
    C_k\coloneqq\mathbb{E}_g\bigl[\theta_{r,m}\theta_{r,m+k}\bigr],
\end{equation}
where $\mathbb{E}_g$ is the expectation over $a_j,b_j$, so that $C_0$ is the variance of the phase of one round.

We first derive when the statistic $S$ separates $\mathsf{H}_n$ from $\mathsf{H}_0$.
By Eq.~\eqref{eq: y_hat} and $y_{r,m}^2=1$,
\begin{equation}
    \hat y_r^2=\frac{1}{M}+\frac{1}{M^2}\sum_{m\neq m'}y_{r,m}y_{r,m'},
\end{equation}
so the signal in $\hat y_r^2$ comes only from the products of outcomes of two different rounds.
Given the random modulation, $y_{r,m}$ and $y_{r,m'}$ are independent with means $e^{-\tau_{\mathrm{evo}}/T_2}\sin\theta_{r,m}$ and $e^{-\tau_{\mathrm{evo}}/T_2}\sin\theta_{r,m'}$ by the same computation as in Eq.~\eqref{eq: conventional_outcome_distribution}.
Writing $\sin\theta\sin\theta'=\tfrac12[\cos(\theta-\theta')-\cos(\theta+\theta')]$ and using $\mathbb{E}[\cos X]=e^{-\mathrm{Var}(X)/2}$ for a Gaussian $X$ with mean zero,
\begin{equation}
    \mathbb{E}_g\bigl[\sin\theta_{r,m}\sin\theta_{r,m+k}\bigr]
    =\tfrac12e^{-(C_0-C_k)}-\tfrac12e^{-(C_0+C_k)}
    =e^{-C_0}\sinh C_k .
\end{equation}
Hence the mean of $\hat y_r^2$ under $\mathsf{H}_n$ exceeds its value $1/M$ under $\mathsf{H}_0$ by
\begin{equation}
    \frac{e^{-2\tau_{\mathrm{evo}}/T_2}}{M^2}\sum_{m\neq m'}e^{-C_0}\sinh C_{m-m'} .
\end{equation}
Under $\mathsf{H}_0$ the outcomes are i.i.d.\ uniform on $\{\pm1\}$, and counting the index patterns with nonzero expectation in $\hat y_r^4$ gives $\mathrm{Var}[\hat y_r^2\mid\mathsf{H}_0]=2(M-1)/M^3\approx 2/M^2$.
Over the $R$ independent random modulations, the mean of $S$ therefore grows by $R$ times this gap, while its standard deviation under $\mathsf{H}_0$ is about $\sqrt{2R}/M$, so $S$ separates $\mathsf{H}_n$ from $\mathsf{H}_0$ when $\sqrt{R}\,s\gg 1$ with
\begin{equation}\label{eq: s_heuristic}
    s\coloneqq\frac{e^{-2\tau_{\mathrm{evo}}/T_2}}{M}\sum_{m\neq m'}e^{-C_0}\sinh C_{m-m'},
\end{equation}
and we take $R\approx\max\{1,1/s^2\}$ random modulations.

We now reduce the total sensing time to the two quantities named above.
Each pair of rounds contributes $e^{-C_0}\sinh C_{m-m'}$ to $s$.
A pair with uncorrelated phases, $C_{m-m'}=0$, contributes nothing, and in particular $M=1$ gives $s=0$.
A pair with coinciding phases, $C_{m-m'}=C_0$, contributes $\tfrac12(1-e^{-2C_0})\approx\min\{C_0,1\}$, which grows with $C_0$ until the phase wraps around.
Suppose that the phases of $\bar M$ consecutive rounds coincide and that rounds farther apart have uncorrelated phases.
Each round then has about $\min\{M,\bar M\}$ partners with coinciding phases, so
\begin{equation}
    s\approx\min\{M,\bar M\}\,\tilde\sigma,
    \qquad
    \tilde\sigma\coloneqq\min\{C_0,1\}\,e^{-2\tau_{\mathrm{evo}}/T_2}.
\end{equation}
Increasing $M$ raises $s$ at a linear cost in time as long as $M\leq\bar M$, while increasing $R$ compensates a small $s$ at a quadratic cost.
Hence the protocol takes $M\approx 1/\tilde\sigma$ and $R=O(1)$ when $\bar M\geq 1/\tilde\sigma$, and $M=\bar M$ and $R\approx 1/(\bar M\tilde\sigma)^2$ otherwise, and the total sensing time is
\begin{equation}\label{eq: our_total_time_scaling}
    RM\tau_{\mathrm{evo}}
    \approx
    \begin{cases}
        \tau_{\mathrm{evo}}/\tilde\sigma, & \bar M\geq 1/\tilde\sigma,\\
        \tau_{\mathrm{evo}}/(\bar M\tilde\sigma^2), & \bar M<1/\tilde\sigma.
    \end{cases}
\end{equation}
The rest of this section computes $C_0$ and $\bar M$ as functions of $\tau_{\mathrm{evo}}$ and minimizes Eq.~\eqref{eq: our_total_time_scaling}.

We first express $C_k$ through the covariance of the modulation function.
By Eq.~\eqref{eq: phase_from_filter}, $\theta_{r,m}$ is the integral over the round of the product of $2g(t)$ and the signal $\epsilon\cos(\omega_n t+\phi)$, and dropping the terms that oscillate at $\omega_n+\omega_j$,
\begin{equation}\label{eq: theta_rm}
    \theta_{r,m}=\frac{\epsilon}{\sqrt N}\sum_{j=1}^{N}\int_{t_{r,m}}^{t_{r,m}+\tau_{\mathrm{evo}}}\Bigl(a_j\cos\bigl((n-j)\Delta t+\phi\bigr)-b_j\sin\bigl((n-j)\Delta t+\phi\bigr)\Bigr)\,dt .
\end{equation}
We compute $C_k=\mathbb{E}_g[\theta_{r,m}\theta_{r,m+k}]$ by multiplying this expression for rounds $m$ and $m+k$ and taking the expectation over $a_j,b_j$.
Since $\mathbb{E}_g[a_ja_{j'}]=\mathbb{E}_g[b_jb_{j'}]=\delta_{jj'}$ and $\mathbb{E}_g[a_jb_{j'}]=0$, only the terms with the same $j$ in both factors survive, and $\cos\alpha\cos\beta+\sin\alpha\sin\beta=\cos(\alpha-\beta)$ combines the cosine and the sine part of each term, so
\begin{align}
    C_k
    &=\frac{\epsilon^2}{N}\sum_{j=1}^{N}\int_{t_{r,m}}^{t_{r,m}+\tau_{\mathrm{evo}}}dt\int_{t_{r,m+k}}^{t_{r,m+k}+\tau_{\mathrm{evo}}}dt'\,
    \nonumber\\
    &\qquad\times\Bigl[\cos\bigl((n-j)\Delta t+\phi\bigr)\cos\bigl((n-j)\Delta t'+\phi\bigr)+\sin\bigl((n-j)\Delta t+\phi\bigr)\sin\bigl((n-j)\Delta t'+\phi\bigr)\Bigr]
    \nonumber\\
    &=\frac{\epsilon^2}{N}\sum_{j=1}^{N}\int_{t_{r,m}}^{t_{r,m}+\tau_{\mathrm{evo}}}dt\int_{t_{r,m+k}}^{t_{r,m+k}+\tau_{\mathrm{evo}}}dt'\,\cos\bigl((n-j)\Delta(t'-t)\bigr)
    \nonumber\\
    &=\int_{t_{r,m}}^{t_{r,m}+\tau_{\mathrm{evo}}}dt\int_{t_{r,m+k}}^{t_{r,m+k}+\tau_{\mathrm{evo}}}dt'\,\Gamma(t'-t),
\end{align}
where $\phi$ and the absolute times have cancelled and
\begin{equation}\label{eq: Gamma_sum}
    \Gamma(u)\coloneqq\frac{\epsilon^2}{N}\sum_{j=1}^{N}\cos\bigl((n-j)\Delta u\bigr).
\end{equation}
We change the variable $t'$ to the difference $u=t'-t$ of the two times.
For fixed $u$ the integrand no longer depends on $t$, so the integral over $t$ gives the length of the set of times $t$ in round $m$ with $t+u$ in round $m+k$.
Round $m+k$ starts $k\tau_{\mathrm{evo}}$ after round $m$ by Eq.~\eqref{eq: window_start}, so this set is the overlap of two intervals of length $\tau_{\mathrm{evo}}$ whose starting points differ by $|u-k\tau_{\mathrm{evo}}|$, and its length is $\Lambda_k(u)\coloneqq\max\{\tau_{\mathrm{evo}}-|u-k\tau_{\mathrm{evo}}|,0\}$.
Hence
\begin{equation}\label{eq: C_k_triangle}
    C_k=\int_{-\infty}^{\infty}\Lambda_k(u)\,\Gamma(u)\,du .
\end{equation}
Thus $C_k$ depends only on $k$, and it is the integral of $\Gamma$ against a triangle of height $\tau_{\mathrm{evo}}$ and half-width $\tau_{\mathrm{evo}}$ centered at $k\tau_{\mathrm{evo}}$, so $C_k$ is determined by the shape of $\Gamma$.
Writing each cosine as the real part of $e^{i(n-j)\Delta u}$ and summing the geometric series,
\begin{equation}
    \Gamma(u)=\frac{\epsilon^2}{N}\cos\Bigl(\bigl(n-\tfrac{N+1}{2}\bigr)\Delta u\Bigr)\frac{\sin(N\Delta u/2)}{\sin(\Delta u/2)} .
\end{equation}
The function $\Gamma$ is even and has period $2\pi/\Delta$.
At every integer multiple of $2\pi/\Delta$ all $N$ cosines equal one, so $\Gamma$ has a peak of height $\epsilon^2$ there, and the first zero of $\sin(N\Delta u/2)$ sets the width of the peak to about $1/(N\Delta)$.
Between the peaks, $\Gamma$ oscillates with an envelope at most $\epsilon^2/(N|\sin(\Delta u/2)|)$.
The integral of $\Gamma$ over one period is exactly $2\pi\epsilon^2/(N\Delta)$, since every term with $j\neq n$ in Eq.~\eqref{eq: Gamma_sum} integrates to zero.

We now read off $C_0$ and $\bar M$ in three cases, set by how the length of a round compares with the width $1/(N\Delta)$ and the period $2\pi/\Delta$ of the peaks of $\Gamma$, and minimize Eq.~\eqref{eq: our_total_time_scaling} within each case.
We first treat the two extreme cases, rounds much shorter than the width of a peak and rounds whose length is close to an integer multiple of the period, and then show that the rounds between them are never optimal.

First, when $\tau_{\mathrm{evo}}\ll 1/(N\Delta)$, the triangle is narrower than the peak at $u=0$.
As long as $k\tau_{\mathrm{evo}}\lesssim 1/(N\Delta)$, the triangle lies inside this peak, where $\Gamma\approx\epsilon^2$, and the triangle has area $\tau_{\mathrm{evo}}^2$, so $C_k\approx C_0=\epsilon^2\tau_{\mathrm{evo}}^2$.
For $k\tau_{\mathrm{evo}}\gg 1/(N\Delta)$ the triangle lies between the peaks and $C_k\ll C_0$, so $\bar M\approx 1/(N\Delta\tau_{\mathrm{evo}})$.
The phase of a round reaches order one at $\tau_{\mathrm{evo}}\approx 1/\epsilon$, which lies in this case only when $\epsilon\gg N\Delta$.
When $\epsilon\gg N\Delta$, the protocol takes $\tau_{\mathrm{evo}}\approx 1/\epsilon$ with $M=O(1)$ and $R=O(1)$, and the total sensing time is $1/\epsilon$, regime 4.
When $\epsilon\ll N\Delta$, every round in this case has $C_0\ll 1$ and $\bar M\tilde\sigma\approx\epsilon^2\tau_{\mathrm{evo}}/(N\Delta)\ll 1$, so the total sensing time $\tau_{\mathrm{evo}}/(\bar M\tilde\sigma^2)\approx N\Delta/(\epsilon^4\tau_{\mathrm{evo}}^2)$ decreases with $\tau_{\mathrm{evo}}$.
The protocol then takes the upper end $\tau_{\mathrm{evo}}\approx 1/(N\Delta)$ with $M=O(1)$ and $R\approx N^4\Delta^4/\epsilon^4$, and the total sensing time is $N^3\Delta^3/\epsilon^4$, regime 3b.

Second, when $\tau_{\mathrm{evo}}\gg 1/(N\Delta)$, we write $\tau_{\mathrm{evo}}=2\pi p/\Delta+\zeta$ with an integer $p\geq 0$ and $|\zeta|\leq\pi/\Delta$, and split $\theta_{r,m}$ in Eq.~\eqref{eq: theta_rm} into the term $j=n$ and the terms $j\neq n$.
The term $j=n$ has an integrand that does not depend on $t$, so it equals $(\epsilon\tau_{\mathrm{evo}}/\sqrt N)(a_n\cos\phi-b_n\sin\phi)$ in every round.
For every $j\neq n$, the product $(n-j)\Delta\cdot 2\pi p/\Delta$ is an integer multiple of $2\pi$, so the $p$ full periods of the round contribute zero, and the terms $j\neq n$ reduce to their integral over a window of length $|\zeta|$ at the end of the round, with a minus sign when $\zeta<0$.
The two parts involve disjoint sets of coefficients and are therefore uncorrelated, so
\begin{equation}\label{eq: C_k_split}
    C_k=\frac{\epsilon^2\tau_{\mathrm{evo}}^2}{N}+D_k,
\end{equation}
where $D_k$ is the covariance of the parts $j\neq n$ of rounds $m$ and $m+k$.
The same computation that led to $C_k$, together with $\bigl|\int_0^{|\zeta|}e^{ix\Delta t}\,dt\bigr|=|\zeta|\,|\operatorname{sinc}(x\Delta\zeta/2)|$ with $\operatorname{sinc}y=(\sin y)/y$, gives
\begin{equation}
    D_0=\frac{\epsilon^2}{N}\sum_{j\neq n}\zeta^2\operatorname{sinc}^2\bigl((n-j)\Delta\zeta/2\bigr).
\end{equation}
When $|\zeta|\ll 1/(N\Delta)$ every sinc factor is about one, and when $|\zeta|\gg 1/(N\Delta)$ the sum over $j$ is a Riemann sum with spacing $\Delta$ of $\int_{-\infty}^{\infty}\zeta^2\operatorname{sinc}^2(\omega\zeta/2)\,d\omega=2\pi|\zeta|$, so
\begin{equation}\label{eq: D0_bound}
    D_0\approx\min\Bigl\{\epsilon^2\zeta^2,\,\frac{2\pi\epsilon^2|\zeta|}{N\Delta}\Bigr\}\lesssim\frac{\epsilon^2}{N\Delta^2},
\end{equation}
where the last step uses $|\zeta|\leq\pi/\Delta$.
By the Cauchy--Schwarz inequality $|D_k|\leq D_0$, so $C_0-C_k=D_0-D_k\leq 2D_0$.
When $p\geq 1$ and $|\zeta|\ll\tau_{\mathrm{evo}}$, Eq.~\eqref{eq: D0_bound} gives $D_0\ll\epsilon^2\tau_{\mathrm{evo}}^2/N$, so $C_k$ stays close to $\epsilon^2\tau_{\mathrm{evo}}^2/N$ for every $k$.
Each pair of rounds then contributes $e^{-C_0}\sinh C_k$, which is at least $e^{-2D_0}$ times the contribution of a pair with coinciding phases of variance $\epsilon^2\tau_{\mathrm{evo}}^2/N$.
Hence $\bar M=\infty$ as long as $D_0\lesssim 1$, while for $D_0\gg 1$ the factor $e^{-(C_0-C_k)}$ suppresses the pairs.
With $\bar M=\infty$ and $C_0\approx\epsilon^2\tau_{\mathrm{evo}}^2/N$, Eq.~\eqref{eq: our_total_time_scaling} gives the total sensing time
\begin{equation}
    \frac{\tau_{\mathrm{evo}}}{\tilde\sigma}\approx e^{2\tau_{\mathrm{evo}}/T_2}\max\Bigl\{\frac{N}{\epsilon^2\tau_{\mathrm{evo}}},\,\tau_{\mathrm{evo}}\Bigr\},
\end{equation}
which without the dephasing factor is minimized where $C_0\approx 1$, namely at $\tau_{\mathrm{evo}}\approx\sqrt{N}/\epsilon$.
When $\epsilon\ll\sqrt{N}\Delta$, this evolution time is much longer than $1/\Delta$, and Eq.~\eqref{eq: D0_bound} gives $D_0\ll 1$ for every $\zeta$, so the evolution time need not lie near a multiple of $2\pi/\Delta$.
If $T_2\gg\sqrt{N}/\epsilon$, the protocol takes $\tau_{\mathrm{evo}}\approx\sqrt{N}/\epsilon$ with $M=O(1)$ and $R=O(1)$, and the total sensing time is $\sqrt{N}/\epsilon$, regime 1.
If $T_2\ll\sqrt{N}/\epsilon$, the protocol takes $\tau_{\mathrm{evo}}\approx T_2$, where $C_0\approx\epsilon^2T_2^2/N\ll 1$, repeats the round $M\approx N/(\epsilon^2T_2^2)$ times with $R=O(1)$, and the total sensing time is $N/(\epsilon^2T_2)$, regime 2.
When $\epsilon\gg\sqrt{N}\Delta$, already the shortest evolution time of this case with $p\geq 1$, $\tau_{\mathrm{evo}}=2\pi/\Delta$, has $C_0\approx\epsilon^2/(N\Delta^2)\gg 1$, so the total sensing time is $1/\Delta$, regime 3a.
In this regime with $\epsilon\ll N\Delta$, Eq.~\eqref{eq: D0_bound} shows that $D_0\lesssim 1$ requires the evolution time to lie within $N\Delta/\epsilon^2$ of $2\pi/\Delta$.

Third, when $p=0$, that is $1/(N\Delta)\ll\tau_{\mathrm{evo}}\leq\pi/\Delta$, the triangle in Eq.~\eqref{eq: C_k_triangle} is much wider than a peak and shorter than the period, and we show that no evolution time in this case is optimal.
For $k=0$, the peak at $u=0$ sits at the apex of the triangle, where $\Lambda_0=\tau_{\mathrm{evo}}$, so the integral of $\Gamma$ over one period gives $C_0\approx 2\pi\epsilon^2\tau_{\mathrm{evo}}/(N\Delta)$.
For $k=1$, the same peak sits at the end point of the triangle, where $\Lambda_1(u)=u$ for $0\leq u\leq\tau_{\mathrm{evo}}$ and $\Lambda_1(u)=0$ for $u<0$, and $\Gamma\approx\epsilon^2$ over the width $1/(N\Delta)$ of the peak, so
\begin{equation}
    C_1\approx\epsilon^2\int_0^{1/(N\Delta)}u\,du\approx\frac{\epsilon^2}{(N\Delta)^2}\ll C_0 .
\end{equation}
For $k\geq 2$, no peak lies under the triangle until it reaches the peak at $2\pi/\Delta$, so $C_k$ is negligible.
Only the $2(M-1)$ ordered pairs of adjacent rounds then contribute to $s$ in Eq.~\eqref{eq: s_heuristic}.
When $C_0\ll 1$, each of them contributes $e^{-C_0}\sinh C_1\approx C_1$, so $s\approx 2C_1\approx\epsilon^2/(N\Delta)^2$ for every $M$, and when $C_0\gtrsim 1$ the factor $e^{-C_0}$ makes $s$ smaller still.
Since $s$ does not grow with $\tau_{\mathrm{evo}}$, lengthening a round beyond $1/(N\Delta)$ adds time but no signal, and the protocol retreats to the first or the second case.
Rounds whose separation is close to the period $2\pi/\Delta$ become correlated again, but a strategy that uses them spends a time $M\tau_{\mathrm{evo}}>2\pi/\Delta$ per random modulation, in which rounds of length $2\pi/\Delta$ have coinciding phases of a larger variance and do at least as well.

We now compare the strategies of the three cases as $\epsilon$ grows.
When $\epsilon\ll\sqrt{N}\Delta$, the protocol takes long rounds that resolve each candidate frequency, and the total sensing time is $\sqrt{N}/\epsilon$ or $N/(\epsilon^2T_2)$, regimes 1 and 2.
When $\epsilon\gg N\Delta$, the round of length $\tau_{\mathrm{evo}}\approx 1/\epsilon\ll 1/(N\Delta)$ places all $N$ candidate frequencies in one frequency bin, so they form an effective single band, and the protocol senses this band like a single frequency in the total sensing time $1/\epsilon$, regime 4.
When $\sqrt{N}\Delta\ll\epsilon\ll N\Delta$, the evolution time $\tau_{\mathrm{evo}}\approx N\Delta/\epsilon^2$ at which $C_0\approx 2\pi\epsilon^2\tau_{\mathrm{evo}}/(N\Delta)$ of the third case reaches order one lies in the third case, where the phases of adjacent rounds are uncorrelated, so the protocol takes either shorter or longer rounds.
The first strategy takes the shorter rounds $\tau_{\mathrm{evo}}\approx 1/(N\Delta)$ and compensates their small signal with $R\approx N^4\Delta^4/\epsilon^4$ random modulations, in the total sensing time $N^3\Delta^3/\epsilon^4$.
The second strategy takes the longer rounds $\tau_{\mathrm{evo}}=2\pi/\Delta$, whose phases coincide in every round, in the total sensing time $1/\Delta$.
The protocol takes the strategy with the shorter total sensing time, and $N^3\Delta^3/\epsilon^4$ equals $1/\Delta$ at $\epsilon\approx N^{3/4}\Delta$.
Hence the protocol takes $\tau_{\mathrm{evo}}=2\pi/\Delta$ for $\epsilon\ll N^{3/4}\Delta$, regime 3a, and $\tau_{\mathrm{evo}}\approx 1/(N\Delta)$ for $\epsilon\gg N^{3/4}\Delta$, regime 3b.
Regime 3a is the only regime that relies on the equal spacing of the candidate frequencies.
Table~\ref{tab: our_regimes} collects the five regimes.

\begin{table}[t]
    \centering
    \small
    \begin{tabular*}{\textwidth}{@{\extracolsep{\fill}}ccccccc@{}}
        \toprule
        Regime & $\epsilon$ & $T_2$ & $\tau_{\mathrm{evo}}^*$ & $M$ & $R$ & Total sensing time \\
        \midrule
        1 & $\epsilon\ll\sqrt{N}\Delta$ & $T_2\gg\sqrt{N}/\epsilon$ & $\sqrt{N}/\epsilon$ & $O(1)$ & $O(1)$ & $\sqrt{N}/\epsilon$ \\
        2 & $\epsilon\ll\sqrt{N}\Delta$ & $1/\Delta\ll T_2\ll\sqrt{N}/\epsilon$ & $T_2$ & $N/(\epsilon^2T_2^2)$ & $O(1)$ & $N/(\epsilon^2T_2)$ \\
        3a & $\sqrt{N}\Delta\ll\epsilon\ll N^{3/4}\Delta$ & $T_2\gg 1/\Delta$ & $2\pi/\Delta$ & $O(1)$ & $O(1)$ & $1/\Delta$ \\
        3b & $N^{3/4}\Delta\ll\epsilon\ll N\Delta$ & $T_2\gg 1/\Delta$ & $1/(N\Delta)$ & $O(1)$ & $N^4\Delta^4/\epsilon^4$ & $N^3\Delta^3/\epsilon^4$ \\
        4 & $\epsilon\gg N\Delta$ & $T_2\gg 1/\Delta$ & $1/\epsilon$ & $O(1)$ & $O(1)$ & $1/\epsilon$ \\
        \bottomrule
    \end{tabular*}
    \caption{Optimal strategy of the random-modulation protocol in each regime, with $\delta=\Theta(1)$. All entries hold up to constant factors, and $T_2\gg 1/\Delta$ throughout. In regime 3a the evolution time must be an integer multiple of $2\pi/\Delta$ to within $N\Delta/\epsilon^2$.}
    \label{tab: our_regimes}
\end{table}

\subsubsection{Improvement factor of the random-modulation protocol}
We now compare $T_{\mathrm{CPMG}}$ and $T_{\mathrm{RM}}$ regime by regime.
Since $\ln N\ll N^{1/4}$, the scales of $\epsilon$ in Tables~\ref{tab: conventional_regimes} and~\ref{tab: our_regimes} are ordered as $\Delta\ln N\ll\sqrt{N}\Delta\ll\sqrt{N}\Delta\ln N\ll N^{3/4}\Delta\ll N\Delta/\ln N\ll N\Delta\ll N\Delta\ln N$.
Dividing the total sensing times of the two protocols regime by regime gives Table~\ref{tab: advantage}, which lists the cases by their improvement factor and separates the range where CPMG spectroscopy is faster.
\begin{table}[h]
    \centering
    \small
    \begin{tabular*}{\textwidth}{@{\extracolsep{\fill}}ccccccc@{}}
        \toprule
        Improvement factor & $\epsilon$ & $T_2$ & \multicolumn{2}{c}{CPMG} & \multicolumn{2}{c}{Random-modulation} \\
        \cmidrule(lr){4-5}\cmidrule(lr){6-7}
         & & & Regime & Total sensing time & Regime & Total sensing time \\
        \midrule
        $\sqrt{N}\ln N$ & $\epsilon\ll\Delta\ln N$ & $T_2\gg\sqrt{N}/\epsilon$ & 1 & $N\ln N/\epsilon$ & 1 & $\sqrt{N}/\epsilon$ \\
        \midrule
        $\epsilon T_2\ln N$ & $\epsilon\ll\Delta\ln N$ & $1/\epsilon\ll T_2\ll\sqrt{N}/\epsilon$ & 1 & $N\ln N/\epsilon$ & 2 & $N/(\epsilon^2T_2)$ \\
        \midrule
        $\sqrt{N}\Delta\ln^2 N/\epsilon$ & $\Delta\ln N\ll\epsilon\ll\sqrt{N}\Delta$ & $T_2\gg\sqrt{N}/\epsilon$ & 3 & $N\Delta\ln^2 N/\epsilon^2$ & 1 & $\sqrt{N}/\epsilon$ \\
        \midrule
        $\Delta T_2\ln^2 N$ & $\Delta\ln N\ll\epsilon\ll\sqrt{N}\Delta$ & $1/\Delta\ll T_2\ll\sqrt{N}/\epsilon$ & 3 & $N\Delta\ln^2 N/\epsilon^2$ & 2 & $N/(\epsilon^2T_2)$ \\
        \midrule
        $N\Delta\ln^2 N/\epsilon$ & $N\Delta\ll\epsilon\ll N\Delta\ln N$ & $T_2\gg 1/\Delta$ & 3 & $N\Delta\ln^2 N/\epsilon^2$ & 4 & $1/\epsilon$ \\
        \midrule
        $\ln N$ & $\epsilon\ll\Delta\ln N$ & $1/\Delta\ll T_2\ll 1/\epsilon$ & 2 & $N\ln N/(\epsilon^2T_2)$ & 2 & $N/(\epsilon^2T_2)$ \\
         & $\epsilon\gg N\Delta\ln N$ & $T_2\gg 1/\Delta$ & 4 & $\ln N/\epsilon$ & 4 & $1/\epsilon$ \\
        \midrule
        $N\Delta^2\ln^2 N/\epsilon^2$ & $\sqrt{N}\Delta\ll\epsilon\ll\sqrt{N}\Delta\ln N$ & $T_2\gg 1/\Delta$ & 3 & $N\Delta\ln^2 N/\epsilon^2$ & 3a & $1/\Delta$ \\
        \midrule
        $\epsilon^2\ln^2 N/(N^2\Delta^2)$ & $N\Delta/\ln N\ll\epsilon\ll N\Delta$ & $T_2\gg 1/\Delta$ & 3 & $N\Delta\ln^2 N/\epsilon^2$ & 3b & $N^3\Delta^3/\epsilon^4$ \\
        \midrule
        \multicolumn{7}{l}{\textit{Improvement factor below $1$ (CPMG spectroscopy is faster)}} \\
        \midrule
        $N\Delta^2\ln^2 N/\epsilon^2\ll 1$ & $\sqrt{N}\Delta\ln N\ll\epsilon\ll N^{3/4}\Delta$ & $T_2\gg 1/\Delta$ & 3 & $N\Delta\ln^2 N/\epsilon^2$ & 3a & $1/\Delta$ \\
        \midrule
        $\epsilon^2\ln^2 N/(N^2\Delta^2)\ll 1$ & $N^{3/4}\Delta\ll\epsilon\ll N\Delta/\ln N$ & $T_2\gg 1/\Delta$ & 3 & $N\Delta\ln^2 N/\epsilon^2$ & 3b & $N^3\Delta^3/\epsilon^4$ \\
        \bottomrule
    \end{tabular*}
    \caption{Improvement factor $\mathcal{I}$ of the random-modulation protocol over CPMG spectroscopy, namely the ratio of their total sensing times at their optimal evolution times, with $\delta=\Theta(1)$. All entries hold up to constant factors, $T_2\gg 1/\Delta$ throughout, and the regimes refer to Tables~\ref{tab: conventional_regimes} and~\ref{tab: our_regimes}.}
    \label{tab: advantage}
\end{table}

The $\sqrt{N}$ improvement factor claimed in the main text is the first line of Table~\ref{tab: advantage}.
The extra factor $\ln N$ is the number of rounds per test that CPMG spectroscopy spends to keep the false positive probability of each of its $N$ tests below $\delta/N$, whereas the random-modulation protocol makes a single decision and pays $O(1)$ rounds for the same purpose, and this factor appears in every line.
The remaining lines show that the random-modulation protocol retains an improvement factor above $1$ when $\epsilon\ll\sqrt{N}\Delta\ln N$ or $\epsilon\gg N\Delta/\ln N$, even when the conditions $\epsilon<\Delta$ and $T_2=\infty$ assumed in the main text are not satisfied, and that CPMG spectroscopy is faster in between.
Three mechanisms shape the table.

The first is the coherence time.
The $\sqrt{N}$ improvement factor requires that the phase of the random-modulation protocol accumulate until $\tau_{\mathrm{evo}}=\sqrt{N}/\epsilon$, where it wraps, whereas the phase of CPMG spectroscopy wraps already at $\tau_{\mathrm{evo}}=1/\epsilon$.
When $1/\epsilon\ll T_2\ll\sqrt{N}/\epsilon$, the evolution time of the random-modulation protocol is limited to $T_2$ and the improvement factor drops to $\epsilon T_2\ln N$.
When $T_2\ll 1/\epsilon$, both protocols are limited to $\tau_{\mathrm{evo}}=T_2$ and their single-round phases are below one radian, so the number of rounds grows as the inverse square of the phase.
The random-modulation protocol tests all $N$ candidate frequencies with one random modulation instead of scanning them one by one, but its single-round phase $\epsilon T_2/\sqrt{N}$ is smaller by a factor of $\sqrt{N}$ compared to the phase $\epsilon T_2$ of CPMG spectroscopy, so it needs $N$ times as many rounds per test and the two factors of $N$ cancel, leaving only the factor $\ln N$.

The second is Fourier broadening on the side of CPMG spectroscopy.
When $\epsilon\gg\Delta\ln N$, one test of CPMG spectroscopy covers a bin of $\epsilon/(\Delta\ln N)$ candidate frequencies, so $N_{\mathrm{bin}}=N\Delta\ln N/\epsilon$ tests suffice instead of $N$ and the improvement factor is reduced by $\Delta\ln N/\epsilon$.
When $\epsilon\gg N\Delta\ln N$, one test covers all $N$ candidate frequencies, both protocols use the same evolution time $1/\epsilon$, and again only the factor $\ln N$ remains.

The third is a range of evolution times that the random-modulation protocol cannot use.
When $1/(N\Delta)\ll\tau_{\mathrm{evo}}\ll 1/\Delta$, Fourier broadening lets the off-resonant modes of the random modulation respond to the signal, and their off-resonant precession within one round leaves consecutive rounds with uncorrelated phases by Eq.~\eqref{eq: C_k_triangle}, while the statistic $S$ detects the signal only through the correlation between the rounds of one random modulation by Eq.~\eqref{eq: s_heuristic}.
The only exceptions are integer multiples of $2\pi/\Delta$, where every off-resonant mode completes a whole number of precession cycles and cancels.
The single sinusoid of CPMG spectroscopy has no off-resonant modes, so it is free to use any evolution time up to $1/\epsilon$.
When $\sqrt{N}\Delta\ll\epsilon\ll N\Delta$, the evolution time of the random-modulation protocol $\sqrt{N}/\epsilon$ falls into this range, and we must retreat to one of its endpoints, either $2\pi/\Delta$ in regime 3a with total sensing time $1/\Delta$ or $1/(N\Delta)$ in regime 3b with total sensing time $N^3\Delta^3/\epsilon^4$, while the total sensing time $N\Delta\ln^2 N/\epsilon^2$ of CPMG spectroscopy keeps decreasing.
The improvement factor therefore falls from $\ln^2 N$ at $\epsilon=\sqrt{N}\Delta$ to $\ln^2 N/\sqrt{N}$ at $\epsilon=N^{3/4}\Delta$, where the two endpoints cost the same, and rises back to $\ln^2 N$ at $\epsilon=N\Delta$, where $1/(N\Delta)$ coincides with $1/\epsilon$ and the range is left from below.

\subsection{End-to-end numerical simulation}\label{app: end_to_end_numerics}
In this subsection we conduct an end-to-end numerical simulation to test the phase diagram of Appendix~\ref{ssec: phase_diagram}.
The analysis behind this phase diagram ignores the nonlinear transformation $\mathrm{sgn}$ in the modulation function, the finite duration $\tau_\pi$ of the $\pi$ pulses during which the signal still acts, and the minimum interval $\tau_{\mathrm{min}}$ between consecutive pulses, and Tables~\ref{tab: conventional_regimes}, \ref{tab: our_regimes}, and~\ref{tab: advantage} hold only asymptotically, up to constant factors.
The end-to-end numerical simulation here accounts for all these effects.
We ask whether the qualitative features of the phase diagram persist, namely whether the regimes of the tables appear in the order given by the tables, and whether the total sensing time scales with $N$, $T_2$, and $\epsilon$ as the tables give.
The simulation also determines the improvement factor including the constant factors that the tables omit.
Appendix~\ref{sssec: optimization_verification} first describes how we optimize the parameters of both protocols and verify the result.
Appendix~\ref{sssec: scaling_N} varies $N$ under the conditions $\epsilon<\Delta$ and $T_2=\infty$ assumed in the main text, and tests whether the improvement factor grows as $\sqrt{N}$.
Appendices~\ref{sssec: scaling_T2} and~\ref{sssec: scaling_eps} scan $T_2$ along a line of fixed $\epsilon$ and $\epsilon$ along a line of fixed $T_2$ in the phase diagram of Fig.~\ref{fig:phase_scan}a, c, and~e.
Appendix~\ref{sssec: max_adv} maximizes the improvement factor over $N$ and $\epsilon$ on a given sensor.

\subsubsection{Optimization of protocol parameters and verification}\label{sssec: optimization_verification}
Every numerical experiment below is built from one optimization unit.
We optimize the protocol parameters of both CPMG spectroscopy and the random-modulation protocol under given signal parameters and sensor parameters, namely $N$, $\Delta$, $\omega_0$, $\epsilon$, $T_2$, $\tau_\pi$, and $\tau_{\mathrm{min}}$, each to minimize its own total sensing time while keeping the false positive probability and the false negative probability at most the target error probability $\delta$.
We repeat this optimization independently for each set of signal parameters and sensor parameters.
This subsection explains how we carry out this optimization and how we verify its result.

The optimization proceeds in two stages.
First, in the design stage, we choose the parameters of the random-modulation protocol by the search of Appendix~\ref{sssec: data_processing} and those of CPMG spectroscopy by the search of Appendix~\ref{ssec: conventional_protocol}, with a restriction that saves computational time.
Then, in the verification stage, we run the end-to-end simulation of both CPMG spectroscopy and the random-modulation protocol at the selected parameters under the hypotheses of all $N$ candidate frequencies, to test whether the false negative probability at the selected parameters exceeds $\delta$ under any of the $N$ hypotheses.
Both stages restrict the initial phase $\phi$ to $N_\phi$ equally spaced values in $[0,\pi)$, since $\phi\to\phi+\pi$ leaves the distributions of $S_r$ and $A_r$ unchanged, and we take $N_\phi=6$ in all numerical experiments below.

We first specify the search of the design stage.
It ranges over $\tau_{\mathrm{evo}}$ and $M$ for both CPMG spectroscopy and the random-modulation protocol, together with $R$ for the random-modulation protocol, while the thresholds $A^*$ and $S^*$ follow from the exact distributions under $\mathsf{H}_0$.
For the random-modulation protocol, we start from three values of $\tau_{\mathrm{evo}}$ with ratio $1.5$ between consecutive values, and add values with the same ratio beyond the end at which the total sensing time is smallest, until the smallest total sensing time is attained at an interior value.
We then add values on both sides of that value with ratios $1.5^{1/2}$, $1.5^{1/4}$, and $1.5^{1/8}$ in turn.
For the random-modulation protocol we also add to these values the integer multiples of $2\pi/\Delta$ by hand, since in regime 3a of Table~\ref{tab: our_regimes} the evolution time must be an integer multiple of $2\pi/\Delta$.
For each value of $\tau_{\mathrm{evo}}$, we compare $M$ and $R$ as in Appendix~\ref{sssec: data_processing} over ranges that we extend until the selected $M$ and $R$ lie in their interior.
For CPMG spectroscopy, for each $M$ we locate to within $1\%$ the shortest $\tau_{\mathrm{evo}}$ at which the estimated false negative probability is at most $\delta$.
For every $M$ whose mean total sensing time at this $\tau_{\mathrm{evo}}$ is within $10\%$ of the smallest one, we then increase $\tau_{\mathrm{evo}}$ in steps of $3\%$ as long as the estimated mean total sensing time decreases.
We extend the range of $M$ until the selected $M$ lies in its interior.
The design stage uses $1000$ samples of the random modulation in the Monte Carlo evaluation of Eq.~\eqref{eq: K_pmf}, and $10^4$ runs in the Monte Carlo simulation of CPMG spectroscopy.

The design stage evaluates every round of the random-modulation protocol by integrating Eq.~\eqref{eq: round_propagator} numerically, as the verification stage does.
To save computational time, the design stage of both protocols takes the maxima of the false negative probability and of the mean total sensing time only over the five hypotheses $\mathsf{H}_n$ with $n=1$, $\lceil N/4\rceil$, $\lceil N/2\rceil$, $\lceil 3N/4\rceil$, and $N$, and the hypotheses of all other candidate frequencies are checked in the verification stage.

In the verification stage, for every hypothesis $\mathsf{H}_n$ with $n\geq 1$ and each of the $N_\phi$ values of $\phi$, we run each protocol $N_{\mathrm{run}}$ times at its selected parameters and record the fraction $\hat P_{n,\phi}$ of these runs that report no signal, which estimates the false negative probability $P_{n,\phi}$ under $\mathsf{H}_n$ and $\phi$.
Every run draws its measurement outcomes, together with the random modulations of the random-modulation protocol and the order of the tested candidate frequencies of CPMG spectroscopy, independently of the design stage.
Because $\hat P_{n,\phi}$ fluctuates binomially about $P_{n,\phi}$ over the finite number $N_{\mathrm{run}}$ of runs, some of the $NN_\phi$ values $\hat P_{n,\phi}$ may exceed $\delta$ even when $P_{n,\phi}\leq\delta$ for all $n$ and $\phi$, so we accept the selected parameters if
\begin{equation}\label{eq: verification_criterion}
    \max_{n,\phi}\hat P_{n,\phi}\leq\frac{k_\alpha}{N_{\mathrm{run}}},
\end{equation}
where $k_\alpha$ is the smallest integer with $\Pr[X>k_\alpha]\leq\alpha/(NN_\phi)$ for a binomial random variable $X$ with $N_{\mathrm{run}}$ trials and success probability $\delta$.
If $P_{n,\phi}\leq\delta$ for all $n$ and $\phi$, each $N_{\mathrm{run}}\hat P_{n,\phi}$ exceeds $k_\alpha$ with probability at most $\alpha/(NN_\phi)$, so by the union bound over the $NN_\phi$ values the selected parameters are rejected with probability at most $\alpha$.
We take $\alpha=0.05$, which, for example, gives $k_\alpha/N_{\mathrm{run}}=0.139$ at $N_{\mathrm{run}}=1000$, $\delta=0.1$, and $N=200$.

\subsubsection{Total sensing time against $N$ at fixed $\epsilon$ and $T_2$}\label{sssec: scaling_N}
We study how the total sensing times of the two protocols scale with $N$ in the regime $\epsilon\ll\Delta\ln N$ and $T_2\gg\sqrt{N}/\epsilon$ of the first line of Table~\ref{tab: advantage}, and how a finite $T_2$ changes this scaling.
We fix $\Delta/2\pi=40$~kHz and $\epsilon=0.05\,\Delta$, and vary $N$ from $13$ to $200$.
The candidate frequencies are $\omega_n=\omega_0+n\Delta$ with $\omega_0=\omega_N-N\Delta$ and $\omega_N/2\pi=10$~MHz for every $N$, so that the assumption $\tau_\pi+\tau_{\mathrm{min}}\ll 2\pi/\omega_N$ holds with the same margin for every $N$.
We use $\tau_\pi=8$~ns and $\tau_{\mathrm{min}}=4$~ns throughout~\cite{Lange2011}, which give $(\tau_\pi+\tau_{\mathrm{min}})\,\omega_N/2\pi=0.12$, and we take $\delta=0.1$.
Fixing $\epsilon/\Delta$ keeps $\epsilon\ll\Delta\ln N$ for every $N$, so CPMG spectroscopy stays in regime 1 of Table~\ref{tab: conventional_regimes}, where each test covers only a single candidate frequency.
The total sensing times of both protocols therefore depend on $N$ only through the number of candidate frequencies, and not through a change in Fourier broadening.
We run the random-modulation protocol with $T_2=\infty$, $4.3$, $2$, $1$, $0.5$, and $0.2$~ms, and CPMG spectroscopy only with $T_2=\infty$.
Setting $T_2=\infty$ for CPMG spectroscopy is a further simplification of the same kind as those of Appendix~\ref{ssec: conventional_protocol}, which can only shorten its total sensing time.
\begin{figure*}[t]
    \centering
    \includegraphics[width=\linewidth]{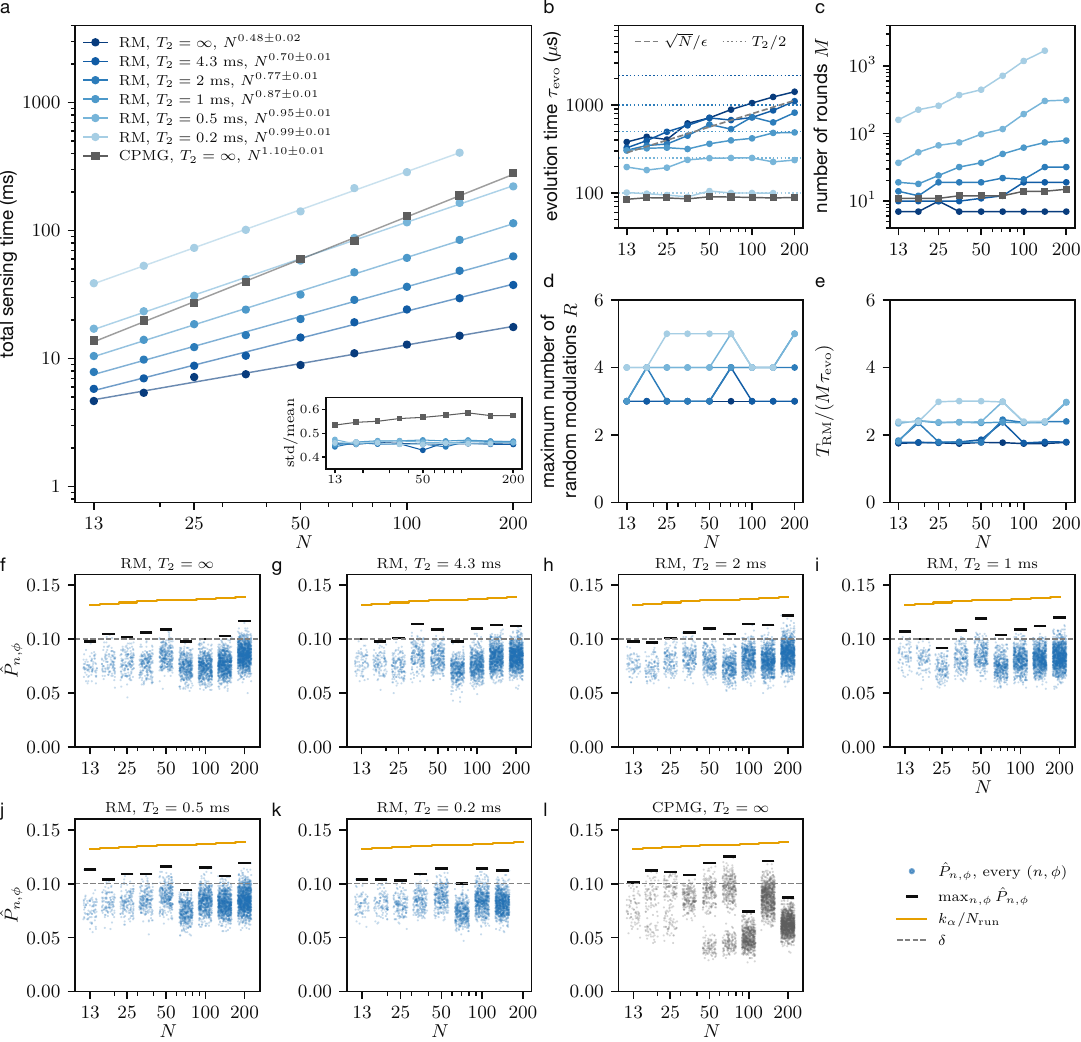}
    \caption{
    \textbf{a.}~Total sensing time $T_{\mathrm{RM}}$ of Eq.~\eqref{eq: T_RM} at six values of $T_2$ and total sensing time $T_{\mathrm{CPMG}}$ of Eq.~\eqref{eq: T_CPMG} at $T_2=\infty$ against the number $N$ of candidate frequencies, for $\Delta/2\pi=40$~kHz and $\epsilon=0.05\,\Delta$.
    The lines are power laws $AN^{\alpha}$ fitted by least squares on logarithmic scales, and the legend gives the fitted exponents $\alpha$ with their standard errors.
    The total sensing time of a single run is random, and the inset shows its standard deviation divided by its mean $T_{\mathrm{RM}}$ or $T_{\mathrm{CPMG}}$ for the curves of the main panel.
    The random-modulation protocol at $T_2=0.2$~ms has no point at $N=200$, which we omit because of its computational cost.
    \textbf{b--d.}~Evolution time $\tau_{\mathrm{evo}}$ (\textbf{b}), $M$ (\textbf{c}), and $R$ (\textbf{d}) selected in the design stage for the curves of panel~a, in the same colors.
    The gray squares in panels~b and~c are the $\tau_{\mathrm{evo}}$ and $M$ of CPMG spectroscopy.
    In panel~b, the dashed line is $\sqrt{N}/\epsilon$, and each dotted line is $T_2/2$ for the finite $T_2$ of its color.
    \textbf{e.}~Ratio $T_{\mathrm{RM}}/(M\tau_{\mathrm{evo}})$ for the curves of the random-modulation protocol in panel~a, which is the mean number of random modulations that the protocol draws under the $\mathsf{H}_n$ and $\phi$ that attain the maximum in Eq.~\eqref{eq: T_RM}.
    \textbf{f--l.}~Fraction $\hat P_{n,\phi}$ of the $N_{\mathrm{run}}=1000$ runs of the verification stage that report no signal, for every hypothesis $\mathsf{H}_n$ and every initial phase $\phi$ on the grid, in panels~f--k for the random-modulation protocol at the six values of $T_2$ and in panel~l for CPMG spectroscopy.
    The points at each $N$ are spread horizontally for visibility.
    The black bars mark $\max_{n,\phi}\hat P_{n,\phi}$ at each $N$, the orange line marks the threshold $k_\alpha/N_{\mathrm{run}}$ of Eq.~\eqref{eq: verification_criterion} with $\alpha=0.05$, and the dashed line marks the target error probability $\delta=0.1$.
    }
    \label{fig:sqrtN_scaling}
\end{figure*}

We first consider $T_2=\infty$.
In Fig.~\ref{fig:sqrtN_scaling}a, the fitted exponent of the random-modulation protocol is $0.48$ and that of CPMG spectroscopy is $1.10$.
For the random-modulation protocol, the selected $\tau_{\mathrm{evo}}$ lies between $1.0$ and $1.3$ times $\sqrt{N}/\epsilon$ at all $N$ (Fig.~\ref{fig:sqrtN_scaling}b), the selected $M$ is $7$ at all $N$ except $N=25$, where it is $10$, and the selected $R$ is $3$ at all $N$ (Fig.~\ref{fig:sqrtN_scaling}c,d).
These values agree with regime 1 of Table~\ref{tab: our_regimes}, which has $\tau_{\mathrm{evo}}^*=\sqrt{N}/\epsilon$, $M=O(1)$, and $R=O(1)$, and hence the total sensing time $\sqrt{N}/\epsilon$.
For CPMG spectroscopy, the selected $\tau_{\mathrm{evo}}$ lies between $1.06/\epsilon$ and $1.14/\epsilon$ at all $N$, and the selected $M$ grows from $11$ at $N=13$ to $15$ at $N=200$ (Fig.~\ref{fig:sqrtN_scaling}b,c).
Regime 1 of Table~\ref{tab: conventional_regimes} has $\tau_{\mathrm{evo}}^*=1/\epsilon$ and $M=\ln N$, with the total sensing time $N\ln N/\epsilon$, and we attribute the excess of the fitted exponent over $1$ to this growth of $M$ with $N$.

At finite $T_2$, the fitted exponent of the random-modulation protocol in Fig.~\ref{fig:sqrtN_scaling}a increases as $T_2$ decreases, from $0.70$ at $T_2=4.3$~ms to $0.99$ at $T_2=0.2$~ms.
Fig.~\ref{fig:sqrtN_scaling}b shows that the selected $\tau_{\mathrm{evo}}$ never exceeds $0.53\,T_2$.
As $T_2$ decreases from $4.3$~ms to $0.5$~ms, the factor by which the selected $\tau_{\mathrm{evo}}$ grows from $N=13$ to $N=200$ falls from $3.4$ to $1.2$, and the factor by which the selected $M$ grows rises from $1.9$ to $8.5$ (Fig.~\ref{fig:sqrtN_scaling}b,c).
At $T_2=0.2$~ms, the selected $\tau_{\mathrm{evo}}$ stays between $0.45\,T_2$ and $0.53\,T_2$ at all $N$, and the selected $M$ grows by a factor of $10.6$ from $N=13$ to $N=141$, while $N$ grows by a factor of $10.8$.
This trend agrees with Table~\ref{tab: our_regimes}, in which a shorter $T_2$ moves the random-modulation protocol from regime 1, with $\tau_{\mathrm{evo}}^*=\sqrt{N}/\epsilon$, $M=O(1)$, and the total sensing time $\sqrt{N}/\epsilon$, toward regime 2, with $\tau_{\mathrm{evo}}^*=T_2$, $M=N/(\epsilon^2T_2^2)$, and the total sensing time $N/(\epsilon^2T_2)$.
At all six values of $T_2$ and all $N$, the selected $R$ stays between $3$ and $5$ and $T_{\mathrm{RM}}/(M\tau_{\mathrm{evo}})$ stays between $1.7$ and $3.0$ (Fig.~\ref{fig:sqrtN_scaling}d,e), so the growth of $T_{\mathrm{RM}}$ with $N$ comes from $\tau_{\mathrm{evo}}$ and $M$.
The inset of Fig.~\ref{fig:sqrtN_scaling}a shows that the standard deviation of the total sensing time of a single run divided by its mean lies between $0.43$ and $0.47$ for the random-modulation protocol at all $N$ and all six values of $T_2$, and between $0.53$ and $0.59$ for CPMG spectroscopy.
This ratio therefore does not depend on $N$ or $T_2$ over the ranges studied, and it is smaller for the random-modulation protocol than for CPMG spectroscopy.

Fig.~\ref{fig:sqrtN_scaling}f--l show the fractions $\hat P_{n,\phi}$ of the verification stage.
At every $N$, $\max_{n,\phi}\hat P_{n,\phi}$ lies below the threshold $k_\alpha/N_{\mathrm{run}}$ for the random-modulation protocol at all six values of $T_2$ and for CPMG spectroscopy.
We therefore expect that the design stage, which checks only five hypotheses, selects parameters whose false negative probability is at most $\delta$ under all $N$ hypotheses.
In the numerical experiments below, we report only the results of the design stage.

\subsubsection{Total sensing time against $T_2$ at fixed $N$ and $\epsilon$}\label{sssec: scaling_T2}
We study how the total sensing times of the two protocols and the improvement factor depend on $T_2$ at fixed $N$.
We take $\Delta$, $\epsilon$, the candidate frequencies, $\tau_\pi$, $\tau_{\mathrm{min}}$, and $\delta$ as in Appendix~\ref{sssec: scaling_N}, with $N=2$, $25$, $50$, and $100$ and $T_2=0.06$, $0.1$, $0.2$, $0.5$, $1$, $2$, $4.3$, $10$, $20$, and $50$~ms and $T_2=\infty$.
As in Appendix~\ref{ssec: conventional_protocol}, CPMG spectroscopy runs at the same $T_2$ as the random-modulation protocol.
\begin{figure*}[t]
    \centering
    \includegraphics[width=\linewidth]{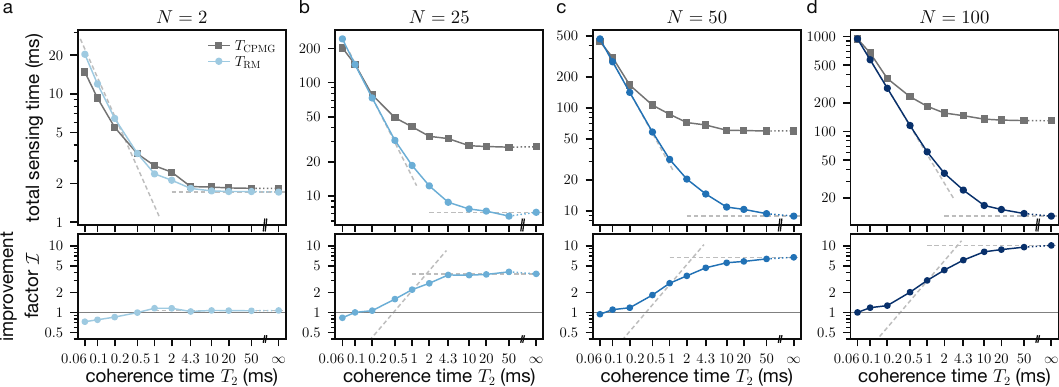}
    \caption{
    Total sensing times $T_{\mathrm{RM}}$ of Eq.~\eqref{eq: T_RM} and $T_{\mathrm{CPMG}}$ of Eq.~\eqref{eq: T_CPMG} against $T_2$, with both protocols at the same $T_2$, for $N=2$ (\textbf{a}), $25$ (\textbf{b}), $50$ (\textbf{c}), and $100$ (\textbf{d}), with $\Delta/2\pi=40$~kHz and $\epsilon=0.05\,\Delta$.
    The rightmost points are at $T_2=\infty$.
    The dashed lines are a line of slope $-1$, with its intercept fitted by least squares on logarithmic scales to the points of $T_{\mathrm{RM}}$ with $T_2\leq 0.5$~ms, and a horizontal line at the value of $T_{\mathrm{RM}}$ at $T_2=\infty$.
    The lower part of each panel shows the improvement factor $\mathcal{I}=T_{\mathrm{CPMG}}/T_{\mathrm{RM}}$ on the same $T_2$ axis.
    Its dashed lines are a line of slope $1$, with its intercept fitted in the same way to the points with $0.5~\mathrm{ms}\leq T_2\leq 2$~ms and omitted for $N=2$, and a horizontal line at the value of $\mathcal{I}$ at $T_2=\infty$.
    The gray line is $\mathcal{I}=1$.
    }
    \label{fig:sqrtN_T2}
\end{figure*}

Fig.~\ref{fig:sqrtN_T2} shows that at every $N$ the total sensing times of both protocols decrease as $T_2$ increases.
The two curves have the same shape.
At small $T_2$ the total sensing time decreases nearly as $1/T_2$, at intermediate $T_2$ the decrease slows down, and at large $T_2$ the total sensing time approaches its value at $T_2=\infty$.
The two curves nearly coincide at small $T_2$ and differ in two respects for $N\geq 25$.
The decrease of CPMG spectroscopy slows down at a smaller $T_2$ than that of the random-modulation protocol, and its value at $T_2=\infty$ is larger.
Therefore, the improvement factor $\mathcal{I}$ is close to $1$ at small $T_2$, then grows with $T_2$, and finally approaches its value at $T_2=\infty$, which grows with $N$.
For $N=2$, the decrease of both protocols slows down at roughly the same $T_2$, and the improvement factor stays close to $1$ at all $T_2$.
This behavior agrees qualitatively with Tables~\ref{tab: conventional_regimes} and~\ref{tab: our_regimes}.
In regime 2 of both tables the total sensing time is proportional to $1/T_2$, and in regime 1 it does not depend on $T_2$.
CPMG spectroscopy passes from regime 2 to regime 1 at $T_2=1/\epsilon$, and the random-modulation protocol at $T_2=\sqrt{N}/\epsilon$, which is larger by a factor of $\sqrt{N}$ and nearly the same for $N=2$.

\subsubsection{Total sensing time against $\epsilon$ at fixed $N$ and $T_2\Delta$}\label{sssec: scaling_eps}
\begin{figure*}[t]
    \centering
    \includegraphics[width=\linewidth]{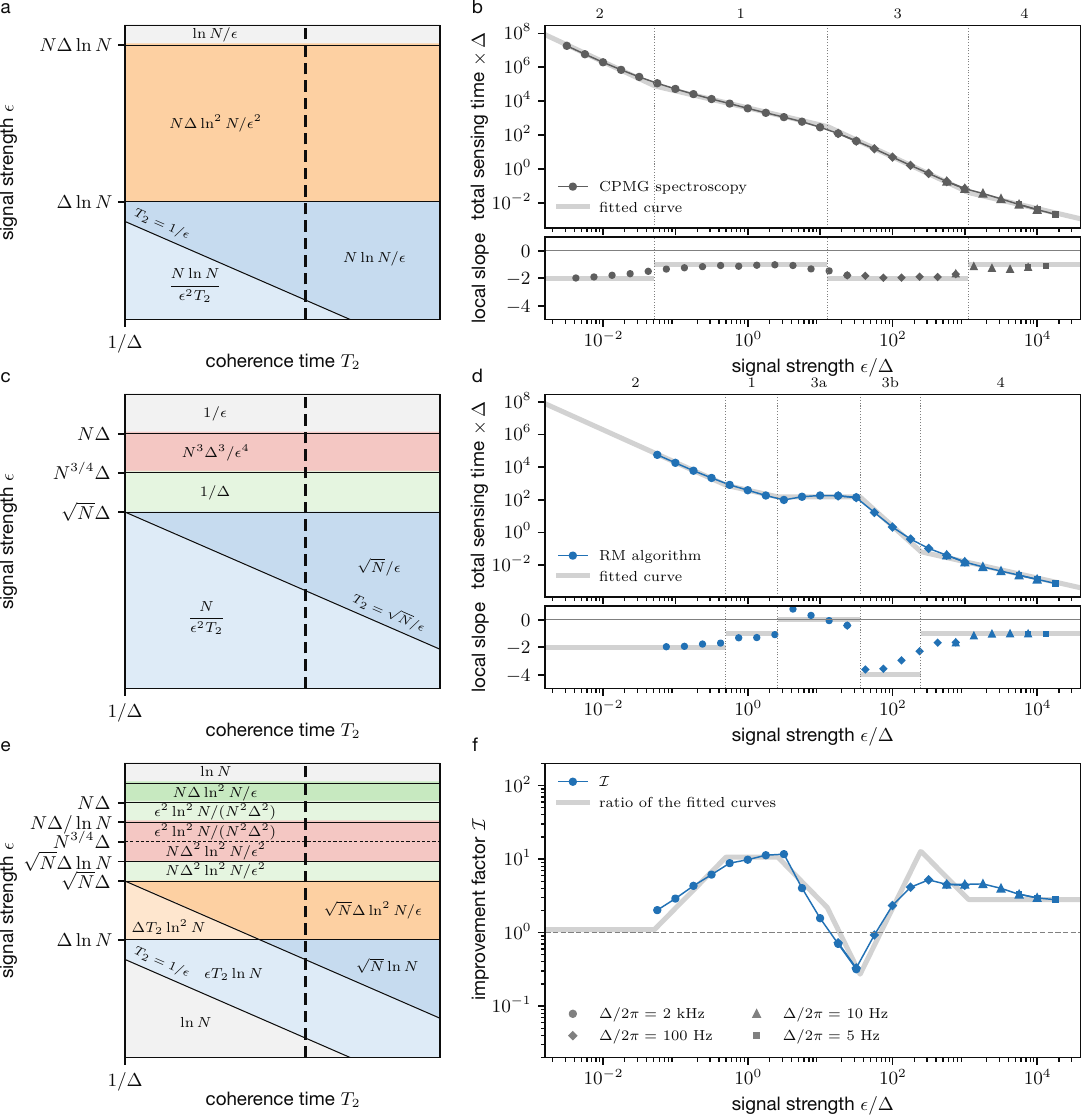}
    \caption{
    \textbf{a,\,c,\,e.}~Schematic phase diagrams of Tables~\ref{tab: conventional_regimes}, \ref{tab: our_regimes}, and~\ref{tab: advantage} in the plane of $T_2$ and $\epsilon$, showing the total sensing time of CPMG spectroscopy in each regime in panel~a, that of the random-modulation protocol in panel~c, and the improvement factor in panel~e.
    The dashed line is a line of fixed $T_2$ with $\sqrt{N}/T_2<\Delta\ln N$, which the scanned line $T_2\Delta=100$ at $N=200$ also satisfies, so both lines cross the regimes of the three tables in the same order.
    \textbf{b,\,d.}~Total sensing times $T_{\mathrm{CPMG}}$ of Eq.~\eqref{eq: T_CPMG} (\textbf{b}) and $T_{\mathrm{RM}}$ of Eq.~\eqref{eq: T_RM} (\textbf{d}) in units of $1/\Delta$ against $\epsilon/\Delta$, along the line $T_2\Delta=100$ at $N=200$.
    The lower part of each panel shows the local slope of $\ln(T\Delta)$ against $\ln(\epsilon/\Delta)$ between neighbouring points of the same $\Delta$.
    The gray lines are a continuous piecewise power law in $\epsilon$ whose exponents are those of the regimes crossed by the line $T_2\Delta=100$ in Tables~\ref{tab: conventional_regimes} and~\ref{tab: our_regimes}, and whose boundaries between regimes and overall constant are fitted by least squares to $\ln(T\Delta)$.
    The dotted vertical lines mark the fitted boundaries between regimes, and the names of the regimes are given above the panels.
    \textbf{f.}~Improvement factor $\mathcal{I}=T_{\mathrm{CPMG}}/T_{\mathrm{RM}}$ against $\epsilon/\Delta$ along the same line.
    The gray line is the ratio of the fits of panels~b and~d, and the dashed line is $\mathcal{I}=1$.
    In panels~b, d, and~f, the markers distinguish the four values $\Delta/2\pi=2$~kHz, $100$~Hz, $10$~Hz, and $5$~Hz.
    }
    \label{fig:phase_scan}
\end{figure*}

We study how the total sensing times $T_{\mathrm{CPMG}}$ of Eq.~\eqref{eq: T_CPMG} and $T_{\mathrm{RM}}$ of Eq.~\eqref{eq: T_RM} depend on $\epsilon$ along a line of fixed $N$ and $T_2\Delta$ in the phase diagram of Fig.~\ref{fig:phase_scan}a and~c.
We ask whether the regimes appear in the order given by Tables~\ref{tab: conventional_regimes} and~\ref{tab: our_regimes}, and whether within each regime the total sensing time scales with $\epsilon$ with the exponent given by the tables.
Since the tables hold only up to constant factors, we only test the qualitative behaviors.
We consider the vertical line $N=200$ and $T_2\Delta=100$.
For CPMG spectroscopy, we scan $\epsilon/\Delta$ from $10^{-2.5}$ to $2\times 10^4$.
For the random-modulation protocol, we scan $\epsilon/\Delta$ from $0.056$ to $2\times 10^4$, since the simulation at smaller $\epsilon$ exceeds our computational resources.

In the numerical simulation, we cover this range of $\epsilon/\Delta$ with four values $\Delta/2\pi=2$~kHz, $100$~Hz, $10$~Hz, and $5$~Hz, with $T_2=100/\Delta$ for each and with overlapping ranges for adjacent values.
A single value of $\Delta$ cannot cover the whole range, since the upper end requires $\Delta$ small enough that $\epsilon\tau_\pi\ll 1$ still holds, while the lower end requires $\Delta$ large enough that the total sensing time, which scales as $1/\Delta$ at fixed $\epsilon/\Delta$, does not exceed our computational resources.
For every $\Delta$ the lowest candidate frequency is $\omega_0/2\pi=10$~MHz, and $\tau_\pi$, $\tau_{\mathrm{min}}$, $\delta$, and the design stage are as in Appendix~\ref{sssec: scaling_N}.
The four values of $\Delta$ scan the same line because, in Tables~\ref{tab: conventional_regimes} and~\ref{tab: our_regimes}, the total sensing time in units of $1/\Delta$ depends only on $\epsilon/\Delta$, $T_2\Delta$, and $N$.
In the simulation, $\tau_\pi$, $\tau_{\mathrm{min}}$, and $\omega_0$ are the same for all four values of $\Delta$ and do not scale with $1/\Delta$, so the total sensing time in units of $1/\Delta$ can depend on $\Delta$, and the overlapping ranges measure this dependence.

Fig.~\ref{fig:phase_scan}a and~c are schematic phase diagrams of CPMG spectroscopy and the random-modulation protocol, drawn from Tables~\ref{tab: conventional_regimes} and~\ref{tab: our_regimes} of Appendix~\ref{ssec: phase_diagram}, and the dashed line in each marks a line of fixed $T_2$.
Fig.~\ref{fig:phase_scan}b and~d show the numerical results along the scanned line, namely $T_{\mathrm{CPMG}}$ and $T_{\mathrm{RM}}$ in units of $1/\Delta$ against $\epsilon/\Delta$.
At the values of $\epsilon/\Delta$ shared by two values of $\Delta$, the two total sensing times in units of $1/\Delta$ differ by at most $0.8\%$ for CPMG spectroscopy and by at most $4.3\%$ for the random-modulation protocol.
We therefore read the four parts as one curve.
The gray lines are continuous piecewise power laws in $\epsilon$ whose exponents are fixed to those of the tables, with only the boundaries between regimes and one overall constant fitted by least squares.

We first examine whether the total sensing time of CPMG spectroscopy agrees with Table~\ref{tab: conventional_regimes}.
Specifically, we compute the local slope of $\ln(T_{\mathrm{CPMG}}\Delta)$ against $\ln(\epsilon/\Delta)$ between neighbouring points of the same $\Delta$, shown in the lower part of Fig.~\ref{fig:phase_scan}b, and compare it with the exponents of the table.
The local slope is $-1.98$ at the smallest $\epsilon$, lies between $-1.16$ and $-1.02$ for $\epsilon/\Delta$ between $0.18$ and $5.6$, lies between $-1.96$ and $-1.78$ for $\epsilon/\Delta$ between $18$ and $5.6\times 10^2$, and lies between $-1.34$ and $-1.09$ for $\epsilon/\Delta$ above $10^3$.
These four ranges are close to $-2$, $-1$, $-2$, and $-1$, the exponents of regimes 2, 1, 3, and 4 of Table~\ref{tab: conventional_regimes} in the order in which the scanned line crosses these regimes.

We next examine whether the total sensing time of the random-modulation protocol agrees with Table~\ref{tab: our_regimes}, using the local slope of $\ln(T_{\mathrm{RM}}\Delta)$ against $\ln(\epsilon/\Delta)$ in the lower part of Fig.~\ref{fig:phase_scan}d.
The local slope is $-1.96$ and $-1.93$ on the two intervals at the smallest $\epsilon$, and changes gradually from $-1.77$ to $-1.08$ for $\epsilon/\Delta$ between $0.18$ and $3.2$.
For $\epsilon/\Delta$ between $3.2$ and $32$ it falls from $0.78$ to $-0.43$, and the total sensing time varies by at most a factor of $1.9$ while $\epsilon$ grows by a factor of $10$.
For $\epsilon/\Delta$ between $32$ and $10^2$ it is $-3.62$ and $-3.57$ on two adjacent intervals, then it changes gradually and lies between $-1.16$ and $-1.00$ for $\epsilon/\Delta$ above $10^3$.
This sequence follows the exponents $-2$, $-1$, $0$, $-4$, and $-1$ of regimes 2, 1, 3a, 3b, and 4 of Table~\ref{tab: our_regimes} in the order in which the scanned line crosses these regimes.
The local slope agrees qualitatively with the phase diagram.

Finally, Fig.~\ref{fig:phase_scan}e is the schematic phase diagram of the improvement factor $\mathcal{I}$, drawn from Table~\ref{tab: advantage}, and Fig.~\ref{fig:phase_scan}f shows $\mathcal{I}=T_{\mathrm{CPMG}}/T_{\mathrm{RM}}$ along the scanned line, computed from the data of Fig.~\ref{fig:phase_scan}b and~d.
It rises from $2.0$ at $\epsilon/\Delta\approx 0.056$ to $11.7$ at $\epsilon/\Delta\approx 3.2$, falls to $0.32$ at $\epsilon/\Delta\approx 32$, rises to $5.2$ at $\epsilon/\Delta\approx 3.2\times 10^2$, and falls to $2.8$ at the largest $\epsilon$.
It is below $1$ at the three values $\epsilon/\Delta\approx 18$, $32$, and $56$.
This behavior agrees qualitatively with the phase diagram of Fig.~\ref{fig:phase_scan}e.
The gray line in Fig.~\ref{fig:phase_scan}f is the ratio of the gray fitted lines of Fig.~\ref{fig:phase_scan}b and~d.

\subsubsection{Maximum improvement factor $\mathcal{I}_{\max}$ against $\xi$}
\label{sssec: max_adv}
We now ask how large the improvement factor $\mathcal{I}$ can be on a given sensor, maximized over the signals that the sensor can probe.
The sensor is described by $T_2$, $\tau_\pi$, and $\tau_{\mathrm{min}}$, and $\xi=T_2/(\tau_\pi+\tau_{\mathrm{min}})$ is the maximum number of pulses that fit within the coherence time.
We choose the width $W=\omega_N-\omega_0$ of the band of candidate frequencies as a compromise between two competing effects.
First, a larger $W$ holds more candidate frequencies at the same spacing $\Delta$, so $N$ can grow without bringing the candidate frequencies closer together, which would let Fourier broadening shorten CPMG spectroscopy.
Second, a larger $W$ requires a higher $\omega_N$ and hence a higher rate of $\pi$ pulses, which increases the dead time and reduces the sensitivity.
In this section we therefore let the candidate frequencies fill the band from $\omega_0=0.1\,\omega_N$ to $\omega_N$, and we choose $\omega_N$ such that $(\tau_\pi+\tau_{\mathrm{min}})\,\omega_N/2\pi=0.24$ for every sensor.
The band then has width $W=0.9\,\omega_N$, and the spacing of the candidate frequencies is $\Delta=W/N$.
Multiplying $\tau_\pi$, $\tau_{\mathrm{min}}$, $T_2$, and every evolution time by the same factor, and dividing $\omega_n$, $\Delta$, and $\epsilon$ by this factor, leaves every outcome distribution unchanged.
This rescaling multiplies the total sensing times of both protocols by the same factor, so their ratio $\mathcal{I}$ is invariant.
With the band chosen as above, the maximum of $\mathcal{I}$ over $N$ and $\epsilon/\Delta$ therefore depends on the sensor only through $\xi$.
We denote this maximum by $\mathcal{I}_{\max}(\xi)$.
Since $\omega_0/\omega_N$ and $(\tau_\pi+\tau_{\mathrm{min}})\,\omega_N$ are fixed rather than optimized, $\mathcal{I}_{\max}(\xi)$ is a conservative estimate of the largest improvement factor achievable on such a sensor.

We estimate $\mathcal{I}_{\max}(\xi)$ from Table~\ref{tab: advantage}.
The largest improvement factor in the table, $\sqrt{N}\ln N$, is that of its first line, where CPMG spectroscopy is in regime 1 of Table~\ref{tab: conventional_regimes} and the random-modulation protocol is in regime 1 of Table~\ref{tab: our_regimes}.
The first line requires $\epsilon\lesssim\Delta\ln N$, so that Fourier broadening does not shorten CPMG spectroscopy, and $\sqrt{N}/\epsilon\lesssim T_2$, so that the evolution time of the random-modulation protocol fits within the coherence time.
Both conditions hold together only if $\sqrt{N}/T_2\lesssim\Delta\ln N$, namely $N^{3/2}/\ln N\lesssim N\Delta T_2=\xi$.
The improvement factor $\sqrt{N}\ln N$ grows with $N$, so the maximum is reached at the largest $N$ with $N^{3/2}/\ln N\lesssim\xi$, which gives $N\sim\xi^{2/3}$ and $\mathcal{I}_{\max}(\xi)\sim\xi^{1/3}$ up to logarithmic factors.
At this $N$ the two conditions leave only $\epsilon\sim\sqrt{N}/T_2\sim\Delta\ln N$, and the evolution time $\sqrt{N}/\epsilon$ of the random-modulation protocol is of order $T_2$.

The analysis in the previous paragraph omits constant factors and assumes the order $\Delta\ln N\ll\sqrt{N}\Delta$.
At finite $N$ the constant factors can reverse this order, and the estimate of $\mathcal{I}_{\max}(\xi)$ then has the exponent $1/2$ instead of $1/3$.
Suppose that the order is reversed once the omitted constant factors are included, which we write as $\sqrt{N}\Delta\lesssim\Delta\ln N$.
CPMG spectroscopy then stays in regime 1 of Table~\ref{tab: conventional_regimes} for every $\epsilon\lesssim\sqrt{N}\Delta$.
The first line of Table~\ref{tab: advantage} therefore requires only $\epsilon\lesssim\sqrt{N}\Delta$, so that the random-modulation protocol stays in regime 1 of Table~\ref{tab: our_regimes}, and $\sqrt{N}/\epsilon\lesssim T_2$ as before.
Both conditions hold together only if $\Delta T_2\gtrsim 1$, namely $N\lesssim N\Delta T_2=\xi$.
The improvement factor $\sqrt{N}\ln N$ again grows with $N$, so the maximum is reached at the largest $N$ with $N\lesssim\xi$, which gives $N\sim\xi$ and $\mathcal{I}_{\max}(\xi)\sim\xi^{1/2}$ up to logarithmic factors.
At this $N$ the two conditions leave only $\epsilon\sim\sqrt{N}/T_2\sim\sqrt{N}\Delta$, and the evolution time $\sqrt{N}/\epsilon$ of the random-modulation protocol is of order $1/\Delta\sim T_2$.
Since $\sqrt{N}/\ln N$ grows with $N$, the reversed order holds only below some value of $N$ if the constant factors do not grow with $N$.
In both estimates the value of $N$ at the maximum grows with $\xi$, so the estimate $\xi^{1/2}$ applies at small $\xi$ and the estimate $\xi^{1/3}$ at large $\xi$.

\begin{figure*}[t]
    \centering
    \includegraphics[width=\linewidth]{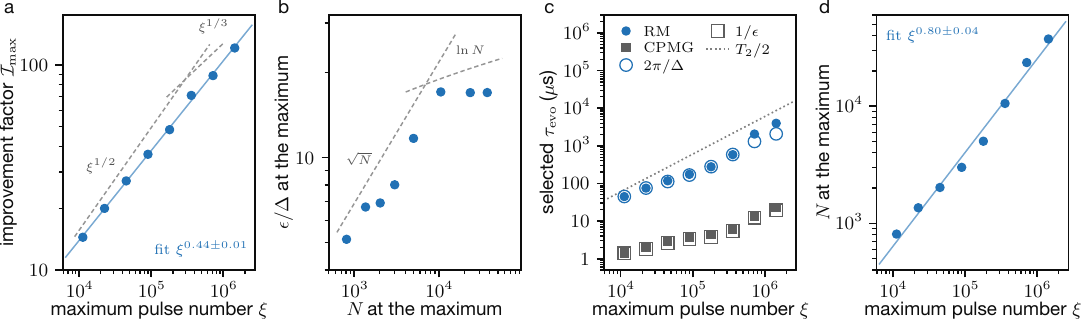}
    \caption{
    \textbf{a.}~Maximum improvement factor $\mathcal{I}_{\max}$ against the maximum pulse number $\xi=T_2/(\tau_\pi+\tau_{\mathrm{min}})$, at eight values of $T_2$ from $0.135$ to $17.28$~ms with $\tau_\pi=8$~ns and $\tau_{\mathrm{min}}=4$~ns.
    The solid line is a power law $K\xi^{\alpha}$ fitted by least squares on logarithmic scales, and the label gives the fitted exponent $\alpha$ with its standard error.
    The dashed lines are guides to the eye proportional to $\xi^{1/2}$ and $\xi^{1/3}$.
    \textbf{b.}~The value of $\epsilon/\Delta$ at which $\mathcal{I}_{\max}$ is attained against the value of $N$ at which it is attained.
    The dashed lines are guides to the eye proportional to $\sqrt{N}$ and $\ln N$.
    \textbf{c.}~The evolution time $\tau_{\mathrm{evo}}$ selected in the design stage at the $N$ and $\epsilon/\Delta$ at which $\mathcal{I}_{\max}$ is attained, for the random-modulation protocol (filled circles) and for CPMG spectroscopy under a signal (filled squares).
    The open circles show $2\pi/\Delta$ and the open squares $1/\epsilon$, both evaluated at the same $N$ and $\epsilon/\Delta$.
    The dotted line is $T_2/2$.
    \textbf{d.}~The value of $N$ at which $\mathcal{I}_{\max}$ is attained against $\xi$.
    The solid line is a power law fitted as in panel~a.
    }
    \label{fig:maxadv}
\end{figure*}

We compute $\mathcal{I}_{\max}(\xi)$ numerically and examine how it and the values of $N$ and $\epsilon/\Delta$ at which it is attained depend on $\xi$ in the range of $\xi$ that we can simulate.
We take $\tau_\pi=8$~ns~\cite{Lange2011}, $\tau_{\mathrm{min}}=4$~ns, and $\omega_N/2\pi=20$~MHz~\cite{Zopes2017, Nomura2023}, so the $N$ candidate frequencies fill the band from $2$ to $20$~MHz.
We take $\delta=0.1$ and use the design stage of Appendix~\ref{sssec: optimization_verification}.
We take eight values of $T_2$ from $0.135$ to $17.28$~ms, each twice the previous one, which give $\xi$ from $1.1\times 10^4$ to $1.4\times 10^6$.
For each $T_2$ we run the design stage on a grid of $N$ and $\epsilon/\Delta$ and extend the grid until the maximum of $\mathcal{I}$ lies in its interior.

The results are shown in Fig.~\ref{fig:maxadv}.
In Fig.~\ref{fig:maxadv}a, we present  $\mathcal{I}_{\max}$ against $\xi$, and a single power law fits $\mathcal{I}_{\max}$ over the whole range, with the exponent $\alpha=0.44\pm0.01$ between the exponents $1/2$ and $1/3$ of the two estimates.
The simulated range of $\xi$ is not wide enough to observe a clear change in scaling.
In Fig.~\ref{fig:maxadv}b, we present the value of $\epsilon/\Delta$ at which $\mathcal{I}_{\max}$ is attained against the value of $N$ at which it is attained.
This value of $\epsilon/\Delta$ first rises with $N$ and then saturates to a plateau.
This is clear evidence that agrees qualitatively with the predicted transition from $\mathcal{I}_{\max}\sim\xi^{1/2}$, where $\epsilon\sim\sqrt{N}\Delta$, to $\mathcal{I}_{\max}\sim\xi^{1/3}$, where $\epsilon\sim\Delta\ln N$.
In Fig.~\ref{fig:maxadv}c, we present the evolution times $\tau_{\mathrm{evo}}$ that the two protocols select at the maximum, together with $2\pi/\Delta$ and $1/\epsilon$.
The random-modulation protocol selects $\tau_{\mathrm{evo}}$ close to $2\pi/\Delta$ at small $\xi$, as in the estimate $\xi^{1/2}$, but not at the two largest values of $\xi$, which also agrees with the predicted transition.
In Fig.~\ref{fig:maxadv}d, we present the value of $N$ at which $\mathcal{I}_{\max}$ is attained against $\xi$, and a single power law fits it with the exponent $0.80\pm0.04$, between the exponents $1$ and $2/3$ of the two estimates.
We observe no clear transition here, which we attribute to the weak dependence of $\mathcal{I}$ on $N$ near the maximum.

\section{Quantum-computing-enhanced Learning in wireless communication}\label{app:qg}
In this section, we apply our quantum-computing-enhanced Learning algorithm to wireless communication, demonstrating that it enables data rates unattainable by any ancilla-free sensing protocol in the weak-signal regime. This provides a concrete new motivation for building quantum computers: they are essential for achieving a practically useful advantage via quantum sensing.

We first define the communication protocol. We consider orthogonal frequency-division multiple access (OFDMA)~\cite{tse2005fundamentals}, where multiple users transmit simultaneously to a base station that decodes their information. The users transmit within a government-allocated frequency band $[\omega_c, \omega_c + W]$ of width $W$ above the carrier frequency $\omega_c$. Since frequencies outside this band are reserved for other uses, the base station must reject out-of-band signals. The sensor qubit downconverts the received signal through quantum frequency mixing with a known bias field at $\omega_c$~\cite{Wang2022}, shifting the signal band to $[0,W]$. For simplicity, each of the $N$ users transmits at a single frequency (its subcarrier), encoding one bit per symbol duration $T$ in the sign of its cosine amplitude.
This encoding is the quadrature phase-shift keying (QPSK) of the main text restricted to the cosine quadrature, which changes every comparison below only by constant factors.
The downconverted signal is therefore
\begin{align}\label{eq:qg-signal}
    f_{\vec v}(t) = \sum_{n=1}^N v_n \cos(\omega_n t),
    \qquad \omega_n = n\Delta,
    \qquad \Delta := \frac{W}{N},
    \qquad v_n \in \{-\epsilon, +\epsilon\},
    \qquad 0 \le t \le T,
\end{align}
where $\Delta$ is the subcarrier spacing, and $f_{\vec v}(t)$ denotes the signal generating the Hamiltonian $H = \gamma f_{\vec{v}}(t)Z$ with $\gamma$ denoting the coupling constant between the receiver hardware and the source of the signal (i.e. we keep the $\gamma$ in Appendix~\ref{sec:sensing-model}, which was previously omitted, explicit).
This is the signal Eq.~\eqref{eq:appendix-signal} with $v_{N+n} = 0$, and the subcarrier frequencies satisfy the separation condition Eq.~\eqref{eq:frequency-separation}.
The base station knows the subcarrier frequencies in advance. Each symbol consists of the signal $f_{\vec v}(t)$ over $0 \le t \le T$ and encodes $N$ bits. The communication protocol is specified by the bandwidth $W$, number of subcarriers $N$, and symbol duration $T$. The \textit{data rate}, defined as the number of bits transmitted per unit time, is
\begin{equation}\label{eq:qg-rate}
R := \frac{N}{T}.
\end{equation}
Thus, the data rate depends only on the protocol parameters, with higher rates being preferable.
Hence $R$ differs from the data rate $2N/T$ of the main text by a factor of 2.

A communication protocol must satisfy two constraints to ensure distinguishable subcarriers and the validity of the signal model Eq.~\eqref{eq:qg-signal}. First, the symbol duration must satisfy $T\ge 1/\Delta$, since a time window $T$ provides frequency resolution $1/T$. For $N\ge 2$, this resolves neighboring subcarriers; for $N=1$, where $\Delta=W$, it suppresses leakage from outside the band into the single subcarrier.
Second, the symbol duration must be at most a time $T^*$. Each user’s carrier arrives with a phase relative to the base station’s local oscillator that varies with user motion. The base station estimates and compensates this phase using periodically transmitted pilot signals (i.e. a known signal), yielding the zero-initial-phase signal Eq.~\eqref{eq:qg-signal}. This compensation requires the relative phase to remain nearly constant over each symbol, limiting its duration to $T^*$, the timescale of negligible phase variation.
Together, these constraints give
\begin{equation}\label{eq:qg-constraints}
\frac{N}{W} = \frac{1}{\Delta} \le T \le T^*.
\end{equation}

A \textit{receiver} processes the signal Eq.~\eqref{eq:qg-signal} over one symbol and outputs $N$ bits. It succeeds at amplitude $\epsilon$ under a given communication protocol if all $N$ bits are correct with probability at least $2/3$ for every $\vec v$ in Eq.~\eqref{eq:qg-signal}. A pair $(\epsilon,R)$ is \textit{achievable} if the receiver succeeds at amplitude $\epsilon$ under some protocol satisfying Eq.~\eqref{eq:qg-constraints} with data rate $R$. The set of such pairs defines the receiver’s \textit{achievable regime}, which we use to compare receivers.
We consider three receivers.
\begin{itemize}
    \item The \textit{time-sampling receiver} divides the symbol into sampling intervals of duration $\pi/W$, estimates the average of $f_{\vec v}(t)$ over each interval by a Ramsey measurement, and recovers every $v_n$ from these estimates by a discrete Fourier transform.
    By the Nyquist theorem, this recovery is possible because the sampling rate $W/\pi$ is twice the highest frequency $W/2\pi$ of the signal Eq.~\eqref{eq:qg-signal}.
    \item The class of \textit{coherent quantum receivers} consists of all sensing protocols in the sense of Appendix~\ref{sec:sensing-model} that use no auxiliary qubits, and it contains the time-sampling receiver.
    \item The \textit{quantum computational receiver} is the Learning algorithm of Appendix~\ref{app:periodic-signal-learning}, which uses auxiliary qubits.
\end{itemize}
As a practical reference, we also consider the \textit{classical antenna}, representing conventional receivers that record the full noisy signal over one symbol:
\begin{equation}\label{eq:qg-antenna}
    y(t) = f_{\vec v}(t) + \xi(t), \qquad 0 \le t \le T, \qquad \mathbb{E}[\xi(t)\xi(t')] = \frac{S_0}{\gamma^2}\,\delta(t - t'),
\end{equation}
where $\xi(t)$ is Gaussian white noise with power spectral density $S_0/\gamma^2$.
We assume that $f_{\vec v}(t)$ is the desired signal and that all noise originates from the antenna's coupling to the field and subsequent electronic amplification, with $S_0$ independent of the bandwidth $W$.

We determine each receiver's achievable regime in three steps.
First, for fixed symbol duration $T$, we vary $N$ from $1$ to $WT$, as permitted by Eq.~\eqref{eq:qg-constraints}.
For each $N$, the minimum amplitude required for successful reception defines an achievable pair $(\epsilon, N/T)$, yielding a curve with data rates from $1/T$ to $W$.
Second, we take the union of these curves over all allowed $T$.
Third, we extend the resulting regime to larger amplitudes and lower data rates: success at amplitude $\epsilon$ implies success at any larger amplitude, while the data rate can be reduced without changing the receiver by allowing some users to transmit bits already known to the base station.
We call the minimum achievable amplitude at each data rate the lower boundary of the achievable regime.
By Eq.~\eqref{eq:qg-constraints}, the data rate $R = N/T$ ranges from $1/T^*$, at $N = 1$ and $T = T^*$, to $W$, at $N = WT$, and we state each lower boundary for this range.
For $R < 1/T^*$, by the third step the lower boundary stays at its value at $R = 1/T^*$.
We focus on scaling with $W$ and $T^*$, using $\sim$ and $\gtrsim$ to suppress constant and logarithmic factors.

\subsection{Time-sampling receiver}

A symbol of duration $T$ is divided into $K = TW/\pi$ sampling intervals, each of duration $\pi/W$.
In each interval, a Ramsey measurement gives a Heisenberg-limited estimate of the average of $f_{\vec v}(t)$ over the interval, with a standard deviation of order $W/\gamma$.
We estimate each $v_n$ by performing a discrete Fourier transform on these sampled points, which averages $K$ independent estimates, so the standard deviation of each $v_n$ is of order $(W/\gamma)/\sqrt{K}$.
The receiver decodes every sign once $\epsilon$ exceeds this standard deviation, so at fixed $T$ it succeeds when
\begin{equation}\label{eq:qg-eps-samp}
    \gamma\epsilon \gtrsim \sqrt{\frac{W}{T}}.
\end{equation}
This smallest amplitude does not depend on $R$, so the curve at fixed $T$ is a horizontal line at $\gamma\epsilon \sim \sqrt{W/T}$ spanning data rates from $1/T$ to $W$.
As $T$ increases, this line moves down and extends to lower data rates, so the line at $T = T^*$ lies below all others and gives the lower boundary
\begin{equation}\label{eq:qg-boundary-samp}
    \epsilon_{\mathrm{samp}} \sim \frac{1}{\gamma}\sqrt{\frac{W}{T^*}},
\end{equation}
shown in Fig.~\ref{fig:qg-sm}a.

\subsection{Coherent quantum receivers}

The signals Eq.~\eqref{eq:qg-signal} are the hard instances of Theorem~\ref{thm:main}, whose proof bounds the sensing time needed to decode every sign of $v_n \in \{-\epsilon, +\epsilon\}$.
Hence every coherent quantum receiver that succeeds at amplitude $\epsilon$ obeys the lower bound on the sensing time in Theorem~\ref{thm:main}, and since the receiver acts during one symbol, its sensing time is at most $T$, so
\begin{equation}\label{eq:qg-time-af}
    T \gtrsim \frac{N}{\gamma\epsilon\left(1 + 2\gamma\epsilon/\Delta\right)}.
\end{equation}
We translate this bound into a condition on $\epsilon$ and $R$.
We consider the two limits $\gamma\epsilon \ll \Delta$ and $\gamma\epsilon \gg \Delta$.
For $\gamma\epsilon \ll \Delta$, the bound Eq.~\eqref{eq:qg-time-af} reduces to $T \gtrsim N/(\gamma\epsilon)$, which gives $\gamma\epsilon \gtrsim N/T = R$.
For $\gamma\epsilon \gg \Delta$, the bound Eq.~\eqref{eq:qg-time-af} reduces to $T \gtrsim N\Delta/(\gamma\epsilon)^2 = W/(\gamma\epsilon)^2$, which gives $\gamma\epsilon \gtrsim \sqrt{W/T}$ independently of $R$.
The two conditions coincide at $\gamma\epsilon \sim \Delta$, so at fixed $T$ every coherent quantum receiver succeeds only when
\begin{equation}\label{eq:qg-eps-af}
    \gamma\epsilon \gtrsim \min\left(R, \sqrt{\frac{W}{T}}\right).
\end{equation}
The curve at fixed $T$ is therefore the line $\gamma\epsilon \sim R$ for data rates from $1/T$ to $\sqrt{W/T}$, followed by the horizontal line $\gamma\epsilon \sim \sqrt{W/T}$ for data rates from $\sqrt{W/T}$ to $W$.
As $T$ increases, the line $\gamma\epsilon \sim R$ stays in place and extends to lower data rates, while the horizontal line moves down, so the curve at $T = T^*$ lies below all others and gives the lower boundary
\begin{equation}\label{eq:qg-boundary-af}
    \epsilon_{\mathrm{coh}} \sim \frac{1}{\gamma}\min\left(R, \sqrt{\frac{W}{T^*}}\right),
\end{equation}
shown in Fig.~\ref{fig:qg-sm}b.
This lower boundary can be reached.
For $R \le \sqrt{W/T^*}$, sequential CPMG spectroscopy reaches it by probing each subcarrier frequency for a time $1/R = T^*/N$.
This probing decodes each sign once $\gamma\epsilon \gtrsim R$, and the probing time $1/R$ resolves neighboring subcarrier frequencies because $R \le \sqrt{W/T^*}$ is equivalent to $R \le \Delta$.
For $R \ge \sqrt{W/T^*}$, the time-sampling receiver reaches it by Eq.~\eqref{eq:qg-boundary-samp}.

\subsection{Quantum computational receiver}

Theorem~\ref{thm:periodic-learning} estimates every $v_n$ to precision $\epsilon/2$ in sensing time $\widetilde O(\sqrt{N}/(\gamma\epsilon) + 1/\Delta)$, and this precision suffices to decode every sign of $v_n \in \{-\epsilon, +\epsilon\}$.
The term $1/\Delta$ is at most $T$ by Eq.~\eqref{eq:qg-constraints}, so at fixed $T$ the quantum computational receiver succeeds when $\gamma\epsilon \gtrsim \sqrt{N}/T$, which reads, after substituting $N = RT$,
\begin{equation}\label{eq:qg-eps-q}
    \gamma\epsilon \gtrsim \sqrt{\frac{R}{T}}.
\end{equation}
The curve at fixed $T$ is therefore the line $\gamma\epsilon \sim \sqrt{R/T}$ for data rates from $1/T$ to $W$, which rises from $\gamma\epsilon \sim 1/T$ at a single subcarrier frequency to $\gamma\epsilon \sim \sqrt{W/T}$ at $R = W$.
As $T$ increases, this line moves down and extends to lower data rates, so the line at $T = T^*$ lies below all others and gives the lower boundary
\begin{equation}\label{eq:qg-boundary-q}
    \epsilon_{\mathrm{QC}} \sim \frac{1}{\gamma}\sqrt{\frac{R}{T^*}},
\end{equation}
shown in Fig.~\ref{fig:qg-sm}c.

Furthermore, by Theorem~\ref{thm:periodic-learning-lower-bound},
every receiver that succeeds at amplitude $\epsilon$ under a communication protocol with $N$ subcarriers and symbol duration $T$ satisfies $T \gtrsim \sqrt{N}/(\gamma\epsilon)$, with or without ancilla qubits.
Since $N = RT$ and $T \le T^*$, this condition reads $\gamma\epsilon \gtrsim \sqrt{R/T} \ge \sqrt{R/T^*}$, so no receiver reaches below the lower boundary Eq.~\eqref{eq:qg-boundary-q} and the quantum computational receiver is optimal among all receivers.

\subsection{Classical antenna}

The classical antenna estimates each $v_n$ by projecting the recorded waveform Eq.~\eqref{eq:qg-antenna} onto its subcarrier frequency,
\begin{equation}\label{eq:qg-antenna-estimate}
    \hat v_n := \frac{2}{T}\int_0^T y(t)\cos(\omega_n t)\,dt,
\end{equation}
and outputs the sign of $\hat v_n$ as the $n$th bit.
We take $T\Delta/2\pi$ to be an integer, so that $\frac{2}{T}\int_0^T \cos(\omega_m t)\cos(\omega_n t)\,dt = \delta_{mn}$ and the signal part of $\hat v_n$ equals $v_n$.
The noise part of $\hat v_n$ is Gaussian with mean zero, and by Eq.~\eqref{eq:qg-antenna} its variance is
\begin{equation}
    \frac{4}{T^2}\int_0^T\!\!\int_0^T \frac{S_0}{\gamma^2}\,\delta(t - t')\cos(\omega_n t)\cos(\omega_n t')\,dt\,dt'
    = \frac{4S_0}{\gamma^2 T^2}\int_0^T \cos^2(\omega_n t)\,dt
    = \frac{2S_0}{\gamma^2 T}.
\end{equation}
The standard deviation of $\hat v_n$ is therefore of order $\sqrt{S_0/T}/\gamma$, and the receiver decodes every sign once $\epsilon$ exceeds it, so at fixed $T$ it succeeds when
\begin{equation}\label{eq:qg-eps-ant}
    \gamma\epsilon \gtrsim \sqrt{\frac{S_0}{T}}.
\end{equation}
This smallest amplitude does not depend on $R$, so the curve at fixed $T$ is a horizontal line at $\gamma\epsilon \sim \sqrt{S_0/T}$ spanning data rates from $1/T$ to $W$.
As $T$ increases, this line moves down and extends to lower data rates, so the line at $T = T^*$ lies below all others and gives the lower boundary
\begin{equation}\label{eq:qg-boundary-ant}
    \epsilon_{\mathrm{ant}} \sim \frac{1}{\gamma}\sqrt{\frac{S_0}{T^*}}.
\end{equation}
The lower boundary Eq.~\eqref{eq:qg-boundary-ant} has the same form as the lower boundary Eq.~\eqref{eq:qg-boundary-samp} of the time-sampling receiver, a horizontal line that decreases as $1/\sqrt{T^*}$, with $S_0$ in place of $W$.
The two lower boundaries differ in their dependence on the bandwidth, since $S_0$ is independent of $W$ by assumption while $\epsilon_{\mathrm{samp}}$ grows as $\sqrt{W}$.

\subsection{Comparison}

\begin{figure*}[t]
    \centering
    \includegraphics[width=\linewidth]{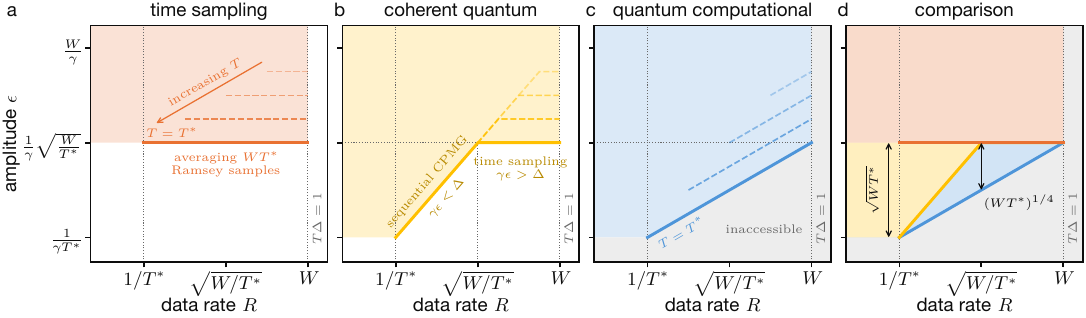}
    \caption{Schematic achievable regimes of the time-sampling receiver, the coherent quantum receivers, and the quantum computational receiver, on logarithmic axes with constant and logarithmic factors omitted.
    \textbf{a}--\textbf{c}~Achievable regime (shaded) of the time-sampling receiver in \textbf{a}, of the coherent quantum receivers in \textbf{b}, and of the quantum computational receiver in \textbf{c}.
    The curves are the smallest amplitude reached at fixed symbol duration $T$, Eqs.~\eqref{eq:qg-eps-samp},~\eqref{eq:qg-eps-af}, and~\eqref{eq:qg-eps-q} with $N = RT$, for several $T < T^*$ (dashed) and for $T = T^*$ (solid).
    Each curve runs from $R = 1/T$, where $N = 1$, to $R = W$, where $T\Delta = 1$.
    The curve at $T = T^*$, extended horizontally to data rates below $1/T^*$, is the lower boundary of the achievable regime.
    In \textbf{b}, no coherent quantum receiver reaches the region below the lower boundary, by Theorem~\ref{thm:main}.
    The lower boundary is reached by sequential CPMG spectroscopy for $R < \sqrt{W/T^*}$ and by the time-sampling receiver for $R > \sqrt{W/T^*}$.
    \textbf{d}~Comparison of the achievable regimes.
    The time-sampling receiver reaches the orange region, the coherent quantum receivers reach the orange and the yellow regions, and the quantum computational receiver reaches the orange, the yellow, and the blue regions.
    Arrows mark the largest improvement factors of the quantum computational receiver over the time-sampling receiver, $\sqrt{WT^*}$ at $R = 1/T^*$, and over the coherent quantum receivers, $(WT^*)^{1/4}$ at $R = \sqrt{W/T^*}$.
    In \textbf{c} and \textbf{d}, no receiver reaches the gray region.}
    \label{fig:qg-sm}
\end{figure*}

Fig.~\ref{fig:qg-sm}d compares the achievable regimes of the time-sampling receiver, the coherent quantum receivers, and the quantum computational receiver.
At the largest data rate $R = W$, the three lower boundaries Eqs.~\eqref{eq:qg-boundary-samp},~\eqref{eq:qg-boundary-af}, and~\eqref{eq:qg-boundary-q} coincide at $\gamma\epsilon \sim \sqrt{W/T^*}$.
Since no receiver reaches below the lower boundary Eq.~\eqref{eq:qg-boundary-q}, no receiver succeeds at a smaller amplitude than the time-sampling receiver at this data rate.
Current communication protocols such as 4G and 5G are operating at $T\Delta = 1$, which gives $R = W$.
At smaller data rates, the coherent quantum receivers and the quantum computational receiver succeed at smaller amplitudes than the time-sampling receiver.
As $R$ decreases below $W$, $\epsilon_{\mathrm{samp}}$ stays constant, $\epsilon_{\mathrm{QC}}$ decreases as $\sqrt{R}$, and $\epsilon_{\mathrm{coh}}$ decreases as $R$ once $R < \sqrt{W/T^*}$.
Both improve on the time-sampling receiver the most at the smallest data rate $R = 1/T^*$, where $N = 1$, by the factor
\begin{equation}
    \frac{\epsilon_{\mathrm{samp}}}{\epsilon_{\mathrm{coh}}} \sim \frac{\epsilon_{\mathrm{samp}}}{\epsilon_{\mathrm{QC}}} \sim \sqrt{WT^*}.
\end{equation}
At this data rate, sequential CPMG spectroscopy and the quantum computational receiver accumulate the phase of the single subcarrier frequency coherently over the whole symbol duration $T^*$, whereas the time-sampling receiver averages $WT^*/\pi$ independent Ramsey measurements of duration $\pi/W$.

Between $R = 1/T^*$ and $R = W$, the quantum computational receiver improves on every coherent quantum receiver by the factor
\begin{equation}\label{eq:qg-adv-af}
    \frac{\epsilon_{\mathrm{coh}}}{\epsilon_{\mathrm{QC}}} \sim \min\left(\sqrt{N}, \sqrt{\frac{WT^*}{N}}\right), \qquad N = RT^*,
\end{equation}
which is largest, $(WT^*)^{1/4}$, at $N = \sqrt{WT^*}$, that is, at $R = \sqrt{W/T^*}$.
The two terms of the minimum in Eq.~\eqref{eq:qg-adv-af} compare the quantum computational receiver with the two receivers that reach the lower boundary Eq.~\eqref{eq:qg-boundary-af}, sequential CPMG spectroscopy for $R \le \sqrt{W/T^*}$ and the time-sampling receiver for $R \ge \sqrt{W/T^*}$.
The factor $\sqrt{N}$ arises because the quantum computational receiver processes all $N$ subcarrier frequencies simultaneously in sensing time of order $\sqrt{N}/(\gamma\epsilon)$, whereas sequential CPMG spectroscopy probes them one at a time in sensing time of order $N/(\gamma\epsilon)$.
The factor $\sqrt{WT^*/N} = \sqrt{T^*\Delta}$ arises because the signal Eq.~\eqref{eq:qg-signal} is periodic with period $2\pi/\Delta$, so a symbol of duration $T^*$ contains of order $T^*\Delta$ periods.
The quantum computational receiver accumulates the phase coherently over these periods, whereas the time-sampling receiver averages them as independent measurements, which gives the factor $\sqrt{T^*\Delta}$.

Finally, by Eqs.~\eqref{eq:qg-boundary-q} and~\eqref{eq:qg-boundary-ant}, the quantum computational receiver outperforms the classical antenna when
\begin{equation}
    R < S_0,
\end{equation}
and in this range it improves on the classical antenna by the factor
\begin{equation}
    \frac{\epsilon_{\mathrm{ant}}}{\epsilon_{\mathrm{QC}}} \sim \sqrt{\frac{S_0}{R}},
\end{equation}
which is largest, $\sqrt{S_0 T^*}$, at $R = 1/T^*$.

\subsection{Finite $T_2$ analysis}

\begin{figure*}[t]
    \centering
    \includegraphics[width=\linewidth]{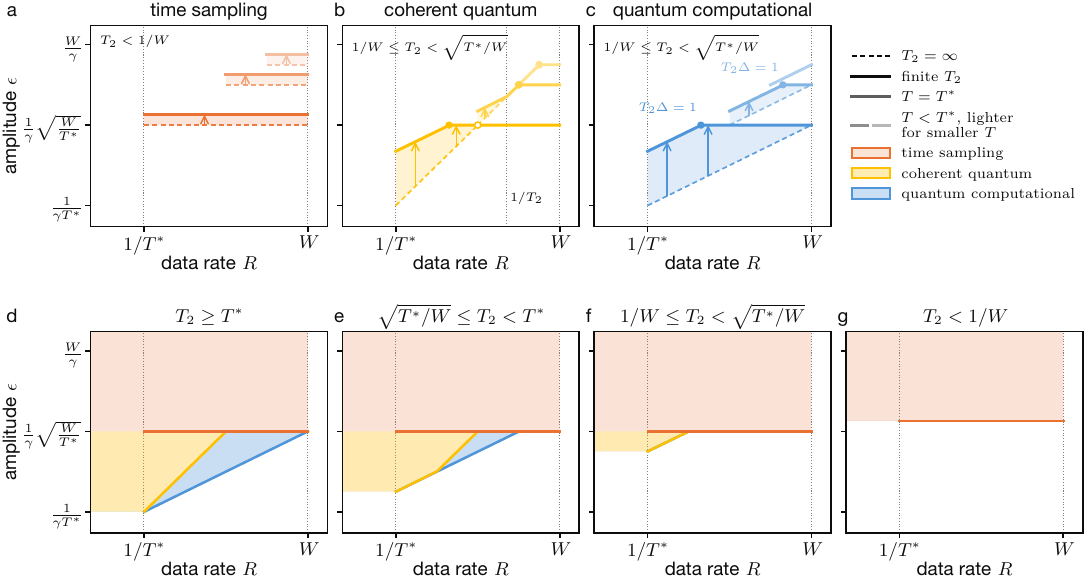}
    \caption{Schematic effect of a finite coherence time $T_2$ on the achievable regimes, on logarithmic axes with constant and logarithmic factors omitted.
    \textbf{a}--\textbf{c}~Curves at fixed symbol duration $T$ of the time-sampling receiver in \textbf{a}, of the coherent quantum receivers in \textbf{b}, taken as the better of sequential CPMG spectroscopy and the time-sampling receiver, and of the quantum computational receiver in \textbf{c}, for three symbol durations $T \le T^*$.
    Dashed curves are at $T_2 = \infty$, and solid curves are at a finite $T_2$ in the range labeled in the panel.
    Lighter curves are at smaller $T$.
    Arrows and shaded bands mark how finite $T_2$ raises each curve.
    In \textbf{b}, open and filled dots mark where the curve switches from sequential CPMG spectroscopy to the time-sampling receiver at $T_2 = \infty$ and at finite $T_2$.
    In \textbf{c}, filled dots mark where the curve at finite $T_2$ meets the curve of the time-sampling receiver, at $T_2\Delta = 1$.
    \textbf{d}--\textbf{g}~Achievable regimes for $T_2$ in the range given above each panel.
    The time-sampling receiver reaches the orange region, the coherent quantum receivers reach the orange and the yellow regions, and the quantum computational receiver reaches the orange, the yellow, and the blue regions.}
    \label{fig:qg-t2}
\end{figure*}

Here we analyze how a finite coherence time $T_2$ of the sensor qubit changes the achievable regimes of the time-sampling receiver, sequential CPMG spectroscopy, and the quantum computational receiver.
We study only how finite $T_2$ changes the performance of these three receivers.
We model $T_2$ heuristically by requiring that a receiver evolve coherently for at most $T_2$.
If a receiver needs a coherent evolution shorter than $T_2$, it performs exactly as for $T_2 = \infty$.
Otherwise, it splits the evolution into rounds of duration at most $T_2$, and we recompute the smallest amplitude at which it succeeds.
For each receiver, we study how finite $T_2$ changes the curve formed by the achievable pairs $(\epsilon, R)$ at fixed $T$ (Fig.~\ref{fig:qg-t2}a--c), and we then compare the resulting achievable regimes (Fig.~\ref{fig:qg-t2}d--g).

We first study the effect of finite $T_2$ on the time-sampling receiver.
Its coherent evolution is a single Ramsey measurement of duration $\pi/W$, independent of $T$ and $R$.
For $T_2 \gtrsim 1/W$, this evolution fits within $T_2$, so no curve at fixed $T$ changes.
For $T_2 \lesssim 1/W$, the receiver instead performs $T/T_2$ Ramsey measurements of duration $T_2$ during one symbol, each with a standard deviation of order $1/(\gamma T_2)$.
The sampling rate $1/T_2$ exceeds $W/\pi$, so the discrete Fourier transform still recovers every $v_n$ and averages the $T/T_2$ independent estimates, and the receiver succeeds when
\begin{equation}\label{eq:qg-eps-samp-t2}
    \gamma\epsilon \gtrsim \frac{1}{\sqrt{T T_2}}.
\end{equation}
Every curve at fixed $T$ thus remains a horizontal line and moves up by a factor of order $1/\sqrt{WT_2}$.

We next study the effect of finite $T_2$ on the coherent quantum receivers, namely the better one between sequential CPMG spectroscopy and the time-sampling receiver.
At $T_2 = \infty$, the curve at fixed $T$ consists of two segments that meet at $R = \sqrt{W/T}$, where $\gamma\epsilon \sim \Delta$.
For $R \le \sqrt{W/T}$, sequential CPMG spectroscopy is better and succeeds when $\gamma\epsilon \gtrsim R$, and for $R \ge \sqrt{W/T}$, the time-sampling receiver is better and succeeds when $\gamma\epsilon \gtrsim \sqrt{W/T}$.
Since the time-sampling receiver was analyzed above, it remains to study how finite $T_2$ changes the segment of sequential CPMG spectroscopy, and then to take the better of the two receivers at each data rate.
Sequential CPMG spectroscopy probes each subcarrier frequency in a coherent evolution of duration $T/N = 1/R$, so at data rates $R \ge 1/T_2$ this evolution fits within $T_2$ and the segment does not change.
At data rates $R < 1/T_2$, the receiver probes each subcarrier frequency in $1/(RT_2)$ rounds of duration $T_2$, each with a standard deviation of order $1/(\gamma T_2)$, and averaging them raises the segment to
\begin{equation}\label{eq:qg-eps-cpmg-t2}
    \gamma\epsilon \sim \sqrt{\frac{R}{T_2}}.
\end{equation}
The shape of the curve at fixed $T$ therefore depends on where $1/T_2$ lies relative to the range $1/T \le R \le \sqrt{W/T}$ of the segment of sequential CPMG spectroscopy.
If $1/T_2 \le 1/T$, that is, $T \le T_2$, no data rate is raised and the curve does not change.
If $1/T < 1/T_2 < \sqrt{W/T}$, only the data rates below $1/T_2$ are raised to Eq.~\eqref{eq:qg-eps-cpmg-t2}, which stays below $\gamma\epsilon \sim 1/T_2 < \sqrt{W/T}$ there, so the two segments still meet at $R = \sqrt{W/T}$.
If $1/T_2 \ge \sqrt{W/T}$, the whole segment of sequential CPMG spectroscopy is raised to Eq.~\eqref{eq:qg-eps-cpmg-t2}, which reaches the height $\sqrt{W/T}$ of the segment of the time-sampling receiver at $R = WT_2/T \le \sqrt{W/T}$, so the time-sampling receiver is better from this lower data rate on.
Fig.~\ref{fig:qg-t2}b shows the three shapes.
These three shapes assume $T_2 \gtrsim 1/W$, for which the segment of the time-sampling receiver does not change.
For $T_2 \lesssim 1/W$, the segment of the time-sampling receiver rises to Eq.~\eqref{eq:qg-eps-samp-t2}, while every data rate satisfies $R \le W < 1/T_2$, so the whole segment of sequential CPMG spectroscopy is raised to Eq.~\eqref{eq:qg-eps-cpmg-t2}.
Since $\sqrt{R/T_2} \ge 1/\sqrt{TT_2}$ for every $R \ge 1/T$, sequential CPMG spectroscopy is never better than the time-sampling receiver, and the curve at fixed $T$ coincides with that of the time-sampling receiver.

Finally, we study the effect of finite $T_2$ on the quantum computational receiver.
We take the ancilla qubits to have the same coherence time $T_2$ as the sensor qubit, which is conservative, since ancilla qubits that do not couple to the signal typically have longer coherence times than the sensor qubit.
At $T_2 = \infty$, the quantum computational receiver divides each period $2\pi/\Delta$ of the signal into short intervals, assigns an ancilla qubit to each interval, and lets each ancilla qubit accumulate the phase of its interval coherently over all periods of the symbol, in a single coherent evolution that lasts the whole symbol duration $T$.
The time scale that matters for $T_2$ is therefore the period $2\pi/\Delta$, since an ancilla qubit must keep its phase coherent from one period to the next, and not just the duration of a single interval.
For $T \le T_2$, this evolution fits within $T_2$, so the curve at fixed $T$ does not change.
For $T > T_2$, the receiver repeats coherent evolutions of duration $T_2$, each covering of order $T_2\Delta$ periods.
If $T_2\Delta \gtrsim 1$, each coherent evolution estimates every $v_n$ with a standard deviation of order $\sqrt{N}/(\gamma T_2)$, as Theorem~\ref{thm:periodic-learning} gives with $T_2$ in place of the sensing time, and averaging the $T/T_2$ evolutions lowers this standard deviation to order $\sqrt{N/(TT_2)}/\gamma = \sqrt{R/T_2}/\gamma$, so the curve rises to Eq.~\eqref{eq:qg-eps-cpmg-t2}.
If $T_2\Delta \lesssim 1$, an evolution covers less than one period, so no phase accumulates over several periods, and the receiver succeeds when $\gamma\epsilon \gtrsim \sqrt{W/T}$, as the time-sampling receiver does.
Unlike for the coherent quantum receivers, finite $T_2$ therefore raises the whole curve for $T > T_2$, and the raised curve meets the curve of the time-sampling receiver at $T_2\Delta = 1$, as shown in Fig.~\ref{fig:qg-t2}c.
For $T_2 \lesssim 1/W$, every data rate has $T_2\Delta \lesssim 1$ since $\Delta \le W$, and an evolution is shorter than even a Ramsey measurement of the time-sampling receiver, so the whole curve coincides with the raised curve Eq.~\eqref{eq:qg-eps-samp-t2} of the time-sampling receiver.

We now compare the achievable regimes of the three receivers at finite $T_2$.
By the analysis above, the achievable regimes behave qualitatively differently in four ranges of $T_2$, separated by $T_2 = T^*$, $T_2 = \sqrt{T^*/W}$, and $T_2 = 1/W$, and shown in Fig.~\ref{fig:qg-t2}d--g.
For $T_2 \ge T^*$ (Fig.~\ref{fig:qg-t2}d), no curve with $T \le T^*$ changes, and the lower boundaries are those of Fig.~\ref{fig:qg-sm}d.
For $\sqrt{T^*/W} \le T_2 < T^*$ (Fig.~\ref{fig:qg-t2}e), the lower boundary of the quantum computational receiver rises to $\gamma\epsilon \sim \sqrt{R/T_2}$ up to $R \sim WT_2/T^*$, while that of the coherent quantum receivers rises to the same value only for $R < 1/T_2$.
The quantum computational receiver therefore improves on the coherent quantum receivers only for $1/T_2 < R < WT_2/T^*$, and its largest improvement factor, still at $R = \sqrt{W/T^*}$, decreases from $(WT^*)^{1/4}$ to $(WT_2^2/T^*)^{1/4}$.
At $R = 1/T^*$, both receivers improve on the time-sampling receiver by $\sqrt{WT_2}$ instead of $\sqrt{WT^*}$.
For $1/W \lesssim T_2 < \sqrt{T^*/W}$ (Fig.~\ref{fig:qg-t2}f), the lower boundaries of the quantum computational receiver and of the coherent quantum receivers coincide.
The coherent quantum receivers still improve on the time-sampling receiver for $R < WT_2/T^*$, by at most $\sqrt{WT_2}$ at $R = 1/T^*$.
For $T_2 \lesssim 1/W$ (Fig.~\ref{fig:qg-t2}g), all three lower boundaries coincide with that of the time-sampling receiver, $\gamma\epsilon \sim 1/\sqrt{T^*T_2}$.
The quantum computational receiver therefore improves on the coherent quantum receivers at some data rate as long as $T_2 \gtrsim \sqrt{T^*/W}$.

\section{Exponential advantage for parity check learning with ancilla qubits}\label{app:parity-learning}

In this section, we show, by embedding a parity check learning problem in the wireless communication signals of Appendix~\ref{app:qg}, that our quantum-computing-enhanced speedup for Learning gives an exponential separation between quantum computational sensing algorithms and ancilla-free sensing algorithms. More specifically, we consider signals whose spectral coefficients have fixed amplitude but unknown sign, where the signs are resampled randomly at periodic intervals, subject to an unknown parity check. The goal is to learn the parity check. This signal construction is identical to that in Ref.~\cite{mele2026exponentialseparationsensingcontinuous} but uses frequency-division multiplexing rather than time-division multiplexing. A quantum-computing-enhanced receiver can use our Learning algorithm to efficiently determine the signs and recover the parity check by simple linear algebra. Meanwhile, we prove that an ancilla-free receiver requires exponentially more sensing time to recover the parity check, regardless of the strategy employed.

We now specify the parity check learning problem more precisely. We use the wireless communication signals of Appendix ~\ref{app:qg}: during each \emph{symbol}, each of the $N$ equally spaced subcarriers has amplitude either $+\eps$ or $-\eps$ for some known $\eps \in (0,1]$. Let $\tau$ denote the symbol duration, and let the subcarrier frequencies be $\omega_n = n \Delta$ for $\Delta =  2 \pi N / \tau$. Each symbol duration therefore consists of $N$ periods, for a data transmission rate of $R = N/\tau$.
A \emph{packet} consists of $L$ consecutive symbols, and therefore transmits $D = L N$ total bits. Finally, the unknown parity check is a nonzero vector $s \in \{0,1\}^D$: its nonzero entries denote the bits which must sum to zero modulo two. The transmitter draws independent packets according to
\begin{align}
    X^{(r)}\sim\operatorname{Unif}(s^\perp),\qquad
    s^\perp=\{x\in\{0,1\}^D:s\cdot x\equiv0 \text{ mod } 2\},\qquad r=1,2,\ldots.
    \label{eq:parity-source}
\end{align}
Write $X^{(r)}=(X_1^{(r)},\ldots,X_L^{(r)})$, with $X_j^{(r)}\in\{0,1\}^N$. In symbol $j$ of packet $r$, the signal is
\begin{align}
    f(t)=\eps\sum_{n=1}^N(-1)^{X_{j,n}^{(r)}}\cos(n\Delta t),\qquad t\in I_{r,j},\qquad I_{r,j}=[((r-1)L+j-1)\tau,\,((r-1)L+j)\tau).
    \label{eq:parity-signal}
\end{align}
This defines the data stream for all $t \ge 0$; a receiver accesses the data stream through the time dependent Hamiltonian $H(t) = f(t) Z$. To reemphasize, consecutive packets are resampled uniformly from $s^\perp$, rather than repeated, and the signal amplitudes change from symbol to symbol.

The parity check learning problem is as follows:

\begin{problem}[Parity check learning]\label{problem:parity-learning}
Given evolution under $H(t) = f(t) Z$ for $f(t)$ of the form Eq.~\eqref{eq:parity-signal}, identify the unknown parity check $s \in \{0,1\}^D$ with probability at least $2/3$.
\end{problem}

Our main result is an exponential separation for parity check learning with and without ancilla qubits:

\begin{theorem}[Exponential separation for parity check learning]\label{thm:parity-learning} Let $C$ be the constant in Lemma~\ref{lem:information}, let $\eps \in (0,1]$, let $R \ge 64C\eps$, and let $L \ge 2$. Set $N=\ceil{c_0(R \ln L / \eps)^2}$ for some constant $c_0$ and let $D = LN$. There is a quantum-computing-enhanced sensing algorithm for Problem~\ref{problem:parity-learning} which uses $O(\log^3 L)$ ancillas and total time
\begin{align}
    T_{\mathrm Q}\le O \left( \frac{D^2}{R} \right).
    \label{eq:parity-quantum-time}
\end{align}
Meanwhile, every ancilla-free sensing algorithm for Problem~\ref{problem:parity-learning} must take expected total time
\begin{align}
    T_{\mathrm{AF}} \ge \Omega \left( \frac{D}{R} 2^L \right).
\end{align}
\end{theorem}
\noindent
At fixed $R,\eps$, the quantum-computing-enhanced receiver takes $O(L^2\log^4 L)$ time, whereas the ancilla-free receiver requires $\Omega(L\log^2 L\,2^L)$ time. The separation is exponential in the number of symbols per packet, $L$. In the full check length $D=\Theta(L\log^2 L)$, the lower bound is $\exp(\Omega(D/\log^2 D)) = \exp(D^{1-o(1)})$.
\\

\noindent\emph{Proof Sketch.}
The quantum-computing-enhanced receiver decodes each symbol using the Learning algorithm of Theorem~\ref{thm:signal_learning_qft}. Each successfully recovered packet gives a linear equation $X^{(r)}\cdot s=0$, so with high probability $O(D)$ packets suffice to determine $s$. $N$ is chosen large enough that the Learning algorithm terminates within a single symbol duration.

For the lower bound, we consider parity checks involving bits from every symbol. Step~1 bounds the information obtained from measurements during one symbol. Step~2 includes the single qubit state (which can be kept coherent from one symbol to the next) and bounds the receiver's ability to distinguish a random symbol parity in a given packet. Step~3 shows that the resulting loss multiplies across the $L$ symbols, making the receiver's outputs for parity-constrained and uniformly random packets exponentially close. Step~4 uses a hybrid argument to show that this implies an exponential lower bound on the sensing time necessary to learn the parity check.

\subsection{Quantum-computing-enhanced parity check learning algorithm}

\begin{proof}[Quantum-computing-enhanced algorithm in Theorem~\ref{thm:parity-learning}]

We use our Learning algorithm to recover $O(D)$ packets, specifically $D+3$. Apply Theorem~\ref{thm:signal_learning_qft} to each symbol with amplitude bound $B=1$, accuracy $\eps/2$, and failure probability $\delta = 1/(16 L (D+3))$. For fixed $R,\eps,c_0$, we have $\log(N/\delta)=2\log L+O(\log\log L)$. Therefore, the duration of the Learning algorithm is bounded as
\begin{align}
    t_{L}\le c\left(\frac{\sqrt N}{\eps}\log L
        +\frac{\log N}{\Delta}\right),
\end{align}
for some universal constant $c$. It follows that $t_L/\tau \le c/\sqrt{c_0} + o(1)$ for sufficiently large $L$; choosing $c_0 = 4 c^2$, the Learning algorithm always concludes within one signal duration. The overall sensing time is at most $(D+3) L \tau = O(D^2/R)$. The total number of ancillas is
$O\bigl(N\log(N/\delta)\log(1/\eps)\bigr)=O(\log^3 L)$.

All packets are successfully decoded except with probability at most $(D+3) L \delta = 1/16.$ Form a binary matrix with the decoded packets as rows. Return a nonzero vector in the nullspace of this matrix, if it exists, and otherwise conclude failure. To see that this works with high probability, observe that the rows of the matrix are independent and uniform on the $(D-1)$-dimensional subspace $s^\perp$. If the rows do not span $s^\perp$, there must be some linear functional which vanishes on every row. There are $2^{D-1}-1$ such functionals, and each has probability $2^{-(D+3)}$ of vanishing on all rows. Hence, the probability that the rows do not span $s^\perp$ is at most $(2^{D-1}-1) 2^{-(D+3)} < 1/16$. Outside of these failure events, the unique nonzero vector in the nullspace is $s$, as desired.
\end{proof}

\subsection{Exponential lower bound for ancilla-free parity check learning}

\begin{proof}[Ancilla-free lower bound in Theorem~\ref{thm:parity-learning}]

Write $s=(s_1,\ldots,s_L)$ for the parity check, and choose each
$s_l$ independently and uniformly from all nonzero bitstrings
in $\{0,1\}^N$. Any sensing algorithm which succeeds for every nonzero check must also succeed on average under this choice. For each $s$, we will compare the output of the sensing algorithm under the true data source (which respects the parity check) with the output under a ``reference source'' for which the data is uniformly random, independent of $s$. We will turn this into a lower bound for parity check learning by demonstrating that replacing the random source by the true source does not substantially improve the probability of identifying $s$.

\emph{Step 1: Bound the classical information obtained from one symbol.}
Consider a new symbol, and fix the measurement record and single qubit state of the algorithm at the start of the symbol. Let $X \in \{0,1\}^N$ be uniformly random independent of the state and record. Let $Y$ denote the measurement outcomes obtained during this symbol. We will bound the total information about $X$ which can be obtained from $Y$, proving that $I(X;Y) \le N/16$.

As in Appendix~\ref{app:ancilla-free-learning-bound}, we introduce a ``dummy,'' noisy copy of the symbol data $V = \eps (-1)^X + 4 \eps G$, for $G$ a standard Gaussian vector independent of all other data. As in Appendix~\ref{app:ancilla-free-learning-bound}, we again have $I(X;V) \le N/32$ (where here and elsewhere in this section entropy is defined in nats). 
We apply Lemma~\ref{lem:information} and use the chain rule of mutual information over different measurements in $Y$, just as in Appendix~\ref{app:ancilla-free-learning-bound}, to obtain
\begin{align}
    I(X;Y) \le I(X;V) + I(X;Y|V) \le \frac{N}{32} + C \eps (1 + \eps /\Delta) \tau \le \frac{N}{16},
\end{align}
as desired. In the final step, we have used $\tau = N/R$, $\Delta = 2 \pi R \ge \eps$, and $R \ge 64 C \eps$.

\emph{Step 2: Convert the information bound to a parity bound} 
As in Step 1, we continue to consider a uniformly random symbol $X \in \{0,1\}^N$. For nonzero $u \in \{0,1\}^N$, compare the output states (including the results of any measurements) conditioned on $u \cdot X = 0$ and $u \cdot X = 1$. We will bound the trace distance between these output states on average of $u$ using the mutual information bound in Step 1.

Let the initial state be denoted $\rho$, with the incoming classical record fixed as in Step 1. Let $\Phi_x$ denote the quantum channel mapping from the input state to the state at the end of the symbol when the symbol data is $x \in \{0,1\}^N$. We again include the single qubit state as well as the results of all classical measurements, so the final states is block diagonal with single qubit blocks. Define
\begin{align}
    \cD_u=2^{-N}\sum_{x\in\{0,1\}^N}(-1)^{u\cdot x}\Phi_x.
    \label{eq:parity-fourier-map}
\end{align}
For $u \neq 0$, $\cD_u$ is half the difference between $\Phi_X$ conditioned on $u \cdot X =0$ and $\Phi_X$ conditioned on $u \cdot X =1$. Our goal is therefore to bound $\| \cD_u(\rho) \|_1$. 

Let $Q$ denote the single qubit system. We can bound the quantum mutual information between $X$ and $YQ$ as
\begin{align}
    I(X;YQ) \le I(X;Y) + \ln(2) \le \frac{N}{16} + \ln(2),
\end{align}
where we have used Step 1, as well as the fact that $X$ and $Y$ are classical, so the additional information in the single qubit $Q$ conditional on any classical record is at most $\ln(2)$. We must convert this mutual information to the mutual information between $YQ$ and the parities of $X$. To this end, take any basis $u_1, \dots, u_N$ of $\{0,1\}^N$; the bits $B_i = u_i \cdot X$ are independent and together determine $X$. Applying the information chain rule and data processing inequality, we have
\begin{align}
    I(X;YQ) =\sum_{i=1}^N I(B_i;YQ\mid B_1,\ldots,B_{i-1}) \ge\sum_{i=1}^N I(B_i;YQ).
\end{align}
Average both sides over uniformly chosen bases of $\{0,1\}^N$; then $u_i$ is uniform over nonzero vectors. Hence, we have
\begin{align}
    \E_u[I(u\cdot X;YQ)] \le \frac{1}{N} I(X;YQ),
\end{align}
where $\E_u$ denotes average over nonzero $u \in \{0,1\}^N$.
Finally, we can use quantum Pinsker's inequality, along with Cauchy-Schwarz, to bound the expected value of $\| \cD_u(\rho) \|_1$:
\begin{align}
   \E_u [\pnorm{1}{\cD_u(\rho)}]^2 \le \E_u[\pnorm{1}{\cD_u(\rho)}^2] \le 2 \E_u[I(u\cdot X;YQ)] \le \frac{2}{N} I(X;YQ) \le \frac{1}{8} + \frac{2 \ln(2)}{N} < \frac{1}{4},
\end{align}
for sufficiently large $N$.

Finally, in the following step, we will compose the maps $\cD_u$ for different symbols in a packet. Since $\cD_u$ need not preserve positivity, the intermediate matrices need not be states. We therefore extend the bound to any block-diagonal Hermitian input $A$ independent of $u$. Decomposing each single qubit block into its positive and negative parts, and applying the bound for states to each part after normalization, gives $\E_u[\pnorm{1}{\cD_u(A)}]\le\frac12\pnorm{1}{A}$.

\emph{Step 3: Compare the output states from one packet.}  Having analyzed the information gained within an individual symbol, we now consider a full packet, say the $r$-th packet. Let $\rho$ denote the initial state. Let $\sigma_s$ denote the state at the end of the packet when the packet data is drawn uniformly from $s^\perp$; let $\sigma_{\mathrm{ref}}$ denote the state when the packet data is instead drawn uniformly from $\{0,1\}^D$. We will bound the expected trace distance between these two states, over random $s$. 

Let $\Phi_{j,x_j}$ and $\cD_{j,u}$ denote the maps from Step 2 associated with symbol $j$ of the packet. If the packet data is $x=(x_1, \dots, x_L)$, the state at the end of the packet is $\Phi_{L,x_L} \cdots \Phi_{1,x_1}(\rho)$. The difference between the state $\sigma_s$ and $\sigma_{\mathrm{ref}}$ is therefore
\begin{align}
    \sigma_s - \sigma_{\mathrm{ref}} &= \sum_{x \in \{0,1\}^D} \Big(\bbP[X^{(r)}=x \mid s] - \frac{1}{2^D} \Big) \Phi_{L,x_L} \cdots \Phi_{1,x_1}(\rho) \\
    &= \sum_{x \in \{0,1\}^D}  \frac{(-1)^{s \cdot x}}{2^D} \Phi_{L,x_L} \cdots \Phi_{1,x_1}(\rho) \\
    &= \sum_{x \in \{0,1\}^D}  \Big( \prod_{j=1}^L \frac{(-1)^{s_j \cdot x_j}}{2^N} \Big)  \Phi_{L,x_L} \cdots \Phi_{1,x_1}(\rho).
\end{align}
Recalling the definition of the maps $\cD_{j,s_j}$, we conclude that
\begin{align}
    \sigma_s - \sigma_{\mathrm{ref}} = \cD_{L,s_L}\cdots\cD_{1,s_1}(\rho).
\end{align}
Finally, applying the result of Step 2 repeatedly, we have the bound
\begin{align}
    \frac{1}{2} \E_s [\| \sigma_s - \sigma_{\mathrm{ref}}\|_1] \le \frac{1}{2^{L+1}}.
\end{align}

\emph{Step 4: Bound the success probability of parity check learning.} Finally, we translate the trace distance bound in Step 3 into a bound on the success probability of an ancilla-free parity check learning algorithm. Let $m$ denote the maximum number of packets before the algorithm outputs its estimate. We use a hybrid argument to bound the final trace distance between the state under the true data stream and the uniformly random reference data stream, replacing true data packets by reference data packets. This shows that the average final trace distance is at most $m/2^{L+1}$.

Let $\hat{s}$ denote the algorithm's estimate for the parity check; we want to bound the expected probability that $\hat s = s$. Under the reference data stream, this success probability is at most $1/(2^N-1)^L$; using the trace distance bound, we conclude that
\begin{align}
    \E_s[ \bbP[\hat s = s]] \le \frac{1}{(2^N-1)^L} + \frac{m}{2^{L+1}}
\end{align}
Let $T$ denote the worst-case elapsed time, so $m = \ceil{R T / D}$. $1/(2^N-1)^L < 1/4$; rearranging the above inequality therefore yields $T \ge \frac{5}{6}(D/R)(2^L - 1)$, as desired. By Markov's inequality, this readily yields the same lower bound (up to constants)  for the expected sensing time $\overline T$.

\end{proof}

\end{document}